\documentclass[12pt]{ociamthesis}  

\usepackage[T1]{fontenc}
\usepackage[utf8]{inputenc}
\usepackage{amssymb}
\usepackage{amsmath} 
\usepackage{graphicx}
\usepackage[colorlinks]{hyperref}
\usepackage[colorinlistoftodos]{todonotes}
\usepackage{verbatim}
\usepackage{xcolor} 
\usepackage{cancel} 
\usepackage{caption} %
\usepackage{subcaption} 
\usepackage{booktabs}
\usepackage{array} 
\usepackage{multirow}
\usepackage{blindtext} 
\usepackage{epigraph}
\usepackage{quotchap}
\usepackage{pdflscape} 
\usepackage{cite}

\title{Bringing\\[1ex]
the Second Quantum Revolution\\[1ex]     
        into High School}   

\author{Filippo Pallotta}             
\college{Department of Science and High Technology}  

\degree{Doctor of Philosophy}     
\degreedate{2022}         

\begin{document}

\baselineskip=18pt plus1pt

\setcounter{secnumdepth}{3}
\setcounter{tocdepth}{3}

\maketitle                  
\include{dedication}        
\include{acknowlegements}   
\setlength{\epigraphwidth}{0.7\textwidth}
\epigraph{<<Tell me one last thing,>> said Harry.\\
<<Is this real? Or has this been happening inside my head?>>\\ 
Dumbledore beamed at him, and his voice sounded loud and strong in Harry’s ears even though the bright mist was descending again, obscuring his figure.\\
<<Of course it is happening inside your head, Harry,\\ but why on earth should that mean that it is not real?>>}{\textit{J.K. Rowling}\\ Harry Potter and the Deathly Hallows} 

\chapter{\label{abstract} Abstract} 

The aim of this thesis is to bridge the gap between the world of physics research and secondary education on contemporary quantum physics.

The second quantum revolution is changing the boundaries of research into the foundations of physics and has enabled the creation of a new generation of technologies that are fundamental to meeting global challenges, now and in the future. Unfortunately, high school curricula are disconnected from these realities and that risks to exclude the new generations from the possibility of participating in a conscious and active way in the construction of the society in which they live.

To address this problem, the first strategy is to strengthen cohesion within the learning ecosystem related to quantum physics. Researchers, teachers and students need to be enabled to carry out activities that allow them to develop knowledge and skills through dialogue, sharing and collaboration.
In this sense, the study of teachers' Pedagogical Content Knowledge (PCK) through the model of Educational Reconstruction for Teacher Education (ERTE) has made it possible to identify those factors that make it possible to achieve the changes necessary to adapt teaching practices to the changes required by the challenges of scientific research.

The second strategic line concerns the creation of an approach to quantum physics that can highlight the revolutionary and disruptive character of the new vision of reality that quantum theory proposes. Moving away from a reflection on the historical origins of quantum physics, the focus is on the informational aspects of the theory and the concept of qubits. Through a "polite informational axiomatic approach", the axioms of quantum theory emerge from the phenomenological analysis of the behaviour of quantum objects representing qubits.

In this approach, quantum technologies are not simply a context in which to see the basic concepts of quantum theory applied, but an opportunity to construct the theory itself. Starting from the problems that these technologies allow to be solved (e.g. the distribution of quantum keys), the focus is on how the information associated with qubits can be manipulated and extracted through quantum laws.

The result of working together with teachers is an understanding of the fundamental role of educational design. During the Professional Development programmes organised, it was possible to provide teachers with the skills to orient their practice towards the development of scientific skills and the building of scientific literacy. The development of the different dimensions of PCK, such as Subject Matter Knowledge for teaching and Educational Structuring allowed teachers to identify strategies to address some of the critical issues in the teaching of quantum physics, such as the role of mathematical formalism and the use of representations.
The use of the developed approach also allowed the creation of learning environments for students that enabled them to grasp the interdisciplinary dimension needed to understand the potential of quantum technologies.


\begin{romanpages} 
\tableofcontents            
\listoffigures              
\listoftables
\end{romanpages}            


\chapter{\label{ch:1-intro}Introduction}

We are in the midst of a scientific revolution. Unfortunately, as is often the case during a revolution, we are not fully aware of it.
Revolutionary in its way of constructing reality, quantum physics has often been narrated as a breakthrough in human scientific thinking. Its disruptive effects have influenced the thinking of scientists, philosophers and writers and in some cases it has managed to break the surface of the common imagination.

The research project aims to explore the educative context related to the teaching of quantum physics.
In view of the cultural relevance of the study of contemporary physical theories and the potential social impact that technological development may have in the coming years, it is important to enable students to understand the contribution that quantum physics is making to the cultural capital of the society in which they live.

This is an enormous horizon in which different research groups have been working for a long time and are producing brilliant results. The present small contribution aims to focus on the possibility of this reflection also becoming part of what happens in the classroom.

It is not just a question of the acquisition of specific notions relating to a given discipline. The tension is directed towards the possibility of making current and future citizens more aware of the impact of the vision of reality introduced by quantum physics.
Often presented as something counter-intuitive or weird \cite{ball2019beyond}, quantum physics is not just something that should remain the preserve of specialists.
In a broader perspective of competence development, the study of quantum physics can and should be an opportunity to reflect on the ways in which science constructs knowledge also through radical changes in the view of reality. In this sense, quantum physics is a perfect example of how certain deep-rooted convictions or dogmatic positions should be revised not with a view to a simple adjustment or revision, but to a transformation. A revolution.\\

Although the starting point is and remains the study of a specific portion of modern physics, the intention is to keep an open eye on issues that are not strictly disciplinary. Studying quantum physics is in fact a way of investigating the evolution of scientific thought.

In particular, dealing with quantum technologies is indeed an opportunity to reflect on the potential of human creativity, on the energy that is released when new ideas shed light on the darkness.

\section{There is an Ichthyosaur at the Zoo}

\epigraph{One never realizes experiments with
a single electron or an atom or a small
molecule. In thought experiments, one
assumes that sometimes this is possible; invariably, this leads to ridiculous consequences \dots.\\
One may say that one does not realize experiments with
single particles, more than one raises ichthyosaurs in the zoo.}{\textit{ E. Schrödinger}\cite{Schrodinger1952}, p.109}

In January 2022 a group of British archaeologists found one of the largest ichthyosaur skeletons on the European continent.

\begin{figure}[hbt!]
    \centering
     \includegraphics[width=\textwidth,height=\textheight,keepaspectratio]{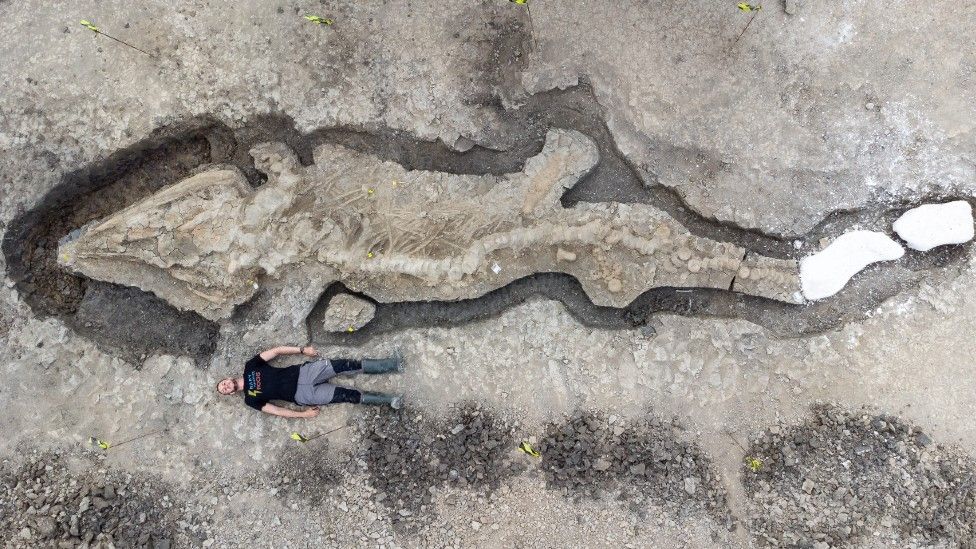}
    \caption[Rutland ichthyosaur fossil]{Palaeontologist Dr. Dean Lomax (being used for scale) near the escavation of the Rutland ichthyosaur fossil. Available at https://www.bbc.com/news/uk-england-leicestershire-59902730}
    \label{fig:Ch3_29_SGAMixed}
\end{figure}

This is probably one of the best results to be achieved in the last 95 million years if the intention is to work with an ichthyosaurus. We agree with Schr\"{o}dinger on that.
But what has happened in the last 40 years has radically changed the possibility for scientists to exploit the essential elements of the known universe.
Today's research has gone beyond the boundaries set by the founding fathers to explore fundamental aspects of the fabric of reality. This new frontier has enabled the creation of a new generation of technologies that will enable new problems to be tackled and new fields of research to be opened up.\\

However, the educational systems have remained fossilised, a bit like an Icthiosaurus.

\begin{quote}
"[...] most modern physics curricula ignore the fact that a second quantum revolution has taken place in the last decades, due to the realization of single-quanta experiments, and a corresponding appreciation of the significance of entanglement
Ideas that were once relegated to the realm of metaphysics are now driving exciting areas of contemporary research, and it is possible to make these developments accessible to introductory quantum physics students" \textit{C. Baily and N.D. Finkelstein} \cite{Baily2015}
\end{quote}

In recent years an important growth in the research on Quantum Technologies (QT) has shown.
Many strategic plans have been presented by various countries to invest in these new technologies
in order to face “the Second Quantum Revolution”: the European Union with the Quantum Flagship
has started an investment of 1 billion Euros in 2018 (https://qt.eu/); China, after several investments in the
past years, has just declared QT one of the new high techs in its 14th Five-Year Plan (2021-2025) \footnote{https://news.cgtn.com/news/2020-10-21/China-to-include-quantum-technology-in-its-14th-Five-
Year-Plan-UM1KUlk80M/index.html}; in the USA the largest companies (IBM, Google, Microsoft and Intel) have been focusing on QT and have already achieved significant results. Thus, today more than ever, alongside this great challenge, there is another very important one: the
educational challenge to prepare the next generations of experts, technicians and, more broadly, citizens because of the wide and deep societal impact that this second quantum revolution is expected to produce.

\section{What it mean to feel part of a revolution?}\label{sec:bepartofrev}


\epigraph{Imagination will often carry us to worlds that never were.\\But without it we go nowhere.}{\textit{ C. Sagan} \cite{sagan2013cosmos}}

Being in a revolution means to be able to start building what is not yet being, it the power of imagination, of creativity.
Teaching the quantum revolution means accompanying future generations in the possibility of feeling empowered to imagine different realities and to perceive how that sense of distance from usual experiences is actually a sense of vertigo and wonder at the new, the unexplored.\\

A scientific revolution has several ways of being perceived. Clearly, its presence can be told through the impact it has on the society in which it develops and expresses itself. A scientific revolution also has the flavour of the future, of a new horizon on which to gaze. The changes brought about by science are not points of arrival, but roads to be travelled that can be glimpsed as you round a bend in the road.

\begin{quote}
    "From the bend, both section of the road are visible. But viewed from a point before the bend, the road seems to run straight to the bend and disappear. \dots And viewed from a point in the next section, after the bend, the road appears to begin at the bend from which it runs straight on" \textit{T.Kuhn} \cite{kuhn1957Copernican}
\end{quote}

A defining aspect of the evolution of scientific practice consists of "communities of people", which gives rise to its "epistemic strength" \cite{oreskes2019why}
Since the days when Galileo, in the pages of the \textit{Dialogo sui Massimi sistemi}, invited people to bring a friend with them before performing experiments with flies and drops of water in the hold of the \textit{gran naviglio}\footnote{G. Galilei, \textit{Dialogo sopra i due massimi sistemi del mondo}, Giornata seconda}, the scientific community has developed its methods through dialogue among researchers and practitioners.

To feel part of the revolution, one must therefore become aware of being part of a \textit{learning ecosystem}. With this term we indicate the close network of connections between the people who make up the living system that is education \cite{robinson2015creative}. 
Often intended as mechanical systems aimed at the transmission of knowledge, learning environments are therefore complex changing interactive systems in which different actors contribute to the activation of different processes linked to the development of competences.

\begin{quote}
    All living systems have a tendency to develop new characteristics in response to changing circumstances. They may have <<emergent features>>, through <<the interaction of small elements forming together to make a larger one>>. In education, there is an abundance of emergent features right now that are changing the context in which schools work and cultures within them. \textit{K. Robinson} \cite{robinson2015creative}
\end{quote}

This ability to adapt and change is therefore an important feature of educational systems that can be exploited so that they can adapt to the challenges and potentialities offered by the second quantum revolution.




\section{Role of education during a revolution}

\epigraph{In most revolutionary times, our times too are caught up in contradictions. Indeed, on closer scrutiny, contradictions in such times often turn out to be antinomies-pairs of large truths, which, though both may be true, nonetheless contradict each other. Antinomies provide fruitful grounds not only for strife, but also for reflection. For they remind us that truths do not exist independently of the perspectives of those who hold them to be so.\\ Educational truths in revolutionary times are also afflicted by antinomies.}{\textit{ J. Bruner} \cite{bruner1996_EduCulture}}


The road was a little wet as the air. This is when the sun rises pale over the lands in northern Honduras.
Groups of men and women from a rural community nearby rolled huge concrete cylinders, \textit{las alcantarillas}, in the middle of the road. These blocks were their last and decisive sign of protest. In a short time, a little stroke hit one of the main freight arteries of Central America.

\begin{figure}[hbt!]
    \centering
     \includegraphics[width=\textwidth,height=\textheight,keepaspectratio]{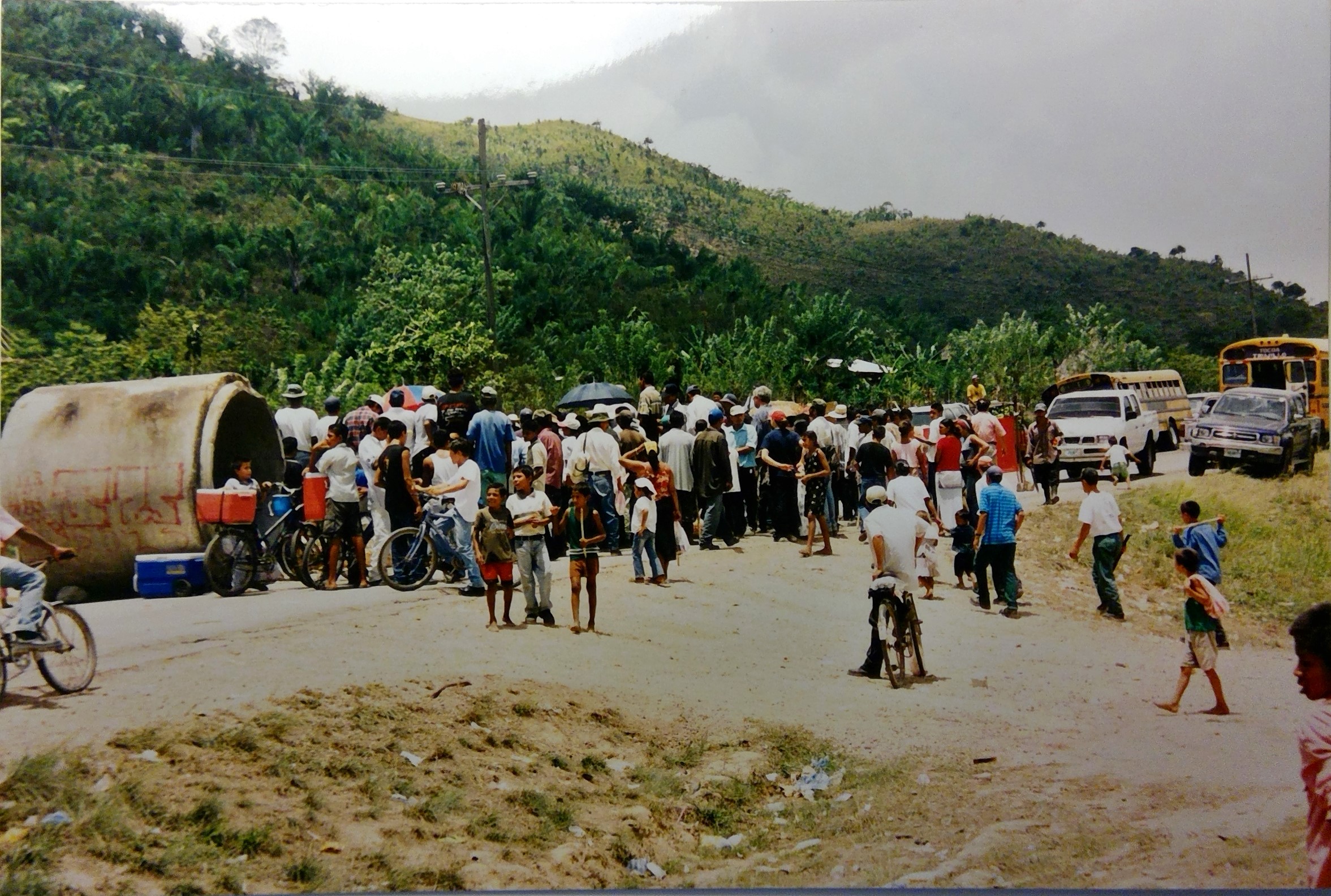}
    \caption[Toma de carrettera]{\textit{Toma de carettera} organized by the  Movimento Capesino del Aguan, 2003. Families from the rural community in northern Honduras block roads in protest}
    \label{fig:Ch1_MCA}
\end{figure}

I was there, 20 years ago. I have been living for about a year in one of the community's \textit{champas} in a large rural area close to that road. The neighbourhood was made up of about 600 \textit{campesinos} families who arrived a few years earlier. They wanted to start a new life on some land allocated to them during the country's last land reform, following the devastation wrought by Hurricane Mitch.
Despite the war in Afghanistan and Iraq was roaring in that same year, at that time Honduras was considered the most violent country in the world. For years, the community had withstood pressure from the local government, which did not want them there and slowed down the process of land appropriation by the \textit{campesinos}' cooperatives in every way possible. The families in the community have been subjected to abuse by a group of local \textit{terratenientes} (land-owners), linked to a local drug cartel. From the top of one of their houses on a hill, someone was used to amuse himself by mowing down the palm trees above the heads of the women who go to collect water from the river with a machine gun.
There was no drinking water or electricity in the community. There was, however, a primary school for more than 700 children who lived there. Built by the families, teachers from neighbouring towns come to that school to teach the children after they return from working in the fields with their parents. One of the rare concessions from the local government.

But the teachers have not been coming for a few days now. They were afraid. They have been threatened on the bus by people who told them not to go to the community, because "there is nothing to do there", there are only \textit{zancudos molestos} (annoying mosquitoes).

There was nothing in the community but the desire for future.

The action taken to block the road was an extreme and dangerous act of protest because it could lead to violent repression by the army, which had immediately arrived on the scene.
Why risk so much for a school? Why now? \dots and for something like this, when there were other priorities anyway?
The reasons came to me from the people who were there on the street and are summed up in the words of one of the elders. "\textit{Sin la escuola, de verdad nos matan}" (If they take away our school, they will really kill us).\\

Any reflection on the real purpose of education necessarily leads to an assessment of the social impact it can have. It is not just a question of perpetuating the transmission of knowledge from past generations but of planning how to overcome future challenges.
\begin{quote}
"The aims of education are to enable students to understand the world around them and the talents within them so that they become fulfilled individuals and active, compassionate citizens." \textit{K. Robinson}  \cite{robinson2015creative} 
\end{quote}

In addition to specific educational activities, the school provides opportunities to weave those social relationships that allow for growth and expression.
By exploiting the dynamic and active nature of educational processes, the learning environments that can be created can promote those meaningful competences, such as \textit{adaptability to change} and the \textit{creativity in generating new ideas}.\footnote{"The Enterprise of the Future" IBM 2008 Gobal CEO Study. Available at https://www.ibm.com/ibm/ideasfromibm/eg/en/ceo/20080505/index1.shtml. Crossreference \cite{robinson2015creative}, pag 19}

Introducing the second quantum revolution into the school curriculum means contaminating schools with the driving force of innovation that comes from deep reflection on fundamental physical theories.
In this scenario, teachers and students must play an active role \cite{FraserMazur2014_Bridging}.
This means creating learning environments in which they are not asked to passively accept results, but to feel part of the science making process.


\begin{quote}
\dots our instruction of science, from start to finish, should be mindful of the lively processes of science making, rather than being an account only of <<finished science>>, as represented in a textbook or demonstration\\ \textit{J. Bruner} \cite{bruner1996_EduCulture} 
\end{quote}





In order to allow the new generation to feel part of such a fundamental change, it is necessary to provide appropriate tools to narrate it.

\begin{quote}
Words can tether actions that were previously invisible, frame them.\\
\textit{A.S. Magnason }\cite{magnason2020TimeWater}   
\end{quote}



In a time of societal changes, transformation also involves language down to its most fundamental elements. At the time when Shakespeare wrote his plays, the English language was changing, eliminating archaic words linked to its past. At that moment, the Bard's creativity generated new words. By mixing terms and meanings he was able to evoke powerful images for his stories \cite{crystal2008think}.
If we want to talk about a revolution it is therefore necessary to activate a rethinking of how we are telling it and whether it is appropriate to transform the tools we are using.\\

The aim here is not simply to propose a critical review of the textbooks in use in schools, but to foster the creation of learning ecosystems that can be used to achieve the educational goals related to the teaching of quantum physics.
In order to do this, it is necessary to promote and support the dialogue within the learning ecosystem and stimulate creativity, i.e. \textit{process of having original ideas that have value}. \cite{robinson2015creative}.

In fact, the choice made is intended to be consistent with the (scientific) revolution that is being narrated. The challenge is not to \textit{re}form the way quantum physics is taught, but to \textit{trans}form it. This educational goal will have an impact on the development of the cultural capital of teachers and students, regardless of their future career choices. It will also be beneficial for the whole community, in the sense that it can help the development of different soft skills and support the awareness of the potential of quantum technologies \cite{EPSRC21_QTreport}.\\

Physics has this transformative power about the way in which reality can be understood. Investigating how to make it part of what happens in secondary school classrooms is the attempt of this project.

\chapter{\label{ch:2-PrFrame}Project framework} 

The aim of this research project is to generate the conditions to bring the second quantum revolution into high school classrooms. We want to investigate how it is possible to teach the core concepts of quantum physics (such as quantum superposition and entanglement) in regular physics classroom activities. The expected outcome is the creation of teaching materials that can be adapted by teachers for their physics lessons and also be used in extracurricular activities with students. All those materials are intended to be the product of the collaboration between physics researchers and teachers. 

In this Chapter we outline the structure of the research project. The analysis of generative needs can help to understand the motivations and enable the project's objectives to be defined. The definition of strategies and actions brings out which processes should be activated in order to achieve these objectives.

\section{Generative needs}

The main research idea is to see if it is possible to rethink the way quantum physics is taught at secondary school level. 
The gap between the list of contents of the ministerial syllabus and the open scenarios of the research world is growing. The indices of school textbooks, which set the pace for curricula completion, appear to be frozen in time when compared to the discoveries of contemporary physics. A starting point is therefore to look at what is happening in the places where physics is taught (both schools and universities) in order to understand how to act in a transformative way.\\


The process of cultural enrichment of society also passes through educational institutions, such as schools. It is therefore important to create meaningful connections between those who manage educational processes in schools (teachers) and those who enable the advancement of research (researchers). Together with students, they are part of the same learning ecosystem (see Section \ref{sec:bepartofrev}). An organic and dynamic vision of the educational processes can facilitate connections between the different members of the ecosystem. These connections can favour the exchange of knowledge elements and the creation of new visions. The attitude is not to "fix what doesn't work", as it with machines. On the contrary, embracing an ecological and organic vision of education orients towards the stimulation of generative processes. For this reason it is necessary to create or consolidate connections between the members of the ecosystem through which it can grow and evolve.\\

This project is therefore designed on the idea of reflecting about the changes that are needed to nurture and stimulate connections within the learning ecosystem. This process can generate the competences that will enable various members of the ecosystem to build an education system that enables the community to rise to present and future challenges. (see Chapter \ref{ch:6-LETeachers} and Chapter\ref{ch:7-LEStudents}).\\

This kind of stimulation requires the creation of a new narrative about quantum physics, a different way of imagining the present and the future.
First of all, it is necessary to create an alternative to the narrative that describes quantum physics as something absurd or strange \cite{ball2019beyond}. There is a need to go beyond the "unprecedented murkiness that has enveloped quantum foundations for the past nine decades" \cite{QBism2014}.  
This narrative has an impact on various aspects of the development of scientific knowledge, because it contributes to conveying contradictory messages about the methods by which science constructs knowledge of reality \cite{Stadermann2020ConnectNOS}. As a consequence, science is populated with nonsensical and contradictory statements that seem to push away the possibility of being able to investigate fundamental issues.\\

The need, therefore, is to try to construct an approach to quantum physics that starts from the profound changes it has brought about in the description and interpretation of reality (see Chapter \ref{ch:3-QApproach}). This effort requires the contribution of the different elements of the learning ecosystem in order to make it effective and applicable at secondary school level. It is not a question of destroying school curricula, since they constitute a formal and normative constraint that guides the work in schools. It is about regenerating and transforming them to make them more coherent and rigorous with the state of the art in quantum physics research.\\
To facilitate experimentation with this approach, it is also necessary to provide teachers with specific tools to reflect on their professional practice in an active and generative way (see Chapter \ref{ch:4-TheorFrame}).The study of teachers' Pedagogical Content Knowledge (PCK) allows to identify those specific competences that enable teachers to deal creatively and flexibly with the complexity involved in teaching quantum physics (see Section \ref{sec:ERTE}). This is complemented by work on building educational design skills to strengthen the coherence between educational objectives and teaching actions (see Section. \ref{sec:LEDesign}).\\

In this way, it is intended to provide teachers and students with the skills to engage in a broader reflection on the role science currently plays in providing the tools to meet the challenges of the present and the future.
With regard to future possibilities, the focus is on the opportunities offered by quantum technologies. They contain the potential to design a different world, together with the risks this may entail. Having the skills to be aware of and manage these aspects is a fundamental element to be included in educational projects related to quantum physics.
Offering teachers and students the opportunity to approach these technologies meets a training need, be it for professional development or university orientation. Exploring the world of quantum technologies is also a question of stimulating reflection and debate to the benefit of society because it helps to create greater awareness of the possibilities of being able to respond to global challenges through the use of revolutionary technologies. This reflection will also support a general debate, not limited to experts, based on coherent and rigorous information. That is beneficial to resolve doubts and avoid misunderstandings about the purpose of quantum technologies. \cite{EPSRC21_QTreport}.
 
The context identified is that of the upper secondary school. It is an environment in which the different elements of the learning ecosystem can contribute to the development of the cultural process by playing different roles. The existing relationships between teachers and researchers created by different professional development projects (e.g. the "Progetto Lauree Scientifiche (PLS)" - Scientific degrees project \cite{PLSWeb}) facilitated the creation of opportunities for cultural exchange and the possibility of organising joint actions.\\

\section{Aim, strategies, actions and Research Questions}

The aim of the project is to generate the conditions for the elements characterising the second quantum revolution to become part of regular classroom activities in upper secondary schools.
A first strategic line is to activate those processes that will generate and strengthen relationships within the learning ecosystem related to quantum physics.
This is accordance with one of the strategic lines of 2018 "recommendations on key competences for lifelong learning"

\begin{quote}
“reinforcing collaboration between education, training and learning settings at all levels, and in different fields, to improve the continuity of learner competence development and the development of innovative learning approaches” \cite{EU_keycompetences2018}
\end{quote}

These processes nurture the ground on which actions for teachers and students can develop.
Related to this are actions that have created opportunities for discussion and exchange between researchers and teachers. 

A second strategic line concerns the activation of processes to create a contemporary approach to quantum physics \cite{Bitz21_fostering}. This approach allows quantum physics to be introduced not as an extension of classical physics \cite{SusskindQT}, but as new approach to reality (see Chapter\ref{ch:3-QApproach}). In this framework, the introduction of quantum technologies becomes significant. From being 'applications' of new concepts in physics, they become a significant context in which to analyse the potential associated with this new vision of reality.

Two types of action are fundamentally linked to this strategy. On the one hand, those aimed at teachers (see Chapter \ref{ch:6-LETeachers}), to develop Pedagogical Content Knowledge (PCK) (see Section \ref{sec:ERTE}) and instructional design skills (see Section \ref{sec:LEDesign}). On the other hand, those addressed to groups of students (see Chapter \ref{ch:7-LEStudents}) interested in deepening some conceptual nodes on the nature of quantum objects (see Section \ref{sec:SumSch_light}) or issues related to the potential offered by quantum technologies (see Section \ref{sec:SumSch_QT}). 

In Table \ref{tab:ProjFrame} the different project strategies related to the general aim are made explicit. The strategies define the processes to be activated and are described by the different actions. For each action the relevant research questions and references in the text are indicated.

\begin{landscape}
\begin{figure}[hbt!]
    \centering
     \includegraphics[width=\textwidth,height=\textheight,keepaspectratio]{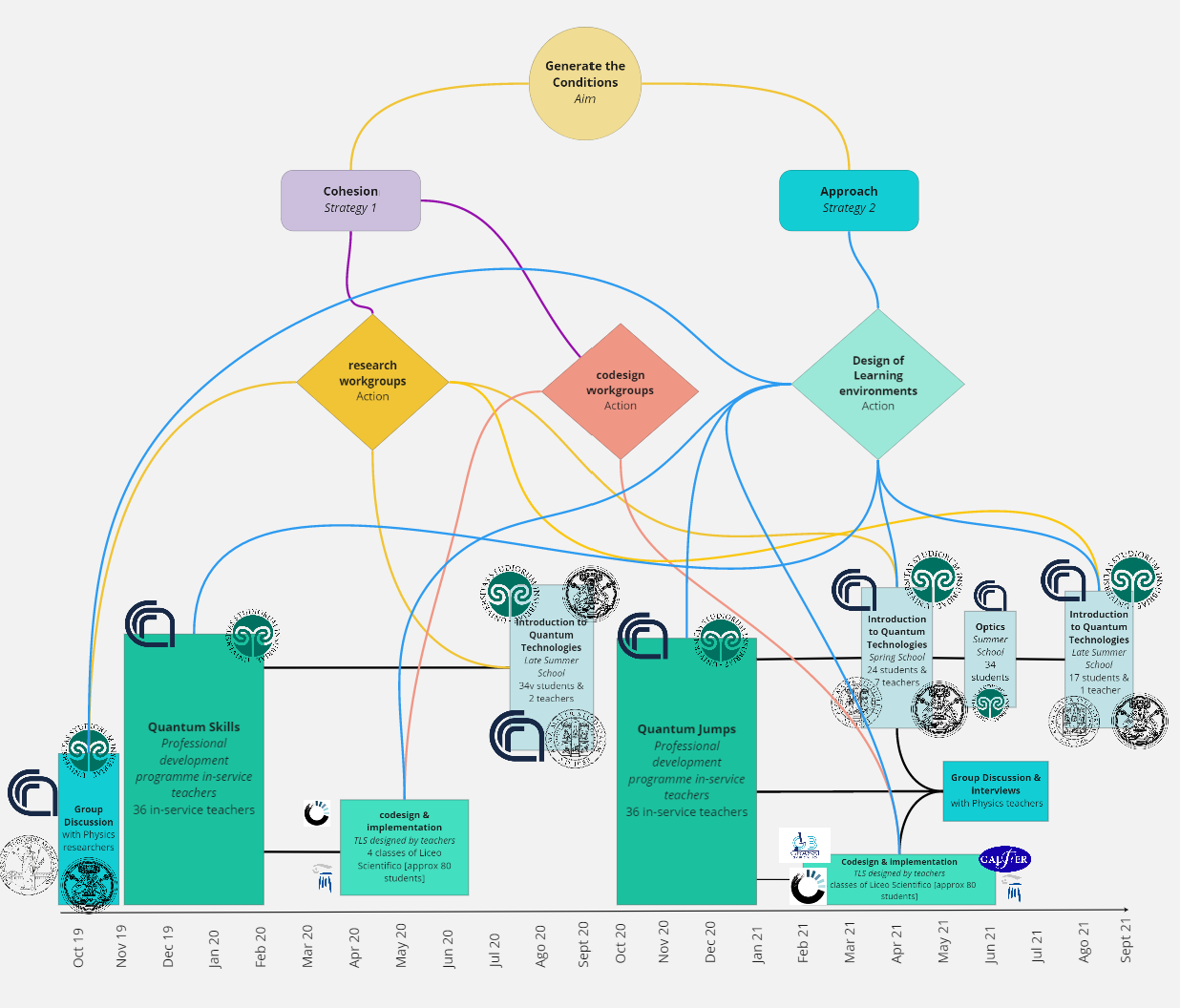}
    \caption{Project framework timeline. Strategies and action are connected to specific moments of the project timeline}
    \label{fig:Ch2_8_NodeFrame}
\end{figure}
\end{landscape}

The connection of the different levels and the related data collection phases is represented by in Figure \ref{fig:Ch2_8_NodeFrame}

\pagestyle{empty}
\begin{landscape}
\begin{table}[hbt!]
\begin{center}
\centering
\begin{tabular}{|l|c|c|r|}
\hline
\multicolumn{4}{|c|}{}\\
\multicolumn{4}{|c|}{\textbf{Aim}}\\
\multicolumn{4}{|c|}{\footnotesize{Generate the conditions to bring the tenets of the second quantum revolution into high school regular classroom activities}}\\
\multicolumn{4}{|c|}{}\\
\hline
\multicolumn{2}{|c|}{} & \multicolumn{2}{c|}{}\\
\multicolumn{2}{|c|}{\textbf{Strategic line 1}} & \multicolumn{2}{c|}{\textbf{Strategic line 2}}\\
\multicolumn{2}{|p{12.5cm}|}{\footnotesize{Promote cohesion within the learning ecosystem oriented towards the identification of the founding elements of education towards quantum physics}} & \multicolumn{2}{p{12.5cm}|}{\footnotesize{Create a contemporary approach to teaching quantum physics at upper secondary school level geared towards building learning environments for teachers and students}}\\
\multicolumn{2}{|c|}{\textbf{\textit{Actions}}} & \multicolumn{2}{c|}{\textbf{\textit{Actions}}}\\
\multicolumn{2}{|p{12.5cm}|}{\footnotesize{$\square$ Organise regular working groups between researchers in physics and education}} & \multicolumn{2}{p{12.5cm}|}{\footnotesize{$\square$ Design and implement learning environmnets for teachers to support PCK development}}\\ 
\multicolumn{2}{|p{12.5cm}|}{\footnotesize{$\square$ Set up educational co-design groups involving researchers and teachers}} & \multicolumn{2}{p{12.5cm}|}{\footnotesize{$\square$ Design and implement learning environments for students about Quantum Technologies}}\\
\multicolumn{2}{|c|}{\textit{Research Questions}} & \multicolumn{2}{c|}{\textit{Research Questions}}\\
\multicolumn{2}{|p{12.5cm}|}{\footnotesize{RQ4.3: What are the key concepts that should characterise quantum physics teaching-learning sequences in upper secondary school? Table \ref{tab:Approach_Researchers}}} & \multicolumn{2}{p{12.5cm}|}{\footnotesize{RQ4.1: How to characterize the educational approach that facilitates the development of the new vision of reality promoted by the second quantum revolution? Table \ref{tab:Approach_RQ}}}\\
\multicolumn{2}{|p{12.5cm}|}{\footnotesize{RQ4.4: What approach to teaching physics can facilitate the introduction of elements of contemporary physics at upper secondary school level? Table \ref{tab:Approach_Researchers}}} & \multicolumn{2}{p{12.5cm}|}{\footnotesize{RQ4.2: How can the educational approach foster awareness of the impact of the second quantum revolution? Table \ref{tab:Approach_RQ}}}\\
\multicolumn{2}{|p{12.5cm}|}{} & \multicolumn{2}{p{12.5cm}|}{\footnotesize{RQ5.1: How is it possible to characterise the quantum nature of physical objects using qubits to encode information? Table \ref{tab:Qubit_RQ}}}\\
\multicolumn{2}{|p{12.5cm}|}{} & \multicolumn{2}{p{12.5cm}|}{\footnotesize{RQ5.2: Through which type of representation is it possible to describe the process by which the information associated with quantum objects can be transformed inside a physical apparatus and extracted through a measurement? Table \ref{tab:Qubit_RQ}}}\\
\multicolumn{2}{|p{12.5cm}|}{} & \multicolumn{2}{p{12.5cm}|}{\footnotesize{RQ5.3: Which experimental contexts are significant for the study of how quantum objects can be used to encode information? Table \ref{tab:Qubit_RQ}}}\\
\multicolumn{2}{|p{12.5cm}|}{\footnotesize{RQ6.1: How can the collaboration with physics researchers foster teachers' PCK about contemporary quantum physics? Table \ref{tab:TLE_RQ_QSkills}}} & \multicolumn{2}{p{12.5cm}|}{\footnotesize{RQ6.3: What are the elements of the informational approach (see Section\ref{sec:ApprPillars}) that can help the development of teachers'PCK? Table \ref{tab:RQ_TLE_QuantumJumps}}}\\
\multicolumn{2}{|p{12.5cm}|}{\footnotesize{RQ6.2: How to support teachers in designing teaching-learning activities about the interpretation of quantum experiments? Table \ref{tab:TLE_RQ_QSkills}}} & \multicolumn{2}{p{12.5cm}|}{\footnotesize{RQ6.4: What role can quantum technologies play in creating learning environments about teaching and learning quantum physics? Table \ref{tab:RQ_TLE_QuantumJumps}}}\\
\multicolumn{2}{|p{12.5cm}|}{} & \multicolumn{2}{p{12.5cm}|}{\footnotesize{RQ7.1: How does the interdisciplinarity between mathematics, computer science and physics can support students' exploration of the tenets of quantum physics and their relation with quantum technologies? Table \ref{tab:RQ_SLE_QT}}}\\
\hline
\end{tabular}
\caption{Project framework: Aim, strategies and actions. Research question are related to each strategic line}
\label{tab:ProjFrame}
\end{center}
\end{table} 
\end{landscape}
\pagestyle{plain}

\pagestyle{empty}

\section{Methods}
To investigate the complex dynamics of teaching, quantitative methods have been chosen. This is to collect practitioners reflections and thoughts using their own words and so facilitate the connection between research and practice.



Qualitative data were collected in the form written responses to questionnaires, focus groups and unstructured interviews.
The choice of using qualitative method was to listen as much as possible to all those who, with different roles, contribute to the evolution of the learning ecosystem. 
The choice of the method of data collection and analysis is focused on the possibility of collecting from the participants in the various actions reflections and arguments from which it is possible to extract information that can be traced back to the specific themes of the research.

\subsection{Data collection and analysis}\label{subsec:datacoll_analys}
By the nature of this study, qualitative data were collected in the form of written responses to questionnaires, focus groups and unstructured interviews.
In some cases, data were collected and then processed quantitatively, such as data from multiple-choice questionnaires. The small sample size of participants in the study does not make this analysis particularly meaningful. Applying a sequential method, the quantitative analysis was essentially used to gather preliminary that was then used to design unstructured interviews and focus groups with the aim of gathering larger quantities of text that was analysed qualitatively.
An important element of data collection was the use of field notes collected during and after the activities. The possibility to interact with course participants also outside the training context allowed the collection of personal comments and reflections. This kind of data is not necessarily more authentic or true, but it contributes to generate a larger number of facets about the evaluations given by the participants to the different training courses. 

From the point of view of data analysis, the methodology used is linked to the Thematic Content Analysis \cite{Braun2006ThemAnalys} which has been developed in the social sciences. In general, the themes are identified inductively by letting them emerge from the data in order to minimise the bias linked to the researcher's choices. The different project actions were structured around specific themes, such as the different dimensions of PCK (see Section \ref{sec:ERTE}). In our case the definition of those dimensions facilitated the process of identification the main themes. The coding of the collected texts (answers to questionnaires, interviews) was done manually as the size of the collected texts allowed it. 
Quotations and anectodes have been used as useful tools to extract themes and connections, and to support arguments \cite{biddle2007QualRes_Quotes}.
Table \ref{tab:Codes} shows the codes for the different text sources. The codes associate the quotation with a particular action (see Figure \ref{fig:Ch2_8_NodeFrame}) in the project and the type of tool used to collect it.

\begin{table}[hbt!]
\centering
\begin{tabular}{cp{12.5cm}}
\toprule
Code & Description\\
\midrule
QSk19EnQ & Quantum Skills 2019 Professional development programme, Entry Questionnaire for teachers\\
QSk19ExQ & Quantum Skills 2019 Professional development programme, Exit Questionnaire for teachers\\
QSk19ObF & Quantum Skills 2019 Professional development programme, Observations and field notes of teachers activities\\
QJs20EnQ & Quantum Jumps 2020 Professional development programme, Entry Questionnaire for teachers\\
QJs20ExQ & Quantum Jumps 2020 Professional development programme, Exit Questionnaire for teachers\\
QJs20ObF & Quantum Jumps 2020 Professional development programme, Observations and field notes of teachers activities\\
TeachRefl & Teachers' reflection collected using field notes in personal or group discussion that happened before, during and after Professional development activities.\\
TeachIntervX & Teachers' reflection collected during semi-structured interviews. Different number "X" have been used to distinguish among different teachers\\
StudentsRefl & Students' reflection collected by questionnaires before, during and after classroom activities or Summer School activities
\end{tabular}
\caption[Codes for Codes identification]{Quotes: Codes to identify the source of quotations collecte during the activities with teachers and students}
\label{tab:Codes}
\end{table}

\subsection{Trustworthiness}
Quotations are included as an element of trustworthiness \cite{Guba1994_Trust} that can be used <<to illustrate the analysis process and  findings>>\cite{Eldh2020Quote}. 
Trustworthiness can also be evaluated by examine some components that characterize the identity of the researcher in the project. In a qualitative research, the role of the researcher is to attempt to make a connection with the thoughts of study participants and so he/she becomes the \textit{human instrument} for data collection \cite{Denzin2013_QualReas}. It is important as a researcher to make as explicit as possible the possible biases, assumptions, experiences that can provide elements of validity to the research. \cite{Greenbank_QualResValues}.

During the research project, I played in an intermediate position. My personal role was in between the identity of researcher in training and that of high school teacher.
On the one hand, I have had the opportunity to be in close contact with high-profile physics researchers whose competence and knowledge of the subject matter have often made up for my shortcomings.
On the other hand, being a lecturer made it easier for me to connect with the participants in the activities, both lecturers and students. Many participants were work colleagues. During the interviews and group meetings with teachers I was positively identified as "one of us". This allowed me to collect many reflections and confidences "out of the records" which were valuable in different phases of the design and evaluation of the training courses. At the same time, it has allowed me to anticipate some of the dynamics that occur when in-service teachers become students during a professional development programme. 
The experience gained in about 15 years of teaching both in Italian public schools and in the IBO curriculum has in fact given me a good familiarity with the relationship and the distance between school curricula and teaching practices. The difficulties those professional development programmes have in reconciling teachers' training needs and the integration with classroom practice are particulary relevant. Research work in education is perceived by those who work in schools as something too distant from what happens in the classroom, too distant from the teachers' real needs \cite{Bitzenbauer2021_Teachers}.

\chapter{\label{ch:4-TheorFrame}Theoretical Framework}

In the context of teacher education, the study of PCK (Pedagogical Content Knowledge) can be used to support teachers in the development of effective teaching practices related to quantum physics. We used the Educational reconstruction for teacher education (ERTE) model to explore secondary school teachers’ different PCK dimensions with the aim of designing learning environment for teacher education programs focused on the core concepts of quantum physics (see Chapter \ref{ch:3-QApproach}).
The ERTE model (see Section \ref{sec:ERTE}) can provide a framework to reconstruct quantum physics from an educational perspective. 
The process of designing learning environments has also been taken into consideration (see Section \ref{sec:LEDesign}) to reinforce the coherence between intended learning outcomes, strategies for effective formative assessment and methodologies for content delivery.

\section{Studing teachers' PCK using ERTE model}\label{sec:ERTE}

Referring to the aims of the present research project, one of the conditions for bringing the second quantum revolution into secondary school is to create learning environments in which physics educational researchers and secondary teachers can collaborate to design and implement teaching-learning activities. Therefore Professional Development Programmes for in-service teachers to support the development of their PCK towards a “modern perspective” \cite{Bitz21_fostering} on quantum physics were organized. The focus was on the analysis of quantum core concepts or  “complex quantum behaviours” \cite{Kim2017} (quantum states, quantum superposition, quantum entanglement and quantum measurement) in relation to the exploration of the basic features of quantum technologies. This action should lead to the creation of teaching materials that could be used for teacher education by educational researchers and that can be then adapted by teachers for regular classroom activities at school.\\

To bring the second quantum revolution into high school is therefore important to support the development teachers’s Pedagogical Content Knowledge (PCK) related to quantum physics core concepts. The definition of PCK evolved in time since Shulman's first introdution as a
\begin{quote}
    "The blending of content and pedagogy into an understanding of how particular topics, problems, or issues may be organised, represented, and adjusted to the diverse interests and abilities of learners, and presented for instruction" \cite{Shulman1987_PCK} 
\end{quote}

Regarding the present study, PCK is to be considered as an "heuristic device for thinking about teacher knowledge"\cite{borko1996a} to gather information about how to support teachers in the process of designing learning environments for their students.\\

\begin{figure}[hbt!]
    \centering
 \includegraphics[width=\textwidth,height=\textheight,keepaspectratio]{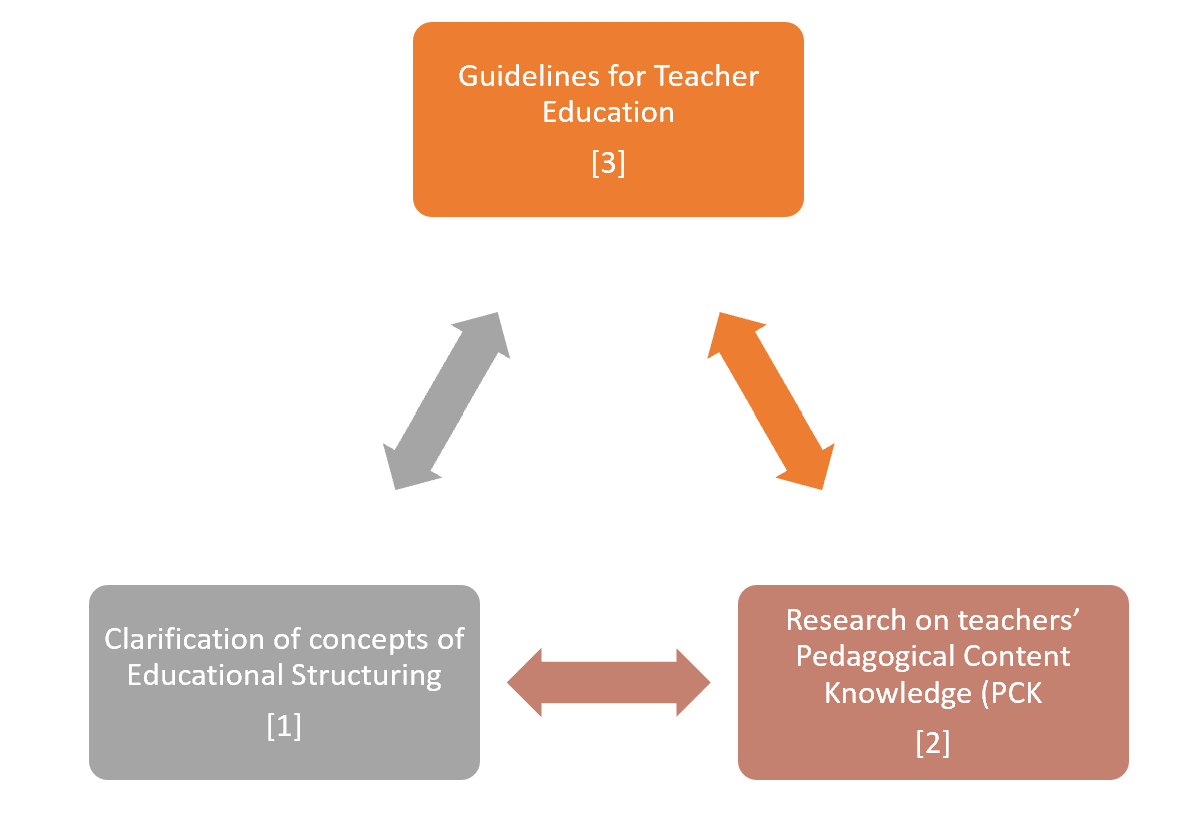}
    \caption{Educational reconstruction for Teacher Education (ERTE). The component [1] comprises the major ideas of the Model of Educational Reconstruction; adapted from \cite{Duit2012_MER}. The component [2] encapsulate the results of PCK studies and component [3] is related to the definition of startegies and actiosn that could be helpful to design learning environments for teachers}
    \label{fig:Ch4_1_ERTE}
\end{figure}

To study teachers' the model of Educational Reconstruction for Teacher Education (ERTE) has been adopted. 
The different elements of ERTE model are shown in Figure \ref{fig:Ch4_1_ERTE}.

\begin{figure}[hbt!]
    \centering
     \includegraphics[width=0.8\textwidth,height=\textheight,keepaspectratio]{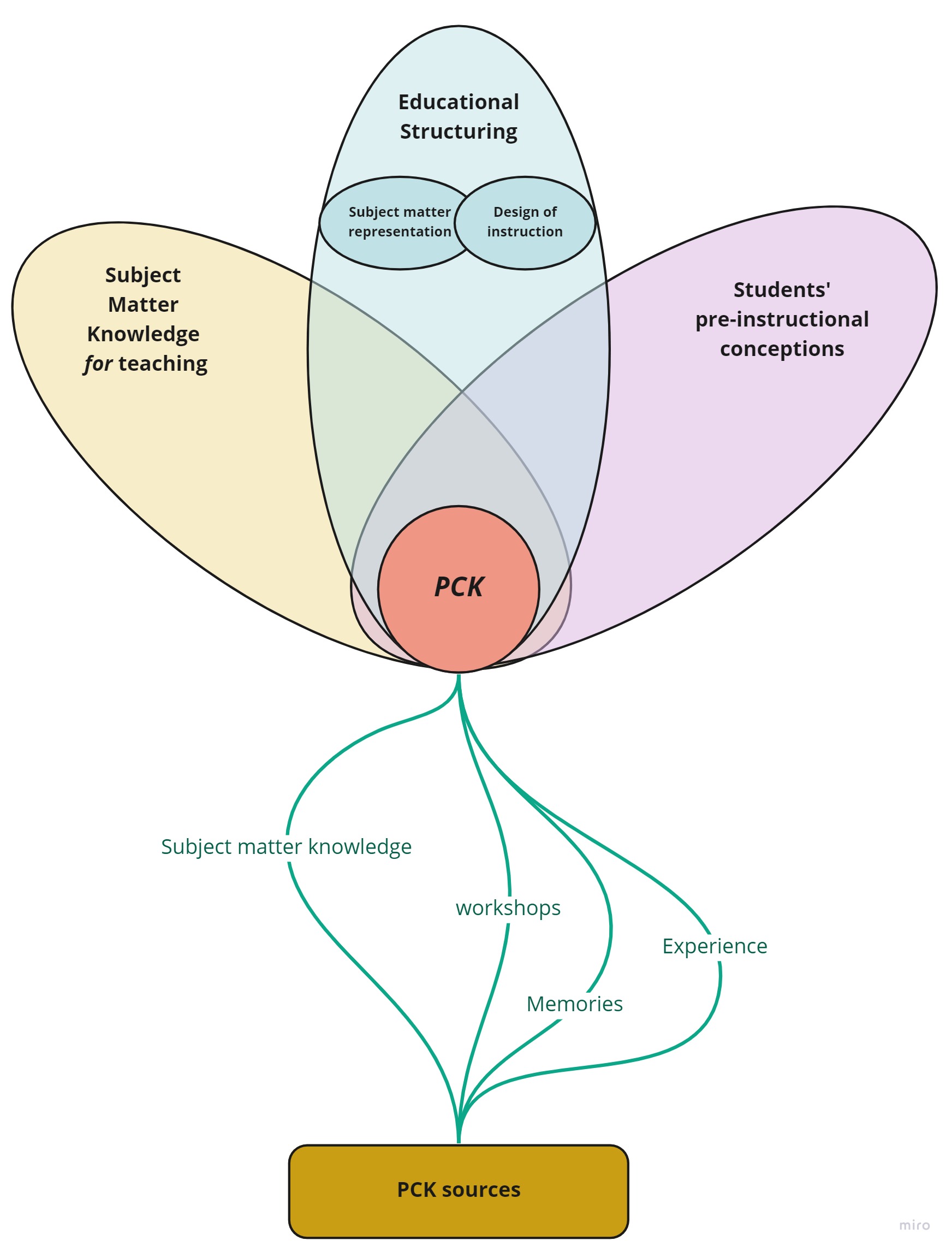}
    \caption{PCK dimensions in the ERTE model: the three main elements of PCK used for the study are shown. Subject matter knowdledge for teaching, Students pre-instructional conceptions and Educational structuring are deeply connected and together with the PCK sources contribute to the PCK development}
    \label{fig:Ch3_PCKflower}
\end{figure}

In the ERTE model, PCK is originated from different sources that contribute to determine the three essential dimensions of the PCK (see Figure \ref{fig:Ch3_PCKflower}). 

One first dimension is the teachers' knowledge and beliefs about students' specific learning difficulties related on a specific topic. These difficulties are related to \textit{students' pre-instructional conceptions} \cite{Dijk2006_ERTE}. A second dimension is \textit{educational structuring} that is related to teachers' knowledge about subject matter representations, in terms of teaching materials (i.e. textbook) and analogies, metaphors, visualizations that can be used to overcome students difficulties. The final dimension is the \textit{subject matter knowledge for teaching} that is related to how teachers use their specific subject matter knowledge in their teaching practice. The development of this last dimension is essential to make teachers able to "handle the complexity of their daily teaching practice flexibly" and "react adequately to different and unanticipated situations" \cite{Dijk2008_ERTEevolution}.

To study how to support the development of teachers’ PCK in all these dimensions, the ERTE model is deeply connected to the Model of Educational Reconstruction (MER). One of the central ideas of MER is that “a certain science content structure has to be transformed into the content structure for instruction” \cite{Duit2012_MER} 
Therefore ERTE model incorporates one of the essential components of the MER, the clarification and analysis of science content. This component is structured around two processes contributing to subject matter clarification and to the analysis of educational significance. The process of \textit{elementarization}, which takes into account aspects linked to the students' experience and language and allows the identification of \textit{elementary ideas} on which to activate the second process of \textit{construction of content structure for instruction} in which these ideas are adapted and organized for teaching activities.\\

\begin{figure}[hbt!]
    \centering
\includegraphics[width=\textwidth,height=\textheight,keepaspectratio]{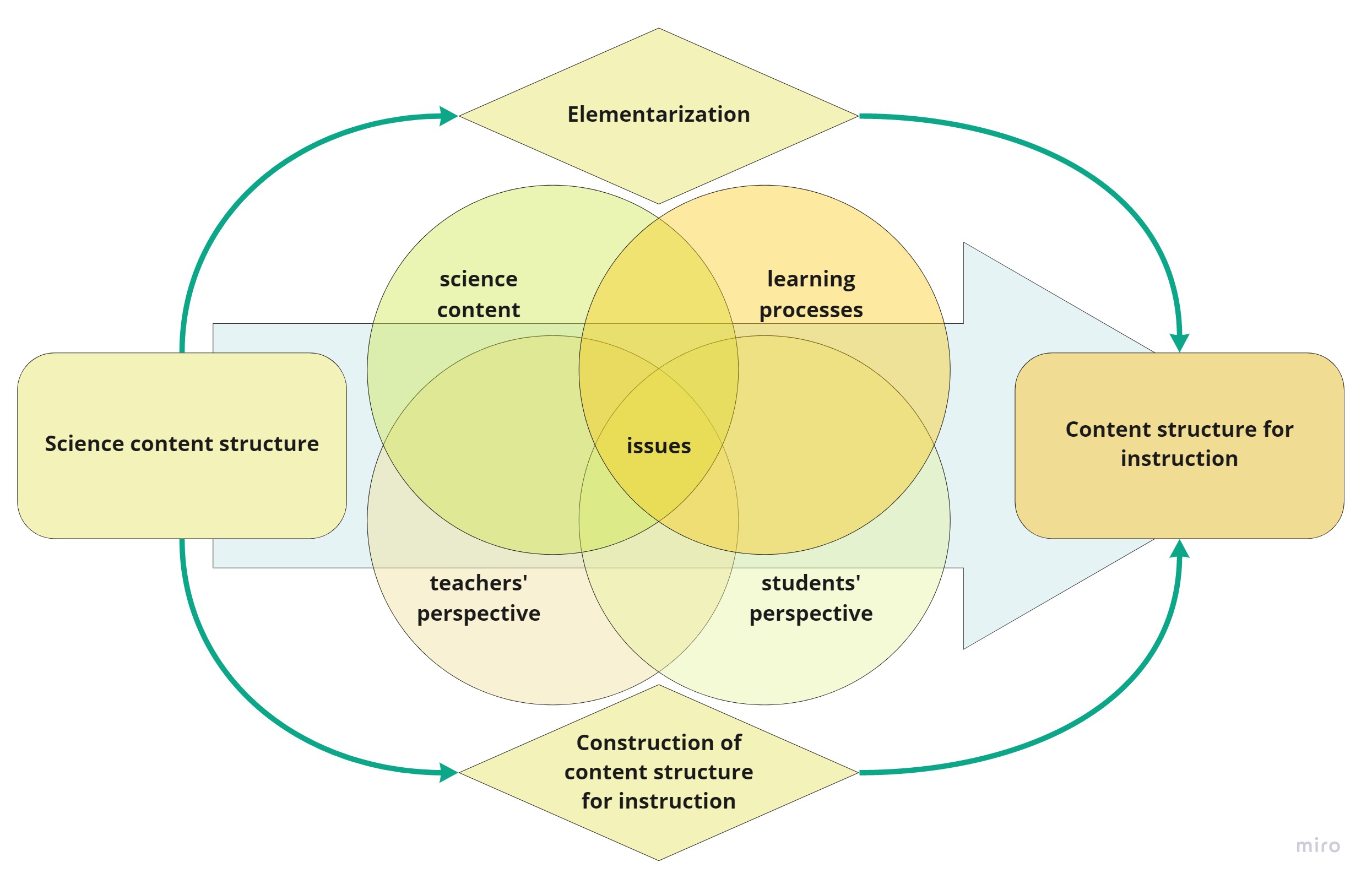}
    \caption[MER Subject matter clarification and analysis of educational significance]{Subject matter clarification and analysis of educational significance: the process of elementarization and construction of content structure for instruction both support the transformation from science content structure to content structure for instruction. Those two processes are not linear and are built around the reflection about different issues related to teaching and learning science.}
    \label{fig:Ch4_1b_MERclarification}
\end{figure}


The actions designed in the present research are deeply related to these processes, which are considered necessary to make the contents of quantum physics accessible at secondary school level. This goal is one of the factors that determined the construction of the approach (see Chapter \ref{ch:3-QApproach}) both in terms of the choice of content and in the attention to creating meaningful links with aspects of contemporary quantum physics.\\ 


    


According to the aforementioned dimensions of the ERTE model, the study of teachers’ PCK can be therefore oriented towards finding the answer to the following question \cite{Duit2012_MER}:

\begin{enumerate}
\item What subject matter knowledge for teaching quantum physics do teachers have at their disposal?
\item What do teachers know about students’ pre-instructional conceptions on the subject matter related to quantum physics and about their learning processes?
\item What conceptions do teachers have of educational structuring (design of instruction, subject matter representation) related to quantum physics?
\item What conceptions do teachers have about the interrelation of subject matter knowledge for teaching, students’ pre-instructional conceptions, and the influence of this interrelation on the process of educational structuring in the context of quantum physics education?
\end{enumerate}

The context for the research project is provided by the Italian programme “Progetto Lauree Scientifiche” (PLS) that includes some specific actions for in-service teachers training. One of the consequences of working with secondary teachers who have been in active service for at least 20 years in public schools [see Chapter \ref{ch:6-LETeachers}] is to consider their teaching experience as the main source of their PCK \cite{Dijk2006_ERTE} and teacher training as an opportunity to learn from it.

A second action is related to the implementation of extra curricular activities for upper secondary students. We used the results of the activities with teachers to design the learning environment for students (see. \ref{ch:7-LEStudents}). Here, the approach has been adapted to emphasise aspects of educational significance, including a reflection on the societal impact of quantum technologies 
\cite{Osborne2001_WhatScienceTeach,Eilks2013_relevanceSE,moraga2020relevance}. 

\begin{figure}[hbt!]
    \centering
     \includegraphics[width=\textwidth,height=\textheight,keepaspectratio]{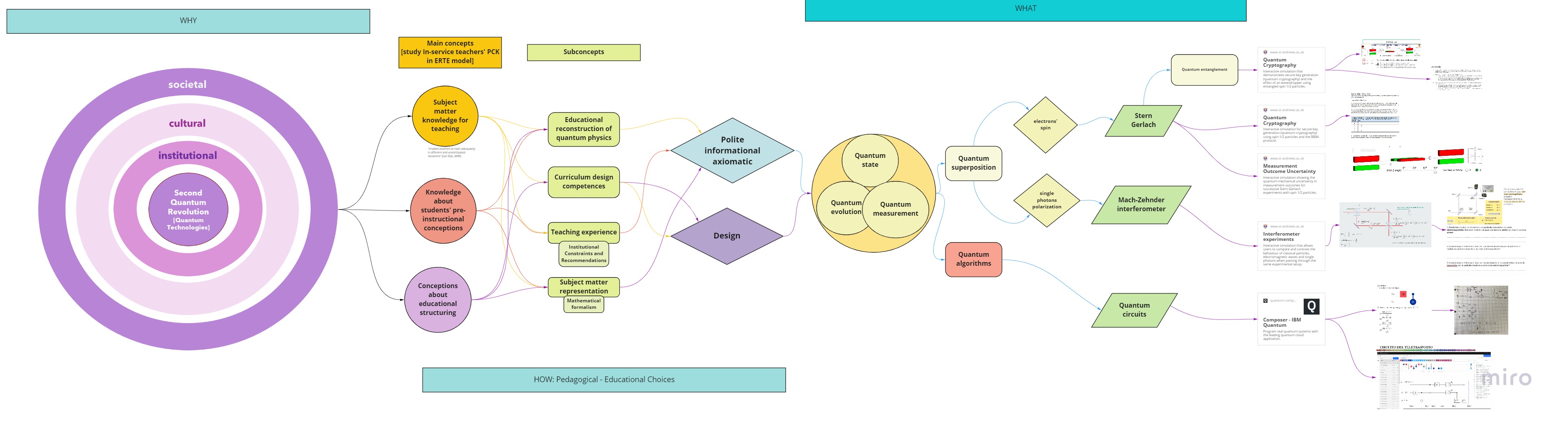}
    \caption[QTEdu PCK Map]{PCK Study map: Starting from multilevel targets (WHY), the maps show the relationship between the three dimensions of ERTE model and the educational choices used to design the learning environment for teachers (HOW). The relationship between the core concepts and the related activities are also shown}
    \label{fig:Ch3_QTEdu_PCKmap}
\end{figure}

In the project all the processes that can contribute to the development of teachers' PCK are fostered. In the Figure \ref{fig:Ch3_QTEdu_PCKmap} an hypothetical map of PCK development is presented. That map can be followed through as guidance of how the informational approach to quantum physics (see Chapter \ref{ch:3-QApproach}) can be used for teachers professional development. This map has been proposed within the working group "QTEdu open school pilot - Quantum Technology PCK for Teachers (PCK)" of the Quantum Flagship (https://qt.eu/about-quantum-flagship/projects/education-coordination-support-actions/)

\subsection{PCK development analysis grid}
\label{sec:PCKdimGrid}

In order to conduct the study of the teachers' PCK and together with them to evaluate the results of the educational design activities, the following grid (see Table \ref{tab:PCK_Analysis_GRID}) was elaborated to monitor the process of constructing learning environments in relation to the three dimensions of PCK development. The idea is to facilitate the process of identifying which of the elements of the elaborated approach to quantum physics (see Section \ref{sec:ApprPillars}) have contributed to the teachers' PCK development. Those same elements have been used in the thematic analysis of teachers interviews (see Section \ref{sec:TLE_focus_interview}) 

\begin{table}[hbt!]
\centering
\begin{tabular}{p{0.3\textwidth}p{0.7\textwidth}}
\multicolumn{2}{c}{\textbf{Study of teachers' PCK}} \\
\hline
& \\
\multicolumn{1}{c}{\textit{PCK Dimensions}} & \multicolumn{1}{c}{\textit{Description}} \\
&  \\
\hline
& \\
\footnotesize{\textbf{\textit{PCK sources}}} &  Subject matter knowledge\\
& Teaching experience\\
& Workshops\\
& Memories\\
&\\
\footnotesize{\textbf{\textit{Subject Matter Knowledge for teaching}}} &  Use of subject matter knowledge\\
& Flexibility in handling daily complexity\\
& Reaction to unanticipated situations\\
& \\
\footnotesize{\textbf{\textit{Students pre instructional conceptions}}} & Awareness of students' prior-knowledge\\
& Ability to follow students conceptions\\
&\\
\footnotesize{\textbf{\textit{Educational structuring}}} & Use of subject matter knowledge representations\\
& Design competences\\
& Familiarity with different teaching methodologies\\
\end{tabular}
\caption{Grid for the PCK study: list of elements used to evaluate different PCK dimensions}
\label{tab:PCK_Analysis_GRID}
\end{table}

\section{Learning environment design process}\label{sec:LEDesign}

An element that can unite the three dimensions of PCK study in the ERTE model is the reflection on the design processes that enable the creation of learning environments.

The process of educational reconstruction is expressed in the way in which learning environments are designed and implemented. The general and specific objectives are in fact constructed by balancing the external design constraints (e.g. institutional guidelines, teachers' experience) with the guidelines emerging from the clarification and analysis of subject matter process. To do this, teachers should organise their action effectively and consistently with the specific educational objectives.
This means promoting the process of transformation from the paradigm of the teacher-lecturer to that of the teacher-designer \cite{sancassani2019progettare}. This involves also a reconstruction of the role of the teacher in terms of "responsibility" no longer linked solely to the correct transmission of content. The teacher is called upon to plan in advance the management of the entire teaching-learning process, which includes reflections on the students' prior-knowledge, methodological choices and assessment actions consistent with learning objectives.
Although important, the explanation of contents exhausts the school's training mandate but not its educational one \cite{DM1392007}, which aims at the complete development of the person as an active and responsible citizen in a society in continuous transformation \cite{eu-2006/962/EC}. 
The active, participatory and creative management of learning experiences is therefore necessary for the development of citizenship competences and must aim to improve the quality of learning.\\
In order to bring a revolution into curricular activities, it is important to support teachers in reflecting about the way they design their learning environments. One of the strategies of the research project is to provide teachers with tools for educational design aimed at the construction of learning environments for the teaching of contemporary quantum physics.
The framework for the development of design competences is based on the coherent structuring of learning experiences starting from learning objectives and outcomes.\\

The stimulus is not to start the design from the list of contents or topics to be addressed.In accordance with the \textit{backward design} framework \cite{wiggins2005understanding}, the suggestion is to reverse the order: starting from the definition of the learning objectives and paying attention to the different learning and assessment strategies \cite{Hattie2016_LearningStrategies}.\\

The Constructive Alignment proposed by Biggs \cite{Biggs2003} \cite{biggs2011teaching} can be indeed useful to foster this practice. Cognitive Alignment starts from the global analysis of the system in which learning takes place and reflects on the elements that favour quality learning. In this framework, the alignment and coherence between Intended Learning Outcomes (ILO), assessment strategies and teaching methodologies is the condition that allows the construction of relevant learning activities that enable the learner to actively construct meaning \cite{FraserMazur2014_Bridging} \cite{Bouchee21} \cite{Malgieri2017}.

Using Biggs' words (2011)
\begin{quote}
\textit{Constructive} comes from the constructivist theory that learners use their own activity to construct their knowledge as interpreted through their own existing schemata. \textit{Alignment} is a principle in curriculum theory that assessment tasks should be aligned to what it is intended to be learned, as in criterion-referenced assessment \cite{biggs2011teaching}, 
\end{quote}

Cognitive alignment can therefore foster the development of teaching - learning strategies related to in-dept learning \cite{Hattie2016_LearningStrategies} that enables students to go beyond the level of notions acquisition. In this framework, the general aim of educational activities is to create a learning environment in which the teaching and learning process is intended "to change the way students see the world and hence to change their behaviour towards it" \cite{biggs2011teaching}.\\

This is not to diminish the role of specific contents and notions, but to clarify how transfer \cite{wiggins2005understanding} can become efficient within a broader learning process.
\begin{quote}
    "Teaching specific topics or skills without making clear their context in the broader fundamental structure of a field of knowledge is uneconomical" \cite{bruner1977} 
\end{quote}

According to Bruner, a teaching activity focused on a covering the course syllabus is \textit{uneconomical} because make it difficult for students to generalize, has little reward in terms of "intellectual excitement" and could easily generate an unconnected set of facts that is likely to be forgotten \cite{bruner1977}.

In that framework one main focus is about what the student does in the learning environment that motivates him or her in learning something. Teachers should be able to support this process. As Shuell pointed out

\begin{quote}
    "If students are to learn desired outcomes in a reasonably effective manner, then the teacher's fundamental task is to get students to engage in learning activities that are likely to result in their achieving those outcomes \dots It is helpful to remember that what the student does is actually more important in determining what is learned than what the teacher does." \cite{Shuell1986CongitiveLearning}
\end{quote}

The definition of the Intended Learning Outcome (ILO) is therefore to be declined by specifying the performance in terms of actions. Those actions refers to different cognitive skills that students have to perform in a specific context. 

A well defined intended learning outcome has the following characteristics \cite{sancassani2019progettare}
\begin{itemize}
    \item it is written in terms of performance ("student will be able to \dots")
    \item it makes explicit the action envisaged by the performance (using specific \textit{verbs})
    \item it defines the object of the action (what)
    \item it makes explicit the field of action (where/when)
    \item it makes explicit the context in which the performance takes place (how) 
\end{itemize}

ILOs should be expressed from the student point of view and should also show what the student will learn (i.e. content), the level of cognitive engagement or what the student is expected to do with the content (i.e. verb), and finally the context in which students will be able to do the “verb” with the “content” \cite{TLTasmaniaILO}
 should foster in-dept comprehension of topics 

\begin{figure}[hbt!]
    \centering
\includegraphics[width=\textwidth,height=\textheight,keepaspectratio]{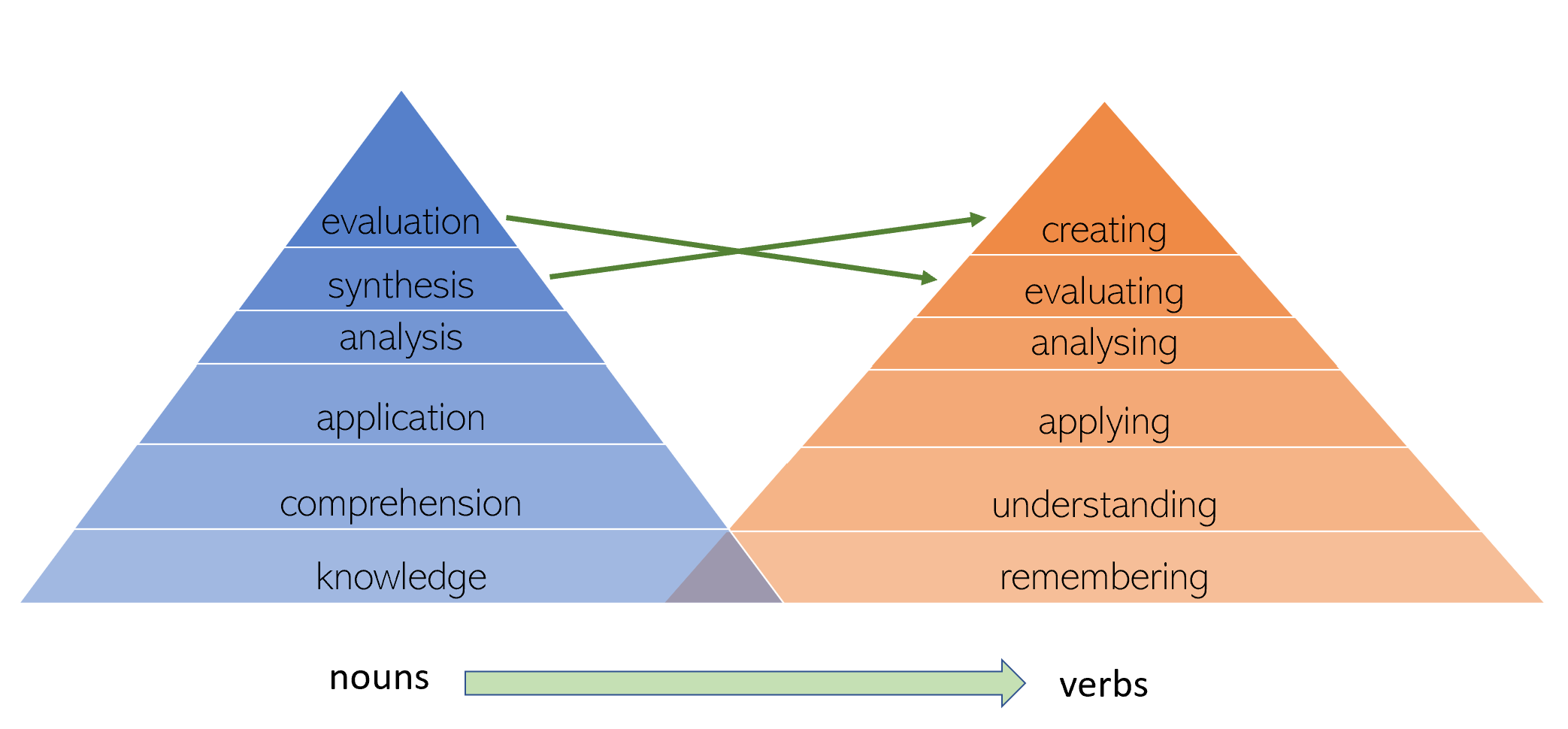}
    \caption[The Bloom taxonomy]{The Bloom taxonomy \cite{bloom1974taxonomy} (left) and the revised Bloom taxonomy (right)\cite{anderson2014_BloomREv}. In the revised version (Anderson 2014) verbs substitute nouns, knowledge becomes remembering and comprehension becomes understanding. Anderson and Krathwohl (2014) introduce the idea of LOTS (Low Order Thinking Skill) and HOTS (High Order Thinking Skills) to differentiate the different types learning goal between lower and higher layers of the pyramid}
    \label{fig:Ch4_4_BLOOM}
\end{figure}

Verbs can refer to different models and types of knowledge. For example, they can be related to different cognitive abilities and levels of abstraction as in the Bloom's Taxonomy \cite{anderson2014_BloomREv} and in its revised version \cite{Krathwohl2002_BloomRevised} (see Figure \ref{fig:Ch4_4_BLOOM})

\begin{figure}[hbt!]
    \centering
    \includegraphics[width=\textwidth,height=\textheight,keepaspectratio]{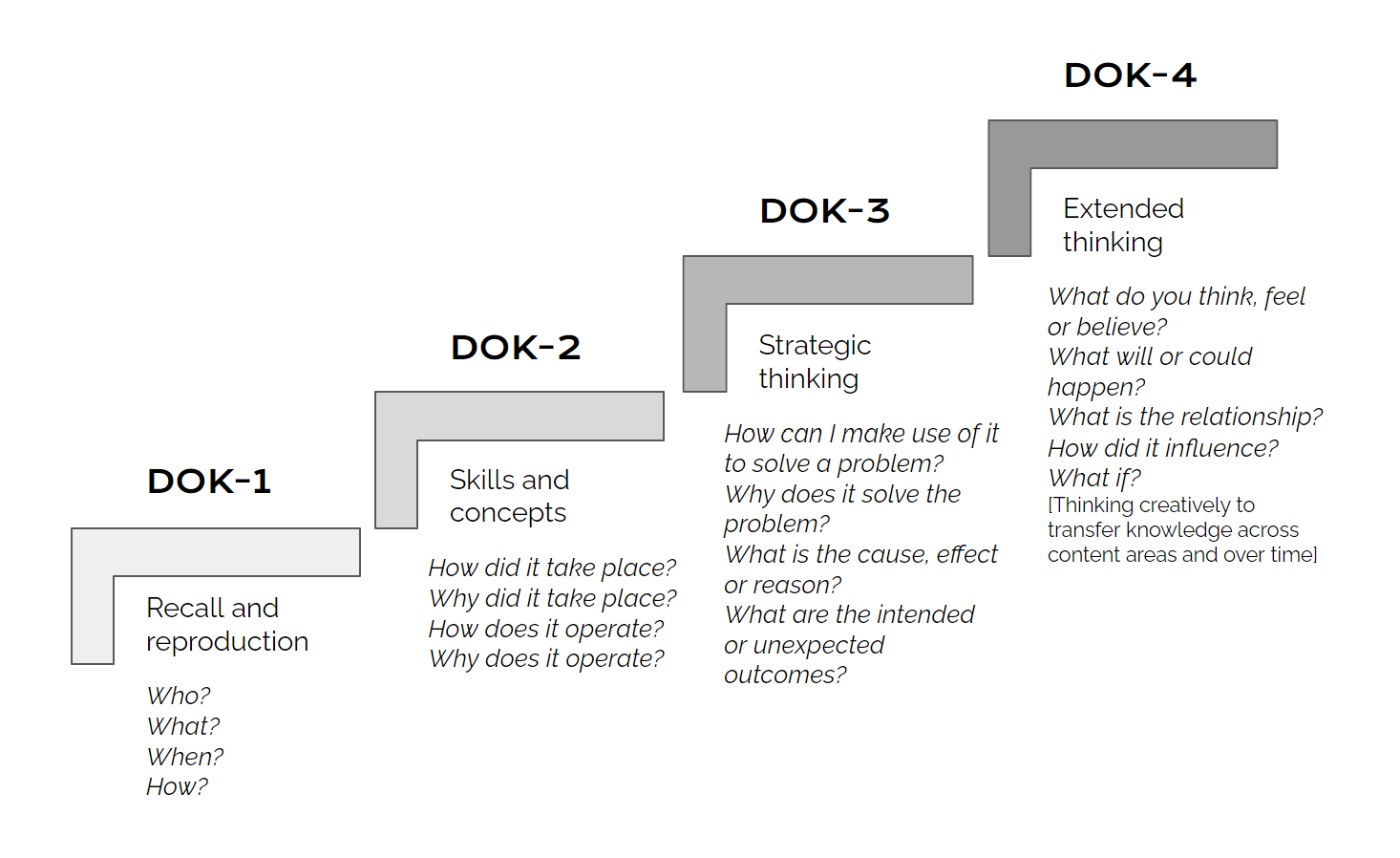}
    \caption[The Depht of Knowledge (DOK) model]{The Depht of Knowledge (DOK) model \cite{Webb1997_DOK}, Randolph Central School Corporation 2020). The Assessment ceilings establish the differences between different levels}
    \label{fig:Ch4_6_DOK}
\end{figure}

It is also possible to refer to different levels of complexity of thought, as in Webb's Depth of Knowledge (DOK) \cite{Webb1997_DOK} (see Figure \ref{fig:Ch4_6_DOK}). 

\begin{figure}[hbt!]
    \centering
    \includegraphics[width=\textwidth,height=\textheight,keepaspectratio]{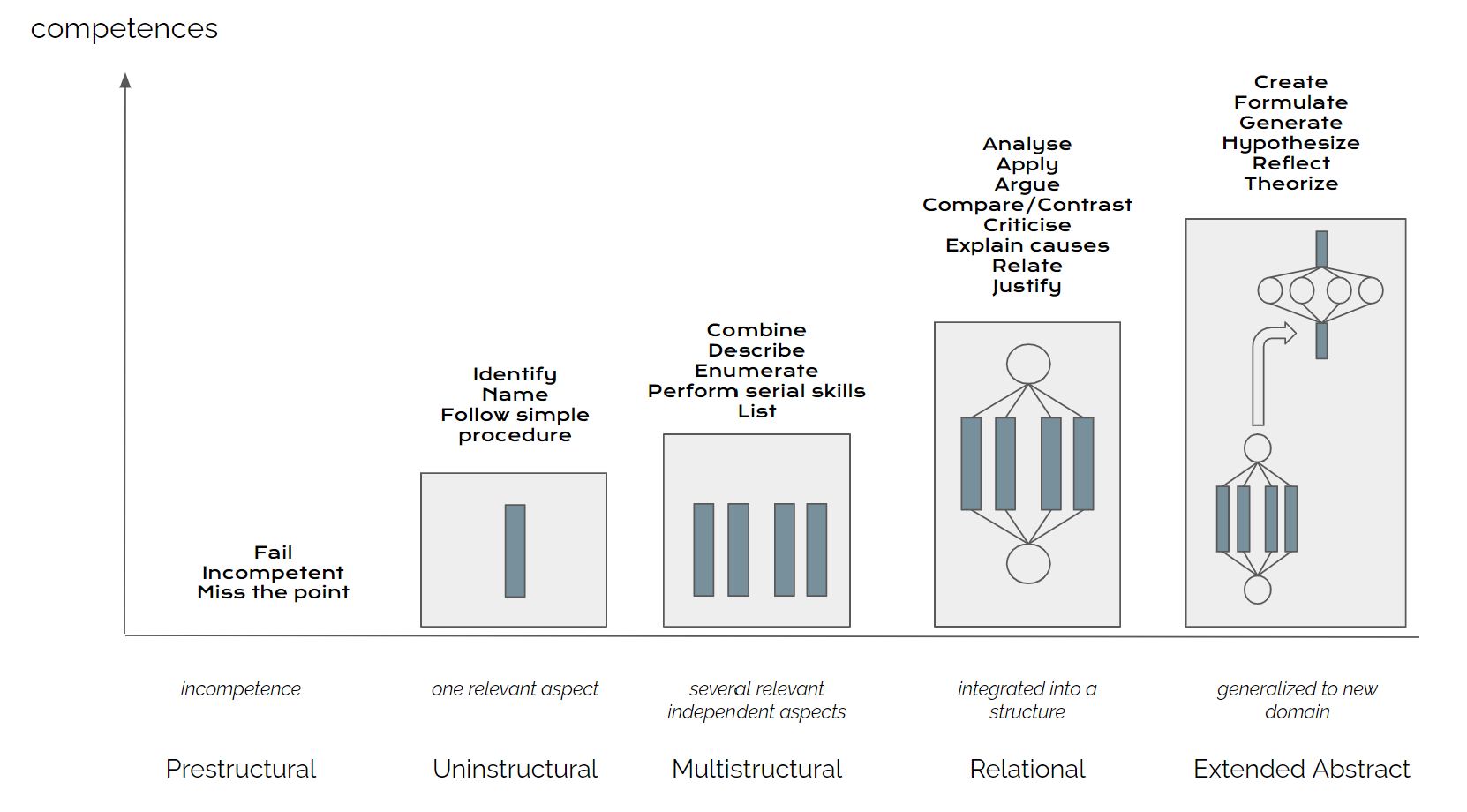}
    \caption{The Structure of the Observal Learning Outcomes (SOLO) taxonomy \cite{biggs1982SOLO}. Sample verbs indicate levels of understanding (adapted from https://www.johnbiggs.com.au/academic/solo-taxonomy/)}
    \label{fig:Ch4_6_SOLO}
\end{figure}

Another possibility in is relation to superficial and deep learning levels \cite{Hattie2016_LearningStrategies}, as in the Structure of the Observed Learning Outcomes \cite{biggs1982SOLO} \cite{Rembach2016_SOLO} (see Figure \ref{fig:Ch4_6_SOLO}).

All those possibilities are taken as reference points in order to create a common language and build shared indicators for the teaching design and for assessment activities.\\

These central elements of backward design \cite{biggs2011teaching, wiggins2005understanding} were shared with the teachers with whom the design activities were carried out and incorporated at various levels in some proposals (see Section \ref{sec:massimo} and Section \ref{sec:carcano}).

The proposed design approach have been used to reflect the about assessment process. In particular formative assessment strategies result as appropriate to monitor the learning process of students during the teaching - learning activities. An effective assessment strategy should allow teachers to grasp lots of evidence. In this scenario, an assessment is more like a "scrapbook of mementos and pictures" rather than a "single snapshot" \cite{wiggins2005understanding, tomlinson2006DiffInstr_UndbyDes}.
Consequently, during the training activities with teachers (see Chapter \ref{ch:6-LETeachers}) and students (see Chapter \ref{ch:7-LEStudents}) the use of formative assessment was promoted to facilitate the process of collecting and exchanging feedback within the different learning environments \cite{SCHILDKAMP2020_FormAssessment}. The most interesting aspect is that the detailed design of formative assessment moments can also help to change the roles of teachers to students in the context of the lesson \cite{Black1998_Assessment}. These moments have therefore been included as a fundamental element of the educational methodology.\\



The restraints to the in-presence activities imposed by the health emergency linked to the COVID-19 had a relevant impact on the implementation of different teaching and learning activities. For this reason, a different lesson setting was identified. That was to allow the meetings to be held at a distance, guaranteeing as much as possible the engagement of the students through the use of online tools during the online lessons. The details are presented in the next section.

\section{Online teaching and learning}\label{sec:online_methodology}

In the process of designing learning environments it is necessary to reflect on the methods of organising activities. This is particularly relevant since most of the activities could only be carried out online. In this section we would like to illustrate the theoretical references used to structure the lessons, both those proposed to the teachers within the Professional development programmes (see Section \ref{sec:QSkills} and Section \ref{sec:QJumps}) and those carried out with the students (see Section \ref{sec:massimo}, Section \ref{sec:carcano}, Section \ref{sec:SumSch_light} and Section \ref{sec:SumSch_QT}).

\begin{figure}[hbt!]
    \centering
   \includegraphics[width=\textwidth,height=\textheight,keepaspectratio]{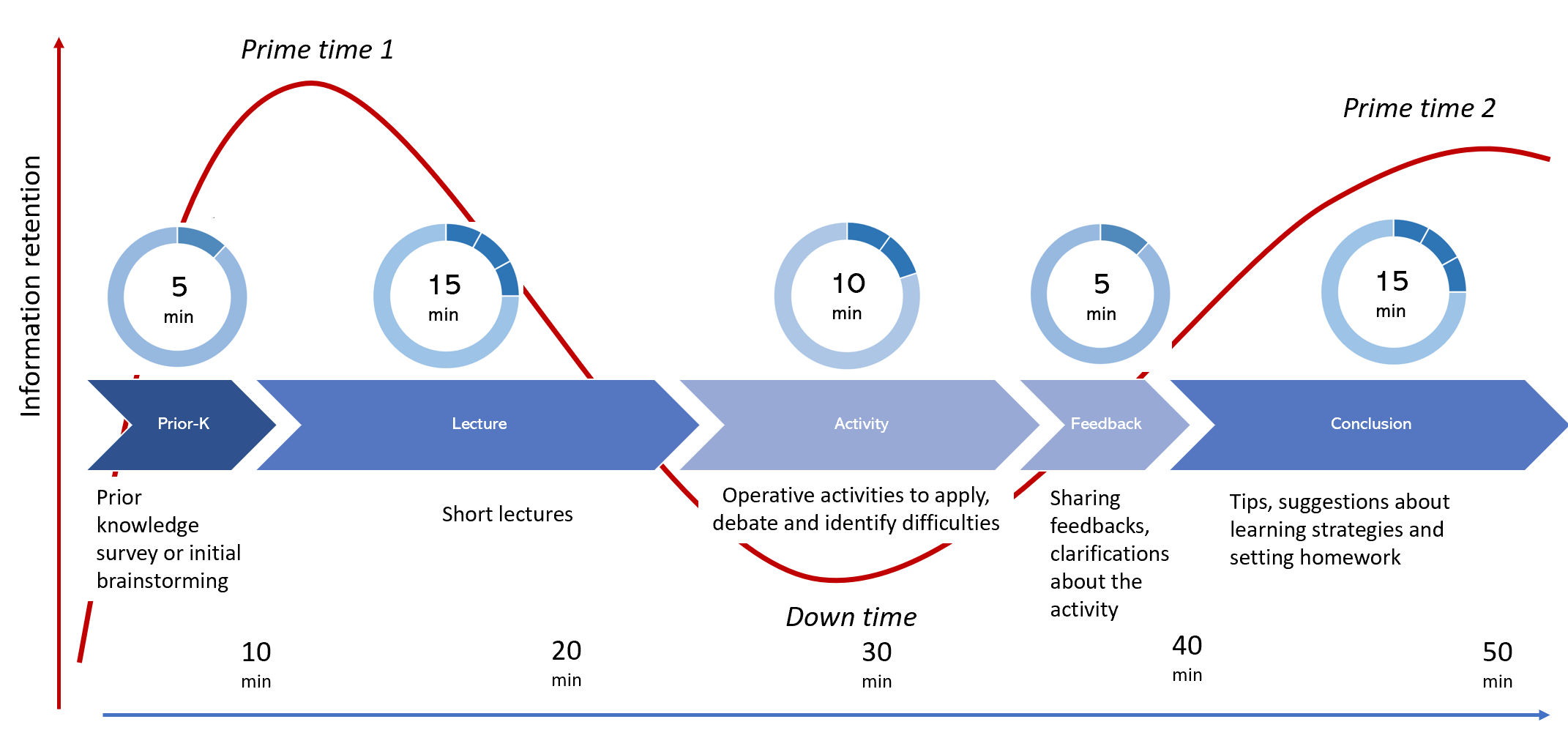}
    \caption{General structure of a Chunked lesson; in relation to the information retention model \cite{sousa2017Brain}. Adapted from \cite{maknouz2021_Chuncked} page 19 and page 22}
    \label{fig:Ch4_4_Chunktime}
\end{figure}

To promote and create a distance learning environment that encourages student participation, the organisation of learning activities for teachers and students was structured using the methodological approach of the \textit{chunked lesson} and the \textit{learning chunk}. \cite{maknouz2021_Chuncked}. 
The chunked lesson is about structuring the lessons (see Figure \ref{fig:Ch4_4_Chunktime}) into 5-15 minutes blocks of time. Moments of "short" explanation of the content are alternated with activities to be carried out independently or in small groups and with moments of sharing and feedback. That structure should facilitate students' information retention \cite{sousa2017Brain} and can be easily adapted to school time slots (that are between 50 and 60 minutes long).


Aware of the difficulty of working at a distance over a long period of time, this mode of time management makes it possible to encourage student participation by keeping them active during moments of synchronous activity. The balance between the teacher's interactive lectures and student activities supports the generation of meaningful learning environments. The use of simulations and worksheet help the creation of participative work groups in which students can learn interacting with their peers and can restructure concepts within an enriched framework.\\

\chapter{\label{ch:3-QApproach}Quantum education towards quantum technologies}

\epigraph{Do you guys just put the word "quantum" in front of everything?}{\textit{Ant-Man (Paul Rudd) to Dr. Hank Pym (Michael Douglas) and Dr. Bill Foster (Laurence Fishburne)}\\Ant-man and the Wasp, Marvel Studios, 2018}

This chapter describes the approach used for the design of the different activities for teaching and learning quantum physics. This is to make explicit the choices made in terms of learning objectives, the selection and organisation of contents and teaching strategies related to those objectives. These choices are the labeled as the \textit{pillars} of our approach. The attempt described in this chapter is to find a way of presenting the fundamental concepts of quantum physics that reflects the reversal of perspectives that is inherent in the way quantum theory describes reality itself.

The term \textit{approach} is referred to the design process. It is the way in which contents related to specific subject elements are chosen and organised as a consequence of making the educational objectives explicit. There are different possible choices through which specific topics, ideas and concepts can be structured. Using the criteria of educational design (see Section \ref{sec:LEDesign}) these choices are shaped by the process of constructing the objectives to be achieved, the strategies to fulfil them and the actions to be taken.   

The reflection presented in this chapter in based on the following guiding questions

\begin{table}[hbt!]
\centering
\begin{tabular}{cp{12cm}}
 & \textbf{Educational approach to quantum technologies} \\
&\textit{Teaching and learning contemporary quantum physics}\\
\hline
\rule[-4mm]{0mm}{1cm}
RQ4.1 & How to characterize the educational approach that facilitates the development of the new vision of reality promoted by the second quantum revolution? \\
 & \\
\midrule
\rule[-4mm]{0mm}{1cm}
RQ4.2 & How can the educational approach foster awareness of the impact of the second quantum revolution?
\end{tabular}
\caption{Research questions: educational approach development}
\label{tab:Approach_RQ}
\end{table}

\section{Shaping the approach}\label{Ch3.1_ApproachShape}
In literature there are different ways of presenting the fundamental concepts underlying quantum physics, categorised in general by the particular quantum phenomenon that is considered emblematic for the introduction of the fundamental concepts \cite{Michelini2021}. Distinctions may be found in contextual aspects, such as selecting a spin, polarization, or two-state systems. Different methodological approaches are also present, aimed to engage the students in the learning process in order to develop specific scientific competences related to the aquisition of important concepts.
There is also a large body of literature on the difficulties encountered in teaching quantum physics at university level. \cite{MarshmanSing2015_QdiffFrame, Singh2008_QStudDiff,Singh2015,Zuccarini2019}. The need to raise awareness of the tenets of quantum physics has stimulated research aimed at identifying effective strategies for the introduction of certain fundamental concepts and contents at secondary school level. \cite{Kim2017,Stadermann2019_Curricula}. The different ways of organising the contents are also influenced by the normative guidelines governing the national curricula \cite{Stadermann2019_Curricula}.

Our work started around a reflection on how to introduce elements of contemporary research in quantum physics into curricular activities. In order to stimulate the educational ecosystem related to the teaching of quantum physics, the first step was to involve a group of researchers in quantum physics in such discussion.

We have therefore organised a focus group with a group of six representatives of present research in quantum physics working in different universities in northern Italy (University of Milan, University of Pavia and University of Insubria in Como). In addition to having a great deal of experience in the field of quantum information in both theoretical and experimental fields, some participants had experience in organising training courses for secondary school teachers and extracurricular activities for high school students.\\
The questions used to guide the discussion with physics researchers are in Table \ref{tab:Approach_Researchers}

\begin{table}[hbt!]
\centering
\begin{tabular}{cp{12cm}}

& \textbf{Key elements for the teaching of contemporary physics}\\
&\textit{Physics researchers' perspective}\\
\hline
\rule[-4mm]{0mm}{1cm}
RQ4.3 & What are the key concepts that should characterise quantum physics teaching-learning sequences in upper secondary school?\\
 & \\
\midrule
\rule[-4mm]{0mm}{1cm}
RQ4.4 & What approach to teaching physics can facilitate the introduction of elements of contemporary physics at upper secondary school level?
\end{tabular}
\caption{Research questions: Physics researchers' reflections about teaching and learning quantum}
\label{tab:Approach_Researchers}
\end{table}

All quotations in this section are taken from the analysis of the recordings of the meeting.
The discussion with the researchers was focused on the identification of a several critical points they had noted in the way the discourse around quantum mechanics is set up at high school level. The main one is the choice of a \textit{historical approach} related to the exploration of the first interpretative models of quantum mechanics \cite{baggott2011Story}.

\begin{quote} {\fontfamily{lmtt}\selectfont
    It is one thing to deal with the history of physics \dots but quite another to deal with physics.
}\end{quote}

The historical approach is seen as not very useful and functional for the general understanding of the role of quantum theories. 

\begin{quote} {\fontfamily{lmtt}\selectfont
    [Presented in a purely historical way] Quantum mechanics seems not to be of much use \dots it only adds a halo of mystery and the people come away thinking that these [quantum] are strange things, a curiosity added without much relevance to the rest and without any connection with the possible applications that are indeed everywhere
}\end{quote}

A historical approach is better suited to introduce paths that have had a more linear development such as relativity (starting from special relativity and passing through the principle of equivalence to arrive at general relativity) or Bell's inequalities (starting from the EPR argument and arriving at Bell's reflection on realism and locality).
However, this kind of approach does not seem to facilitate a coherent analysis of quantum theory, exposing the students to the 
\begin{quote} {\fontfamily{lmtt}\selectfont
    \textit{tortuosity} in which the founding fathers have been trapped for years [\dots] Bell's inequalities are a result of the 1960s and if you follow a historical approach you will never get there.
}\end{quote}

The group's reflections then converged around the observation that contemporary quantum theory is axiomatic in nature. In their experience, the structure of courses at university level introduces the basic principles from the outset by means of a structure made up of postulates presented using a very rigorous formal language.
The advantage of this type of argumentative structure is that it allows one to overcome \textit{tortuosity} and provides effective interpretative tools.
However, this type of content presentation is not feasible at secondary school level.
\begin{quote} {\fontfamily{lmtt}\selectfont
    I don't think our learners' minds are ready to digest such a thing [the axiomatic structure] and still consider it as physics.
}\end{quote}
Although some linear algebra content is part of the mathematics curriculum \cite{MIUR2010_211}, the level of confidence in using that kind of language in non-mathematical contexts is difficult, particularly in the process of linking formal mathematical description and the experiences with physical objects \cite{Bouchee21}.

\begin{quote} {\fontfamily{lmtt}\selectfont
Until that moment [of the school course, at the end of electromagnetism] there is the phenomenon and the mathematics captures the intuitive characteristics of that phenomenon \dots. Our intention is to reduce the gap between the teaching of electromagnetism and the teaching of modern quantum mechanics, which look like different things \dots but are still physics and the paradigm is the same. By introducing an axiomatic approach we do the opposite service.
}\end{quote} 

However, an interesting starting point to introduce the quantum vision can be traced back to quantum information. Quantum physics is therefore a theory that describes how reality is perceived on the basis of the information that it is possible to derive from it, and "how" it is possible to derive this information \cite{maccone2008}.
It is therefore considered necessary to identify how to make this idea accessible to those who do not have a complex and structured formal language. The use of the axioms \cite{nielsen_chuang_2010} must be \textit{educated} and \textit{polite} in order to help understand the relationships between the main concepts through a language and representations that are sufficiently rigorous.

The key concepts to activate the use of an informational approach with students are:

\begin{itemize}
    \item \textbf{quantum state} as a conceptual construct \cite{turchi2007MADIT} representing all the information it is possible to possess on the object \cite{maccone2008}
    \item \textbf{quantum superposition} as one of the peculiar aspects of quantum states and marking a precise boundary between the informational nature associated with classical and quantum objects. Quantum entanglement is considered as superposition applied to multipartite systems
    \item \textbf{quantum measurement} as an action on a quantum state that can be associated with a specific measuring apparatus. The probability of obtaining a given result is provided by Born's rule
\end{itemize}

These concepts can be introduced via meaningful connections with those phenomena related to physical quantum objects. This is to replicates the argumentation schemes that students are already accustomed to use in other contexts of the study of physics (such as Newtonian mechanics).
In this sense, axioms can be introduced by formulating them from the description of some specific experimental contexts. The axioms thus become \textit{rules of the game} that allow the description of observed phenomena.\\

An example of such a context is the polarisation of light. Also in the classical context, polarisation can be described as a vector state. This should facilitate the transition process between classical and quantum objects (see Section \ref{sec:qubit_polarization}). Once the quantum nature of light has been clarified with a suitable definition of the single-photon state (see Section \ref{sec:SingPhot}), it is possible to reconstruct the different axioms by linking them with fundamental phenomenological aspects (such as experiments with Mach-Zehnder interferometers, see Section \ref{subsec:MZIBS_sim} and Section \ref{subsec:MZIexper})

\begin{quote} {\fontfamily{lmtt}\selectfont
The polarisation of light is a classical degree of freedom \dots but it is "vectorial" and therefore we can say superposition states "exist" \dots because the 45$^{\circ}$ polarisation is not just a mixture of horizontal and vertical polarizations.\dots There is the postulate of reduction that corresponds to the action of a polaroid \dots There is Born's rule that corresponds to Malus's law.   
}\end{quote}

Bringing out the axiomatic structure from the experimental evidence facilitates the construction of a general theoretical framework.

\begin{quote} {\fontfamily{lmtt}\selectfont
    the logical passage to be smuggled in is that any physical entity behaves like this \dots follows these rules of the game, the axioms \dots Even those objects that at a macroscopic level do not seem to behave like this \dots when we go to a microscopic level we find these behaviours.
    But this is the \textit{hiatus}, this is the \textit{discovery}.
}\end{quote}

An informational approach can be based on the study of physical systems that can represent qubits (see Chapter \ref{ch:3b-Qubits}). This makes it possible to deal with quantum phenomena using a simpler formalism that is at the same time effective and rigorous, while also avoiding the \textit{tortuosity} associated with the first elements of quantum mechanics development. 

\begin{quote} {\fontfamily{lmtt}\selectfont
    The Schr\"{o}dinger equation is an operator differential equation with partial derivatives [\dots] It all started from that, but do we really need all this crazy mechanism? No, because from a conceptual point of view it equivalent to use unitary 2 x 2 matrices and two-dimensional vectors. The Schr\"{o}dinger equation can easily be shown to be equivalent to unitary evolution, only that while the former is difficult to understand even at university level, the latter can be approached even with a basic knowledge of linear algebra.
}\end{quote}

Finally, an informational approach is seen as essential to build meaningful links with technological applications. Modern quantum information theory is in fact the basis of the second quantum revolution.

\begin{quote} {\fontfamily{lmtt}\selectfont
    If the students are already clear about what a qubit is, talking to them about topics such as quantum cryptography is easier, and perhaps they will understand more about the theory presented to them.
}\end{quote}

Based on the results of this dialogue, a reflection on the construction of an informational approach has been built up over time. The elements that emerged can be found in the next section.

\section{Polite Informational Axiomatic Approach}\label{sec:PIAA}
The research project starts from the observation that a scientific revolution is taking place and this process is opening new possibilities and it is changing our vision of reality (see Chapter \ref{ch:1-intro} and Chapter \ref{ch:2-PrFrame})

The second quantum revolution does not only concern the production of new technologies, but it is about the generation of a new vision of reality as we know it. So the aim of the project is also to engage students (and their teachers) in the exploration of the quantum realm through the analysis of quantum technologies.
The work on the approach to quantum physics is not limited to the choice of a particular selection of contents around which to build a specific teaching-learning sequence. It indeed concerns a broader reflection on the specific didactic objectives linked to the teaching of quantum physics in the context of secondary school. This reflection serves as a guide to design educational activities that allows these objectives to be achieved.
The direction has therefore been to understand how to convey the sense of rupture that the emergence of quantum physics has produced in the usual way of knowing reality and how the study of quantum technologies can facilitate this process..\\ 
The approach is therefore the way to design educational activities in terms of the definition of learning outcomes, educational strategies and content organization (see Chapter \ref{sec:LEDesign}).\\
In general, the study of quantum technologies is proposed following the study of the concepts and fundamental principles that describe the behaviour of quantum objects through the evolution of interpretative models originating in the classical world.\\

The construction of the theoretical framework necessary to understand the behaviour of elementary microscopic entities (such as electrons or photons) precedes the analysis of its application context (in which electrons or photons are, for example, seen as a resource for implementing protocols).

\begin{figure}[ht]
    \centering
    \includegraphics[width=\textwidth,height=\textheight,keepaspectratio]{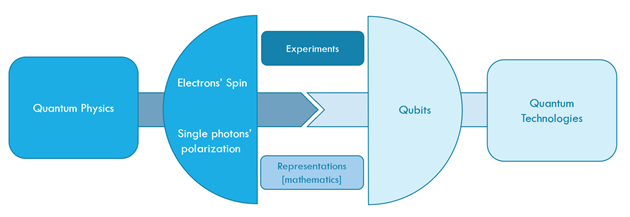}
    \caption{General approach to Quantum Physics: technologies as applications}
    \label{fig:Ch3_1_TraditionalApproach}
\end{figure}

Our choice is to reverse the approach and introduce the study of quantum physics starting from the problem of how information can be encoded in a physical system and the difference between choosing a classical or a quantum system.
We have therefore chosen to introduce the concept of a qubit, as a minimum unit of quantum information, and to approach the study of the basic principles of quantum physics through those physical systems that can be used to represent the concept of a qubit, such as electrons’ spin and single photons polarization (see Chapter \ref{ch:3b-Qubits}).\\
Reversing the approach and focusing on the informational process opens the possibility to introduce the axioms that are the tenets of quantum theory. Those axioms will emerge from the analysis of the algorithms as the “rules of the game”  that quantum states follow. And those rules provide tools to interpret experiments about complex quantum behaviours, such as quantum superposition and entanglement.

\begin{figure}[ht]
    \centering
    \includegraphics[width=\textwidth,height=\textheight,keepaspectratio]{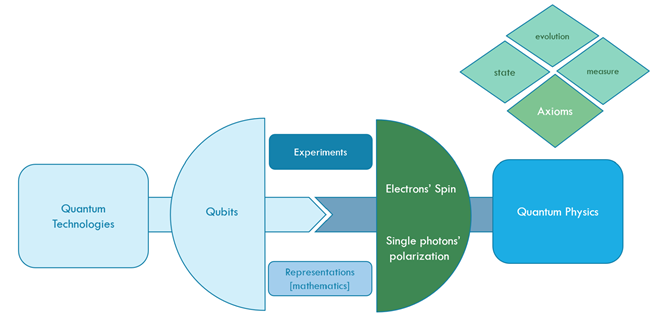}
    \caption{Informational approach to Quantum Physics: technologies as a context to learn quantum}
    \label{fig:Ch3_2_InformatApproach}
\end{figure}

\subsection{The rules of the game}\label{sec:QAxioms}
The axioms referred to are those that enable to make sense of the concept of state and its evolution and that can clarify the meaning of the measurement process in quantum mechanics. In particular, the relationship between the axioms and the phenomenological aspects of the quantum objects that can best represent them have been taken into consideration. In the following paragraphs these connections will be made explicit.

\begin{itemize}
    \item[Rule 1:] Each isolated physical system is associated with a Hilbert space $H$, which is called the system's state space. The system is completely described by a state vector (or wave function) $| \psi \rangle$, that is a unit vector of the space of the states.
\end{itemize}

In this context, the state is not a physical entity but is a conceptual construct \cite{turchi2007MADIT} that represents all the information one has on a physical system.
This is undoubtedly an abstract concept, and this kind of abstraction is part of the repertoire of physics education, even in classical physics (point-masses, inertial reference frames, fields, etc.). What is important is what these kinds of abstractions can help one to say about the nature of quantum objects. The process of educational reconstruction therefore wants to start from this abstract concept to see how it can be used in the interpretation and description of quantum phenomena.\\

The first axiom establish the mathematical background that can be used to describe the state of the physical system. One crucial idea is that if $|\psi_1\rangle$ and $|\psi_2\rangle$ are vectors in $H$, the vector 
\begin{equation}
|\phi \rangle = c_1 |\psi_1\rangle + c_2 \psi_2\rangle \qquad c_1,c_2 \in \mathbb{C}
\end{equation}
is also a vector of $H$.
Another important feature is related to the normalization of the state vector 
$$ || \psi ||^2 = 1 $$
that is necessary for the statistical interpretation of a general state.\\
In general, the vector $|\psi\rangle$ and $e^{i\alpha}|\psi\rangle$ represent the same state.\\
Any vector $|\psi\rangle$ can be written as the sum of the element of a base $e_n$ using a specific set of complex numbers $c_n$.

\begin{equation}
    |\psi\rangle \; = \; \sum_n c_n |e_n\rangle \qquad \langle e_i|e_j \rangle = \delta_{i,j}
\end{equation}


\begin{itemize}
    \item[Rule 2:] The time evolution of a closed quantum system is described by a unitary transformation $\hat{U}$. 
\end{itemize}

This second rule describes the dynamic of the physical system. The equation
\begin{equation}
|\psi (t') \rangle \; = \; \hat{U} (t,t')\,|\psi(t)\rangle   
\end{equation}

is used to describe how a generic $|\psi(t)\rangle$ state evolves in the state $|\psi(t')\rangle$ , with $t'>t$
This type of representation is particularly useful to describe the action on the state of apparatus within an experiment (see Section \ref{sec:seqSGA_matrix} and Section \ref{subsec:Polar_MathInterplay}) or of logic gates in a circuit.
The time dependent Schr\"{o}dinger equation

\begin{equation}
    i\hbar \frac{\partial}{\partial t}|\psi(\mathbf{r},t) \rangle = \hat H |\psi(\mathbf{r},t)\rangle
\end{equation}

describes the evolution of the state.\\
The Hamiltonian operator $\hat{H}$ is obtained from the classical
energy expression
\begin{equation}
    H \; = \; \frac{p^2}{2m} + V(x)
\end{equation}
by replacing the momentum $p$ and position $x$ by their corresponding
operators $\hat{p}$ and $\hat{x}$.
This type of description has been used to analyse the relationship between classical and quantum descriptions in particular experimental contexts, such as electron diffraction (see Section \ref{QJTEresult2}).

\begin{itemize}
    \item[Rule 3:] A single measurement of the observable $A$ on the state $|\psi\rangle$ results in one of the eigenvalues $a_n$ of $\hat{A}$ with a probability equal to $|\langle \lambda_n | \psi \rangle|^2$. Immediately after the measurement, the system is in the state $\frac{\hat{P}_n |\psi\rangle}{||\hat{P}_n |\psi\rangle ||}$, where $\hat{P}_n$ is the projector operator in the subspace of the eigenstates of $\hat{A}$ with eigenvalue $a_n$ 
\end{itemize}

This final rule expresses the intrinsically probabilistic character of quantum measurement. That is  essential to understand the different nature of measurement between classical and quantum physics.
In classical physics, the physical system is the "bearer of the properties" and the act of measurement only makes the value of those properties explicit. In such context, every (macro)state encodes "the way things are" since the reference to a given microstate defined in an appropriate space leads to "unequivocal truth values" for all the propositions about the system \cite{DoringIsham2010}.\\
In quantum physics, any physical quantity $A$ that can be measured (i.e. an observable) is represented in the Hilbert space by a Hermitian operator\footnote{linear, self-adjoint ($\hat{A}^{\dagger} = \hat{A}$) differential operator} $\hat{A}$ that acts on the state functions.
Considering physical quantities that can only assume a discrete number of values (e.g. a discrete spectrum), the only possible result of measuring an observable $A$ are the eigenvalues $a_n$ of operator $\hat{A}$ that represents the observable $A$.
So if $\hat{A}$ has only a discrete set of eigenvalues $a_l , a_2,\dots$, with corresponding eigenvectors $|\lambda_1\rangle, |\lambda_2\rangle,\dots$ then, for a given state $|\psi\rangle$ the probability that a measurement of $A$ will yield a particular eigenvalue $\lambda_n$ is


\begin{equation}
    P(A=a_n; \psi) \; = \; |\langle \psi| \lambda_n \rangle|^2 \; = \; |\alpha_n|^2
\end{equation}

where $\alpha_n \in \mathbb{C}$ is one of the complex coefficients with which the state can be expressed as a linear combination of the basis formed by the eigenvectors di $\hat{A}$.

\begin{equation}
    |\psi\rangle \; = \; \sum_n \alpha_n |\lambda_n\rangle
\end{equation}

Another way to see it: if the system is in the eigenstate $|\lambda_n\rangle$, the result of a measurement of $A$ is \textit{guaranteed} to be $\psi_n$ \cite{SusskindQT}.

A key aspect of this rule also lies in the characterisation of the measurement process as the attribution of a property to the physical object.

As in fact emerges from the analysis of experiments with qubits (see Section \ref{sec:qubit_electron}, \ref{sec:qubit_polarization}), when considering the action of measuring devices on quantum objects represented by states, it is necessary to distinguish between observables and operators \cite{SusskindQT} 
The operators $\hat{A}$ acts on state vectors and not on physical objects. When $\hat{A}$ acts on a state vector, it generates a new state vector and not a numerical output (measurement result).\\
The use of sequences of Stern Gerlach apparatus (see Section \ref{subsec:SGAsims} and Section \ref{sec:seqSGA_matrix}) and Mach Zehnder interferometers (see Section\ref{subsec:MZIBS_sim} and Section \ref{subsec:MZIexper}) will highlight this fact by comparing what happens if you swap the order of the measuring apparatus or if you choose to perform or not to perform a measurement within the apparatus sequences. In other words, the measurement result cannot be properly described without taking into consideration the apparatus.\\

Similarly, the choice of the basis associated with the observable turns out to be a highly relevant aspect in the process of representing the states associated with quantum objects (see Section \ref{sec:seqSGA_matrix}). In general the same state can be written as a combination of different bases. So one of the operations through which we want to express the meaning of rule 3 in the analysis of experiments is the possibility of being able to rewrite the state in the basis associated with the observable being considered.

\section{Approach pillars}\label{sec:ApprPillars}
The approach is based on specific \textit{changes} that are the pillars upon which the design process is based. This pillars guided the construction of the basic ingredients that have been used in the educatonial reconstruction process (see Chapter \ref{ch:4-TheorFrame}), orienting the definition of the learning outcomes, the assessment strategies and the content definition both for teachers and students education activities.
The reflections generated are inevitably strongly intertwined. In the following paragraphs will be explained some of the general characteristics of each of the pillars, highlighting some key words.

\begin{figure}[ht]
    \centering
    \includegraphics[width=\textwidth,height=\textheight,keepaspectratio]{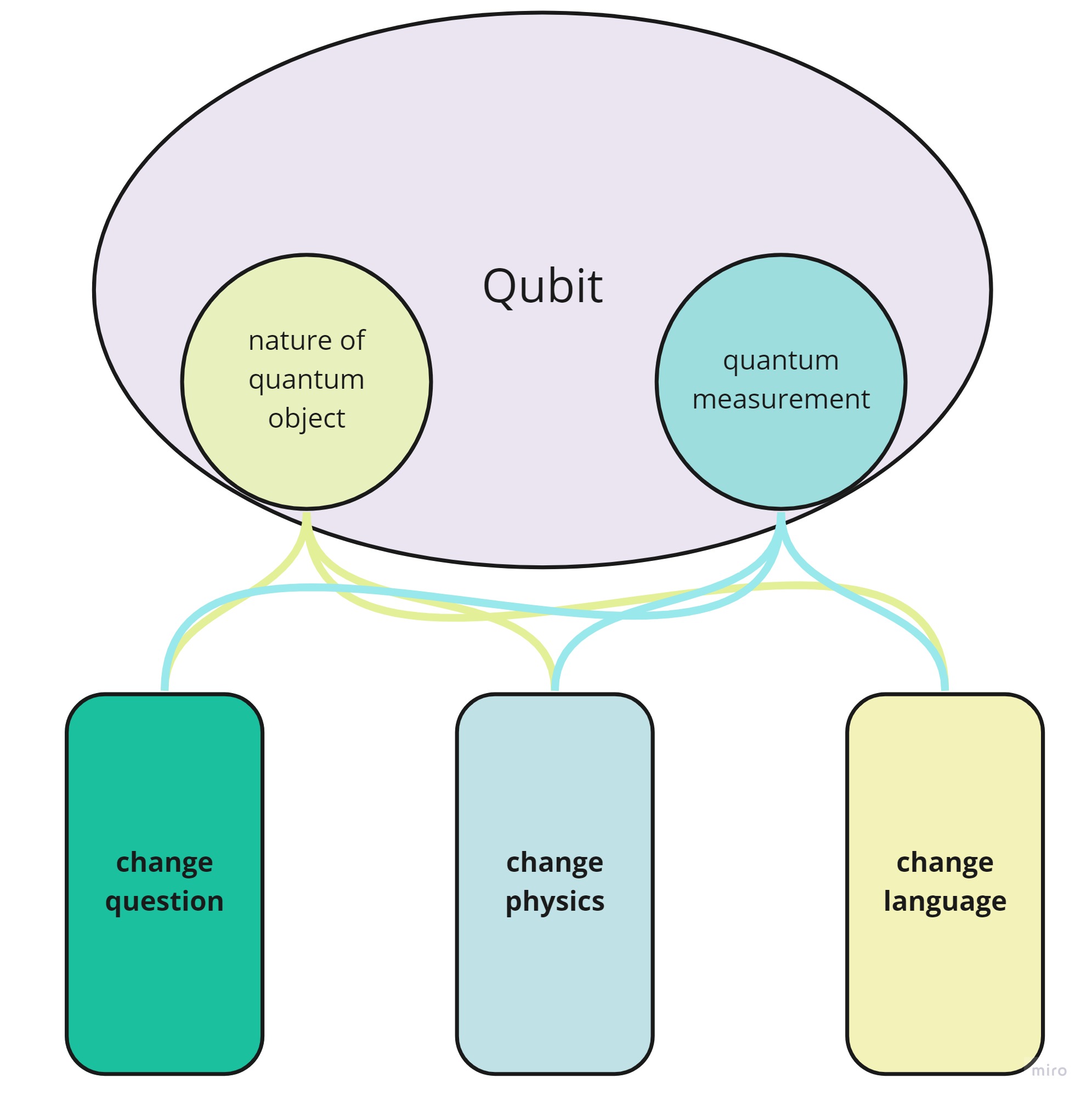}
    \caption{Polite informational axiomatic approach pillars}
    \label{fig:Ch3_7_PIAApproach_pillars}
\end{figure}

\subsection{Change the question}\label{sec:ChangeQ}
One of the goal of most quantum physics curriculum (including the Italian one) is to find how to construct a model to describe the behaviour of microscopical physical object using the analysis of the phenomenological evidence that can be gathered from experiments (as in the case of early stages of the development of quantum mechanics).\\
Starting from an informational perspective, we think about the problem of encoding information by associating it to a physical object. The question is to what extent the choice between associating the information to a 'classical' or 'quantum' object can revolutionize the way this information can be used.
This is a central question at the heart of quantum information and that triggered the research on quantum technologies.

The aim is to introduce quantum physics not as an extension of classical physics born out of the need to explain new phenomena, but as a completely different way of approaching the description of elementary physical phenomena.

One of the critical points noted in other ways of introducing quantum physics is that of justifying the transition from classical to quantum description by using a series of expedients that manage to adapt the classical paradigm to the evidence of quantum experiments. Quantum physics is basically classical physics <<with a couple of new gimmicks thrown in>>  \cite{SusskindQT}. However, this does not resolve many of the interpretative difficulties that make the process of reconstructing the theoretical framework and meaning-making difficult for students, such as the consistency between the corpuscular and wave nature of elementary quantum objects such as electrons or photons. \cite{Bouchee21,Malgieri2017}.\\
The choice was made to build a quantum theory on the basis of another problem, namely that of using quantum physical objects to encode information. The design of the activities therefore started by looking for \textit{new questions} to ask in the discovery of the quantum world. The formulation of the questions was then adapted to the objectives of the teachers' professional development programmes (see Chapter \ref{ch:6-LETeachers}) and the activities with the students (see Chapter \ref{ch:7-LEStudents}).

\subsection{Change the physics}\label{sec:ChangeP}
The choice of an approach centred on the problem of information is closely linked to a reflection on the physical nature of objects. Through the study and analysis of the quantum state, the focus is on what is possible "to say or not to say" about the object and on the possibility that it is not possible to fully capture all its properties. 
While in classical physics <<knowing the state of a system implies knowing everything that is necessary to predict the future of that system>>, in quantum physics <<knowing the state of a system means as much as can be known about how the system was prepared>>\cite{SusskindQT}. In this scenario the role played by the meaning given to the term \textit{knowing} a physical object is fundamental, especially in relation to the measurement process.
The necessary educational reconstruction inevitably brings with it an ontological confusion \cite{BrookesEtkina2007} since that purely quantum objects do not fit ontologies students are familiar with \cite{Bouchee21}. The change in the way one looks at the problem should lead to a more gradual transition towards a description of quantum objects that is unrelated to the classical framework. The use of classical representations such as those of "particle" or "wave" makes it difficult to interpret these counter-intuitive aspects \cite{Bouchee21}, \cite{Baily2015} that almost spontaneously emerge when analysing the behaviour of quantum objects such as electrons from a phenomenological point of view.

The most relevant aspect in the description of quantum objects is that of measurement. This is a characteristic element of the new quantum physics. In classical physics, measurement attributes properties to the object that common sense considers to be pre-existing to the measurement process itself.

In the quantum context, measurement is a probabilistic process of distinction between two or more mutually exclusive states. Through this process it is possible to associate properties with the physical object that encodes the information related to the state itself.
By choosing a quantum object to encode information, not only does the way in which this information is encoded change, but also the type of information that can be processed and how this information can be extracted by a measurement process.

This radically changes the way of approaching knowledge of the physical world and students must be accompanied in this transition by careful reflection on the difference between state and quantum object.

\subsection{Change the language}\label{sec:ChangeL}

\epigraph{When a system collapses, language is released from its moorings. Words meant to encapsulate reality hang empty in the air, no longer applicable to anything. Textbooks are rendered obsolete overnight and overly complex hierarchies fade away. People suddenly find it difficult to hit upon the right phrasing, to articulate concepts that match their reality.\\
\textit{A.S. Magnason}\cite{magnason2020TimeWater}}

The representation of physical objects therefore needs to be supported by an adequate mathematical formalism to compensate for the inability to visualise the behaviour of quantum objects through representative schemes based on the senses or common language. This is a new but necessary level of abstraction.

\begin{quote}
In physics, the formal structure of theories is not just a tedious frill that can be circumvented.\cite{BessonMalgieri2018}
\end{quote}

Quantum states are representable by a rigid formal structure that allows for the intrinsically probabilistic nature of the states themselves to be taken into account. The abstraction required to represent quantum objects also necessitates a new logic and a definition of probability that is distinct from what is usually used for classical systems in terms of Boolean logic \cite{DoringIsham2010}.
From the perspective of educational reconstruction it is therefore necessary to include certain formal aspects in order to preserve the objective of constructing a coherent interpersonal theory based on a rigorous formal language.
Such is the language used to describe operations on vector states using operators and matrices (see Section \ref{sec:seqSGA_matrix}).\\
Educational research shows that it is difficult for students to establish a link between mathematical representation and physical phenomena. \cite{Bouchee21}.
For this reason, various learning environments built around quantum state and qubit concerts are always structured with constant reference to the physical objects representing the qubits and make constant use of simulations that can be used in conjunction with mathematical representations (see Section \ref{sec:qubit_electron} and Section \ref{sec:qubit_polarization}).

Mathematical language is to be understood as a tool for reasoning around concepts. The mere acquisition of formal mechanisms may not, however, be a guarantee of personal reworking and acquisition at conceptual level \cite{Singh2015,MazurPeer1997}. The use of the formalism makes it possible to highlight difficulties in the transitions between different epistemological frameworks \cite{Modir2019, Modir2020}. In the different paths we have always tried not to limit ourselves to the symbolic register of mathematics, but to ensure that the use of formulas and equations is always accompanied by explanations and comments using ordinary language. \\
The construction of the training courses has therefore always foreseen moments of formative assessment (see Section \ref{sec:LEDesign}) to stimulate those cognitive processes suitable for the reconstruction of the theoretical framework by the students. Through the collection of closed and open answers, listening to the reflections in moments of shared reflection, it was possible to collect the text produced by the participants and support the process of reconstruction of the theoretical framework by providing useful feedback to link the physical world and mathematical representation.

\section{Conclusions}\label{Ch4_Conclusions}
As for the educational approach to quantum technologies (see Table \ref{tab:Approach_RQ}), the proposal is not to merely present quantum physics as an adjustment of classical physics. The main suggestion is to turn the problem around by starting with questions about the meaning of reality ("what is an object", "what can I say or not say about its properties") and focusing on the way quantum physics constructs its answers.
In this way, quantum technologies become the context in which to search for such answers because they allow us to deal directly with the problem of information related to quantum states. The impact that this way of talking about quantum physics can have in the educational context comes from the fact that the basic principles of quantum theory emerge during the exploration of this context and turn out to be the necessary and indispensable tool not only to explain the results of experiments but also to understand the functioning of quantum technologies.\\

With respect to the key elements for the teaching of contemporary physics (see Table \ref{tab:Approach_Researchers}), it is strategic to move away from a purely historical approach and focus on those concepts that are at the basis of the gap between the classical quantum world. The advantage of introducing the concept of information linked to the quantum state is that it allows us to go directly to the heart of quantum theory and to include in the teaching reflection the fundamental axioms of quantum theory that emerge from the study of physical phenomena.
This process of transformation must be mediated by an appropriate revision of the language and representations of physical phenomena that are useful for the construction of models. In particular, the role of mathematical formalism is central in order to rigorously address the process of constructing the theoretical framework in the delicate relationship with the knowledge to be provided to students.

The elements discussed in this chapter have been useful in the process of educational reconstruction of the central concepts (see Chapter \ref{ch:3b-Qubits}) and in the work with teachers (see Chapter \ref{ch:6-LETeachers}) and students (see Chapter \ref{ch:7-LEStudents}).




\chapter{\label{ch:3b-Qubits}Qubits and their representations}

The general aim of this chapter is to create a meaningful link between quantum qubits and the physical systems that represent those concepts. The characteristics of the quantum states associated with these objects can be used to encode information. This information is transformed as the physical systems interact with the experimental apparatus (e.g. Stern-Gerlach apparatus, polarizers or beam splitters). Information can finally be extracted through measurements on the physical system itself.\\

The choice of the qubit as the central concept around which to develop the educational reconstruction of the elements characterising quantum physisc has certain advantages \cite{Dur2013qubitlearn, Ford1966keyideas, Schneider2010qm4beginners, Muller2002, Mermin2003CbitQbit,Muller2021penny}.\\
Firstly, the qubit is the simplest physical system that encompasses all the characteristics of a quantum system.
Secondly, the necessary mathematical tools to describe them can be linked to mathematical skills that can be introduced at secondary school level. In particular in the context of the Italian high school where the study of the basic elements of linear algebra are part of the mathematics curriculum \cite{MIUR2010_211}.\\

A critical point in the process of educational reconstruction (see Section \ref{sec:ERTE}) is identifying a proper representation to describe the concept of qubit. The educational context proposed to implement this reconstruction is to analyse the results of experiments performed on quantum physical systems describing qubits through a mathematical representation of states and apparatus in terms of vectors and matrices.\\

From this analysis, it is possible to construct a theoretical framework that allows the quantum nature of physical systems to be characterised in comparison with classical ones, providing a key to interpret "complex quantum behaviours" \cite{Kim2017}.\\

The electron spin and polarisation states of individual photons are in fact equivalent to each other in the context of quantum information and quantum technologies \cite{Fox_QuantumIntro}. 
Using qubits is also particularly effective when dealing with purely technological aspects, such as the analysis of cryptographic protocols. (see. Section \ref{subsec:crypto}).
In fact, the principles behind quantum cryptography performed using single photons polarization states are exactly the same as those of performing Stern–Gerlach experiments with the magnet axes at different angles.
It is therefore clear how the abstract concept of qubits can be associated with real, measurable physical entities and how quantum gates can, for instance, be created using specific experimental apparatus \cite{DiVincenzo2000QCphysicalImplement}.\\

A characteristic of the approach (see Chapter \ref{ch:3-QApproach}) is to represent the way in which quantum objects interact and change in interaction with the experimental setup. This representation reflects the computational and algorithmic structure of qubits within a circuit or protocol in the context of quantum technologies.\\
The use of mathematical representations in terms of vectors and matrices is particularly effective. As the quantum state evolution in the interaction with an experimental apparatus can be described using the action of specific matrices on the state vectors, so the information associated with the qubit state changes in the interaction with quantum gates within a circuit. This is the case of quantum spin states interacting with a Stern-Gerlach apparatus (see Section \ref{sec:qubit_electron}) or a light polarization state interacting with polarizers and beam splitters in a Mach-Zehnder interferometer (see. Section \ref{sec:qubit_polarization}).\\

The study of the physical systems that represent qubits is shaped around the following research questions

\begin{table}[hbt!]
\centering
\begin{tabular}{cp{12cm}}
 & \textbf{Qubit respresentation} \\
 & \textit{Exploit the nature of quantum objects to encode information}\\
\hline
\rule[-4mm]{0mm}{1cm}
RQ5.1 & How is it possible to characterise the quantum nature of physical objects using qubits to encode information?\\
 & \\
\midrule
\rule[-4mm]{0mm}{1cm}
RQ5.2 &  Through which type of representation is it possible to describe the process by which the information associated with quantum objects can be transformed inside a physical apparatus and extracted through a measurement?\\
 & \\
\midrule
\rule[-4mm]{0mm}{1cm}
RQ5.3 & Which experimental contexts are significant for the study of how quantum objects can be used to encode information?
\end{tabular}
\caption{Research questions: qubit representation}
\label{tab:Qubit_RQ}
\end{table}


\section{Qubits as electrons' spin states}\label{sec:qubit_electron}
The main advantage of introducing the study of quantum systems through the \textit{spin first approach} \cite{mcintyre2016quantum}
\cite{SakuraiModern}, \cite{Sadaghiani2015} \cite{Corsiglia2020Spinfirst}, \cite{Sadaghiani2016} is that it favours the detachment from classical interpretative schemes \cite{Rodriguez2017_SGA} and the use of a simple mathematical formalism that can be firmly linked to the analysis of elementary experiments \cite{Muller2021penny}.\\ 

\begin{figure}[hbt!]
    \centering
     \includegraphics[width=\textwidth,height=\textheight,keepaspectratio]{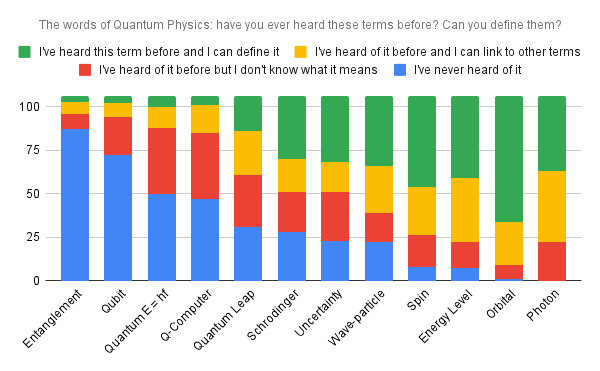}
    \caption{Results of the survey "the words of Quantum Physics": this survey have been designed in collaboration with a group of three teachers that participated in the Quantum Skill project in 2019. The survey was shared with several high school teachers from the Como school district.}
    \label{fig:Ch3_26_QWords}
\end{figure}

The introduction of spin is not conceptually neutral at secondary school level. Students have already been exposed to the idea of spin in the study of the atomic model in Chemistry and General Science courses. Spin is presented as a property of electrons associated with magnetic moment. It is therefore something that the electron \textit{possesses} as an "intrinsic characteristics" \cite{amaldi2020il} . 
This definition is part of the students' narrative repertoire, as emerged from a survey carried out involving physics teachers in the Como school district. A questionnaire was proposed to several schools in the area to be given to students before they started studying quantum physics. The aim was to get a picture of the familiarity students have with some quantum physics terms before any conceptual and formal introduction in a curricular course. Answers from 106 questionnaires were collected (see Figure\ref{fig:Ch3_26_QWords}and the analysis obtained shows that students have a fair knowledge of the vocabulary of quantum mechanics and are able to identify areas where the different terms are connected.\\

Specifically, of the 81 people who said they knew the physical meaning of spin, 71\% of cases (58 responses) wrote that

\begin{quote} {\fontfamily{lmtt}\selectfont
    spin is the direction of rotation of the electron
}\end{quote}

\begin{figure}[hbt!]
    \centering
     \includegraphics[width=\textwidth,keepaspectratio]{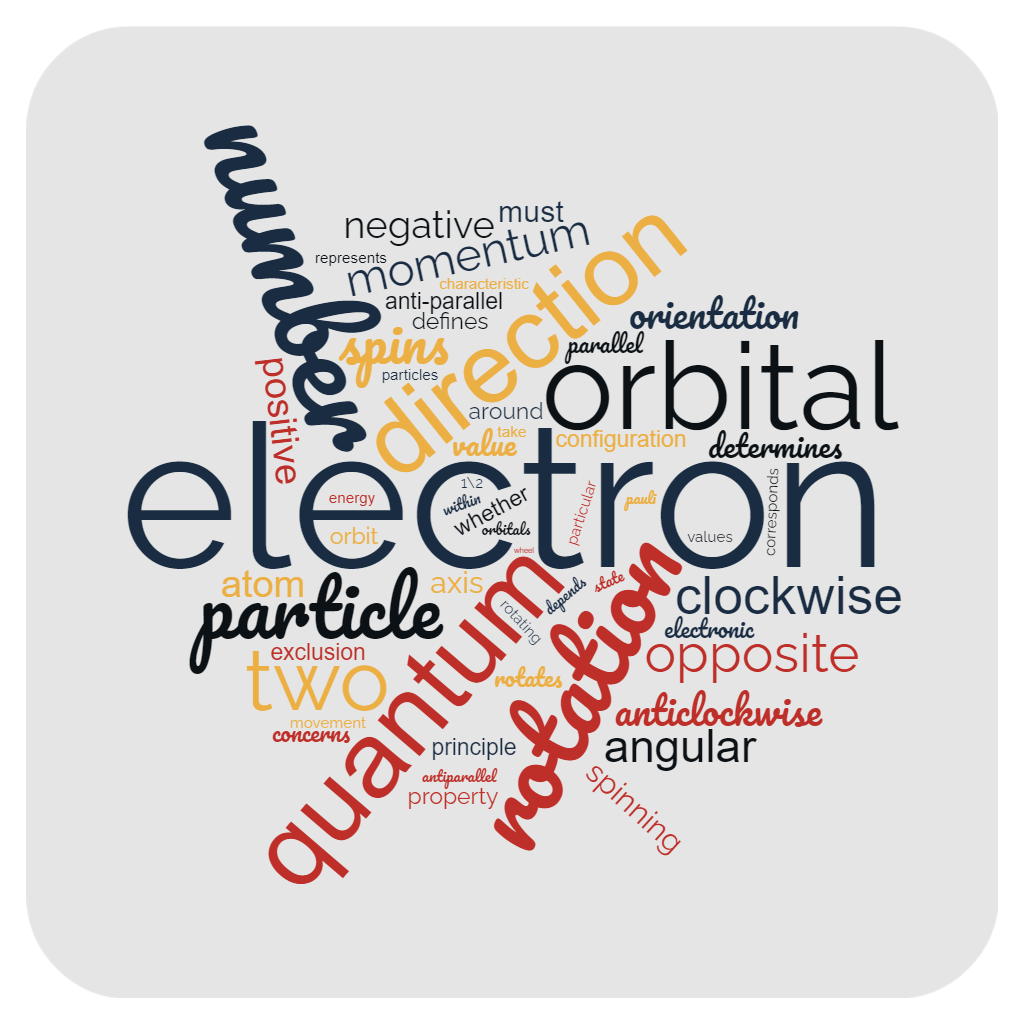}
    \caption{Definition of spin: word cloud built with the definition provided by students. Students related spin to electrons as something related to their rotation motion around the atom. It is also an atomic number that can take only two values. Created using https://www.wordclouds.com/}
    \label{fig:Ch3_27_spin}
\end{figure}

Although the textbooks emphasise that spin has nothing to do with the motion of the electron \cite{amaldi2020il,romeni2017fisica}, this representation is what permeates the students' description.
It is not always specified around what the rotation takes place. In only one case, rotation is ``around the axis of the electron''. In 10 cases out of 58 ``rotation of the electron around the atom'' is mentioned and in 6 cases out of 58 rotation ``within the orbital'' is mentioned.\\
In 6 cases out of 71, reference is made to Pauli's esclusion principle \cite{Kaplan2020_PauliPEP,Scerri1995_Pauli} and the fact that spin makes it possible to determine "how many electrons can fit in the orbital".
This is a "property" or "quantum number" which can only take on two values for 28\% of the total number of answers collected.

These results highlight the need to provide students with more rigorous and consistent ways of approaching the specifically quantum characteristics of classically introduced objects, such as electrons.
This idea has also been taken up in teacher-designed courses (see Chapter \ref{ch:6-LETeachers}) on the nature of microscopic objects.
In the next section (see Section \ref{sec:qubit_characterization}) the chosen way to introduce the concept of spin within an informational framework based on the concept of quantum state will be shown.\\

\subsection{Change the characterization of the physical object}\label{sec:qubit_characterization}
The aim of teaching quantum physics using spin states is to make the quantum nature of physical objects recognizable using the analysis of quantum experiments. To do this, the following definition is adopted:

\begin{quote}
    \textit{Spin is an internal degree of freedom of the electron and has no classical analogue}.
\end{quote}

This statement recalls Pauli's description of spin \cite{PauliSpin} in terms of 
\begin{quote}
    a classical non-describable two valuedness
\end{quote}
according to which a quantum-mechanical degree of freedom is claimed to exist without a corresponding classical one \cite{Giulini2008}.
\begin{quote}
  the cautions expression "classically non-describable two-valuedness" experienced a certain verification during later developments, since Bohr was able to show on the basis of wave mechanics that the electron spin cannot be measured by classically describable experiments (as, for instance, deflection of molecular beams in external electromagnetic fields) and must therefore be considered as an essential quantum-mechanical property of the electron.(\cite{PauliSpin}p.4)
\end{quote}

The justification for the use of the previous definition comes from showing how the results of experiments related to the measurement of spin are totally incompatible with predictions made using classical paradigms.\\

The representation of the spin degree of freedom is therefore made using a classical image, that of a compass oriented in a magnetic field. This is not to highlight the link between magnetic moment and intrinsic angular momentum \cite{mcintyre2016quantum}, \cite{benenti2019principles} (the electron possesses a magnetic moment antiparallel to the angular momentum) and so to avoid the visualisation of spin as something related to a rotation (see Figure \ref{fig:Ch3_27_spin}).\\

Pauli itself emphasised the inadequacy of the classical models for interpreting the behaviour of the electron, referring in particular to specific inconsistencies with special relativity \cite{Giulini2008}.

\begin{quote}
    Emphasising the kinematical aspects one also speaks of the ‘rotating electron’ (English ‘spin-electron’). However, we do not regard the conception of a rotating material structure to be essential [\dots] \cite{pauli1964}
\end{quote}
In the introduction to the concept of spin we have therefore chosen to follow the advice attributed to Max Born:

\begin{quote}
 “the idea of the rotating electron is not be taken literally” (\cite{born1989atomic}, p. 188)    
\end{quote}

The Stern Gerlach experiment is normally conducted silver atoms \cite{Gerlach1922} that are electrically neutral particle. This is so to avoid the large deflection of charged objects while moving through a magnetic field. Moreover, for atoms such as silver, the electronic configuration is such that the magnetic moment is equivalent to the spin of the outer electron \cite{benenti2019principles}.\\
The dynamics of silver atoms inside the Stern-Gerlach apparatus is then introduced with a parallelism between the behaviour of a magnetic dipole in a non-uniform magnetic field and the behaviour of an electric dipole in a non-uniform electric field.

The study of electric dipoles is part of the students' general experience of electromagnetic phenomena.
This additional element of familiarity facilitates the reconstruction process because it makes the link with students' prior knowledge explicit \cite{Duit2012_MER,Bouchee21, Dijk2008_ERTEevolution}.\\

The Stern Gerlach experiment allows measuring the information connected to the electron spin by establishing a biunivocal correspondence between the deflection of the beam of silver atoms and the value of the magnetic moment \cite{Singh2009_SGA}.\\

The interpretation of the experimental results is done by focusing on those aspects that show the hiatus between classical and quantum behaviour.\\
In order to bring out these aspects, different sequences of Stern Gerlach apparatuses obtained using Quvis simulations were analysed \cite{Kohnle15_QuVIS_2states}. The possibility to change measurement conditions and easily compare data in different situations is one of the main advantages of using these simulations, especially in helping students to maintain the link between concepts and physical reality \cite{Bouchee21}.\\
The methodology involves the use of simulations to verify the predictions made on the results of the experiments. For the detailed description of the activities and the materials used please refer to Appendix \ref{appx:SGA}.\\

\begin{figure}[hbt!]
    \centering
     \includegraphics[width=\textwidth,keepaspectratio]{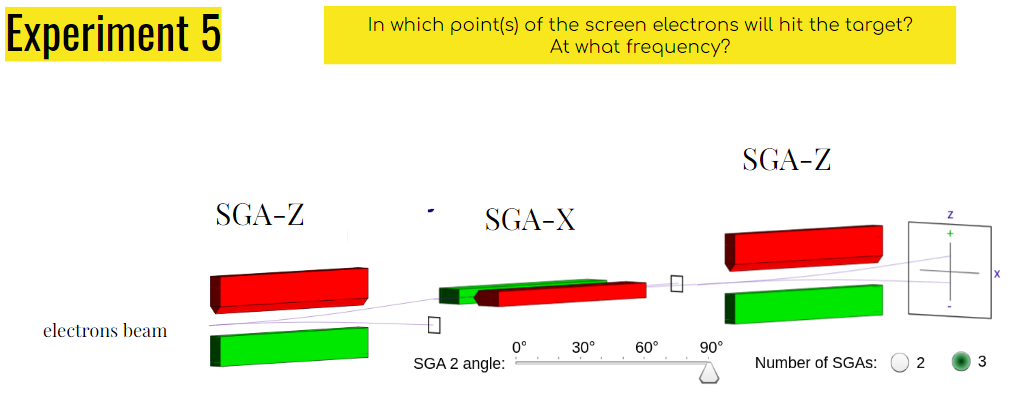}
    \caption{Analysis of simulations using sequences of Stern-Gerlach experiments: we used simulations to verify the predictions to enhance reflection about the difference between classical and quantum behaviours. Figure adapted from \cite{QuVisSim}}
    \label{fig:Ch3_28_SGAexp}
\end{figure}

\subsubsection{Simulations with sequences of Stern Gerlach apparatus}\label{subsec:SGAsims}
Guided reflection sheds light on important aspects of the nature of quantum states and their associated physical systems.\\
Firstly, there is no possibility that the silver atom can continue without deflection. This deflection is related to the direction of the magnetic field regardless of the choice of the spatial orientation of the magnets. The quantization of spin is expressed by the two distinct output states, two possibilities that are classically alternative and incompatible. This applies whatever direction one chooses for the magnetic field. If the field is directed along the z-axis, the two possible states will be $|\uparrow \rangle$ and $|\downarrow \rangle$.  If the field is directed along the x-axis, the states will be $|+ \rangle$ and $|- \rangle$. \\


The existence of quantized physical quantities is not in itself new in the classical world. What is relevant, however, is that the atoms in the furnace have no knowledge of the particular direction in which the field they are about to enter has been directed. Somehow the property that can be associated with the physical object does not pre-exist the measurement. This aspect is relevant when analysing the evolution of the qubit: it is possible to know and determine the shape of the state according to the specific sequence of instruments/operators with which it interacts, but not to predict in a deterministic way the value assumed by the property connected to the state. Since the atoms emerging from the furnace are identical, there must be a relationship between the two different ways of representing the same qubit, one using $|\uparrow \rangle$ and $|\downarrow \rangle$ and the other using $|+ \rangle$ and $|- \rangle$. \\

Using sequences of two SGAs it is therefore possible to highlight the relationshipp between the representations of the same qubit using different basis. The classical prediction for an electron prepared in the spin state $|+ \rangle$ along the $X$-direction is that it will not be deflected once passing through an apparatus with a field directed in the $Z$-direction. This is because the field is orthogonal to the intrinsic magnetic moment. However, this is not compatible with the experiments. When the spin measurement along $Z$-direction is performed, two possible distinct values are still found. This result, also extended to the case of the single atom entering the apparatus, allows us to highlight how the state $|+ \rangle$ must be expressed as a superposition state of the states $|\uparrow \rangle$ and $|\downarrow \rangle$.

\begin{equation}
    | + \rangle \; = \; \frac{1}{\sqrt{2}} \, | \uparrow \rangle \, + \, \frac{1}{\sqrt{2}} \, |\downarrow \rangle 
\end{equation}

This result shows how a total determination of the input state (e.g. prepared in a pure state $|\uparrow \rangle $) corresponds to a total indetermination of the output state.  In other words, the experiment shows how possessing all the information about an object cannot predict all its possible characteristics. This result is linked to the way in which observables are represented by operators that do not commute (see Section \ref{sec:HeisenCommut}).

\begin{figure}[hbt!]
    \centering
     \includegraphics[width=\textwidth,height=\textheight,keepaspectratio]{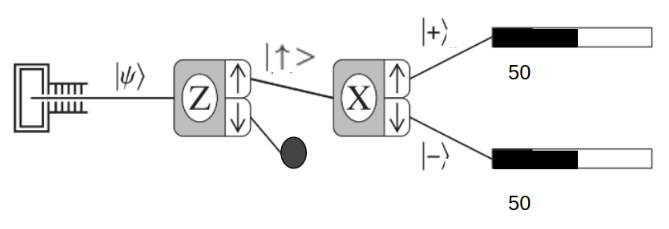}
    \caption{Interpretation of experimental outcomes: the information we have on the spin input state entering the Stern Gerlach apparatus oriented in $x$-direction is complete, hover the output value is totally undetermined. Figure adapted from \cite{mcintyre2016quantum}}
    \label{fig:Ch3_29_SGAMixed}
\end{figure}

Another possible interpretation of this result could be made using the concept of a mixture of states, i.e. the electrons in the sequence have different and equally distributed spin values. That kind of model is related to the idea of physical properties as part of the object and so it is a useful context to inquiry the quantum nature of the qubit \cite{mcintyre2016quantum} \cite{benenti2019principles} and useful to bringing out the classical interpretative patterns \cite{Bouchee21}.\\
To test this interpretative hypothesis, the experiment can be performed with three Stern Gerlach apparatus. At the output of the third apparatus the spin state is such that the measurements are incompatible with the idea of a mixture. This evidence shows instead the validity of the representation of the states in terms of superposition.\\

\begin{figure}[hbt!]
    \centering
     \includegraphics[width=\textwidth,height=\textheight,keepaspectratio]{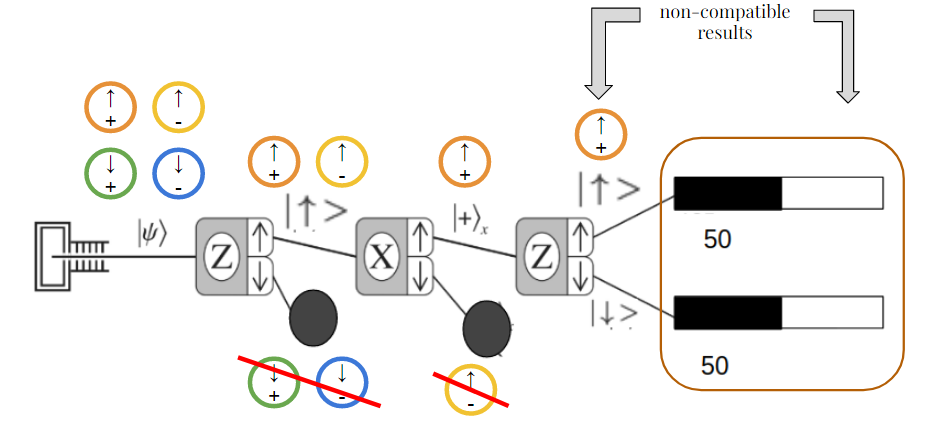}
    \caption{Interpretation of experimental outcomes: mixed quantum states cannot be used to reproduce the measurement outcomes of three apparatus sequence. Figure adapted from \cite{mcintyre2016quantum}}
    \label{fig:Ch3_29_SGAMixed}
\end{figure}

It is evident that the experimental result that attributes a certain property to the object (e.g. spin-up or spin-down) is linked to the fact that the state can be expressed in a basis that depends on the particular choice of the measuring apparatus (in this case, the field orientation of the Stern-Gerlach apparatus).\\

The way in which state and observables are represented is therefore fundamental to provide useful tools to analyse the experimental results. This representation should also help how to describe quantum states and their relationship with measurement outcomes.

\subsection{Change the representation of electrons' spin}\label{sec:spin_repres}

The characterisation of the object, i.e. the description of its characteristics that determine its nature, passes through the necessity of describing using the concept of quantum state. Therefore it is necessary to develop a language suitable for representing those states. Such language can be abstract as long as it can be manipulated through tools that can be made accessible to the students.\\

The choice is to represent the spin states by two dimensional vectors and the quantities that are measured (observables) using matrices that operate on these vectors by transforming them \cite{Dur2013qubitlearn, nielsen_chuang_2010}. The matrix that is associated with each observable thus contains all the information about the a priori possibilities that a measurement of that observable will give certain results.\\

\begin{figure}[hbt!]
    \centering
     \includegraphics[width=\textwidth,height=\textheight,keepaspectratio]{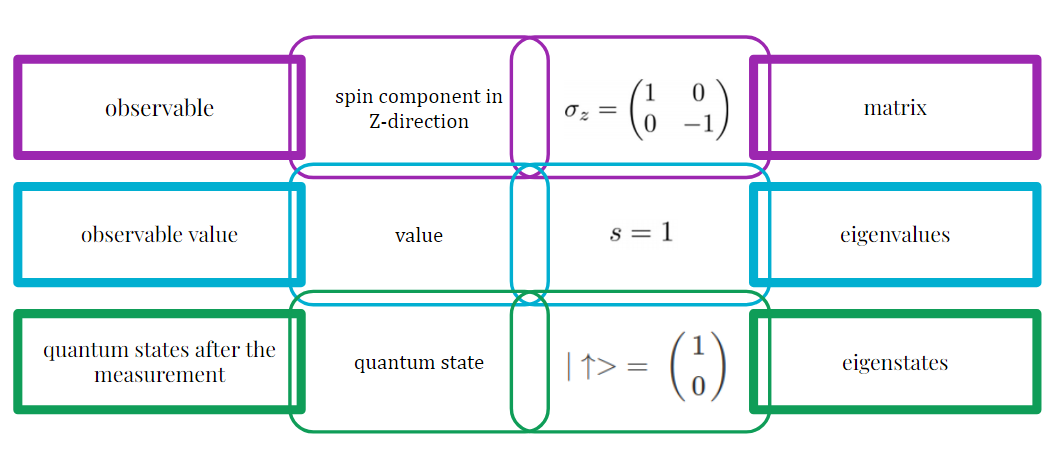}
    \caption{Quantum measurement elements: matrix represent observable and can be used to relate states to the corresponding values.}
    \label{fig:Ch3_29b_Qmeasure}
\end{figure}

Given the mathematical nature of matrices and the relationship between eigenvalues and eigenvectors, 
the eigenvalues of the matrix can be interpreted as the possible (quantized) values of the observable, and determine those particular coefficients that define the corresponding quantum states. Associating a certain observable (e.g., the spin component in the $Z$-direction) with a specific matrix (e.g., $\sigma_z$) allows encoding in a single mathematical entity both the rules of quantization, i.e. how the physical observable is quantized (the eigenvalues), and which states correspond to the possible values of the observable (the eigenstates). In Figure \ref{fig:Ch3_29b_Qmeasure} those relationship are presented. Note that the possible values of spin are 

\begin{equation}
    s = + \frac{\hbar}{2} \qquad s = - \frac{\hbar}{2}
\end{equation}

and the expression of the matrix is 

\begin{equation}
    \sigma_z \; = \; \frac{\hbar}{2}\,\begin{pmatrix}
    1 & 0\\
    0 & -1
    \end{pmatrix}
\end{equation}

The coefficient $\frac{\hbar}{2}$ has been neglected for simplicity.

This is related to the action of the apparatus on the state and to the process of measurement of the observable in terms of the action of a unitary operator \cite{nielsen_chuang_2010} (see. Section \ref{sec:QAxioms}).\\

This way of describing the measurement process and the fact that it is intrinsically probabilistic also leads to the problem of how two states can be distinguished \cite{SusskindQT}.
In fact, it is possible to unambiguously distinguish between a prepared $|\uparrow \rangle$ and $|\downarrow \rangle$ state by a measurement performed with a $Z$-oriented apparatus. On the contrary, the two states $|\uparrow \rangle$ and $|+\rangle$ are not \textit{unambiguously distinguishable} since a measurement performed along $Z$-direction of a prepared $|+\rangle$ state has anyway a probability of 50\% to obtain as result the eigenvalue relative to the $|-\rangle$ state.\\
That is a key aspect about the nature of quantum objects that can emerge also from the analysis of cryptographic protocols \cite{QUVIS_Kohnle2017_Qkeydistrib} (see also Chapter \ref{ch:7-LEStudents}).\\


To increase the role of the interplay between physical phenomenon and mathematical formalism as a powerful thinking tool to analyse quantum objects, the consistency of the mathematical structure with what is observed is shown. This procedure will also be implement in the case of the polarisation of light (see Section \ref{eq:PHbasic}). These exercises have been proposed during the Professional Development Programme for teachers (see Chapter \ref{ch:6-LETeachers}) as examples of didactic tools to support students in the acquisition of a useful language to describe the quantum nature of physical objects.\\

On the one hand, it is possible to \textit{verify} that if the two spin states in the $Z$-direction are the two vectors $|\uparrow \rangle$ and $|\downarrow \rangle$ then the two corresponding eigenvalues of the matrix $\sigma_z$ are precisely + 1 and - 1

\begin{figure}[hbt!]
    \centering
    \includegraphics[scale=0.5]{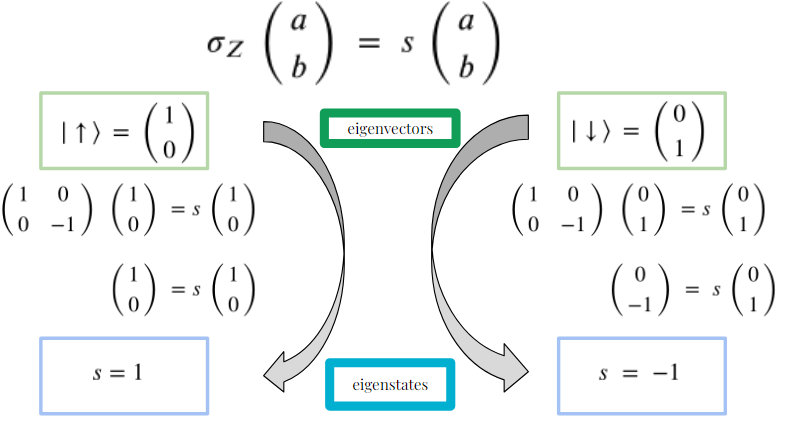}
    \caption{Quantum measurement elements: showing the relationship between eigenvectors and eigenstates}
    \label{fig:Ch3_29a_QmeasElements}
\end{figure}

\begin{figure}[hbt!]
    \centering
     \includegraphics[width=\textwidth,height=\textheight,keepaspectratio]{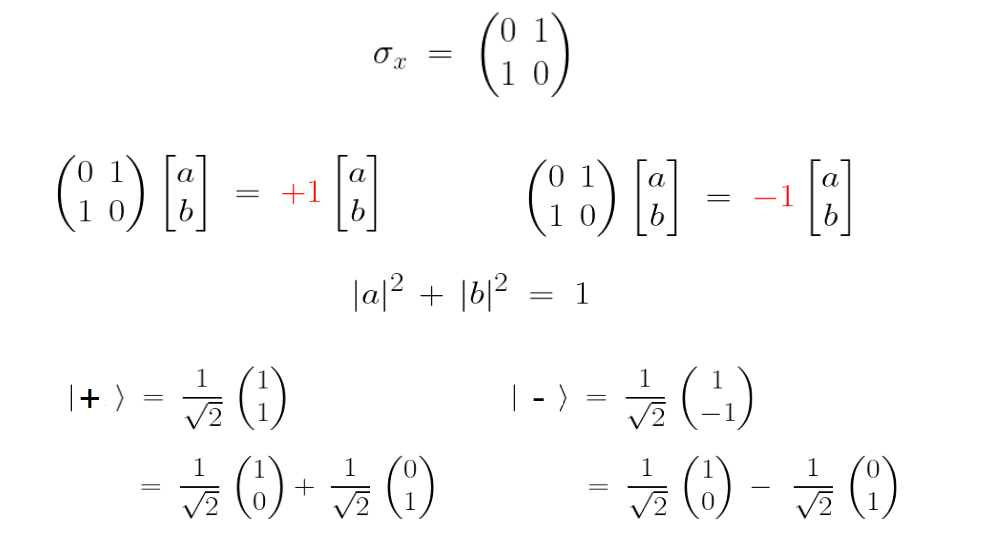}
    \caption{Quantum measurement elements: showing the relationship between eigenvectors and eigenstates for spin in the $x$-direction}
    \label{fig:Ch3_29c_SigmaX}
\end{figure}

The same procedure is applied to the study of the spin along the X-direction. Since the values that the spin are quantized and are the same as those obtained for the $Z$ component (i.e. $s= 1$ and $s= -1$), it can be verified that the spin state $|+ \rangle$ is actually written as a linear combination (with equal weights) of the states $| \uparrow \rangle$ and $| \downarrow \rangle $ (see Figure \ref{fig:Ch3_29c_SigmaX})

This result can be easily generalised to a generic angle $\theta$ with respect to the direction $Z$. The corresponding matrix is

\begin{equation}
    \sigma_\theta \; = \; \cos \theta \sigma_z \, + \sin \theta \sigma_x \,=\, \begin{pmatrix}
    \cos \theta & \sin \theta\\
    -\sin \theta & \cos \theta
    \end{pmatrix}
\end{equation}

and it can be shown that the eigenvectors of $\sigma_z$ corresponding to the eigenvalues $s= 1$ and $s=-1$ are

\begin{equation}
    |\theta\rangle_{s=+1} \; = \; \cos\frac{\theta}{2} \begin{pmatrix} 1 \\ 0 \end{pmatrix} + \sin\frac{\theta}{2} \begin{pmatrix} 1 \\ 0 \end{pmatrix}
\end{equation}
\begin{equation}
    |\theta\rangle_{s=-1} \; = \; \sin\frac{\theta}{2} \begin{pmatrix} 1 \\ 0 \end{pmatrix} - \cos\frac{\theta}{2} \begin{pmatrix} 1 \\ 0 \end{pmatrix}
\end{equation}

\begin{figure}[hbt!]
    \centering
     \includegraphics[width=\textwidth,height=\textheight,keepaspectratio]{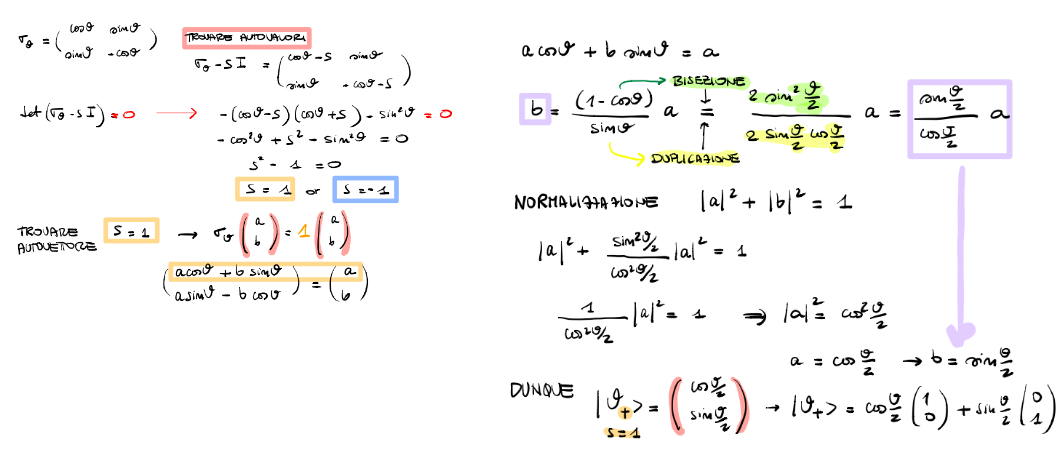}
    \caption{Eigenstates and eigenvectors at a generic angle $\theta$: copy of an exercises carried out by a teacher during the Professional Develoment programme}
    \label{fig:Ch3_29g_SigmathetaEX}
\end{figure}

This demonstration was proposed as an exercise (see Figure \ref{fig:Ch3_29g_SigmathetaEX}) during the professional development programme for teachers.

Through an appropriate reconstruction by the teachers, the expression of the states prepared along $\theta$ has been proposed also in the activities proposed in class (see Section \ref{sec:massimo}) as a tool to study the EPR problem and to introduce quantum entanglement.\\

Using this mathematical representation with matrices and vectors, the general procedure to perform a measurement of a certain observable is defined. If the spin is in fact described by a vector
    $\begin{pmatrix}
    a\\b
\end{pmatrix}$
to perform a measurement of the spin component in a certain direction $\theta$ it is necessary to write the state on the basis of the eigenvectors of the matrix $\sigma_{\theta}$. The square modulus of the coefficients represent the probability that the measurement will result in the corresponding eigenvalue.
After performing the measurement, the state will be represented by the eigenstate that corresponds to the measurement result, which is therefore to be considered certain.

\subsubsection{Heisenberg's uncertainty inequality}\label{sec:HeisenCommut}
Another element that can be derived from the study of states in Stern Gerlach sequences concerns the relation between the observables $\sigma_z$ and $\sigma_x$ \cite{mcintyre2016quantum} and can be use to introduce Heisenberg uncertainty \cite{heisenberg1927a}. 
What we have found about the eigenvectors of ${sigma_z}$ and ${\sigma_x}$ shows that these constitute complete bases of the space, since any state of the quantum object can be represented completely in each of the two bases.
This is particularly significant because in the description of the transformation of the state of the same quantum object the two bases refer to two different observables.\\

The use of this representation makes it possible to interpret the concept of quantum uncertainty without relating it to measurement accuracy \cite{Brumfiel2012_Heisen,Hadzidaki2008_HeisnMicr, Velentzas2011_HeisenMicrosc} or the wave-like nature of the object \cite{Boe2021_HeinsenWave}.
Uncertainty is instead presented in relation to the non commutativity of the operators \cite{Kennard1927Hein, Robertson1929_Hein} representing observables, through a relation that can be verified by operating with matrices.

\begin{figure}[hbt!]
    \centering
     \includegraphics[width=\textwidth,height=\textheight,keepaspectratio]{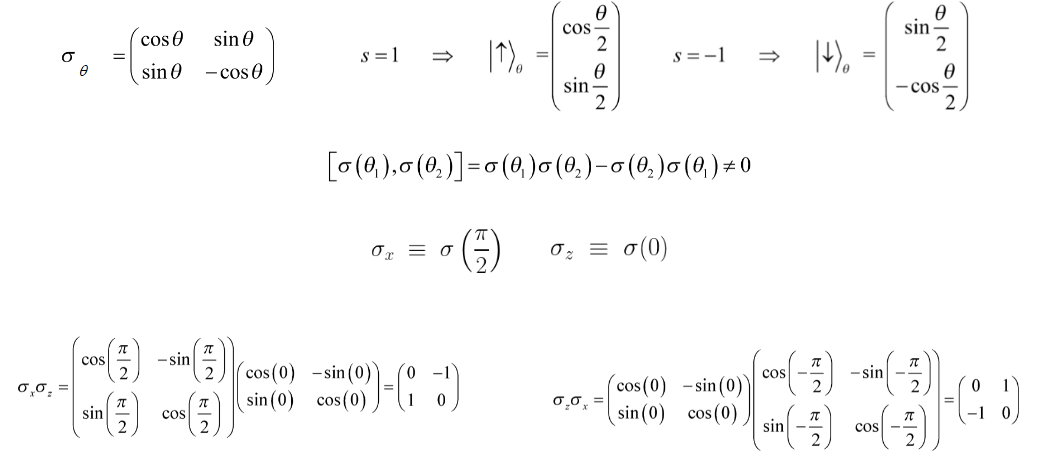}
    \caption{Heisenberg' uncertainty related to the state preparation and measurement related to non commutative operators. The operators $\sigma_z$ and $\sigma_x$ do not commute; this property of the operators associated to two different observables can be use to interpret the fact that a single measurement cannot extract a complete information about all the physical properties of that object}
    \label{fig:Ch3_29f_NonCommMatrix}
\end{figure}

Uncertainty is not presented as an abstract principle, but is derived as an intrinsic feature of quantum objects that can be revealed by the analysis of experiments (see Section \ref{subsec:SGAsims}) and interpreted using a mathematical representation. If we prepare a pure state (e.g. $|\uparrow \rangle$ coming out of a Stern Gerlach oriented in the direction $Z$), that state has therefore a well defined value for the observable represented by $\sigma_z$: the eigenvalue associated to the state $|\uparrow \rangle$ is surely +1. To such a state, which represents one specific physical object, experiments show it is not possible to associate at the same time the value of the observable represented by $\sigma_x$. In mathematical terms, the two operators $\sigma_z$ and $\sigma_x$ do not commute.
 
Non-commutativity expresses the fact that the two operators $\sigma_z$ and $\sigma_x$ do not have the same system of eigenvectors and eigenvalues: changing the order of the two operators makes the outcomes of the measurement different. In the example above, considering a $Z$-$X$ oriented apparatus (see Section \ref{sec:seqSGA_matrix}), the state resulting at the output of the first element of the sequence is not an eigenstate of the second operator and will therefore be modified by the second operator into a different eigenstate. Exchanging the order X-Z generate a different state.\\ 
 
Referring to what is expressed in terms of axioms (see Chapter \ref{ch:3-QApproach}), the average value of an operator $\hat{A}$ on the state $|\psi\rangle$ is given in Dirac notation by

\begin{equation}
    \langle A \rangle \;  = \; \langle \psi | \hat{A} | \psi \rangle
\end{equation}
 
and the standard deviation is
\begin{equation}
\Delta \hat{A} \, = \, \sqrt{\langle \hat{A^2} \rangle - \langle \hat{A} \rangle^2 }    
\end{equation}

In general, if two physical quantities $A$ and $B$ are represented by operators that do not commute $[\hat{A}$,$\hat{B}] \neq 0$, there exists a relation about the minimum uncertainty with which they can be measured
In fact, if a system is prepared to that the uncertainty with respect to the first observable is given by $\Delta A$ then the uncertainty on the second observable is given by 

\begin{equation}
    \Delta A \, \Delta B \; \geq \; \frac{1}{2} \left | \left \langle \left [ \hat{A},\hat{B} \right ] \right \rangle \right |
\end{equation}

where $\left \langle \left [ \hat{A}, \hat{B} \right ] \right \rangle$ is the expectation value of the commutator $[\hat{A} , \hat{B}]$.

In our example, having prepared the state in one of the eigenstates of $\sigma_z$ it means that the standard deviation $\Delta \sigma_z$ is zero and hence $\Delta \sigma_x$ represents the maximum possible uncertainty about the state, so we have 50$\%$ probability of getting eigenvalue + 1 and 50$\%$ probability of getting eigenvalue - 1 by measuring the spin along the $x$-direction.

\subsubsection{Analysis of Stern Gerlach sequences}\label{sec:seqSGA_matrix}
The description of experiments on quantum systems carried out using sequences of Stern Gerlach apparatus is presented by associating information about the system with a quantum state and the action of the apparatuses on the physical system with the action of the operators on the states. The measurement process through which information is extracted from the state is represented by the calculation of the probabilities related to the coefficients of the specific eigenstates in which the state is represented.\\
In this context, it is important to take into account that the state associated with a single object should be written on a different basis depending on the type of apparatus that constitutes the sequence. The relationship between state and measurement also emerges by introducing different sequences in which one chooses to perform a measurement or not between one apparatus and the next one.

\begin{figure}[hbt!]
    \centering
     \includegraphics[width=\textwidth,height=\textheight,keepaspectratio]{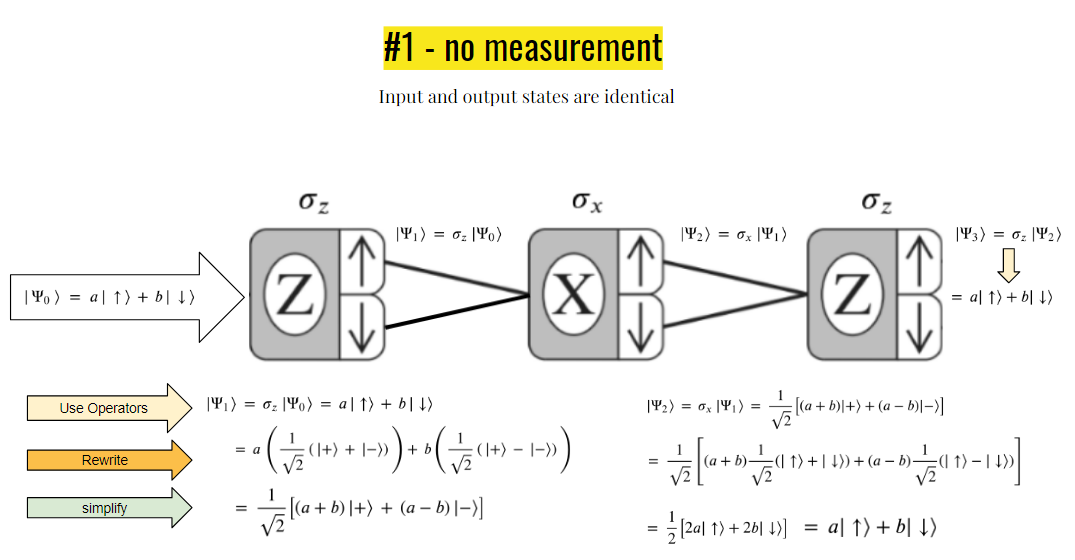}
    \caption{sequence of Stern Gerlach apparatus: no measurement is perfomed, no information is estracted. The output state is identical to the input state}
    \label{fig:Ch3_30_SGAmathseq1}
\end{figure}

As shown in Figure \ref{fig:Ch3_30_SGAmathseq1}, the operator $\sigma_z$ acts on the input state which is not modified since it is prepared as a superposition of the eigenstates $|\uparrow \rangle$ and $|\downarrow \rangle$ of $\sigma_z$. In order to understand how the same state is transformed by the next operator $\sigma_x$ it must first be rewritten on the basis $|+ \rangle$ and $|- \rangle$ of the operator $\sigma_x$.

The focus here is on the abstract concept of state as information carrier. To know the properties of the object, a measurement must be performed and so the state needs to be modified.\\

\begin{figure}[hbt!]
    \centering
     \includegraphics[width=\textwidth,height=\textheight,keepaspectratio]{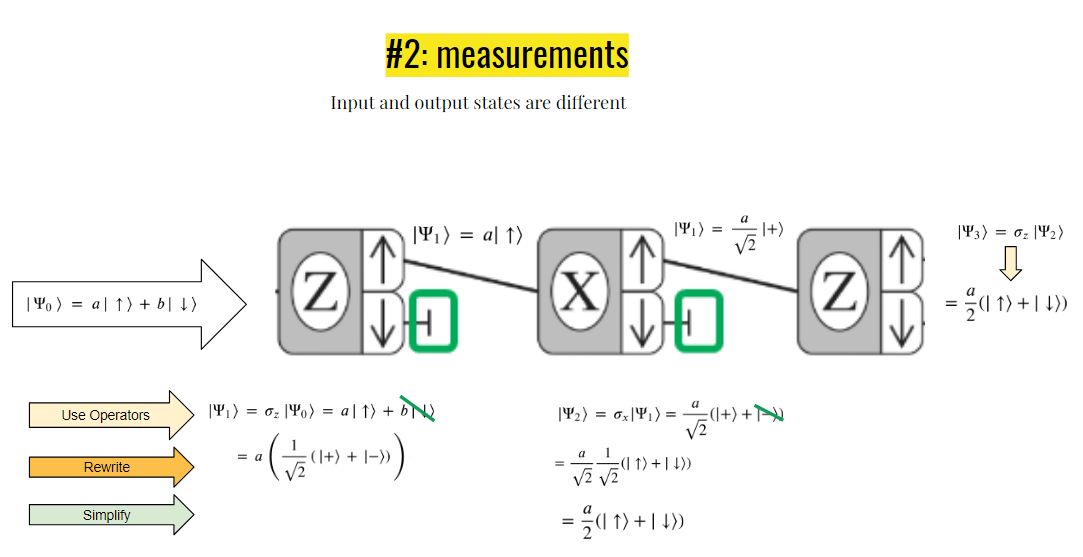}
    \caption{Sequence of Stern Gerlach: a measurement is performed at each step. That act of measurement modifies the state.}
    \label{fig:Ch3_31_SGAmathseq2}
\end{figure}

In the second example (see Figure \ref{fig:Ch3_31_SGAmathseq2}), measurements are performed after each step.
This is the situation presented in the simulator (see Section\ref{subsec:SGAsims}) and is equivalent to ponder the question: "what is the path the object follows within the sequence of apparatus?". That this is equivalent to the problem of figuring out what is the path of the single photon inside a Mach-Zehder interferometer (see Section\ref{subsec:MZIexper}). This is a central question in the process of reconstructing the nature of quantum objects \cite{Maries2020_WhichPath}, and in this context the answer is constructed through a calculation showing that the output state is a superposition of $|\uparrow\rangle$ and $|\downarrow \rangle$ different from that at the input.

In this context is therefore important to bring out that the measurement modifies the state. The state preparation is therefore different from the previous case (see Figure \ref{fig:Ch3_30_SGAmathseq1}) and so will be the output state.\\
This result, counter-intuitive but in perfect agreement with the simulations, is an illustration of a fundamental property of quantum mechanics and of the properties of quantum states. It can be interpreted by considering the way in which the state has been rewritten on the basis of the various operators. Having had to rewrite the output state from the first device using the second operator's eigenstate basis, it is as if the second device had "destroyed any information" \cite{benenti2019principles} about the value that was measured as output of the first apparatus oriented in the $Z$-direction. This is because the two basis sets needed to describe the same physical system are orthogonal to each other and therefore refer to states which, although representing the same physical system, cannot be known at the same time.

In the third example \begin{figure}[hbt!]
    \centering
     \includegraphics[width=\textwidth,height=\textheight,keepaspectratio]{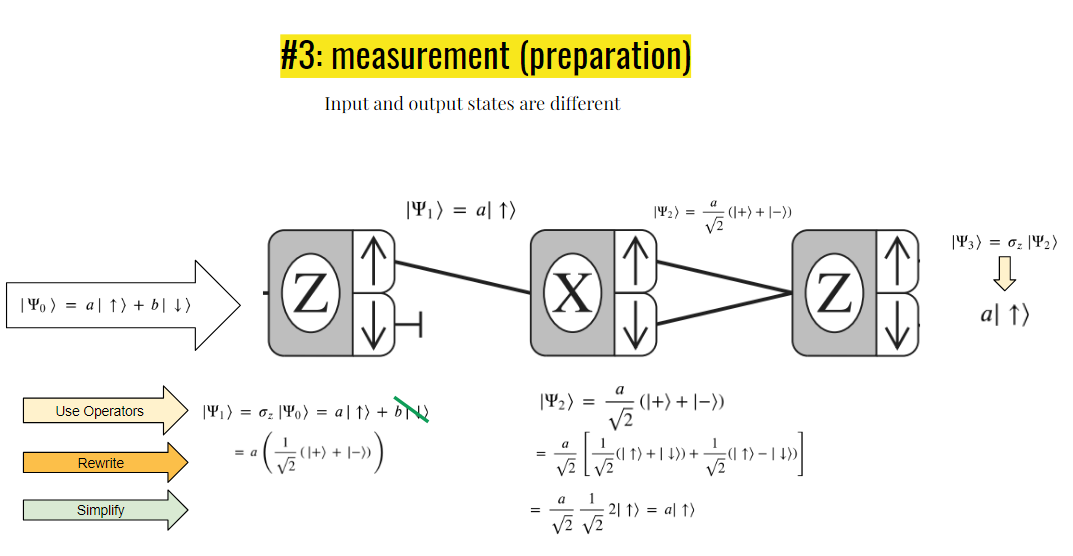}
    \caption{sequence mixed}
    \label{fig:Ch3_32_SGAmathseq3}
\end{figure}
(see fig \ref{fig:Ch3_32_SGAmathseq3}) an intermediate situation is considered, in which the input state in the second apparatus is prepared on the specific $|\uparrow \rangle$ state but no measurement is performed in the $X$-direction.
In this case the action of the second apparatus does not change the output state of the sequence. Nothing has happened between the first and the last apparatus in the sense that no measurement has been performed and therefore the state has not been modified, but only rewritten.\\

\section{Qubits as single photon polarization states}\label{sec:qubit_polarization}

All basic physics phenomena are quantum, and so is light. Therefore, exploring the quantum nature of light is a fruitful way to have access to the core concepts of quantum physics (see Section \ref{Ch3.1_ApproachShape}). The central idea is to use mathematical concepts (such as complex numbers, vectors and matrices) as a powerful tool to help teachers and students interpret experimental evidence and derive explanations of physical phenomena in the quantum theoretical framework. Our department has also a long tradition in teaching quantum optics to high school physics students in different learning context \cite{Bondani2014_SinglePhoton}. Therefore we decided to include single-photon polarization states as one the possible representations of a qubit.

\subsection{Classical and quantum light}\label{sec:ClQtlight}

In high school physics courses, the word \textit{light} is used to describe a wide range of physical phenomena \cite{MIUR2010_211}. 
The value is to show in which context the quantum nature of light can emerge and so how to help students understand the difference between classical and quantum interpretation of light.

In the first two years of italian \textit{scientific oriented} high schools (Liceo scientifico) \cite{MIUR2010_211} \footnote{The same structure is also common to other italian school educational pathways which, however, differ in terms of timing and levels of detail. The scientific high school route is generally the one that allows for more in-depth study of concepts and content, even using a more complex mathematical formalism.}, the study of light is focused on geometric optics and phenomena related to the interaction of light with mirrors and lenses. In the ''corpuscular- newtonian" theoretical framework, lights moves from a source along rays following geometrically describable trajectories.

\begin{figure}[hbt!]
    \centering
     \includegraphics[width=\textwidth,height=\textheight,keepaspectratio]{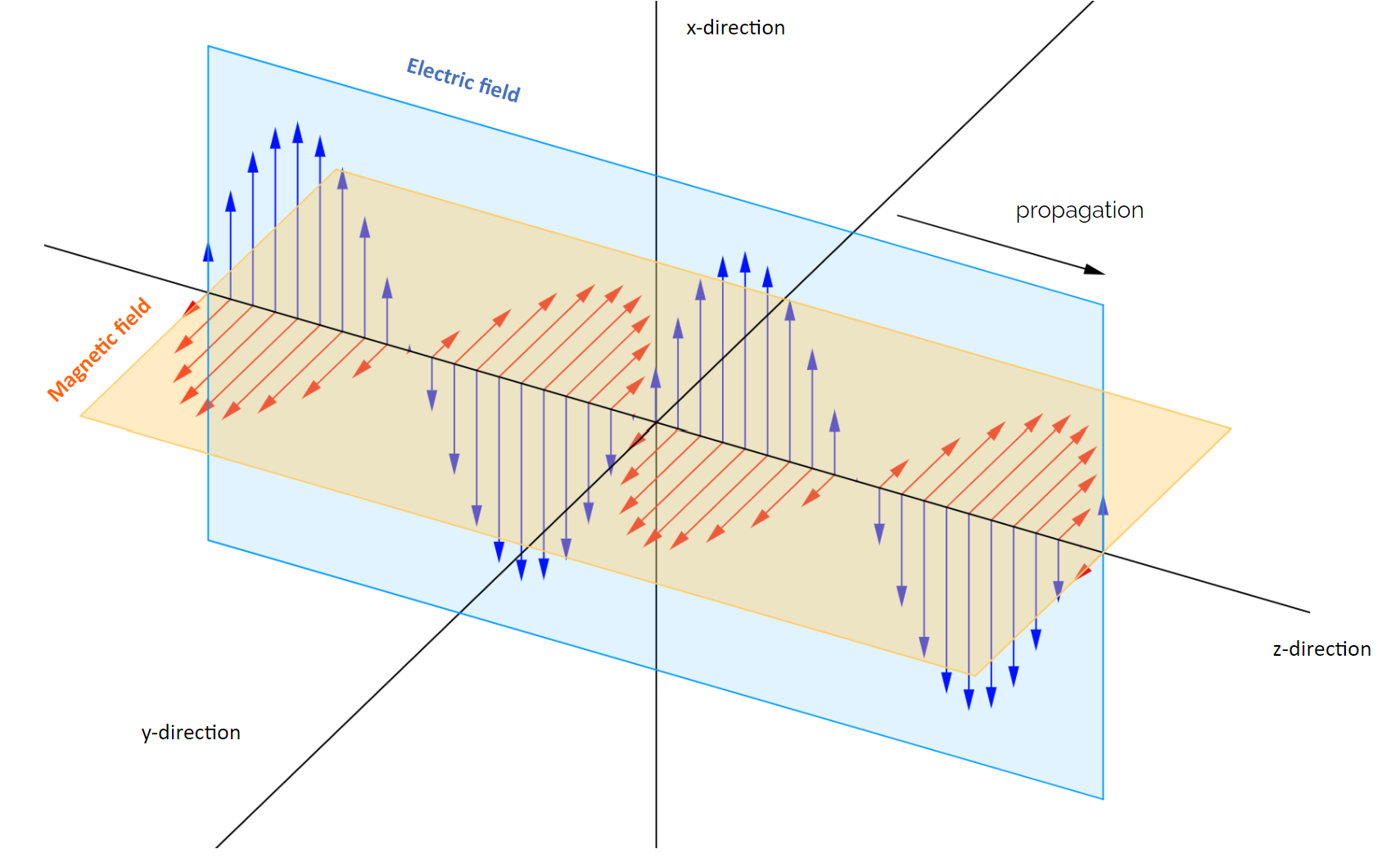}
    \caption{A representation of a linearly polarized sinusoidal electromagnetic wave, propagating in the direction +z in vacuum. Adapet from Tom Walsh - Geogebra Simulation https://www.geogebra.org/m/xhYwXSsH}
    \label{fig:Ch3_3_EMwave}
\end{figure}

In the following two years, the wave nature of light is analyzed, studying diffraction and interference. To do so, students use what they have learned about mechanical waves (i.e. sound waves) in terms of  mathematical description using concepts like wavelenght, frequency and amplitude. During the final year of study, light is presented within Maxwell's theory of electromagnetic waves based on the concept of electromagnetic field. Students became familiar with the idea that from the Maxwell's equation a wave equation for the electric and magnetic field can be derived. The idea is that fields can propagate in space and time like mechanical waves (fig. \ref{fig:Ch3_3_EMwave}).\\

Knowing the strong relation between electric and magnetic field and considering a frequency domain where the magnetic effects are negligible, it is possible to omit to describe the magnetic field part of the solution. Using complex notation, the following equations provide a description of a monochromatic electromagnetic wave that moves in the $z$ direction in complex notation.

\begin{eqnarray} \label{eq:Efield}
E(z,t) \, &=& \,E_0 \cos(kz-\omega t+\delta)\\ &=&  E_0 \frac{e^{i(kz-\omega t+\delta)} + e^{-i(kz-\omega t+\delta)}}{2}\\ &=& \frac{E_0}{2} e^{i(kz-\omega t + \delta)} \,+\, c.c.
\end{eqnarray} 

The same field that propagates can be described using a vectorial representation

\begin{equation}\label{eq:Evector}
\begin{split}
\vec{E}(z,t) & =\,\hat{x}\;E_x(z,t) + \hat{y}\;E_y(z,t)\\
&  = \hat{x}\;E_{0x}\, e^{i(kz-\omega t + \phi_x)} + \hat{y}\;E_{0y}\,e^{i(kz-\omega t + \phi_y)} \\
& = \begin{bmatrix}
E_{0x}e^{i\phi_x} \\
E_{0y}e^{i\phi_y} 
\end{bmatrix}
\, e^{i(kz-\omega t)}  \\
& = \begin{bmatrix}
\widetilde{E}_{0x}\\
\widetilde{E}_{0y}
\end{bmatrix}
e^{i(kz-\omega t)} 
\end{split}
\end{equation}

Light is an electomagentic field that propagates in space time. The previous equation show the two field amplitudes ($\tilde{E}_{0x}$ and $\tilde{E}_{0y}$) can be used to describe the propagate of that field. So light can be represented using as a two-dimensional state vector $\begin{bmatrix}
\widetilde{E}_{0x}\\
\widetilde{E}_{0y}
\end{bmatrix}$
that propagate in space and time. 

This representation create an strong link between the mathematical representation of light (state vector) and the result of the measurement of light that is light intensity
The approach we propose is focused on the link between knowledge of the characteristics (information) of a physical object and what can be obtained through experimental measurements.\\

From an experimental point of view, it is not possible to get direct access to the information about light that is enclosed in the vector state components. In other words we cannot measure the vector field itself. However, we can measure the intensity of light that is related to light vector by the following equation
\begin{equation}\label{eqLightIntensity}
I_0 \; =\; \left| E_{0x} \right |^2 \; + \;\left | E_{0y} \right |^2 
\end{equation}

So we focus our attention on light intensity because is the physical variable we can have access to as the result of an actual act of measurement. Measuring light intensity is therefore the way we use to get information about how light behaves. 

Based of the different kinds of interaction and energy exchange, we can distinguish between the different types of light, each of which can be represented using different light states.  
This approach is indeed useful to properly introduce the concept of single photon and to distinguish it from a classical or semi-classical description of light \cite{Fox_QuantumIntro, gerry2005introQOptics, loudon2000_QthoeryLight}.

\subsection{Single photon}\label{sec:SingPhot}

To introduce the quantum nature of light, our approach is to consider how electromagnetic field interacts with matter while exchanging energy. That is the same phenomena described while analyzing the photoelectric effect but we want to address that problem without introducing any \textit{particle-wave duality} interpretative model.\\
The concept of photon can indeed easily introduce difficulties in its representation especially in relation to the classical model of light that associate behaviours that could be related to particle and wave models \cite{Jones1991_photonMiscon}.\\
Students have absorbed from different contexts the notion that the nature of light has behaviours that can be associated with both waves and particles, but they often do so uncritically \cite{Henriksen2018_Whatislight}.
In the previously mentioned questionnaire (see Section \ref{sec:qubit_electron}) about the applications of quantum mechanics, it can be seen that students show great confidence in defining what a photon is (see Figure \ref{fig:Ch3_26_QWords}).

\begin{figure}[hbt!]
    \centering
     \includegraphics[width=\textwidth,height=\textheight,keepaspectratio]{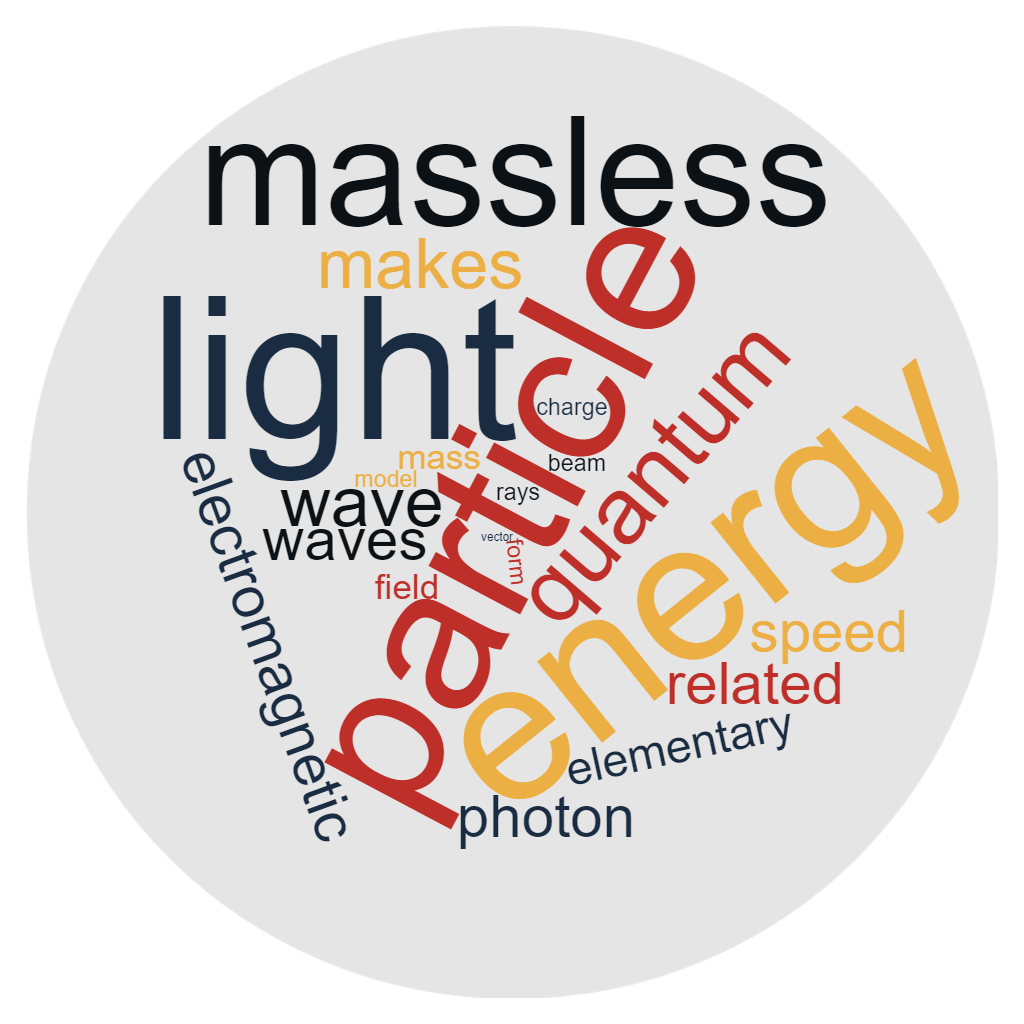}
    \caption{Definition of photon: word cloud built with the definitions provided by students. Students relate the idea of a photon to a massless particle carring energy}
    \label{fig:Ch3_34_WordPhoton}
\end{figure}

Only one student said that he had only heard of photons. For the rest, the analysis of the answers shows that for more than half of the students (45 out of 81)
\begin{quote} {\fontfamily{lmtt}\selectfont
    the photon is a particle of light
}\end{quote}

In the rest of the sample, the most common response relates the photon to the concept of how much and more specifically to the energy carried by light (18 out of 81).\\

In physics school textbooks, the concept of single photon is introduced by the idea of reducing the intensity of light \cite{romeni2017fisica}. In this framework, light is considered to be made by tiny entities, quantized energy grains, called photons. The intensity of the wave is therefore proportional to the number of photons hitting the detector. Reducing the intensity reduces the number of photons hitting the detector. One can think of the single photon as the limit case when the intensity is so low that it is possible to think of the detector being hit by a single photon and thus distinguish individual detections.\\

A better description of this process is to relate the states of light to the the intensity that is measured \cite{Mitchell2016_QOptImpatien}.
In the interaction between matter and light, 
single quantities of energy $E= h \nu$ are exchanged. The concept of \textit{single photon} is related to that of indivisible quantity of energy. A single photon is not the elemental "particle of light", but the basic unit that can be used to describe how light exchanges energy with matter.\\
This concept can be used to interpret the photoelectric effect but not to give a full picture of what light is when moving in space.
In the following discussion the field, described by the state vector and its amplitudes, is the physical entity that ''moves through space" like a wave. To know where that field is, an interaction (measurement) is required. The energy of that interaction is related to the 
square of the amplitudes (see Equation \ref{eqLightIntensity}). That is true for all different types of light. \\
What makes classical wave description of light different from quantum description of light is the nature of the field that is used to represent the information about the light.

The attention has to be focused on the physical processes that can generate specific light states. Single photon can be generated for example by the emission of single atoms or molecules or using the Parametric Down Conversion (PDC) process \cite{gerry2005introQOptics}. The light generated by PDC is called twin beam, a bipartite state described by a superposition of perfectly correlated number of photons. 


\begin{equation}
| \Psi_{twin} \rangle \; = \; \sum_{n=0}^{\infty} \, c_n \, |n,n\rangle \; \approx c_0 |0,0\rangle + c_1 |1,1\rangle + c_2 |2,2 \rangle \dots
\end{equation}

where $|n,n\rangle $ represents the state of a pair of $n$ photons.
If the process of pair production is not efficient, the probability of having pairs of single photons $|1,1\rangle$ is significantly higher that the other pairs \cite{Mitchell2016_QOptImpatien}. When selecting those pairs, \textit{single photons} can be obtained. 

\subsection{Polarization and the interplay between maths and physics}\label{subsec:Polar_MathInterplay}

The behaviour of quantum objects is arduous to depict and represent using everyday language. Describing the trajectory of a thrown ball seems easy because we can use our ordinary language based on a classical perception of reality. Doing the same thing with an electron or a photon is indeed quite challenging. It is possible to describe light using a mathematical formalism we presented in the previous section (see Section \ref{sec:ClQtlight} and Section \ref{sec:SingPhot}). Mathematics acts as the proper language that can be used to describe quantum object and predict the outcomes of experiments.

To try to avoid or reduce ontological confusion (see Section \ref{sec:ChangeP}), light is not said to \textit{be} a wave or to \textit{be} a particle. Light states moving through space and time are described using a mathematical language which is derived from wave mechanics \cite{schwichtenberg2019no-nonsense}. The interaction between light and detectors can be used to extract information from that specific physical system. In the measurement process, light interacts with the detector in the same way as a particle does, being localized and exchanging a specific amount of energy.

That formal mathematical language can support the understanding of quantum objects behaviour and give students proper tools to interpret the results of quantum experiments. The link between physical phenomenon and mathematical formalism is expressed in creating a parallel between a given physical instrument (i.e. polarizers and beam splitters) and the mathematical tool (matrices) used to describe how the interaction between the quantum states of light and the device change that state.\\
The idea is to build these tools almost from scratch with students and use them to describe the action of physical objects on light states.\\

Polarization is a fundamental property of light and its study can  provide deep information about light states that can be studied in different experimental setups.


The representation previously introduced (see Section \ref{sec:ClQtlight}) can be very useful when the students will start studying it.
Students can learn how to manipulate the polarization state of light using vectors and matrices. If light polarization states are represented as vectors, the physical objects that can transform the polarization states can be seen as matrices that transform the state vectors.\\
In teaching optical polarization, the strong interplay between mathematics and physics can be an opportunity. In fact, complex numbers and matrices are in principle included in the mathematics curriculum of scientifically oriented High Schools [3] but seldom taught due to the abstractness of the subject and the lack of connections to meaningful physical examples. On the other hand, the description of light in terms of vectors, required to handle polarization properties, is perceived as too difficult to explain because of the lack of mathematical foundations of formalism. If taught together, both mathematical and physical learning outcome can be achieved.
The central idea is to build the mathematical tools that can be used to describe the transformations of light states. Those transformations are general and the differences are related to the measurements outcomes that depends on the specific light states we consider (classical light or single photon). \\

\begin{figure}[hbt!]
    \centering
     \includegraphics[width=\textwidth,height=\textheight,keepaspectratio]{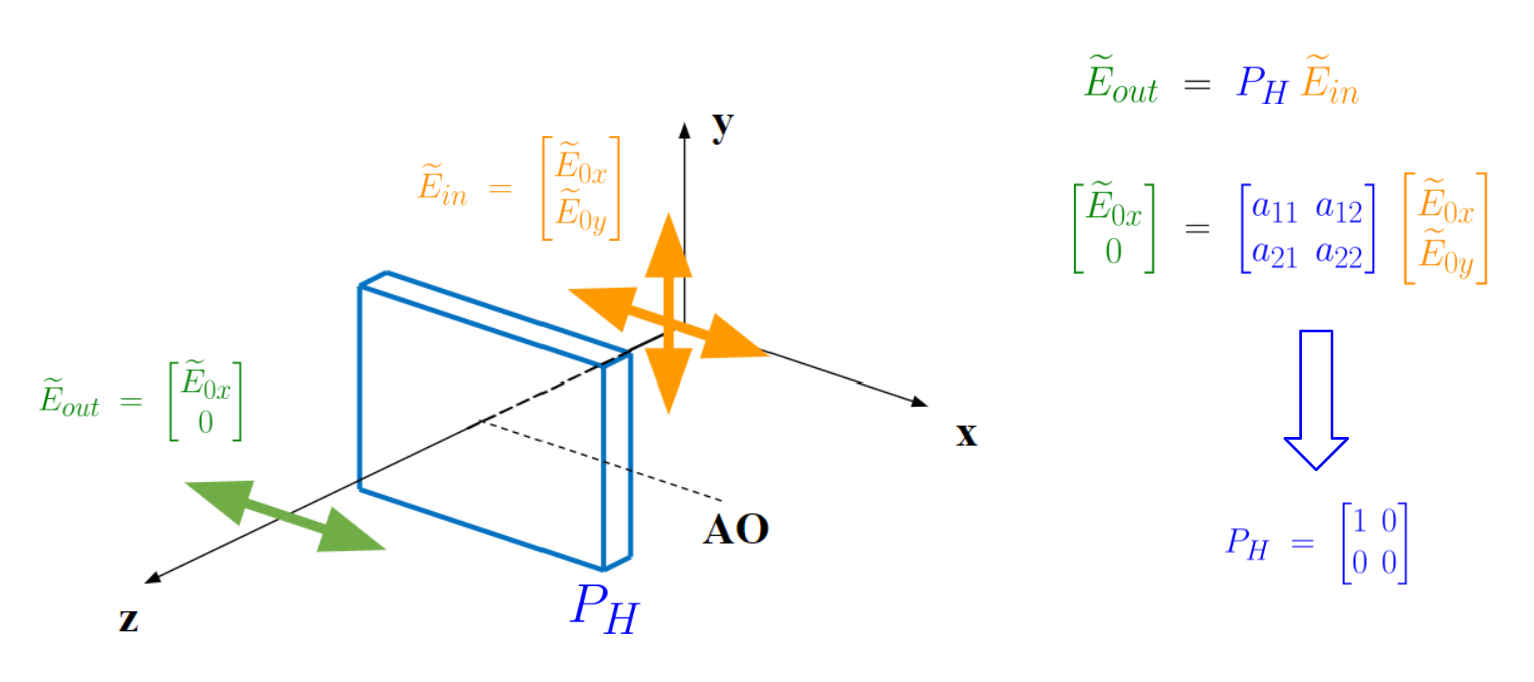}
    \caption{Scheme to infer the matrix representing a polarizer aligned horizontally (along x-axis).}
    \label{fig:Ch3b_6_PolarMatrixH}
\end{figure}

To build proper mathematical tools we started from the description of a polarized light field passing through a polarizer having the optical axis oriented along the x-axis. Based on simple observations, we know how polarizers work: the action of a polarizer is to transform the input field 
$\tilde{E}_{in}$ into the output field $\tilde{E}_{out}$, horizontally aligned (see Figure \ref{fig:Ch3b_6_PolarMatrixH}).

This mathematical formalism has been chosen because the calculation can be easily performed after being introduced on the rules to multiply matrices and vector. This kind of mathematical tool is also part of the Italian maths curriculum \cite{MIUR2010_211}.

As seen with Stern-Gerlach apparatus (see Section \ref{sec:seqSGA_matrix}) the transformation introduced by polarizer on the light state can be mathematically described by the action of the matrix $P_H$ on the vectors so that 

\begin{equation}\label{eq:PHbasic}
    \widetilde{E}_{out} \; = \; P_H \, \widetilde{E}_{in}
\end{equation}


\begin{equation}
    \begin{bmatrix}
        \widetilde{E}_{0x}\\
        0
    \end{bmatrix} \; = \;
    \begin{bmatrix}
        a_{11} & a_{12}\\ 
        a_{21} & a_{22}
    \end{bmatrix} \; \begin{bmatrix}
\widetilde{E}_{0x}\\ 
\widetilde{E}_{0y}
\end{bmatrix}
\end{equation}

and so we obtain
\begin{equation}
    P_H \; = \;  \begin{bmatrix}
        1 & 0\\ 
        0 & 0
    \end{bmatrix}
\end{equation}

%

\begin{figure}[hbt!]
    \centering
     \includegraphics[width=\textwidth,height=\textheight,keepaspectratio]{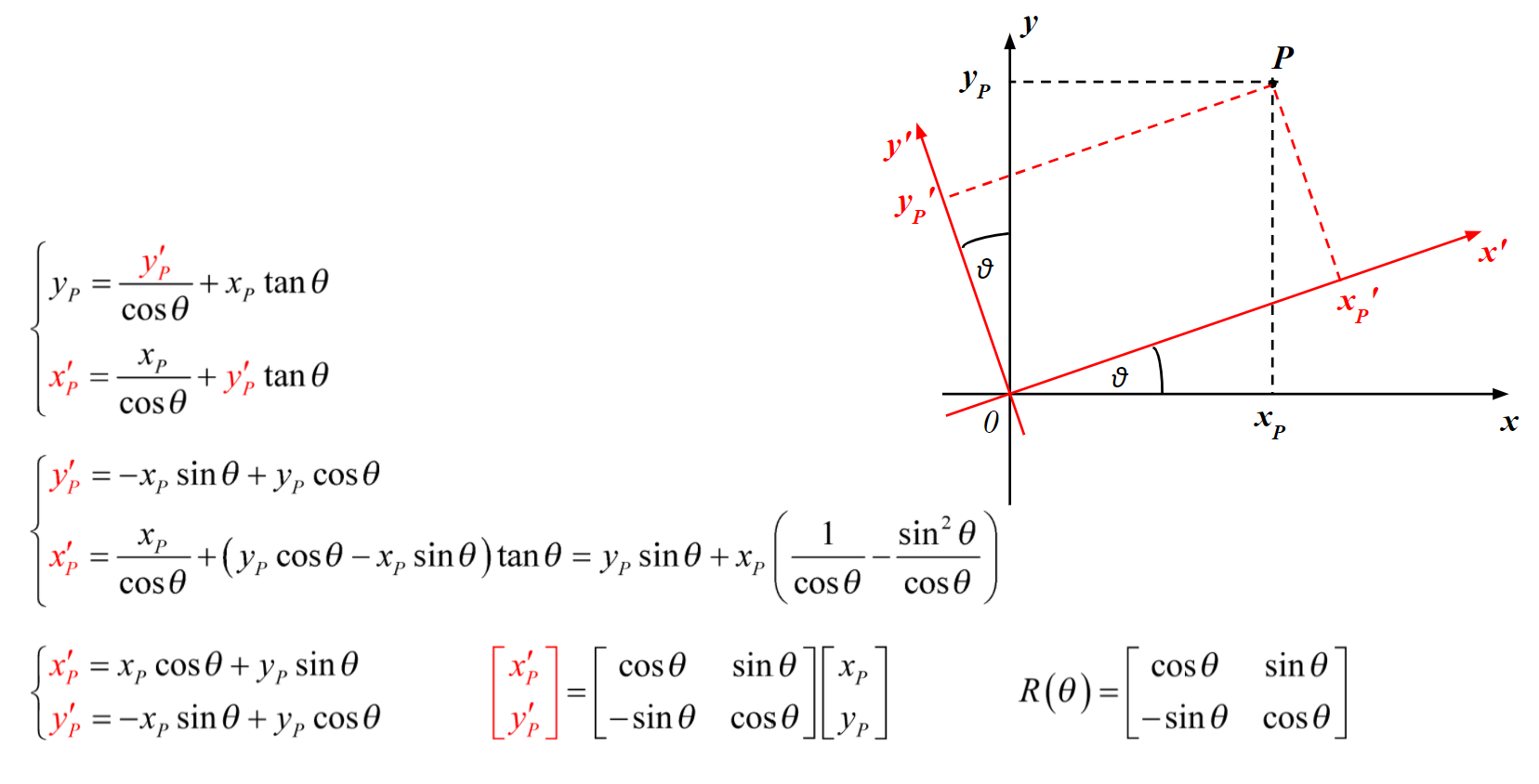}
    \caption{Rotation matrix can be defined writing the coordinated of point $P$ in reference frame that have been rotated counter-clockwise by $\theta$}
    \label{fig:Ch3_4_RotMatrix}
\end{figure}

This result can be generalized to a polarizer oriented at a generic angle $\theta$. The matrix associated to that polarizer $P_{\theta}$ can be obtained by rotating the matrix $P_H$ through the rotation matrix (see Figure \ref{fig:Ch3_4_RotMatrix})
\begin{equation}
R(\theta) \; = \; \begin{bmatrix}
    \cos \theta & \sin \theta\\
    - \sin \theta & \cos \theta
    \end{bmatrix}
\end{equation}    

taking into account that the transformation of a matrix is given by

\begin{equation}
P_{\theta} \;  = \; R(-\theta)\,P_H\,R(\theta) \; = \; \begin{bmatrix}
    cos^2 \theta & \cos \theta \sin \theta \\
    \cos \theta \sin \theta & \sin^2 \theta
\end{bmatrix}\\
\end{equation}

\begin{figure}[hbt!]
    \centering
     \includegraphics[width=\textwidth,height=\textheight,keepaspectratio]{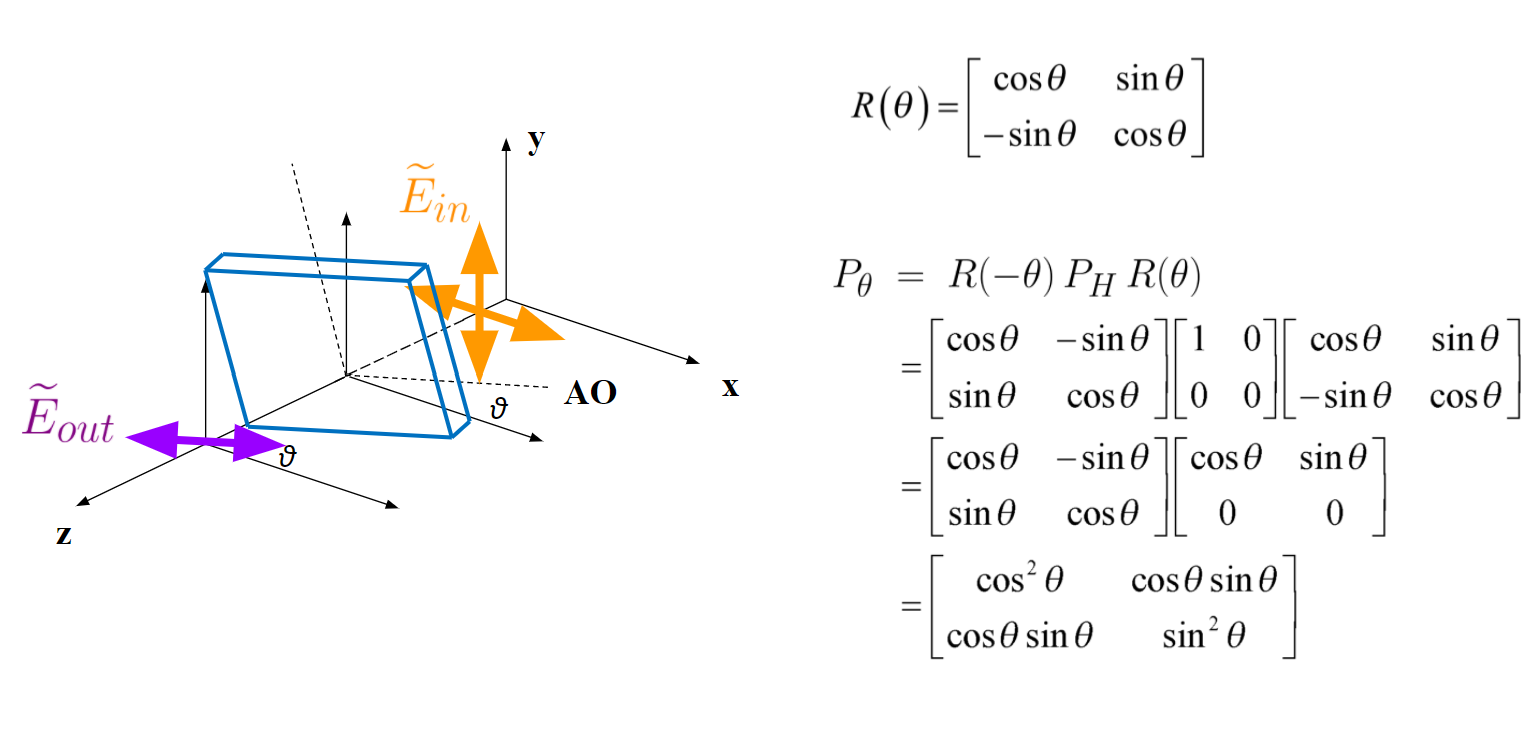}
    \caption{The matrix representing a polarizer at a generic angle $\theta$}
    \label{fig:Ch3_5_Ptheta}
\end{figure}

To support the meaning-making process it is important to explicitly connect the mathematical formalism and the physical object this formalism has been used to represent \cite{Bouchee21}.
The use of simulations and laboratory experiences can support the understanding such connection, fostering an active reflection about the phenomenological aspects and theoretical models.\\
The aim of this section is to show how the use of a rigorous mathematical formalism is an extremely effective tool for the construction of an effective interpretative model of experiments with light. This should help the process of reconstructing the contents, showing how the process of preparing and transforming the state of a quantum object in its interaction with the measuring apparatus is actually useful for the analysis of behaviour and nature of the quantum object. 

\subsubsection{Malus' law}
These powerful mathematical tools describe above can be easily applied to describe the actions of polarizers on light states in more general situation, as for example considering sequences of polarizers.

Consider a field polarized along the angle $\theta_0$ with respect to the horizontal axis $x$
\begin{equation}
    E_0 \; = \; A \begin{bmatrix}
        \cos \theta_0 \\
        \sin \theta_0
    \end{bmatrix}
\end{equation}

The intensity of this field is $I = |A|^2$

\begin{figure}[hbt!]
    \centering
     \includegraphics[width=\textwidth,height=\textheight,keepaspectratio]{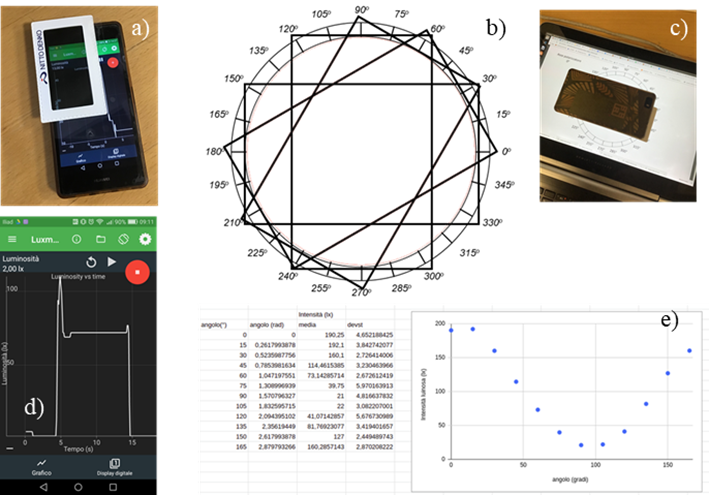}
    \caption{Experimental setup for the measurement of the Malus’ Law. a) plastic polarizer stuck onto the light sensor of a smartphone; and b) goniometer template; c) simple polarizers; d) light intensity measured with a smartphone e) sample of experimental results.}
    \label{fig:Ch3_14_Smartp1}
\end{figure}

\begin{figure}[hbt!]
    \centering
     \includegraphics[width=\textwidth,height=\textheight,keepaspectratio]{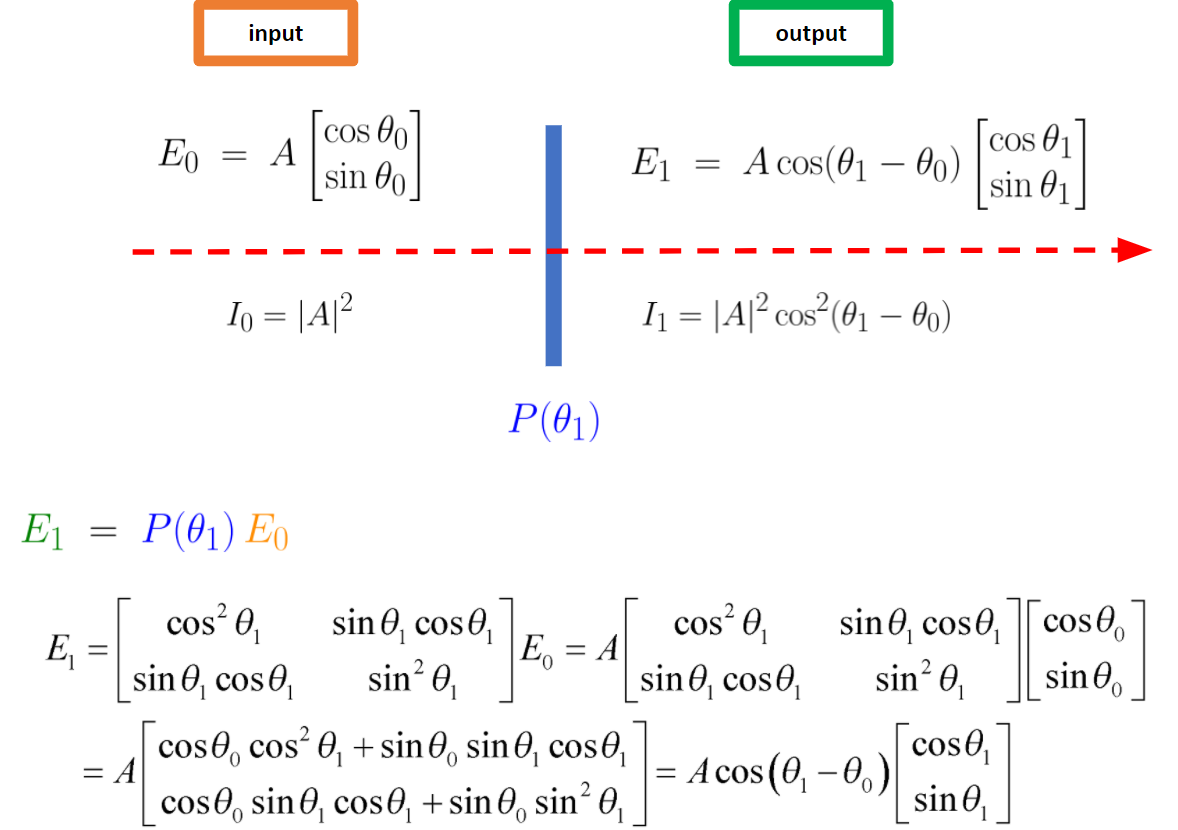}
    \caption{Malus' law using light vector fields}
    \label{fig:Ch3_12_MalusEfield}
\end{figure}

If the field passes through a polarizer oriented at angle $\theta_1$ with respect to the horizontal axis, the output field is linearly polarized along $\theta_1$ (see Figure \ref{fig:Ch3_12_MalusEfield}) and its expression $\widetilde{E}_{1}$ can be obtained using the same sequence we have seen previously (see fig \ref{fig:Ch3_12_MalusEfield})

\begin{equation}
\begin{split}
    \widetilde{E}_{1} \; &= \; P_{\theta_1} \, \widetilde{E}_{0}\\
    & = \begin{bmatrix}
        cos^2 \theta_1 & \cos \theta_1 \sin \theta_1 \\
    \cos \theta_1 \sin \theta_1 & \sin^2 \theta_1
    \end{bmatrix}
    \, A \begin{bmatrix}
        \cos \theta_0 \\
        \sin \theta_0
    \end{bmatrix} \\
    & = A \cos (\theta_1 - \theta_0) \begin{bmatrix}
        \cos \theta_1 \\
        \sin \theta_1
    \end{bmatrix}
\end{split}
\end{equation}

and its intensity is:

\begin{equation}
    I_1 = |A|^2 \, \cos^2(\theta_1 - \theta_0)
\end{equation}
This equation is the Malus' law for polarizers. In this way it is easy to show how phenomenological experiments with polarizers can be interpreted using a model of light based on fields. The advantage is that this equation can be applied to any light field and it connects the abstract light vector state that describes the physical propagating object to the light intensity that describes the information about that same physical object as a result of a measurement process.\\
The experimental context to verify the Malus' law can be performed using rather simple materials and equipment (see Figure \ref{fig:Ch3_14_Smartp1}) and can easily create a stimulating and rich learning context (see Chapter \ref{ch:6-LETeachers}).

This learning experience was proposed to the teachers during the Professional Development programme (see Section \ref{sec:QJumps}) and it has been included in the teaching-learning sequences designed for students (see Chapter \ref{ch:7-LEStudents}).

\begin{figure}[hbt!]
    \centering
     \includegraphics[width=\textwidth,height=\textheight,keepaspectratio]{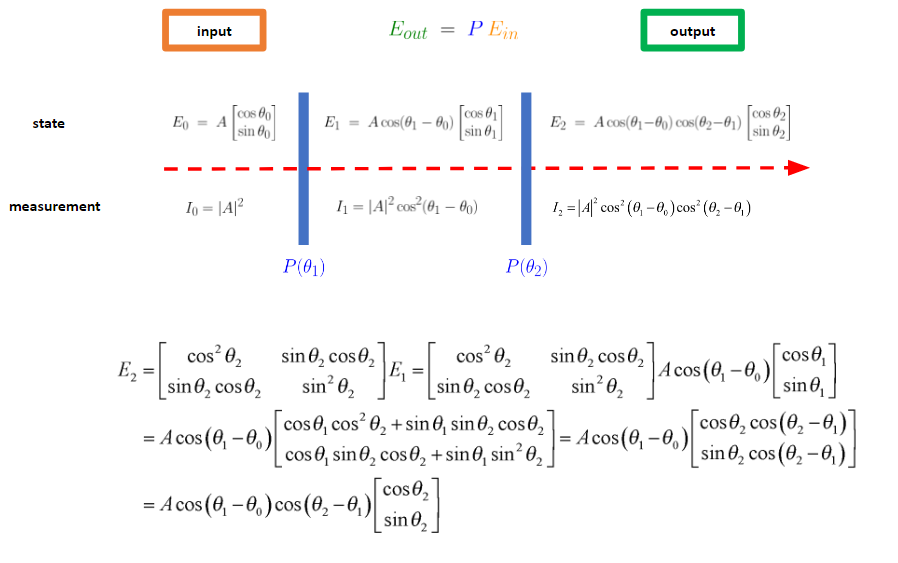}
    \caption{Computational structure for processing information represented using a sequence of two polarizers}
    \label{fig:Ch3_13_2polarz}
\end{figure}

This structure can be also generalized to any computational process (see Figure \ref{fig:Ch3_13_2polarz}). The relationship between the input and output states that encode the information about the system depends on the specific sequence of interactions with polarizing devices that determine the evolution of the field state. The output measurement in this case corresponds to the light intensity. In the case of single photons, the measurements consist in counting the detection events and results are given in term of detection probability.


\subsubsection{Generic sequence of polarizers}
\begin{figure}[hbt!]
    \centering
     \includegraphics[width=\textwidth,height=\textheight,keepaspectratio]{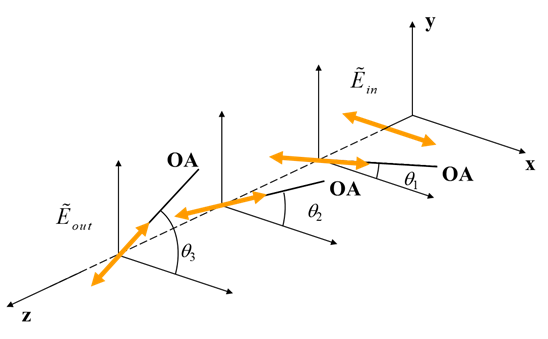}
    \caption{Representation of a generic sequence of polarizers at different angles}
    \label{fig:Ch3_11_PolSeq}
\end{figure}

This result in the previous section can also be easily generalized for a generic sequence of $N$ polarizers, oriented at different angles $\theta_j$ with respect to the horizontal axis (see Figure \ref{fig:Ch3_11_PolSeq}).\\

By iterating the calculation of above, the output field becomes
	\begin{equation}
	    E_N \; = \; A \prod_{j=0}^{N-1} \cos (\theta_{j+1} - \theta_j) \begin{bmatrix}
	        \cos \theta_N \\
	        \sin \theta_N
	    \end{bmatrix}
	\end{equation} 	
whose intensity is
	 	\begin{equation}
	    I_N \; = \; I_0 \prod_{j=0}^{N-1} \cos^2 (\theta_{j+1} - \theta_j)
	\end{equation} 

An interesting physical situation is given by the case in which the angle between the first and the last polarizer is $\frac{\pi}{2}$, thus preventing the transmission of light. If other polarizers are inserted between the two so that the angles between subsequent polarizers is the same $\Delta \theta \; = \; \theta_{j+1} - \theta_j \; = \; \frac{\pi}{2N}$, the intensity becomes
	\begin{equation}
	    I_N \; = \; I_0 \left [\cos^2 \left(\frac{\pi}{2N}\right) \right]^2
	\end{equation} 

\begin{figure}[hbt!]
    \centering
     \includegraphics[width=\textwidth,height=\textheight,keepaspectratio]{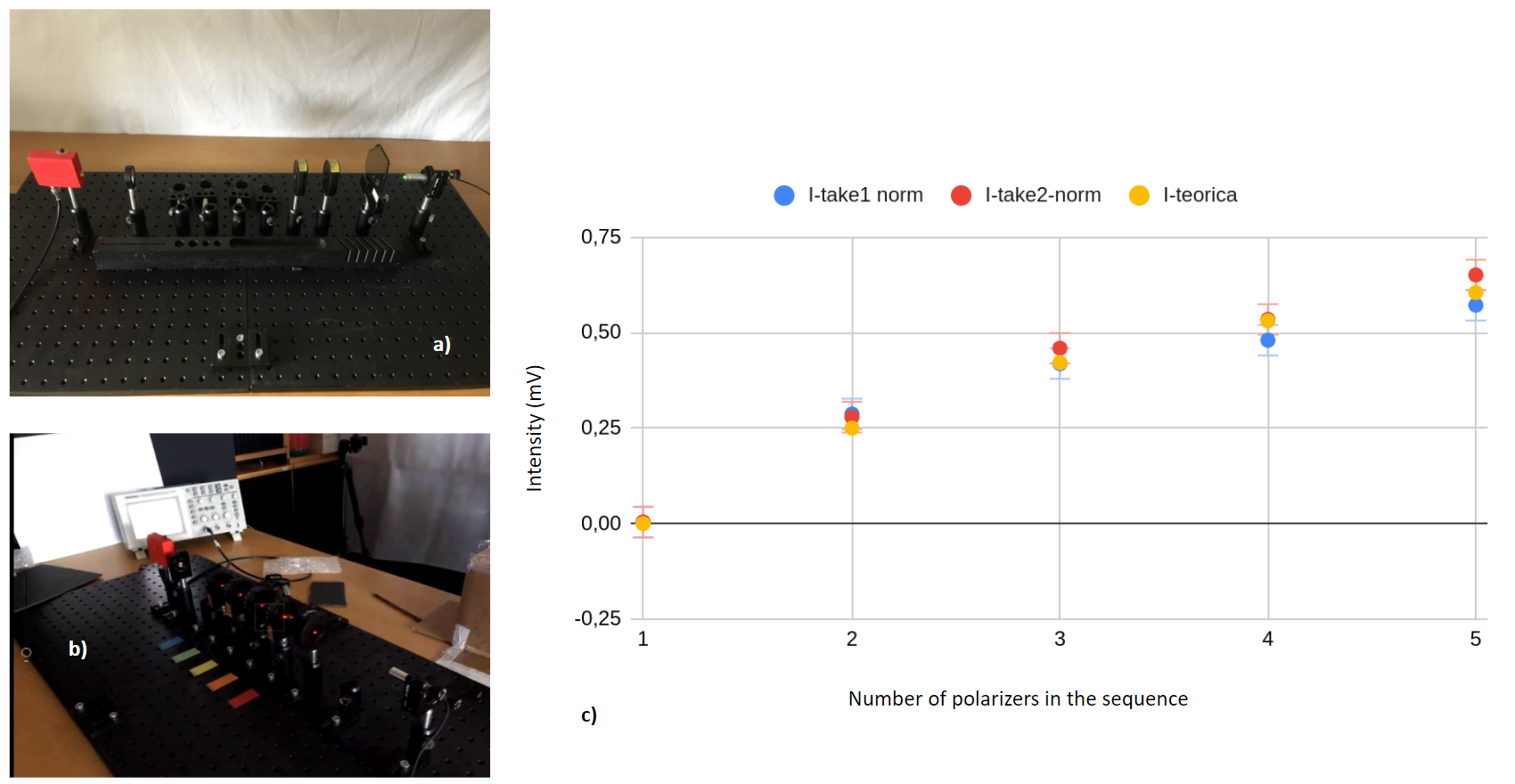}
    \caption{Experimental setup for the measurement of the intensity of a light exiting a sequence of N polarizers. a) and b) professional equipment on a portable breadboard; c) light intensity measured with a photodiode as a function of the number of polarizers in the sequence}
    \label{fig:Ch3_15_LabSeqPolar}
\end{figure}

Note that $I_N \to I_0$  for $N \to \infty$, a result that is often addressed to as paradoxical \cite{dirac1981_QMprinciples}, since for very large $N$ all the light is transmitted entirely. 



\subsection{The Nature of light, beamsplitters and Mach-Zehnder interferometer}\label{subsec:MZIexper}

The opportunity to study the quantum nature of light is provided by the use of Mach-Zehnder interferometers.
Using simulations \cite{Kohnle15_QuVIS_2states} it is possible to analyze how light states interact with the apparatus. The investigation is focused on comparing wave and particle interpretative models of light (see Section \ref{subsec:BSsim} and \ref{subsec:MZIBS_sim}) and introducing superposition states (see Section\ref{subsec:MZIexper}).


Using the QuVis simulations \textcolor{red}{[REF QUVIS simulation]}, interpretative models of light can be compared based on what is the expected output. By reading the intensity of light measured at the output of an interferometer, very small differences in optical paths (related to phase shifts) can be made evident. The reconstruction of the field at the interferometer output can be used to identify the characteristics of different states of light. So it is the analysis of the measurements results that creates a model about the light behaviour inside the interferometer.\\
In particular, the concept of single photon can be enriched from different perspectives.
Single photons cannot be described as electromagnetic waves, as the coincidence counter never fires (see Section \ref{subsec:BSsim}) and cannot be described as classical particles, as they do not show interference between the beams when two beamsplitters are in place (see Section \ref{subsec:MZIBS_sim}). Single photons are also not \textit{sometimes} classical particles and \textit{sometimes} electromagnetic waves, as can be seen from the experiments with the two beamsplitters and the phaseshifter (see Section \ref{subsec:MZIBS_sim}): photons show interference like electromagnetic waves (the counts depend on the phase), but never trigger the coincidence counter like particles. Thus, single photons must be something different to both electromagnetic waves and classical particles. These conclusions can further be analyzed by the development of a theoretical model.\\
In the next sections we illustrate how an interpretative model of light can be built up analysing the data collected from the simulation. An example of worksheet for this activity can be found in Appendix \ref{appx:QuVisMZI}

\subsubsection{Light through a unpolarized beam splitter} \label{subsec:BSsim}
In the simulation, the measurements of light with a single beam splitter show no coincidences when firing a single photon.\\
As seen in the case of polarizers (see Figure \ref{fig:Ch3b_6_PolarMatrixH}), we can build a mathematical model that represents the action of a beamsplitter on the light state vectors. Because the transformation associated to the beam splitter is linear, it can be represented by a matrix.

\begin{figure}[hbt!]
    \centering
 \includegraphics[width=\textwidth,height=\textheight,keepaspectratio]{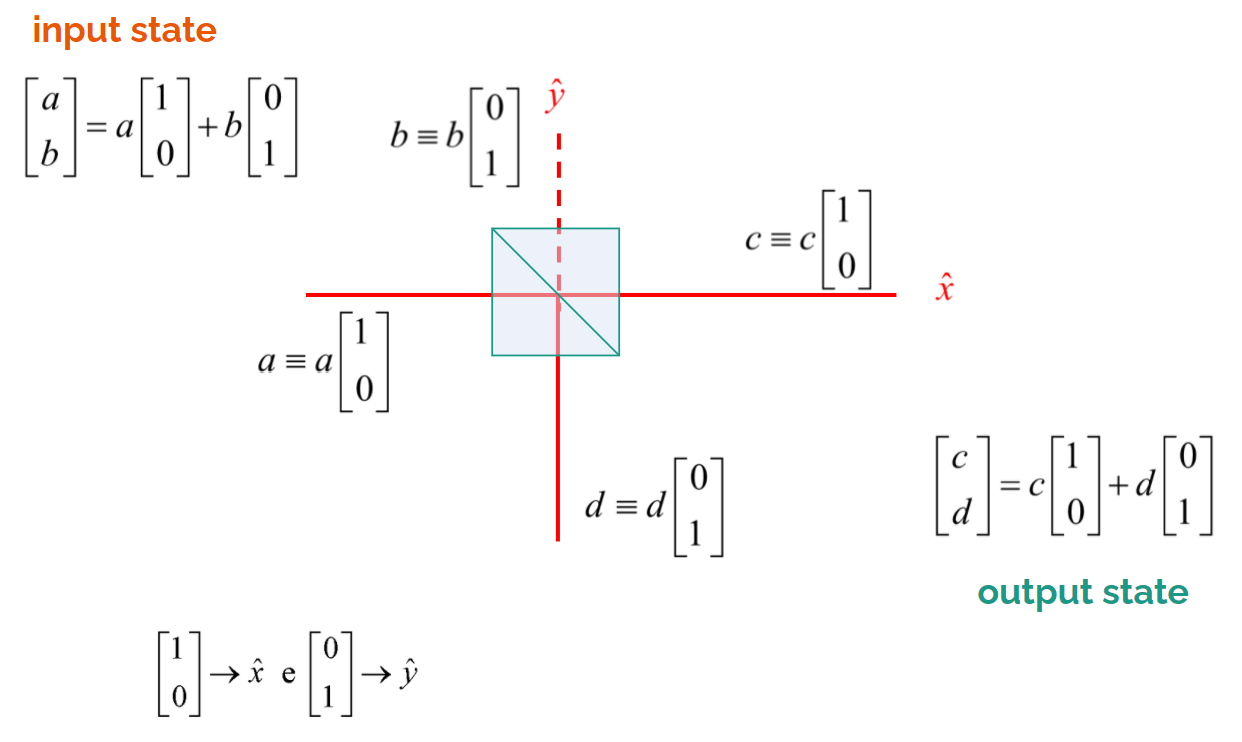}
    \caption{Non polarizing beam splitter; vectors $\begin{bmatrix}
        1\\
        0
    \end{bmatrix}$ and $\begin{bmatrix}
        0\\
        1
    \end{bmatrix}$ represent the direction of propagation}
    \label{fig:Ch3_16_NpBS}
\end{figure}

\begin{figure}[hbt!]
    \centering
    \includegraphics[scale=0.6]{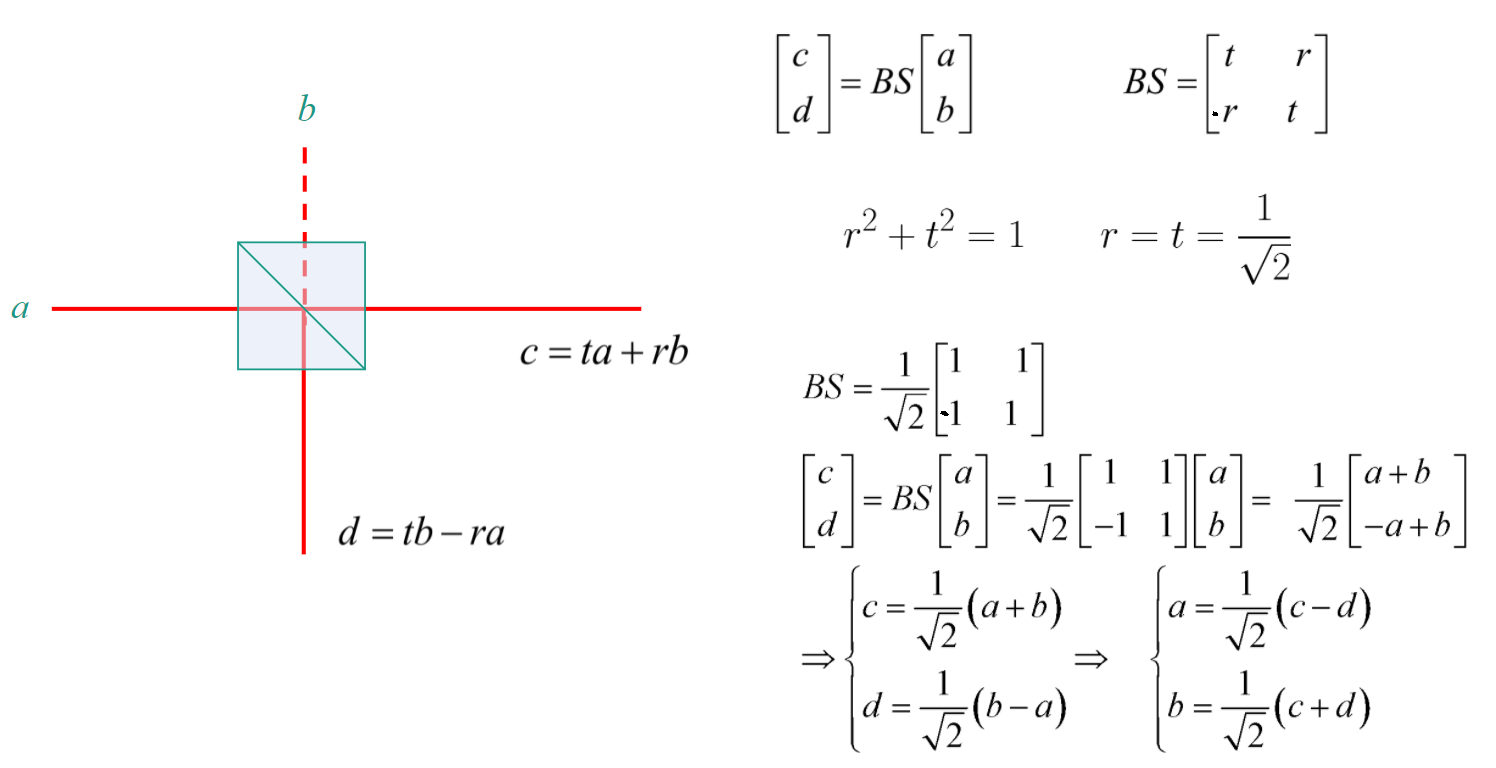}
    \caption{Non polarized beam splitter: matrix and output fields}
    \label{fig:Ch3_17_BSnpOUT}
\end{figure}

Considering a balanced beam splitter ($r = t$) we can define its matrix as
\begin{equation}
BS \; = \; \frac{1}{\sqrt{2}}\begin{bmatrix}
1 & 1 \\
-1 & 1
\end{bmatrix}    
\end{equation}
In this case it is possible to determine the general component of the light state outputs vectors (see Figure \ref{fig:Ch3_17_BSnpOUT})

In the simulation, a particle can follow only one path at the time, while a classical monochromatic wave that passes through the beam splitter can travel on both paths (like in a double slit experiment). The presence of coincidences can be explained relating the firing of the detector to the statistical process of promoting electrons in a conducting band at the detector surface.
Comparing the outcomes, the behaviour of a single photon is identical to that of particles. This could be explained thinking that the state of a single photon after the beam splitter can be written as a superposition of states (see Equation \ref{eq:singlePh_NPBS})

\begin{equation}\label{eq:singlePh_NPBS}
    |1\rangle_a |0\rangle_b \; = \; \frac{1}{\sqrt{2}} |1\rangle_c |0\rangle_d \, + \, \frac{1}{\sqrt{2}} |0\rangle_c |1\rangle_d
\end{equation}

Where $|0\rangle_b$ represent a vacuum state entering the beam splitter port $b$. The nature of superposition and the difference with a mixed states interpretation can be showned using a polarized beam splitter (see Section \ref{subsec:MZIexper})\\
This kind of mathematical representation explain how it is possible to generate a qubit, realized by a single photon, in a superposition state. From a computational perspective, the beam splitter transformation is represented by an Hadamard gate \cite{Adami97OpticGates, Barenco95Gates}.

\subsubsection{Light in a Mach-Zehnder interferometer with unpolarized beam splitters} \label{subsec:MZIBS_sim}

The analysis of the Mach-Zehnder interfometer experiment can be performed using rigorous mathematical tools to show the nature of light states and the relationship between the output light intensity and apparatus parameters, such as the phase shift $\delta$.

\begin{figure}[hbt!]
    \centering
     \includegraphics[width=\textwidth,height=\textheight,keepaspectratio]{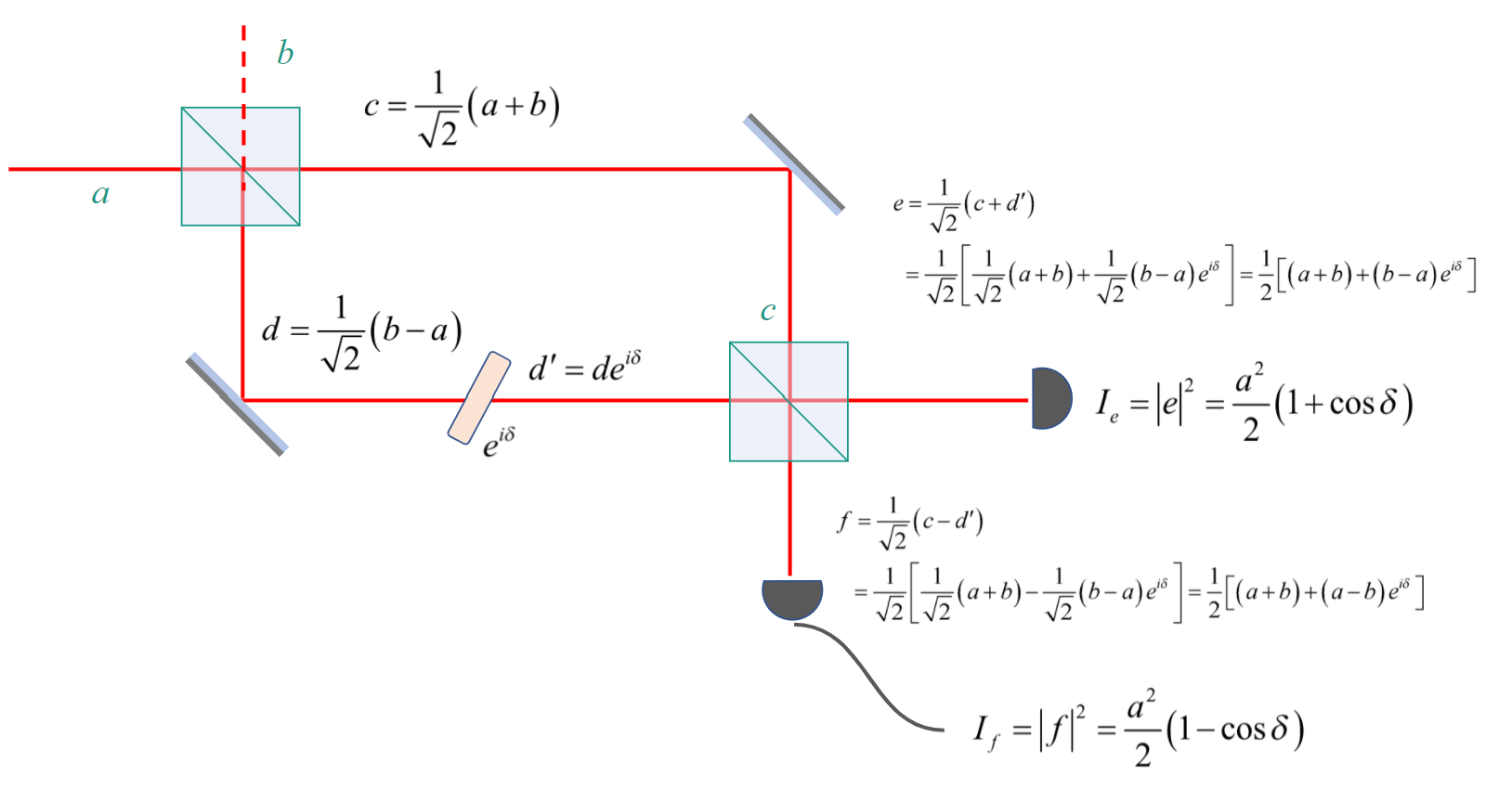}
    \caption{Mach-Zender interferometer: formal mathematical description}
    \label{fig:Ch3_17b_MZImath}
\end{figure}

The light field states $e$ and $f$ are described for both paths and so their intensities $I_e$ and $I_f$ that are the quantities that are actually measured and that can be used to extract information from light and to model its behaviour.\\ Comparing the value of light intensity we can interpret the results in the simulation.

With $\delta = 0$ only one detector fires both using waves and single photons in the simulation. The particle model cannot be used to interpret the measurements because in that case we aspect 50\% chance to see both detectors firing.\\
So a single photon must be interpreted again as superposition of states. The outcomes can be interpreted using the same mathematical description used with waves. The description in terms of a photon that "interfere with itself" is not presented to not trigger the idea that the elemental indivisible photon is somehow "splitted in two" inside the interferometer. The idea of describing the light inside the interferometer using mathematical concepts like fields is instead supported because it still allows the interpretation the measurement outcomes. 

\begin{figure}[hbt!]
    \centering
    \includegraphics[scale=0.6]{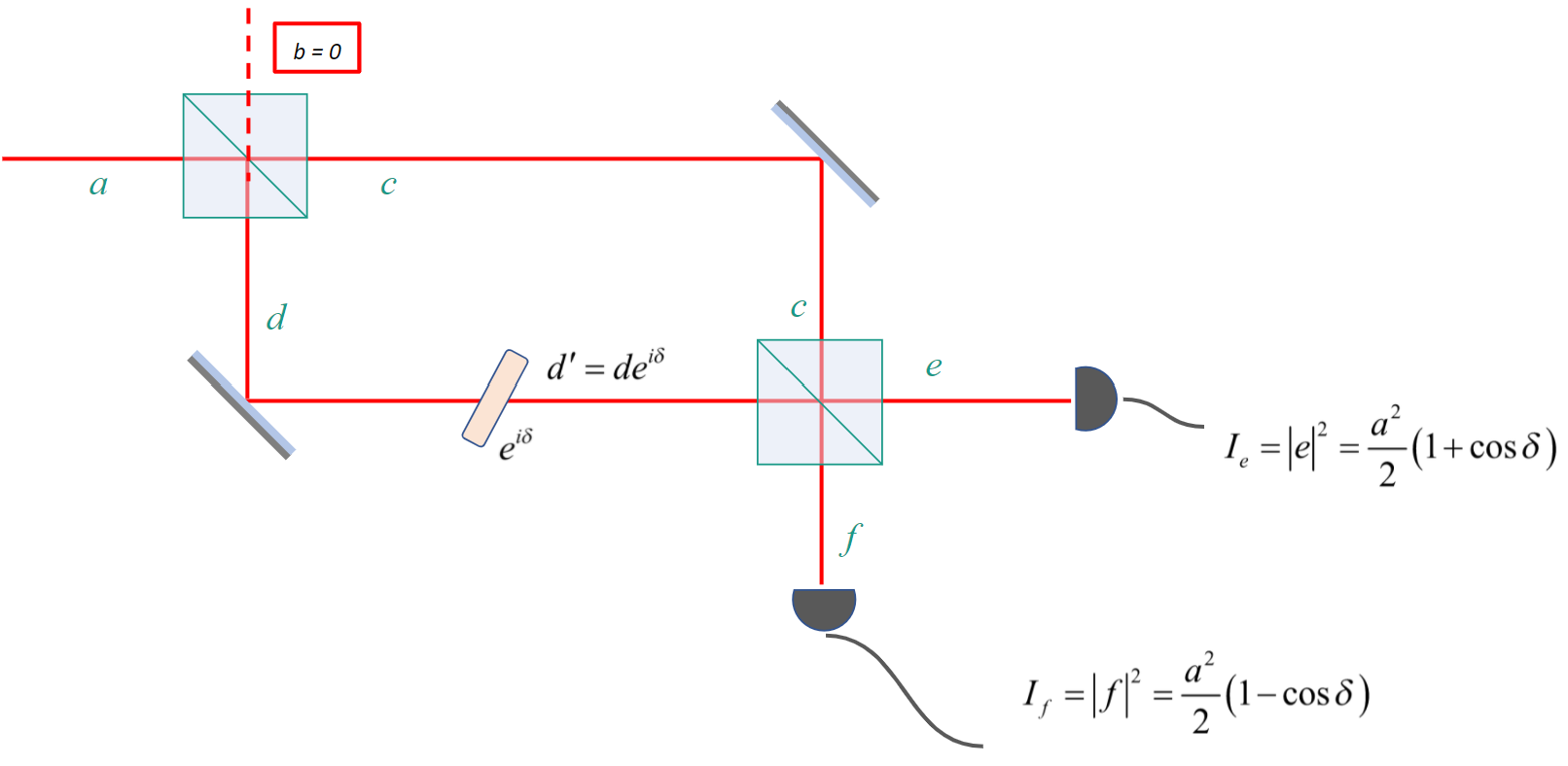}
    \caption{Mach-Zender interferometer: wave-light interpretative model}
    \label{fig:Ch317c_MZIwave}
\end{figure}

Considering the case with $\delta \ne 0$ 
in a simplified condition where light enters only the $a$-channel ($b=0$), the output of the interferometer is similar to what is revealed in the double slit experiment: light that has travelled two different paths is collected and an interference pattern that depends on $\delta$ is obtained. (see Figure \ref{fig:Ch317c_MZIwave}). The energy is conserved (the sum of the output intensities is equal to the intensity of the input signal) and by varying $\delta$ all cases are obtained, even those where light is detected in both detectors ($\delta = \frac{\pi}{2}$).

\begin{figure}[hbt!]
    \centering
 \includegraphics[width=\textwidth,height=\textheight,keepaspectratio]{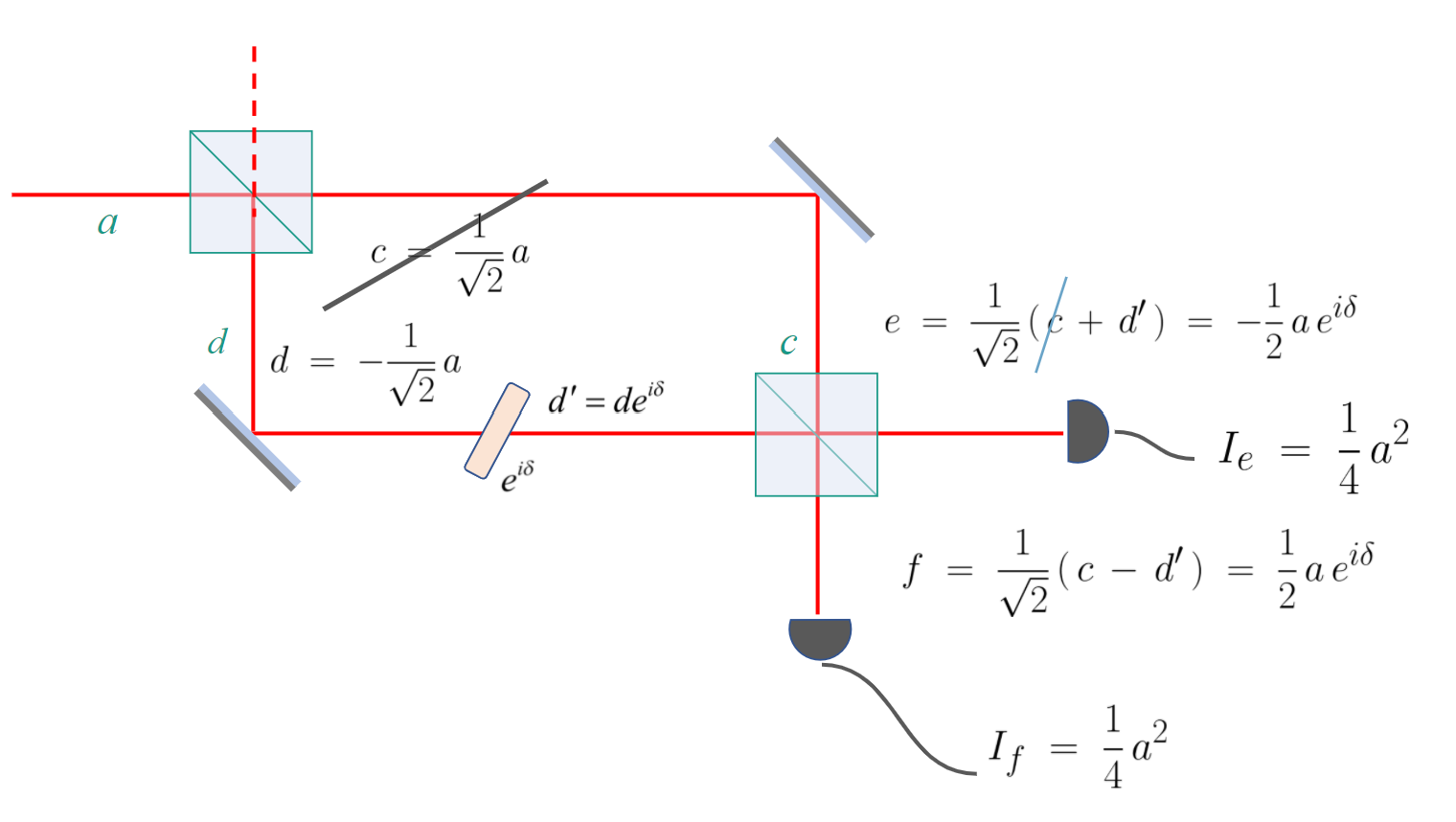}
    \caption{Mach-Zender interferometer: particle-light interpretative model}
    \label{fig:Ch3_17d_MZIpart}
\end{figure}

The particle-like model case presented in the simulator can be recovered by considering that the light passes only on one of the two branches at a time, e.g. only in $d$.
From a mathematical point of view, this means eliminating the field term $c$ from the equations (see Figure \ref{fig:Ch3_17d_MZIpart}) and interpreting the field intensities as numbers counts.

In this case all terms containing the phase shift $\delta$ have disappeared in the output intensities. So the particle model excludes interference. The comparison with the simulation with the single photon, in which a dependence of the measurement results on $\delta$ is detectable, leads to the conclusion that the single photon cannot be considered as a classical particle.\\

The possible conclusion is that what moves inside the interferometer is a field that propagates like a wave and therefore presents a typical wave-like behaviour such as superposition.\\
The question to be clarified is what it means to send a single photon into the interferometer and in particular how to answer the question about the path or trajectory followed by this object. When a single photon is measured, it is localised by the click in the detector, so it is theoretically possible to tell where it is. Once again, the focus is not on the physical object and its properties (where \textit{is} the photon) but on the possibility of being able to describe the single photon state, how does it evolve inside the interferometer and what happens when the state is measured.
This approach to experimental description have been used also in a different context (electron diffraction experiments) in one of the teaching-learning sequence designed by teachers (see Section \ref{sec:carcano}).

\subsubsection{Light inside a Mach-Zehnder interferometer using a polarized beam splitter}\label{subsec:MZIexper}

This section deald with the use the Mach-Zehnder interferometer to introduce the idea of quantum superposition. The different concepts and representation are orgaized in a way similar to the introduction of superposition using Stern-Gerlach apparatus (see Section \ref{sec:seqSGA_matrix}).

\begin{figure}[hbt!]
    \centering
 \includegraphics[width=\textwidth,height=\textheight,keepaspectratio]{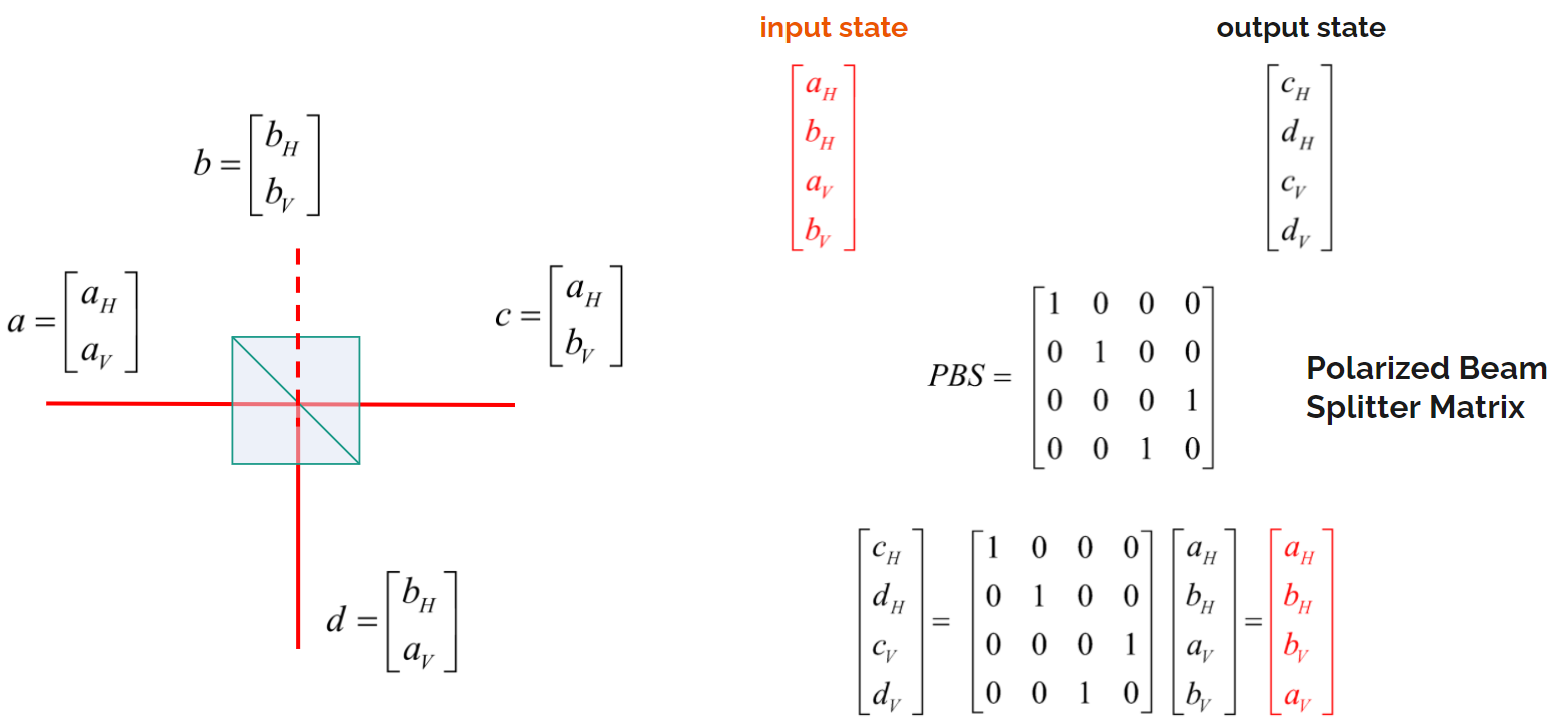}
    \caption{Polarized Beam Splitter matrix and outputs}
    \label{fig:Ch3_18_PBSMatrix}
\end{figure}

The focus is on how it is possible to describing the state of a single photon and what is the most appropriate model to do it while interacting with the experimental device. By using a polarised beam splitter (PBS) \cite{ghirardi2015_carte} and an appropriate mathematical representation the evolution of the state of the single photon at the moment it passes through the PBS is described.
A lossless symmetric polarizing beam splitter is a device that transmits only the horizontal component and reflects all and only the vertical component of polarisation (see Figure \ref{fig:Ch3_18_PBSMatrix}). Consequently, we can know what the state of the single photon is at the moment it has passed through the PBS.

Now consider a complete Mach-Zehnder interferometer with an input state that is linearly polarized in the $\theta$ direction
\begin{equation}
    a \; = \; A \begin{bmatrix}
        \cos \theta\\
        \sin \theta
    \end{bmatrix} \; = \; A \left ( \cos \theta \begin{bmatrix}
        1\\
        0
    \end{bmatrix} + \sin \theta \begin{bmatrix}
        0\\
        1
    \end{bmatrix} \right) \qquad I_a = |A|^2
\end{equation}

\begin{figure}[hbt!]
    \centering
 \includegraphics[width=\textwidth,height=\textheight,keepaspectratio]{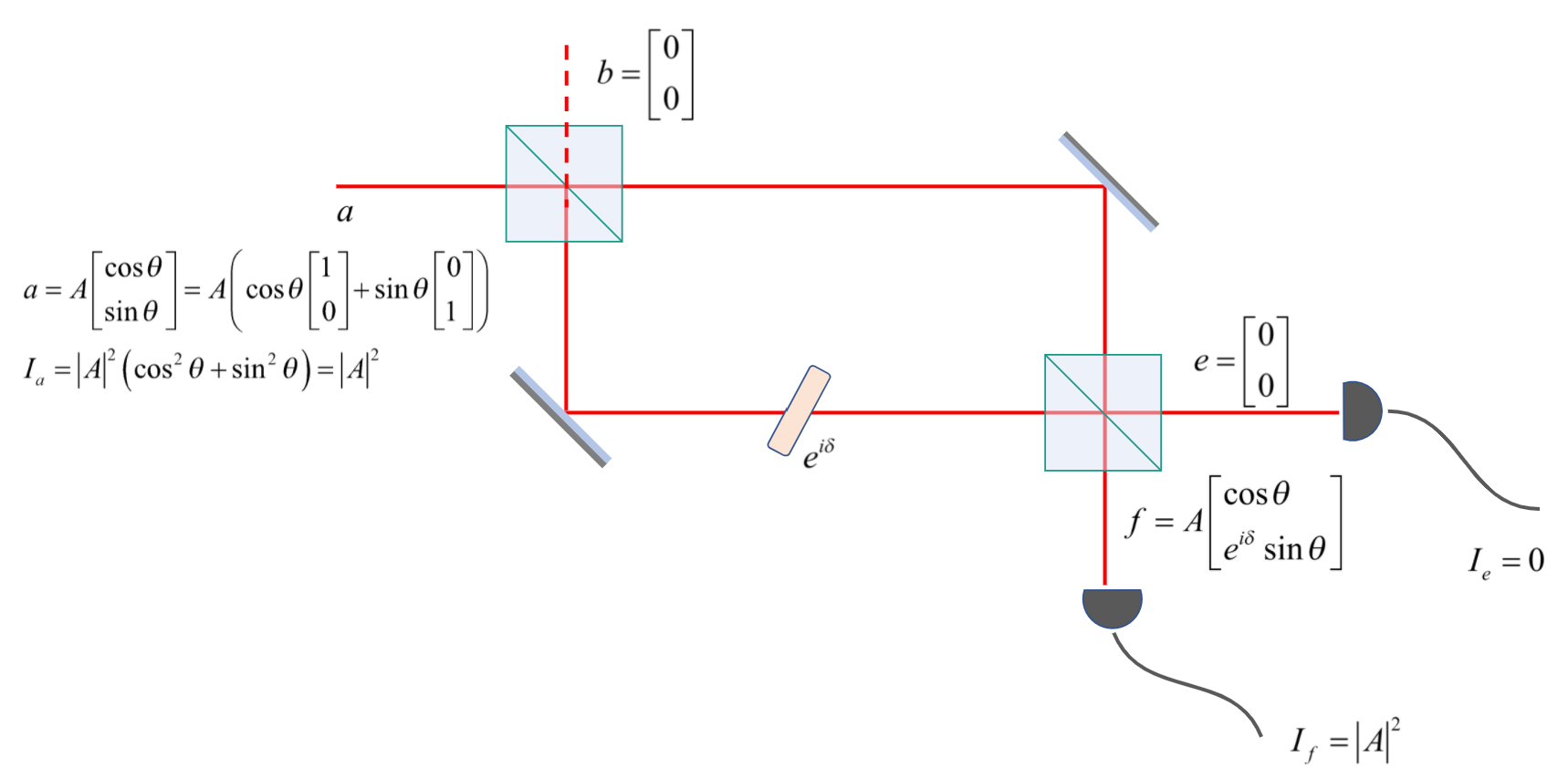}
    \caption{Mach-Zehnder interferometer with polarized beam splitter. $b$ is the vacuum state b = $\begin{bmatrix}
        0\\
        0
    \end{bmatrix}$  The output state component depends on the phase shift $\delta$. The light intensity (the property of light we can measure) does not show any dependency from $\delta$}
    \label{fig:Ch3_19_PBSMZI_f}
\end{figure}

The measured input and output intensities are the same $I_f = I_a = |A|^2$ but the output state is different. The output state is elliptically polarized to its second component depends the phase $\delta$ (see Figure \ref{fig:Ch3_19_PBSMZI_f})\\

\begin{figure}[hbt!]
    \centering
 \includegraphics[width=\textwidth,height=\textheight,keepaspectratio]{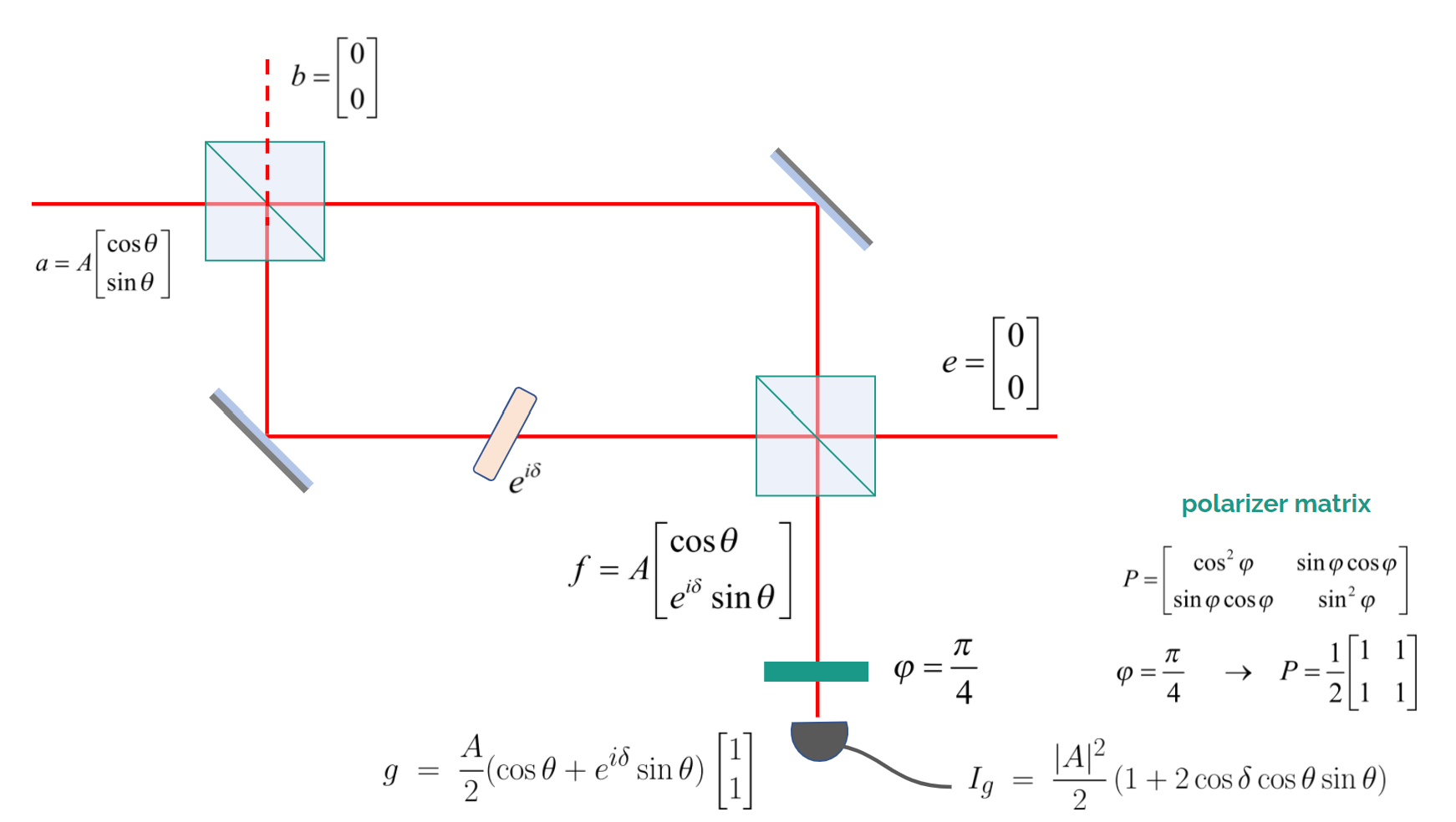}
    \caption{Mach-Zehnder interferometer with polarizing beam splitter. $b$ is the vacuum state b = $\begin{bmatrix}
        0\\
        0
    \end{bmatrix}$.}
    \label{fig:Ch3_20_PBSMZI_g}
\end{figure}

To investigate the inner structure of the output state and its dependency on the phase $\delta$, a polarizer with $\phi = \frac{\pi}{4}$ can be used to analyze output state $g$ (see Figure \ref{fig:Ch3_20_PBSMZI_g}). As seen in Equation \ref{eq:I_g}, in this case the intensity measured depends on the phase $\delta$ and it must be concluded that, as a wave, the single photon have travelled on both paths because the properties we can measure from its state (its intensity) depends on a characteristic of one of the paths.

\begin{equation}\label{eq:I_g}
    I_g \; =\; \frac{|A|^2}{2}\left ( 1 + 2 \cos \delta \cos \theta \sin \theta \right )
\end{equation}

If the light in the input $a$ is polarized at 45$^{\circ}$, the output intensity shows an interference pattern.\\
\begin{figure}[hbt!]
    \centering
 \includegraphics[width=\textwidth,height=\textheight,keepaspectratio]{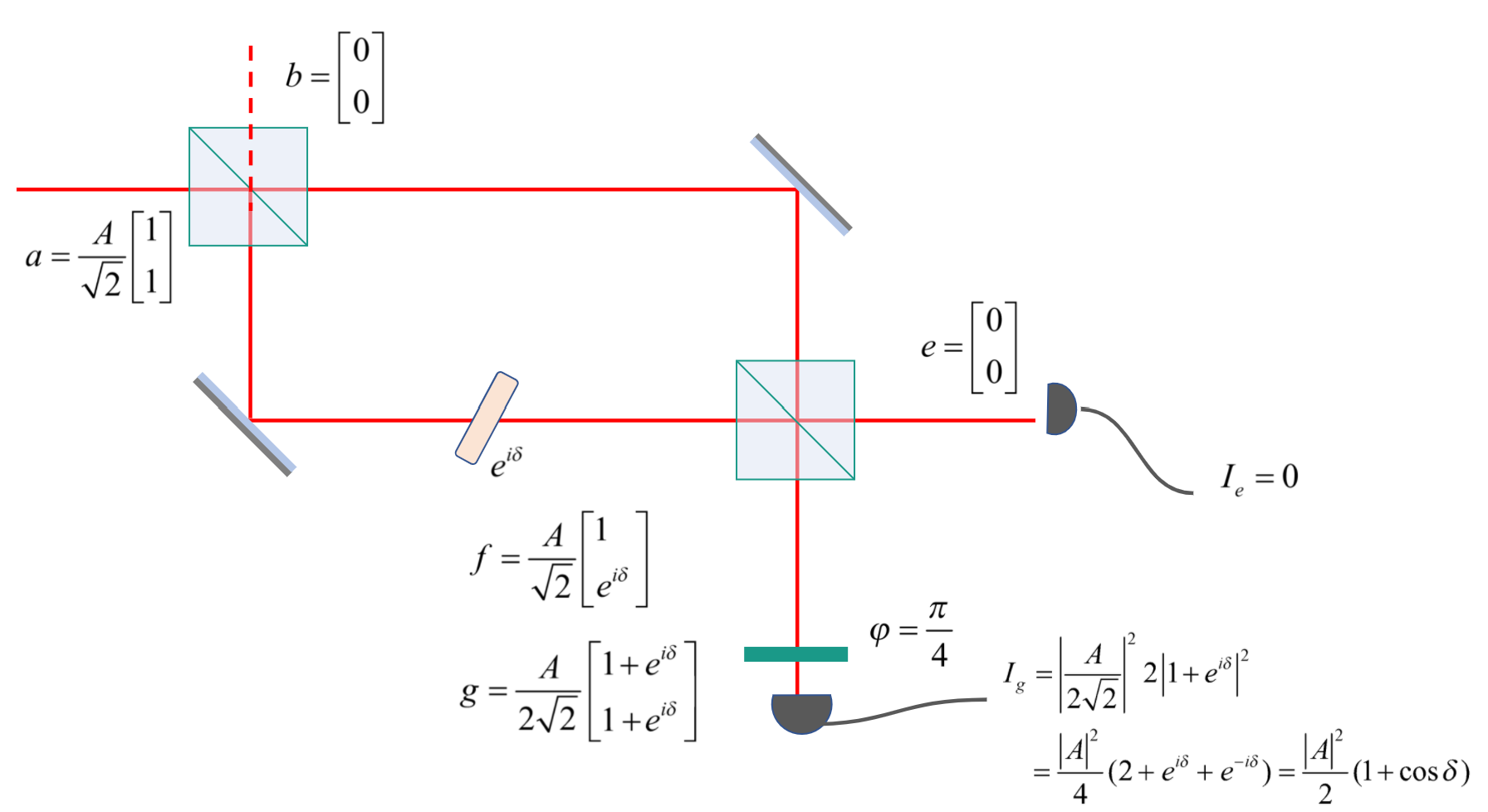}
    \caption{Mach-Zehnder interferometer with polarized beam splitter. The input state is 45$^{\circ}$ polarized}
    \label{fig:Ch3_21_PBSMZI_45input}
\end{figure}

This result is not a surprise if we consider light as a wave. The light at 45$^{\circ}$ can be thought of as a superposition of orthogonal polarisation  components: each component can travel along the two paths and then superimpose again at the end.\\

To extend the inquiry it can be useful to analyse what happens when sending a single photon (see Equation \ref{eq:singlePh_NPBS}) polarized in the 45$^{\circ}$ direction. That is similar to considering a spin state prepared in the Z-up direction that is sent through a X-oriented Stern-Gerlach apparatus (see Section \ref{subsec:SGAsims}). The problem is then approached figuring out whether the state of a sequence of $N$ photons polarised at 45$^{\circ}$

\begin{equation}
    |1\rangle_a |0\rangle_b \qquad \mbox{with} \qquad |1\rangle_a  \equiv |1\rangle_{45^{\circ}} 
\end{equation}

must be considered a mixture or a superposition.
A first interpretative model is to think of the set of $N$ photons as a mixture of $N/2$ horizontally polarized $|1\rangle_H$ photons and $N/2$ vertically polarized $|1\rangle_V$ photons.

\begin{equation}
   N( |1\rangle_{45^{\circ}}|0\rangle_b) \qquad \mbox{with} \qquad N(|1\rangle_{45^{\circ}}) \approx \frac{N}{2} |1\rangle_H \, + \, \frac{N}{2} |1\rangle_V
\end{equation}

\begin{figure}[hbt!]
    \centering
 \includegraphics[width=\textwidth,height=\textheight,keepaspectratio]{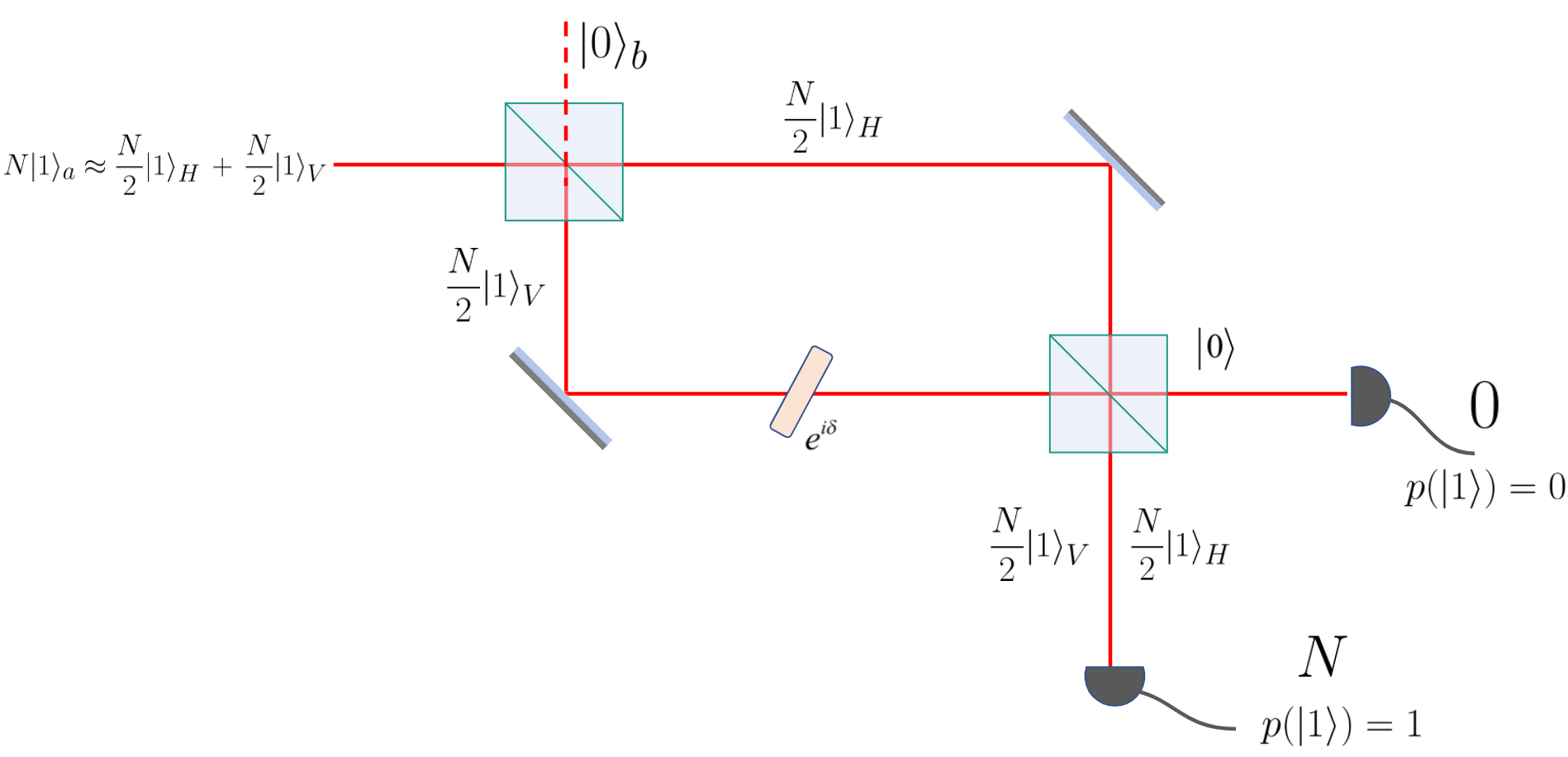}
    \caption{Mach-Zehnder interferometer with polarized beam splitter. Interpretative model using a mixed state. Without the use of polarizer in the output channel the mixed state model is coherent with measurements}
    \label{fig:Ch3_22_PBSMZI_mixed1}
\end{figure}

Without the ${\pi}/{4}$ polarizer after the $f$-output the predictions are consistent with the results of the experimental measurement: the counts reproduce the measured intensities (see Figure \ref{fig:Ch3_22_PBSMZI_mixed1}). All clicks will be detected on one detector (in this case $f$). The phase shift is not detectable since counting photons is equivalent to summing energies that are conserved.\\

Consider now the case with the ${\pi}/{4}$ polariser after the  $f$-output. The experimental results show a phase shift dependence and thus an interference figure similar to that revealed with waves. The mixture-interpretative model considered does not predict that the counts depend on the phase shift.

\begin{figure}[hbt!]
    \centering
 \includegraphics[width=\textwidth,height=\textheight,keepaspectratio]{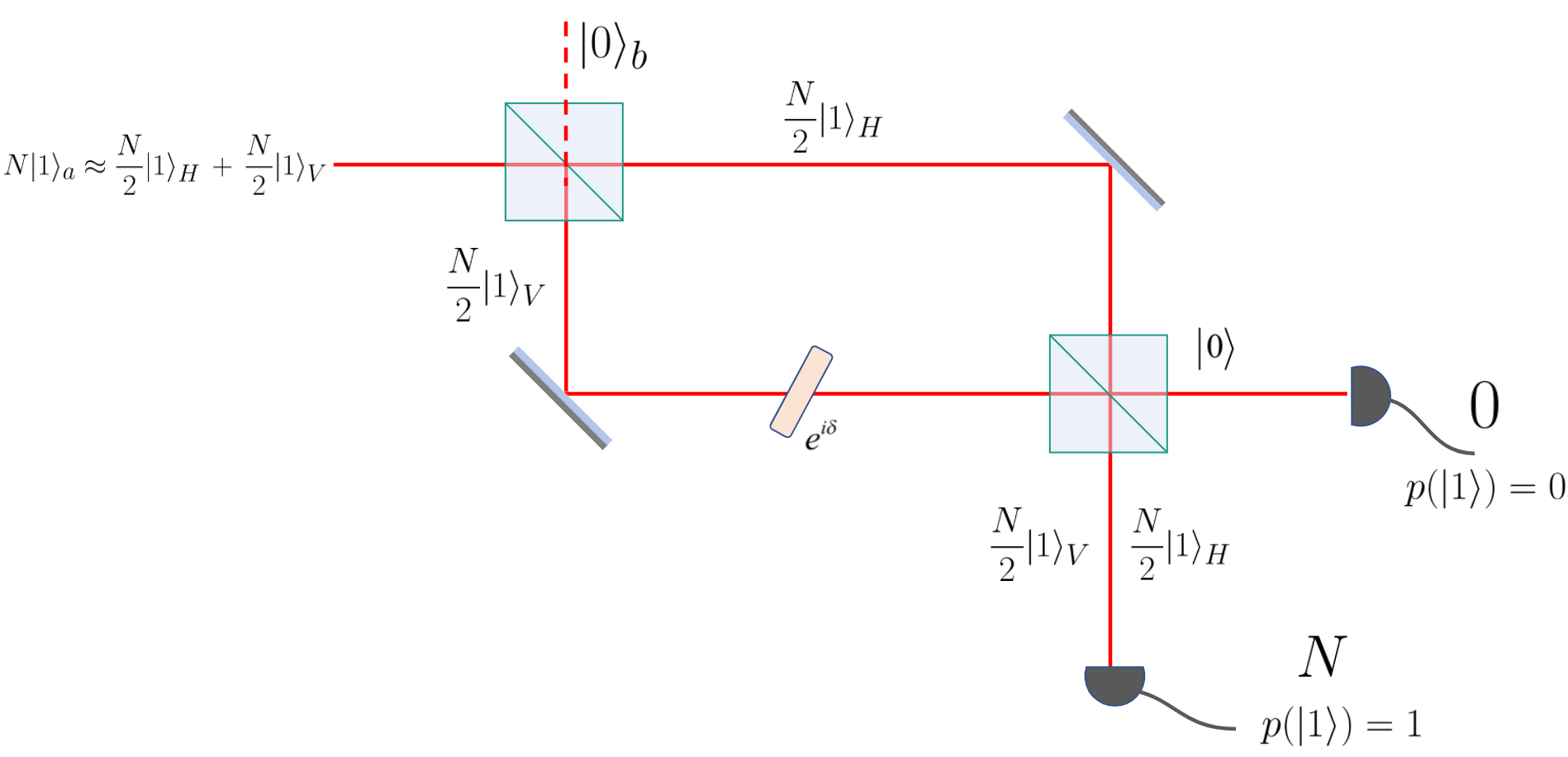}
    \caption{Mach-Zehnder interferometer with polarized beam splitter. Interpretative model using a mixed state. With the use of polarizer in the output channel the mixed state model is not coherent with measurements}
    \label{fig:Ch3_22_PBSMZI_mixed2}
\end{figure}

As in the case of the Stern-Gerlach sequence, the mixed-state model is not consistent with the experimental results. This demonstration should lead us to consider the possibility that the 45$^{\circ}$-polarised single-photon state can be represented as a superposition of horizontally and vertically polarised states. The need of a different interpretative model seems reasonable and that support the process of creating a new meaningful quantum interpretative model \cite{Bouchee21}.\\

A photon polarized at a 45$^{\circ}$ angle can therefore be considered as a superposition of photons horizontally and vertically polarized.\\

\begin{equation}
    | 1 \rangle_{\frac{\pi}{4}} \; = \; \frac{1}{\sqrt{2}} | 1 \rangle_H + \frac{1}{\sqrt{2}} | 1 \rangle_V \; = \; | 1 \rangle \, \otimes \, \frac{1}{\sqrt{2}}\begin{bmatrix}
        1 \\ 1
    \end{bmatrix} 
\end{equation}

With this definition, the outcomes using one single PBS are identical to the ones previously obtained. The probability of detecting a single photon in the two detectors are equal.

\begin{figure}[hbt!]
    \centering
 \includegraphics[width=\textwidth,height=\textheight,keepaspectratio]{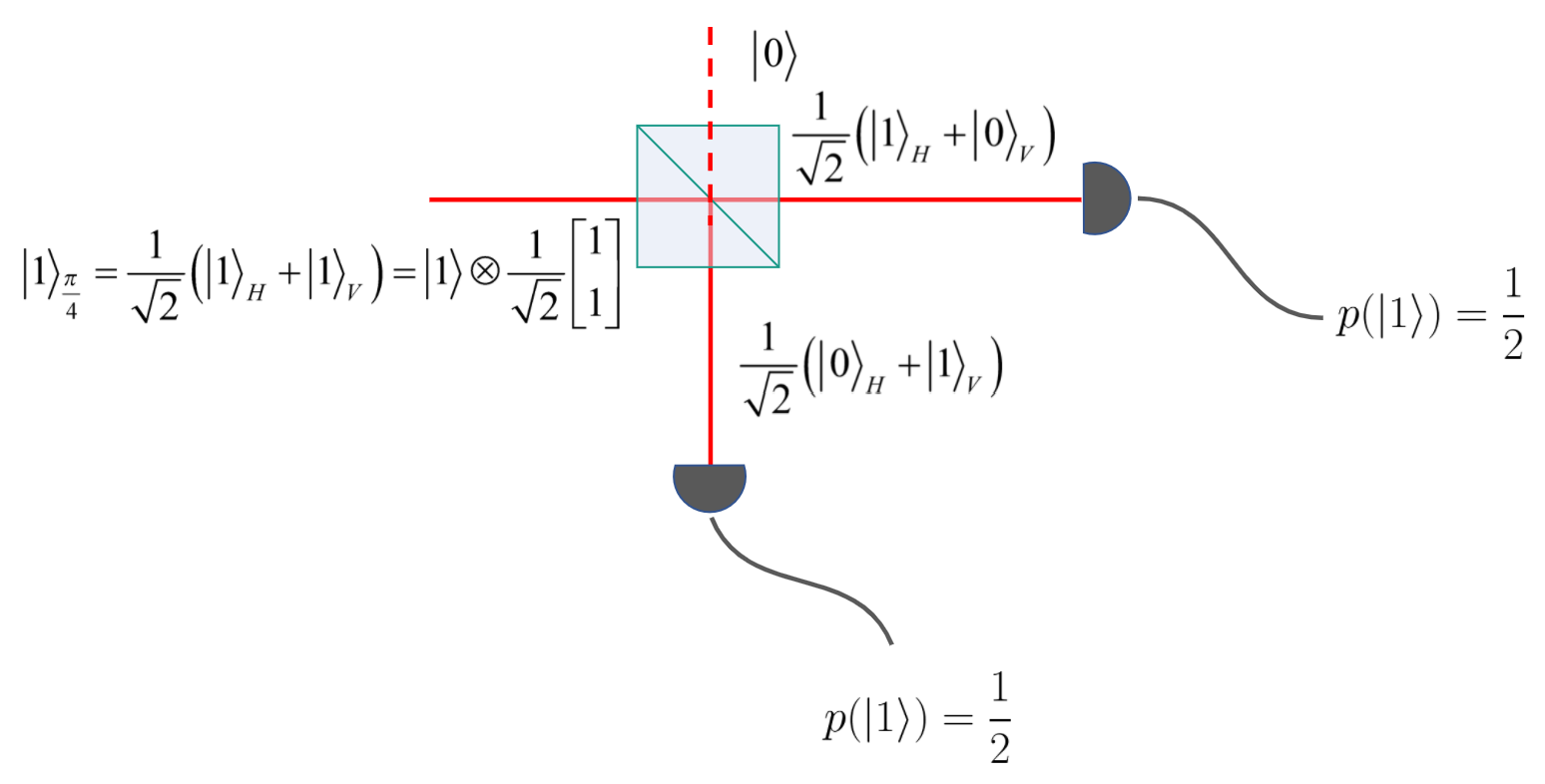}
    \caption{Polarized beam splitter (PBS) with the input state is a superposition state. The PBS sends the horizontal and vertical component along different paths}
    \label{fig:Ch3_35a_PBSsingle45}
\end{figure}

Completing the interferometer with the second PBS, the state at the two outputs can be measured. Also in this case (see Figure \ref{fig:Ch3_19_PBSMZI_f}) the photons are all collected at one output and the state depends on the phase shift $\delta$.

\begin{figure}[hbt!]
    \centering
 \includegraphics[width=\textwidth,height=\textheight,keepaspectratio]{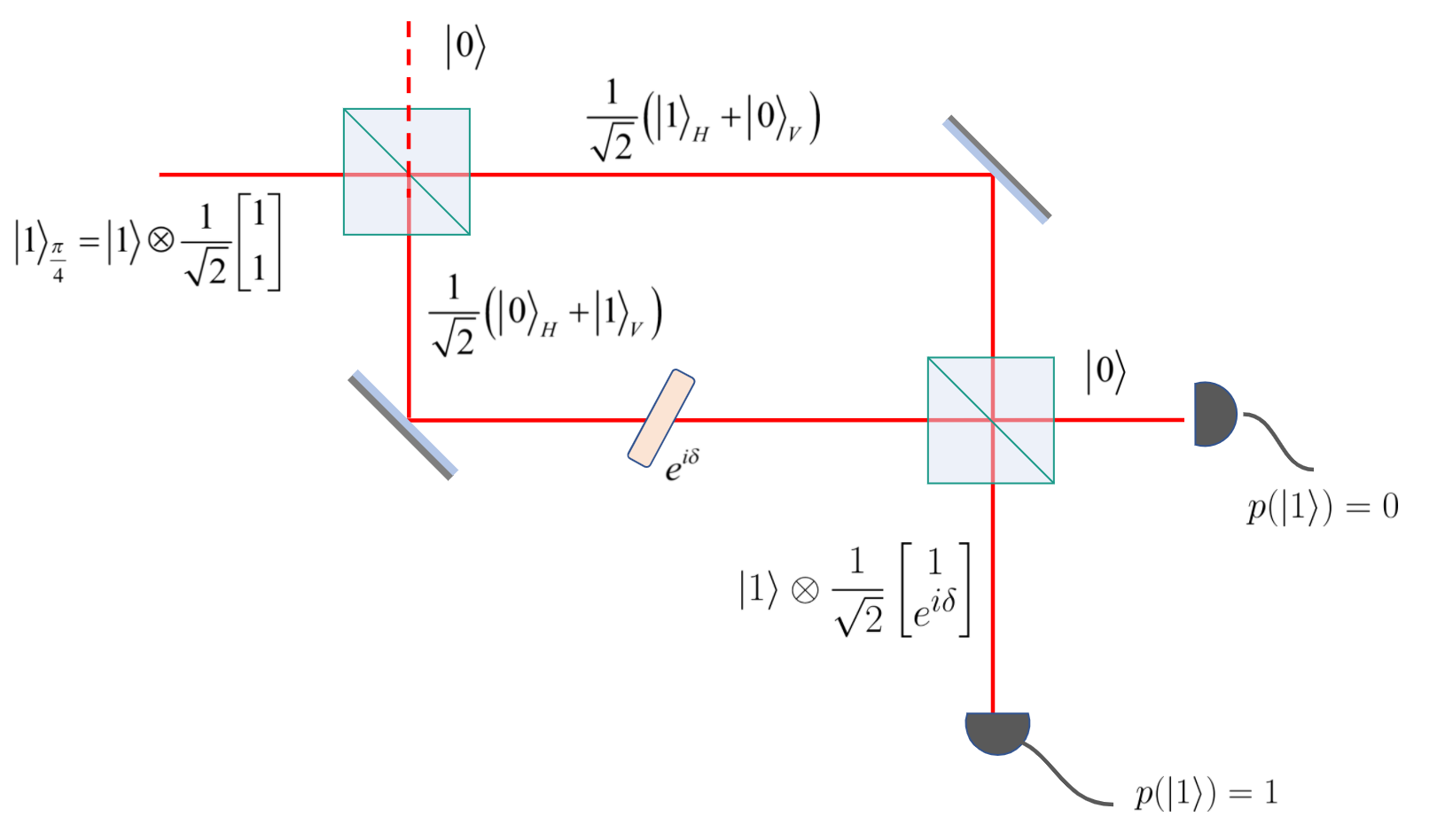}
    \caption{Mach-Zehnder interferometer with polarized beam splitter. The input state is a superposition state. The ouput state depends on the phaseshift $\delta$. All the detections are registeret at the same output $f$}
    \label{fig:Ch3_35b_PBSMZI45f}
\end{figure}
If we want to analyse the state we first use a 45$^{\circ}$ polariser and what we get is a phase-shift dependent intensity, consistent with the measurements.
\begin{figure}[hbt!]
    \centering
 \includegraphics[width=\textwidth,height=\textheight,keepaspectratio]{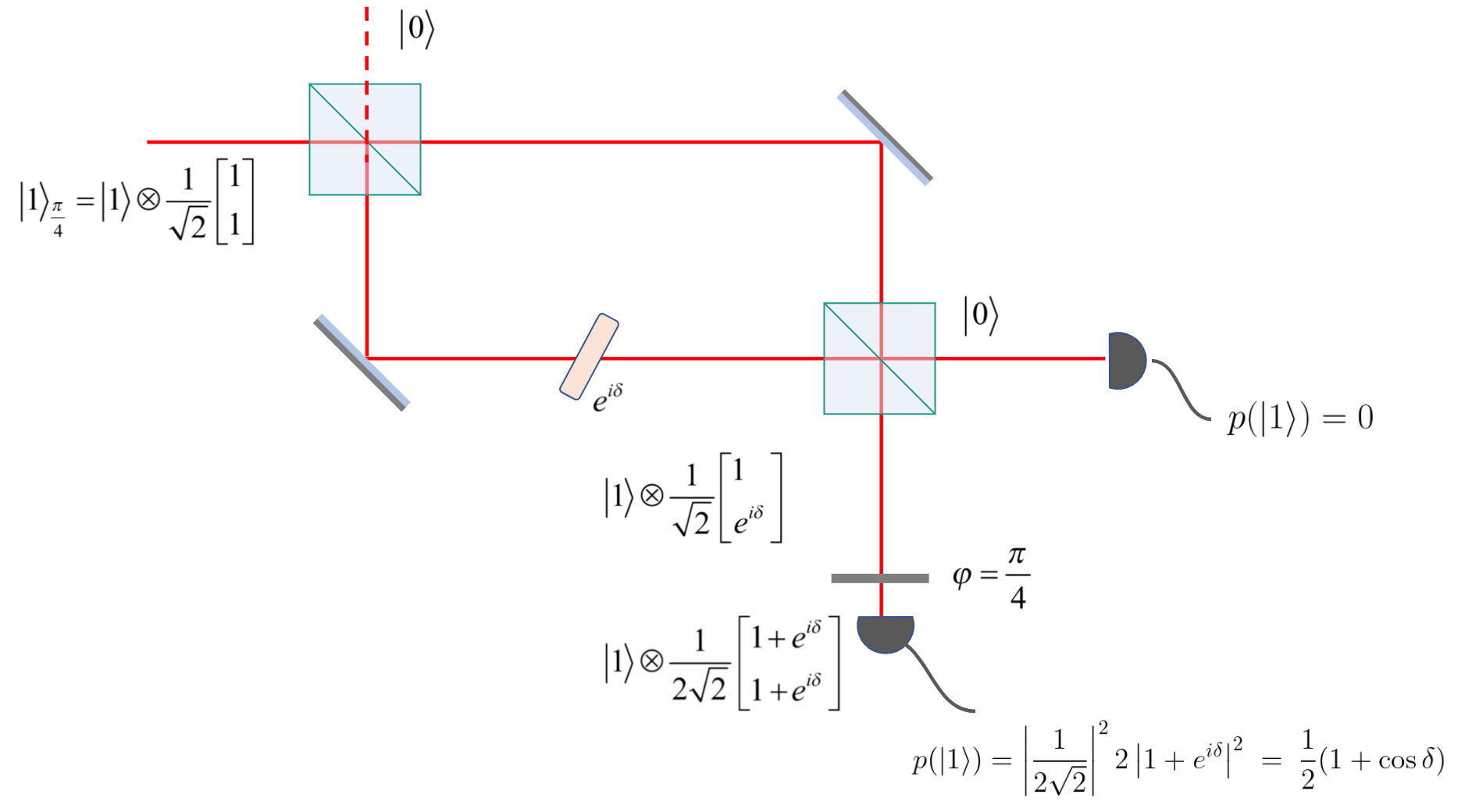}
    \caption{Mach-Zehnder interferometer with polarized beam splitter. The input state is a superposition state. Using the polarize show how the output state depends on $\delta$}
    \label{fig:Ch3_35c_PBSMZI45g}
\end{figure}

In conclusion the results of the experiments performed with  a Mach-Zehnder interferometer with polarized beam splitters are analogous for classical monocromatic light with constant amplitude (i.e. laser beam) and for a single photon.\\
The point is that the single photon has to be thought of in a state of superposition of polarizations that coexist until the photon is measured.
The difference between the two cases is related to what has been measured to extract information from the two different states: in the classical case, the quantity that shows interference is the intensity of the light; in the quantum case, it is the number of counts which, in the case of single photons, represents the probability of observing the photon.\\
The properties of the interferometer (sending light only to one side and showing interference) depend on the properties of the polarisation and not on the type of state of light being considered.

\section{Conclusions}
The use of the qubit as a minimal piece of information about a physical system can be studied by referring to the quantum physical systems that can represent it. Our choices have been the electron spin states and the polarization states of the single photon.

Referring to the research questions at the beginning of the chapter.

\begin{itemize}
  \item[RQ5.1] How is it possible to characterise the quantum nature of physical objects in order to encode information?
\end{itemize}  
It possible to use quantum objects to encode information by associating such information to the quantum state of that object instead of its properties. In this way we can study the process of state evolution in the interaction with devices using the specific properties of these states, such as superposition. The information can be then extracted through a measurement process. The present chapter presents how both spin and single photon polarization states can be used to illustrate such processes.

\begin{itemize}
  \item[RQ5.2] Through which type of representation is it possible to describe the process by which the information associated with quantum objects can be transformed inside a physical apparatus and extracted through measurement?
\end{itemize}

Representing states as vectors and observables as matrices it is possible to describe the process by which the information associated with quantum objects can be transformed inside a physical apparatus and extracted through measurement. The way in which this formalism is used is the same, regardless of the particular type of physical object (electron or single photon). The way in which states are represented allows a transposition to the way in which qubits are processed within quantum computation processes. It is possible to parallel the action of operators on states and quantum gates on qubits.
\begin{itemize}  
    \item[RQ5.3] Which experimental contexts are significant for the study of how quantum objects can be used to encode information?
\end{itemize}

The significant experimental contexts identified are related to the use of the Stern-Gerlach apparatus and the Mach-Zehnder interferometer. Familiarity with these devices can support the exploration of the operating principles of quantum technologies, for example in cryptographic protocols (see Section \ref{subsec:crypto}).
The use of simulators \cite{QuVisSim} makes the use of the qubit as the spin state of the electron particularly exploitable in educational terms.Although these experiments cannot be performed directly in a school laboratory, simulations can facilitate the active learning processes of students \cite{CrouchMazur2004_classroomdemo} by enabling them to make predictions or test explanatory hypotheses. 
The use of polarisers and the Mach Zehnder interferometer opens up the possibility of exploring the quantum nature of quantum objects such as light through the use of experimental apparatus. These kinds of instruments can be used directly by the students or be part of a guided exploration in a well-equipped universitary laboratories \cite{Bondani2014_SinglePhoton}, \cite{Bitzenbauer2021_herald}.
\chapter{\label{ch:6-LETeachers} Learning environments for teachers}

This chapter presents the work done in collaboration with local teachers. The general objective is to analyse processes that can support teachers in developing the necessary competences to implement a contemporary approach to quantum physics learning in their classroom activities.

The strengthening of cohesion within the learning ecosystem (Strategy 1 see Table \ref{tab:LET_RQIncipit}) is an essential strategy to foster dialogue, to enrich the cultural capital of the community and to create opportunities for the design of activities that improve educational practice.


At the same time, it is necessary to develop an approach to teaching (Strategy 2 see Table \ref{tab:LET_RQIncipit}) that can combine the institutional constraints of the national physics curriculum with recent developments in research related to quantum technologies. (see Table \ref{tab:LET_RQIncipit}).

The elements for the study of PCK and the development of design tools become the "indicators" through which these strategic processes are monitored during the project and provide the basis for the evaluation of the effectiveness of the actions undertaken.

\begin{table}[hbt!]
\centering
\begin{tabular}{p{0.1\textwidth}p{0.4\textwidth}p{0.4\textwidth}}
\multicolumn{3}{c}{\textbf{Learning environments for Teachers}} \\
\hline
& &\\
&\multicolumn{1}{c}{\textit{Strategy 1}} & \multicolumn{1}{c}{\textit{Strategy 2}} \\
&\textbf{Generating cohesion in the learning ecosystem} & \textbf{Creating a teaching approach to quantum physics}\\
\hline
& & \\
\footnotesize{\textbf{\textit{PCK study}}} & RQ6.1 How can the collaboration with physics researchers foster teachers' PCK about contemporary quantum physics? &
RQ6.3 What are the elements of the informational approach  (see Section \ref{sec:ApprPillars})  that can help the development of teachers'PCK? \\
&  & \\
\footnotesize{\textbf{\textit{Design tools}}} & RQ6.2  How to support teachers in designing teaching-learning activities  about the interpretation of quantum experiments? & RQ6.4  What role can quantum technologies play in creating learning environments for teaching and learning quantum physics?
\end{tabular}
\caption{Research Questions related to the design of learning environments for teachers. Each question is related to the project strategies. The study of teachers' PCK and the development of design competences are the "indicators" have been use to moitor the processes}
\label{tab:LET_RQIncipit}
\end{table}


The specific actions planned and carried out in the last two years concern the creation of moments of dialogue between teachers and researchers to promote reflection on the different approaches and methodologies for teaching quantum physics and the co-design of teaching activities for students.
The project actions specifically concern the realisation of two editions of Professional Develompent Programmes (PD) used to detect criticalities and elaborate proposals about teaching quantum physics. The first PD programme "Quantum Skills" (see Section \ref{sec:QSkills}) was implemented between October 2019 and May 2020. The second PD "Quantum Jumps" (see Section \ref{sec:QJumps}) was implemented between November 2020 and June 2021.

In the following Sections we present a brief description of the context in which the project took place (see Section \ref{sec:TLE_context}) followed by a description of the two Professional Development programmes (see Section \ref{sec:QSkills} and \ref{sec:QJumps}) and the other actions for teachers (see Section \ref{sec:TLE_focus_interview}).

\section{Elements from the context}\label{sec:TLE_context}

In Italian schools, quantum physics aims to study the microscopic world, culminating in the construction of an atomic model capable of justifying a series of experimental results. This discussion has many parallels with what students have already seen in Chemistry. The standard sequence \cite{MIUR2010_211, MIUR_Esami21} starts from the study of light quantization through the study of thermal radiation and Planck's hypothesis (addressed in a qualitative way). This is followed by the study of the photoelectric effect and its interpretation by Einstein, and the discussion of the theories and experimental results showing the presence of discrete energy levels in the atom (through the Frank-Hertz experiment). The analysis of the experimental evidence for the wave nature of matter (e.g. de Broglie hypothesis) and the uncertainty principle are indicated as the conclusion of the course.
Once the model to describe the behaviour of the microscopic elements of matter has been formed, the programme suggets to link this model to technological applications. In some textbooks \cite{romeni2017fisica,amaldi2020il} there are references to solid states of matter, such as the study of the function of semiconductors and their link with LEDs or the function of solar panels.\\

This type of organisation of the contents favours the resumption of some salient stages in the historical evolution of quantum physics. Some of the limitations of this choice have already been identified (see. Section \ref{Ch3.1_ApproachShape}) mainly related to showing the inadequacies of classical physics rather than building a complete theory \cite{SusskindQT}. The focus is on the \textit{process of affirmation} of new ideas rather than promoting the investigation around a coherent set of phenomenomena that are the fundations on which contemporary quantum physics is built \cite{BessonMalgieri2018}.

The activities carried out with teachers have therefore had the general aim of defining how a more meaningful link can be built between what is taught in school and the theoretical framework underpinning quantum physics and its recent technological applications.
It is not the intention, however, to update the curriculum by simply trying to find ways of adding previously neglected topics. As seen in the previous Chapters, the aim is to provide the elements to redefine the approach to the physical description of reality that the quantum revolution has generated.(see Ch.\ref{ch:3-QApproach}).
The following is a description of the PD programmes designed for in-service secondary teachers (see Section \ref{sec:QSkills} and Section \ref{sec:QJumps}) and teaching-learning sequence proposals designed by teachers from the reflections and ideas collected in the Professional Development programmes (see Section\ref{sec:Teachersdesign}).


\section{Quantum Skills 2019 Professional Development programme}\label{sec:QSkills}
In 2019 we have designed and implemented Quantum Skills, a continuous professional development program for teachers to enable in-service physics teachers in secondary schools to introduce the superposition principle, quantum entanglement, and their technological applications into regular classroom activities.
As first edition of this PD programme, one of the main focus was to collect information about teachers' practice related to quantum physics.

\subsection{Activities design}
The need to develop scientific competences related to the key concepts of quantum physics can foster the creation of meaningful teaching and learning sequences. These educational activities should be coherent with current physics research strands and be also integrated in regular instruction activities at school.
In the project framework, we address the following research questions:


\begin{table}[hbt!]
\centering
\begin{tabular}{cp{12cm}}

 & \textbf{Learning Environments for Teachers} \\
 & \textit{Collaboration to design teaching and learning sequences} \\
\hline
\rule[-4mm]{0mm}{1cm}
RQ6.1 & How can the collaboration with physics researchers foster teachers' PCK about contemporary quantum physics? \\
\midrule
\rule[-4mm]{0mm}{1cm}
RQ6.2 & How support teachers in designing teaching-learning activities about the interpretation of quantum experiments?
\end{tabular}
\caption{Research questions: Learning Environment for teachers, Quantum skills 2019}
\label{tab:TLE_RQ_QSkills}
\end{table}

\subsection{Methods}
\begin{figure}[hbt!]
    \centering
 \includegraphics[width=\textwidth,height=\textheight,keepaspectratio]{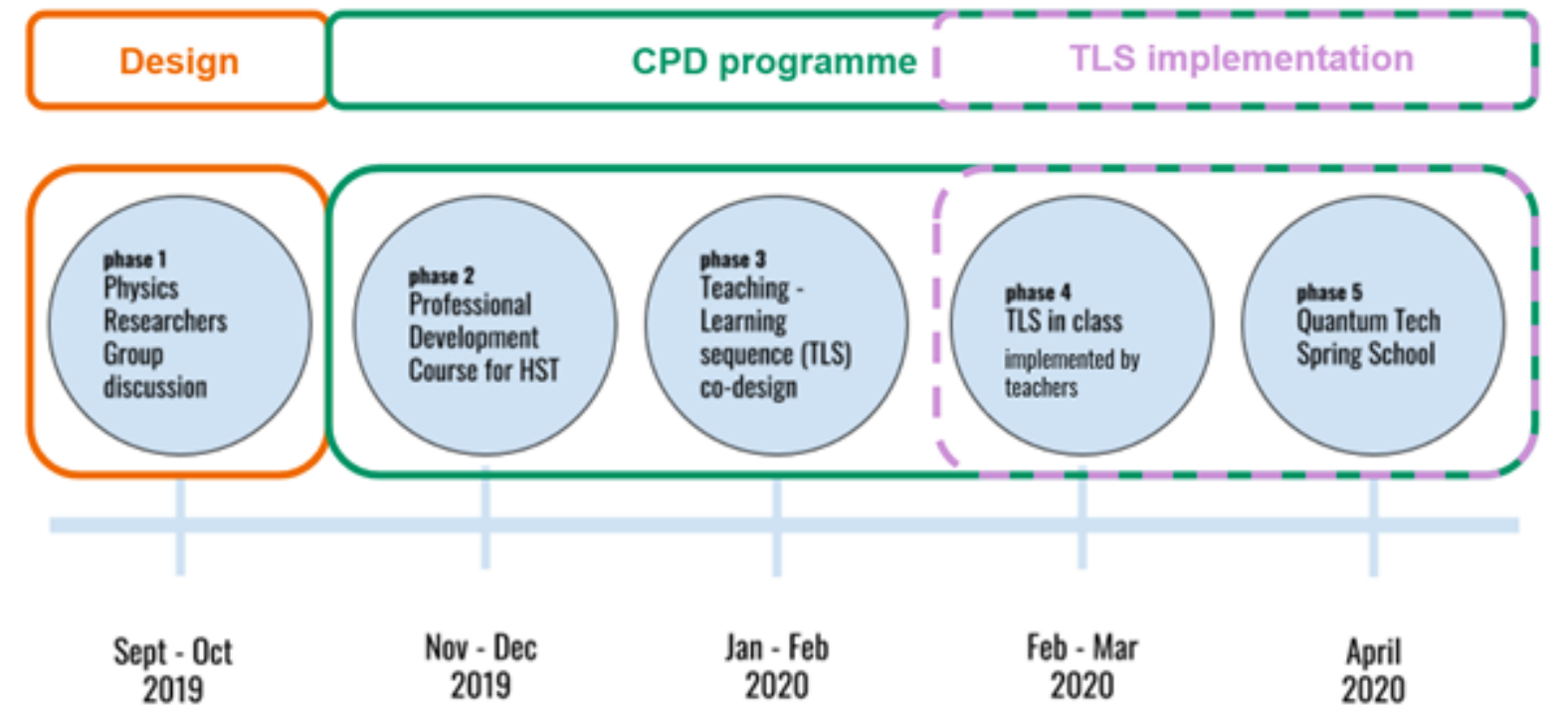}
    \caption{Quantum Skills Professional Development programme timeline. Phase 1 have been dedicated to the definition of the core concepts. In Phase 2 and 3 the programme was implemented, followed by a co-design activity of teaching learning sequences to be proposed in the classes. Phase 4 and 5 relate to actions carried out with students}
    \label{fig:Ch6_1_QSK_sequence}
\end{figure}

The PD programme aim is based on a analysis of the difficulties and the possibilities in teaching quantum physics at the high school level followed by a classroom testing of a teaching-learning sequence (TLS) based on the results of the first part (see Figureure  \ref{fig:Ch6_1_QSK_sequence}).

\begin{figure}[hbt!]
    \centering
 \includegraphics[width=\textwidth,height=\textheight,keepaspectratio]{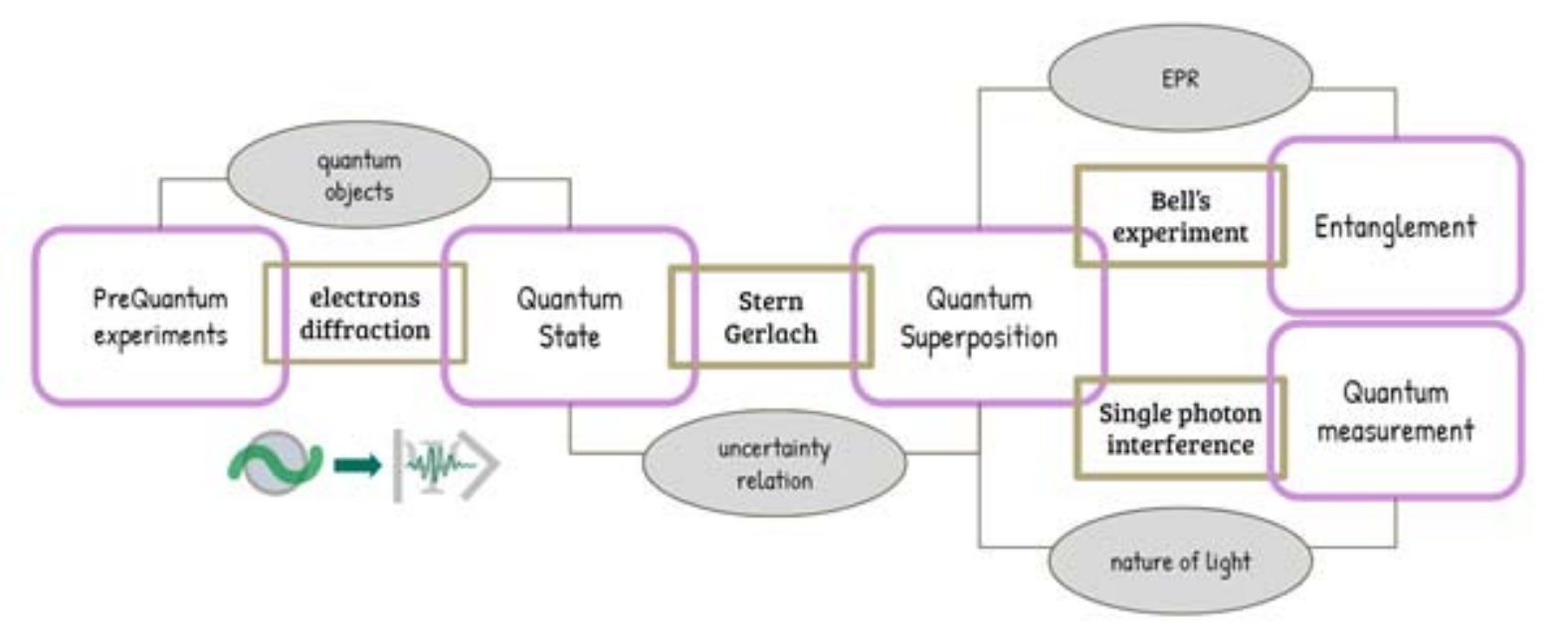}
    \caption{The main structure of Quantum Skills Professional Development programme. Linking the mandatory topics, the central concept of quantum state have been introduced. Superposition and entanglement have been explored using simulations}
    \label{fig:Ch6_2_QSKPDstructure}
\end{figure}

The Quantum Skills project began in September 2019 in the context of the Piano Lauree Scientifiche (PLS). The design process started as the result of a discussion within a group of active quantum researchers from the University of Insubria, the University of Pavia and the University of Milan (see Section \ref{Ch3.1_ApproachShape}).
During the PD, a group of 29 teachers discussed the conditions for these concepts to be part of regular teaching activities with their students. The PD was structured into 5 weekly afternoon meetings of 2 hours each (see Figureure \ref{fig:Ch6_2_QSKPDstructure}). Starting with early historical quantum physics experiments (blackbody radiation spectrum, photoelectric effect), teachers moved on to discuss the tenets of contemporary quantum physics. Through the interpretation of quantum experiments results, the participants reflected on what characteristics of quantum objects are worth to be taught at high school level and how to support their students in the transition from classical to quantum vision of reality.

To facilitate the teacher participation, we used an online platform for live-streamed lessons. Course materials and video recordings of the lessons were also shared with teachers. Meanwhile, we collected teachers’ reflections on the lesson topic and its pedagogical impact using online surveys.\\
Then, a group of three teachers participated in a series of 6 meetings to discuss ideas and define different phases of learning activities to implement with their students. In the end, these teachers tested the teaching-learning sequence they designed and the materials they prepared in their Grade 12 classrooms online activities between March and June 2020 during the Covid19 school lockdown.\\
At the beginning of PD, an online questionnaire was used to collect teachers' expectations of the course and what they thought would be helpful in making quantum physics part of their classroom activities.\\
During each PD class, the analysis of students' prior knowledge and learning difficulties as found in the literature \cite{BessonMalgieri2018, Malgieri2017,Kim2017} was shared and discussed with teachers. The starting point was the about how to introduce the concepts they normally teach in school \cite{MIUR2010_211}, such as duality or the Heisenberg uncertainty relation. At the end of each meeting, participants were invited to write in an online form their reflections on the topic of the lesson, focusing on the subject-specific knowledge they need to introduce the main concepts to their students and the pedagogical approaches they think will facilitate students' understanding.\\
A final questionnaire was used to collect teachers' evaluation of the course with a focus on how well the PD met their expectations. Only 18 teachers out of 29 completed the exit questionnaire. All of those teachers had participated to all the lesson of Phase 2.\\
For the second part of the PD about the teaching-learning sequences design and implementations,teachers’ thoughts and considerations were collected using field notes and post-activity reflections written by the teachers. All the results for this specific phase are reported in Section \ref{sec:Teachersdesign}.

\subsection{Teachers' participation}
\begin{figure}[hbt!]
    \centering
  \includegraphics[width=\textwidth,height=\textheight,keepaspectratio]{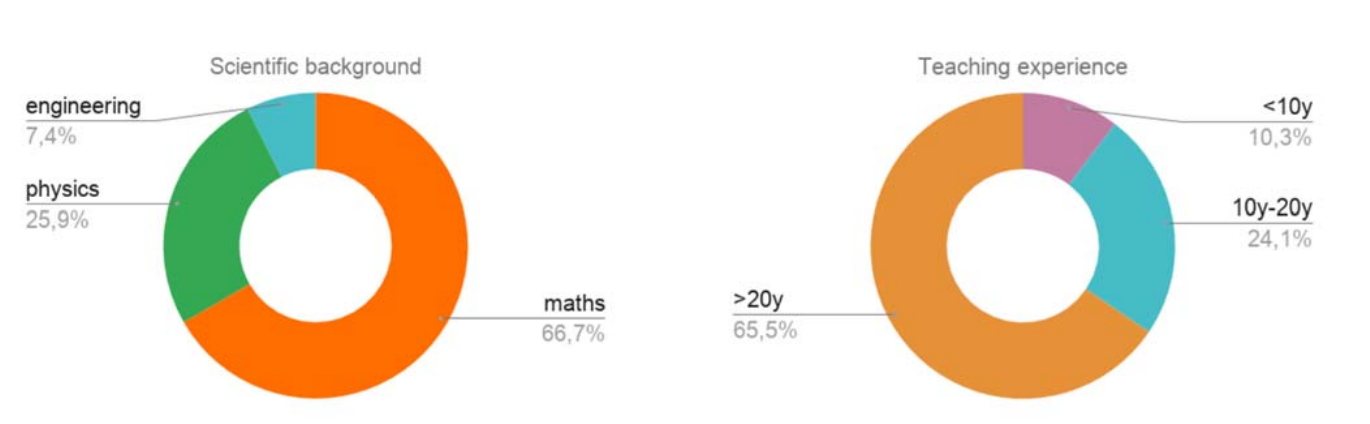}
    \caption{Quantum Skills teachers' general profile}
    \label{fig:Ch6_3_QSKPDTeachersProfile}
\end{figure}

In-service teachers who participated in the program have extensive experience in teaching physics at the high school level. Most of them have a master’s degree in mathematics (see Figureure \ref{fig:Ch6_3_QSKPDTeachersProfile}). This fairly closely resembles the general profile of physics teachers in the Como School District.

\begin{figure}[hbt!]
    \centering
 \includegraphics[width=\textwidth,height=\textheight,keepaspectratio]{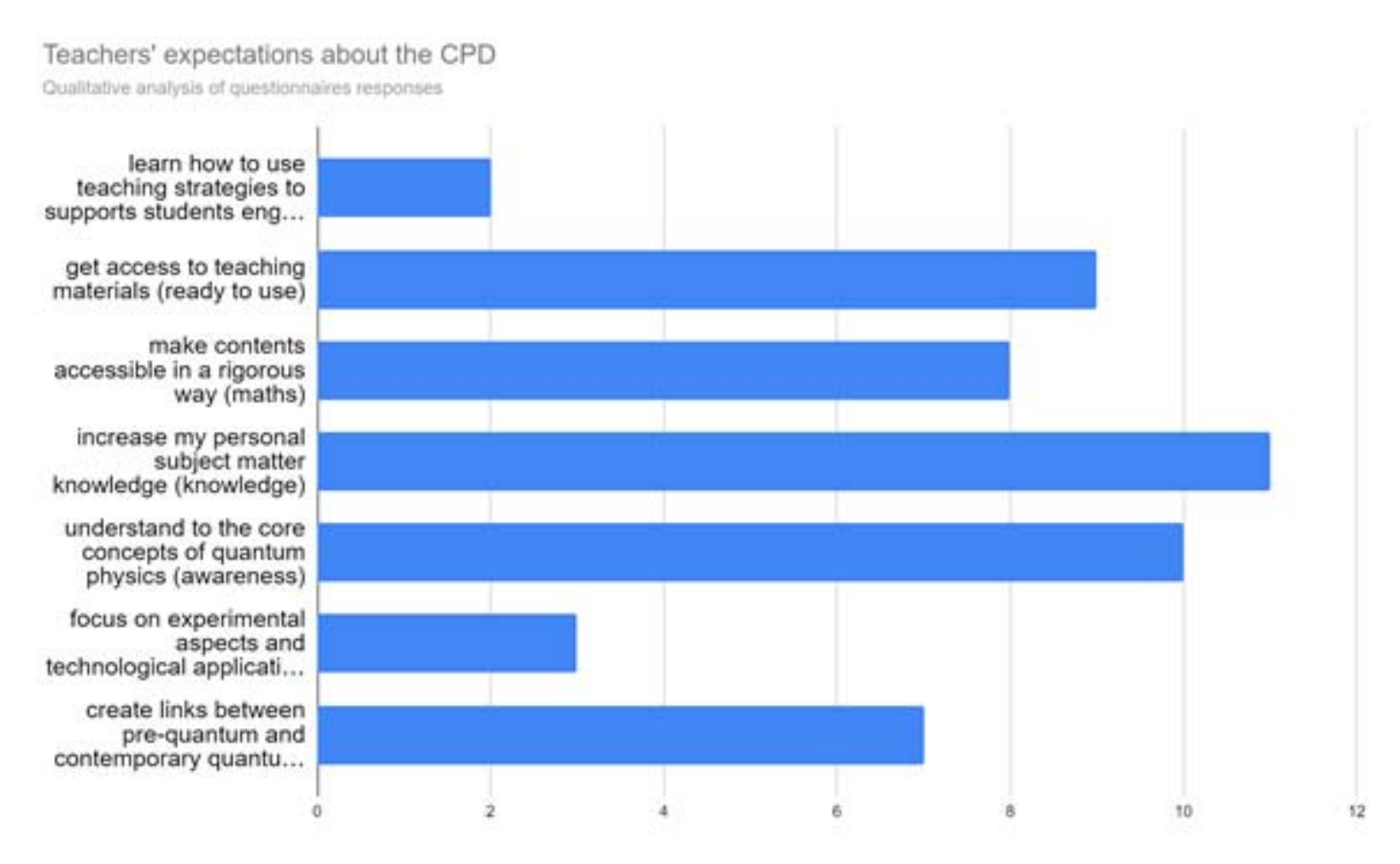}
    \caption{Quantum Skills teachers' expectations}
    \label{fig:Ch6_4_QSKPDTeachersExpectations}
\end{figure}

This is a group of "self-selected" teachers, in the sense that they chose to participate in the course voluntarily, motivated by a desire to enrich their professional practice. This element has to be taken into account when assessing the applicability of the results of educational design to a general school context. These are teachers who want to bring about change in their daily practice and this attitude is not always easy to generalize to other teachers.
Teachers' expectations were mostly focused on improving their specific knowledge of quantum topics and getting materials ready to use in the classroom with their students (see Figureure \ref{fig:Ch6_4_QSKPDTeachersExpectations}).

Their main request was about increasing personal awareness of the essential core concepts of quantum physics. In the words of one of the participants:
\begin{quote} {\fontfamily{cmtt}\selectfont
to go beyond the experiments described as "the crisis in classical physics" to finally see "what happens next"… what quantum physics looks like now.[Qsk19EnQ]
}\end{quote}

That is to create a meaningful relationship between the quantum physics topics taught at school and the actual research activities in that field.

A second emerging expectation was to have access to materials that can be used in classroom activities. Teachers reported the lack of teaching resources on quantum physics and the difficulty they had with their students in creating active learning activities related to the experimental aspects of quantum physics. As one teacher wrote in the entry questionnaire:

\begin{quote} {\fontfamily{cmtt}\selectfont
Quantum is very theoretical \dots I have difficulties getting my students access to quantum experiments [QSk19EnQ]
}\end{quote}

\subsection{Results}
The main results of this activity with teachers concern the initiation of a reflection on a way of introducing quantum physics in secondary school lessons that departs from traditional activities. These ways present some criticalities that have been pointed out by the teachers about the difficulties students may encounter. At the same time, however, the fundamental contribution of being able to discuss the results of experiments was highlighted.

\subsubsection{RQ6.1: Collaboration to support PCK development}

The main result of the collaboration was to initiate a reflection with the teachers on how to organise the contents starting from the identification of some central concepts linked to the choice made on the teaching objectives. The aim of the activities is to make students understand the specific nature of quantum objects and the need to introduce a new interpretative paradigm \cite{DM769_18Quadri}. The discussion promoted with teachers therefore revolved around the question of what fundamental concepts should be introduced in lessons with students so that they can understand the difference between classical and quantum physical objects. A first element is to have made explicit the centrality of the concept of quantum state. A second element comes from the opportunity to use qubits to explore the specifically quantum behaviour of simple physical systems. The main topic of discussion was how this can be integrated into teaching practice.

As for the subject matter knowledge for teaching (see Table \ref{tab:PCK_Analysis_GRID}), the choice of the qubit approach (see Chapter \ref{ch:3-QApproach}) proposed by quantum researchers helped teachers to directly address the non-classical nature of quantum objects. Researchers used the idea of quantum state as a starting point to address the problem of interpreting quantum experiments such as Stern-Gerlach measurements of the electron spin and quantum entanglement.

The need expressed by the teachers was to include, however, those contents that are linked to the first phase of the development of quantum mechanics, such as the photoelectric effect. As expressed by one of the participants

\begin{quote} {\fontfamily{cmtt}\selectfont
    Certain things (e.g. photoelectric effect, Bohr's atom) cannot be excluded from what you do in class. Maybe you can try to reduce the time you spend on them, but you can't eliminate them.[QSk19ExQ]
}\end{quote}

However, what was proposed stimulated reflection on how it is possible to show students how the theory has evolved from those first steps and what developments can be covered in class.

\begin{quote} {\fontfamily{cmtt}\selectfont
    Certain types of experiments (e.g. MZI) are really useful for understanding the problem \dots as it relates to theory.[TeachRefl]
}\end{quote}

A second relevant element are related to the student's conception about the use of mathematics. In agreement with the literature on physics education \cite{Pospiech1999_EPRatHS, Bouchee21} (see Section \ref{sec:ChangeL}), teachers pointed out that quantum physics is challenging at the high school level because of the mathematical formalism required. One of the teachers put it this way:

\begin{quote} {\fontfamily{cmtt}\selectfont
It is too demanding for the students and exceeds their skills![TeachRefl]
}\end{quote}

Collaboration with quantum physics researchers enables teachers to identify the minimum mathematical tools needed to link the interpretation of quantum experiments to core quantum concepts: a selection of linear algebra tools to describe the evolution of quantum states interacting with the experimental apparatus (see Section\ref{sec:qubit_characterization} and Section \ref{sec:seqSGA_matrix}). The teachers expressed concerns about the use of such mathematics. Even though these tools are part of the Italian mathematics curriculum \cite{MIUR2010_211}, they argued that they are not part of how such curriculum is "put into practice" and that it is not possible to add these topics to what is covered in school. As one teacher wrote in the meeting's feedback questionnaire,

\begin{quote} {\fontfamily{cmtt}\selectfont
    we can't add such a piece of math to a physics course: they have some notion about vectors, but no clue about matrices...eigenvectors and eigenstates are completely off the spectrum! [QSk19ExQ]
}\end{quote}

With regard to educational structuring aspects, the main contribution was to reinforce the idea of always showing a strong parallelism between theoretical interpretative structures and experimental results..
The idea was to use quantum experiments as a context in which those pieces of mathematics are useful for making predictions about the experimental results. The main point was not to "add content" but to create a meaningful relationship between the physical phenomena and the mathematical model and representation that could be used to interpret the phenomena \cite{Bouchee21}.
For example, the description of the Stern-Gerlach experiments was based on the use of Dirac notation and the linear algebra formalism: the evolution of the quantum state is represented by the action of the operator (matrix) on the vector state (see Section \ref{sec:qubit_characterization}. The same approach has been used to describe the interaction between light and polarizers in a polarizer sequence or in a Mach-Zehnder interferometer (see Section \ref{subsec:MZIexper}).


This was seen as an opportunity for teachers to be trained and enriched primarily on a personal level.

\begin{quote} {\fontfamily{cmtt}\selectfont
    Honestly, this is the first time I've seen quantum mechanics treated in these terms, it's definitely not the way I'm used to telling it at school or found in textbooks. For me it was enriching.[TeachRefl]
}\end{quote}

\subsubsection{RQ6.2: Interpretation of quantum experiments}
The proposed idea of structuring the introduction of quantum concepts through meaningful experiments \cite{DM769_18Quadri} was viewed positively by teachers. As reported by one of the participants

\begin{quote} {\fontfamily{cmtt}\selectfont
    I think that using experiments facilitates the explanation a lot. It gives students a sense of reality. It is also the way in which these things are introduced, even in other areas of physics (e.g. mechanics). Students are used to this. [TeachRefl]
}\end{quote}

Teachers identified a guided exploration of quantum experiment as a possible learning sequence to introduce the main characteristics of quantum objects to their students. In this sense, the use of simulations has been identified by many as an important resource.

\begin{quote} {\fontfamily{cmtt}\selectfont
[\dots] in fact, one of the main reasons I attended the course was to receive suggestions on how to experiment, at least virtually, with what was explained, which is impossible to do in high school laboratories.[QSk19ExQ]
}\end{quote}

This reduced the impact of mathematical formalism on the teaching-learning sequence and allowed students to be actively engaged in online lessons. In one of the teachers' words:

\begin{quote} {\fontfamily{cmtt}\selectfont
    I am convinced that the mathematics of quantum physics is too difficult, at least for the students I work with\dots Simulations, on the other hand \dots I think are more accessible and allow you to understand how things work without putting too many obstacles in the way. [QSk19Obf]
}\end{quote}

Another teacher pointed out how the interplay between mathematics and physics can become a useful tool but under certain conditions

\begin{quote} {\fontfamily{cmtt}\selectfont
    Before I can introduce that kind of formalism with my students \dots I definitely need to review how to introduce these things. With a little more work the students can learn how to use it. [QSk19ObF]
}\end{quote}

In general, course participants evaluated the course positively, mainly as an opportunity for personal cultural enrichment.

\begin{figure}[hbt!]
    \centering
    \includegraphics[width=\textwidth,height=\textheight,keepaspectratio]{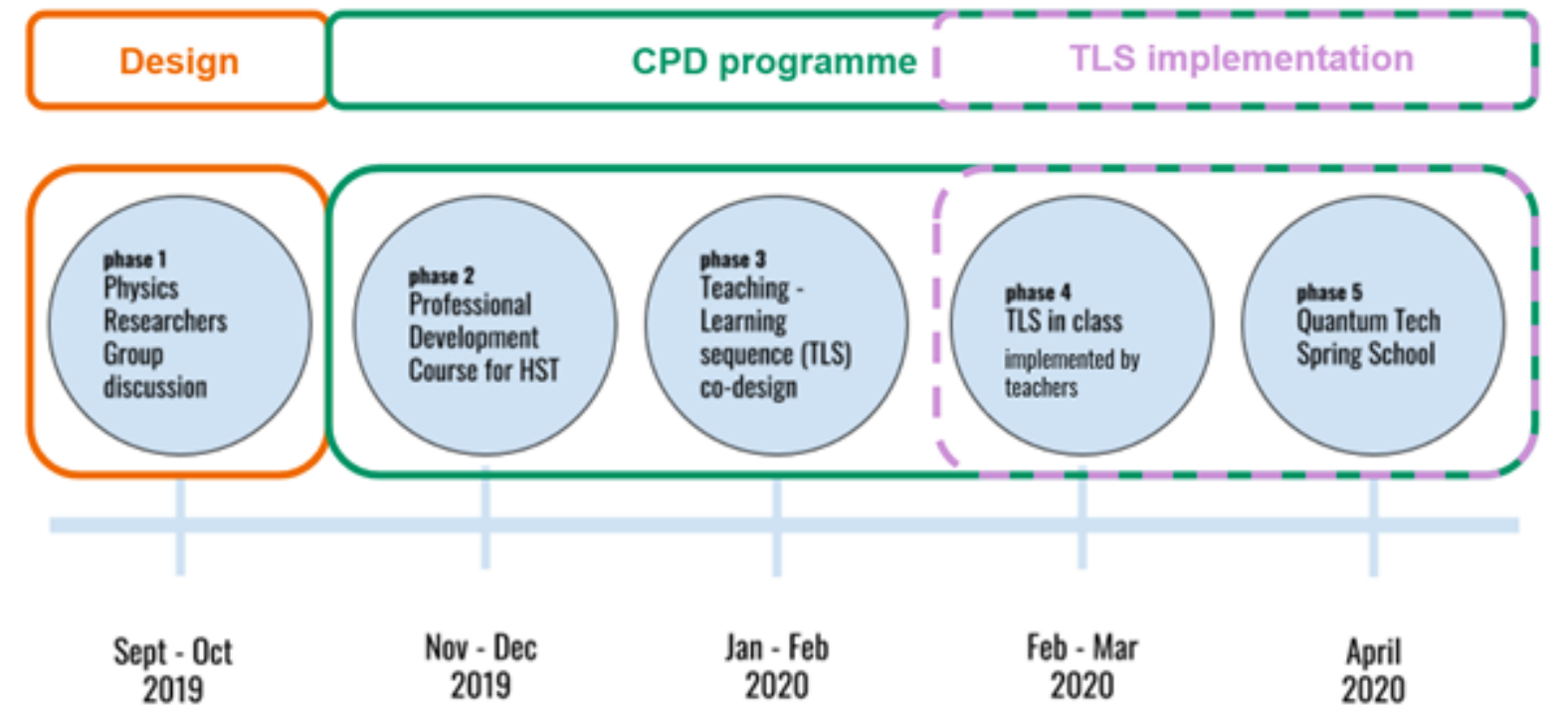}
    \caption{Quantum Skills Professional Development programme: course evaluation.}
    \label{fig:Ch6_123QKsEvaluation}
\end{figure}

\begin{quote} {\fontfamily{cmtt}\selectfont
    I have noted that the approach to quantum physics is not so straightforward even for a graduate teacher (at least not a physics graduate). Therefore, I personally will need a preliminary phase of study and in-depth analysis in order to transmit the contents to the students. [QSk19ExQ]
}\end{quote}

The main doubts concern the possibility of transferring what has been learned directly into the classroom, with the students.
\begin{quote} {\fontfamily{cmtt}\selectfont
It [PD] gave me the opportunity to review the pre-quantum topics in the fifth grade curriculum, pointing out some key aspects, sometimes not well highlighted in the current textbooks. It illustrated some aspects of later developments (entanglement, EPR paradox, Bell's inequality, polarisation...), which are useful and significant in order to have a broader conceptual frame of reference in our teaching interventions, even if we choose not to propose them in detail to the students. I hoped - even if I know that objectively both the difficulty of the contents and the time available to deal with them in fifth grade do not allow it - to get some didactic ideas from them that could be more directly transferred to classroom lessons. [QSk19ExQ]
}\end{quote}

\begin{quote} {\fontfamily{cmtt}\selectfont
\dots At several points I noticed a disconnection between what I was following and what I know I will have to deal with in a class (year 4 and year 5) of a scientific high school. I thought I would be able to get more 'teachable' ideas directly in the classroom, which was not the case. I'd like to be able to follow a course designed for teachers, but with the aim of being able to transfer the concepts I've learnt according to the times and methods of high school, which are very different from those of a university course. [QSk19ExQ]
}\end{quote}

The criterion of those contents to be teachable remains central for teachers. It is an important indicator for the work of reconstructing the content and development process of the subject matter for teaching.
The collected reflections have been used for the design of the PDt programme Quantum Jumps which is presented in the next section.

\section{Quantum Jumps 2020 Professional Developmewnt programme}\label{sec:QJumps}

The second edition of the PD programme was built on the results of the first edition. The second edition of the PD programme was built on the basis of the results of the first edition. During an open meeting between researchers and teachers, an effort was made to investigate some of the critical points that had emerged, in particular the possibility of a greater adherence between institutional teaching objectives and implementation in the classroom using an information-based approach.
The proposal is centred on the idea that the reflection made on the concept of qubit should in fact be more closely linked to the school curriculum. The introduction of the "historical experiments" and of the "indispensable contents" \cite{DM769_18Quadri} is therefore an essential constraint if we wish to maintain a high level of teacher involvement.
A first proposal was then to try to revise the content planning by acting on the topics planned for the last two school years. The possibility to use the polarization state as one of the possible ways to introduce the qubit (see Section \ref{sec:qubit_polarization} ) allows already in the fourth year of high school to introduce the formal and conceptual elements that can then be used for the treatment of quantum physics (usually provided only in the fifth and final year).
In general, therefore, this second edition of Pd has tried to make more explicit to the students the useful ingredients for a treatment of quantum physics with an informational approach, proposing them two strands: one on the nature of light based on the study of polarization, and one on the nature of quantum objects based on the study of electron spin states.
The proposal they were given was then to design learning environments for their students using the materials and reflections proposed by the PD and adapting them to their teaching style.

\begin{table}[hbt!]
\centering
\begin{tabular}{cp{12cm}}
 & \textbf{Informational approach to support teaching quantum}\\
 & \textit{Use of the polite informational axiomatic approach in Professional Developmet programme for teachers}\\
\hline
\rule[-4mm]{0mm}{1cm}
RQ6.3 & What are the elements of the informational approach (see Section\ref{sec:ApprPillars}) that can help the development of teachers'PCK?\\
 & \\
\midrule
\rule[-4mm]{0mm}{1cm}
RQ6.4 & What role can quantum technologies play in creating learning environments about teaching and learning quantum physics?
\end{tabular}
\caption{Research questions: Learning Environment for teachers, Quantum Jumps 2020}
\label{tab:RQ_TLE_QuantumJumps}
\end{table}


\subsection{Activities design}
The design of the PD activities started from the results of the questionnaire given to participants at the beginning of the course.
The aim of this first activity was to understand how participants were used to organising the different contents, making explicit their choices in relation to the physics curriculum.

\begin{figure}[hbt!]
    \centering
  \includegraphics[width=\textwidth,height=\textheight,keepaspectratio]{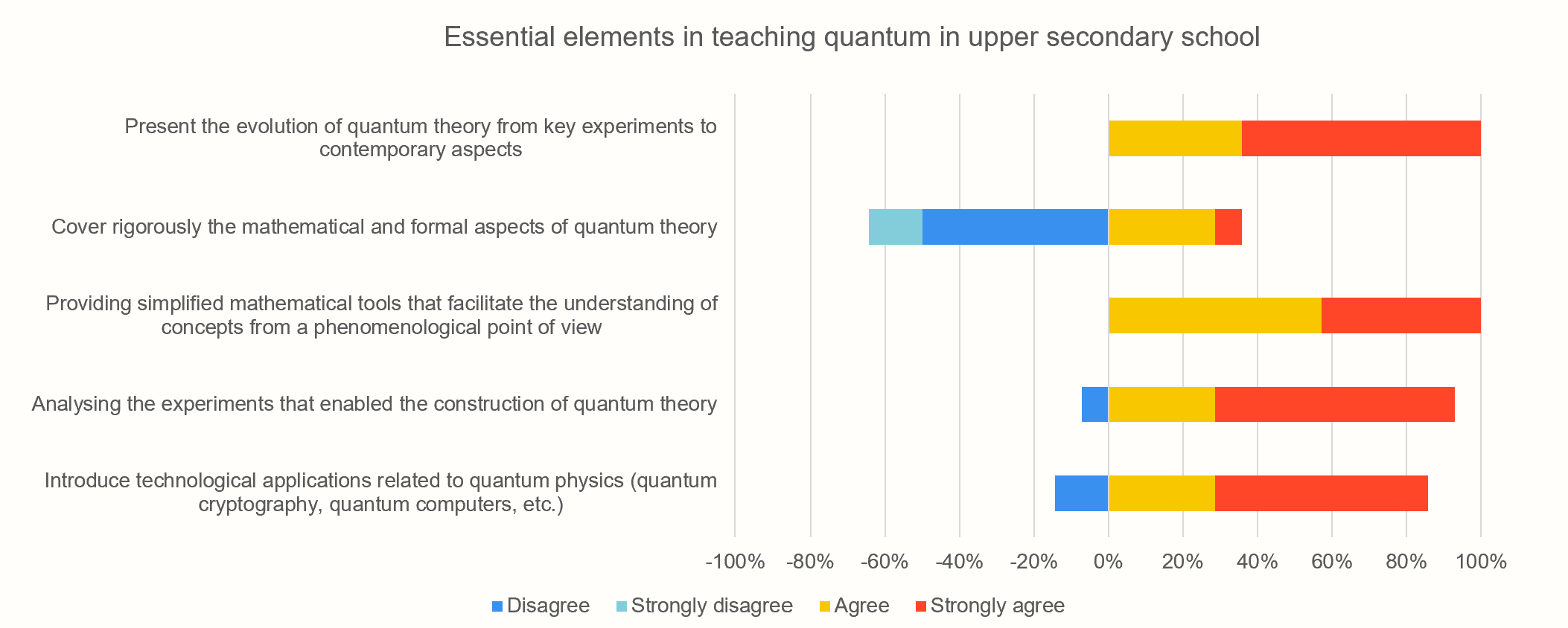}
    \caption[Quantum Jumps: teachers' expectations]{Teachers expectations at the beginning of the PD Quantum Jumps focused on improving their subject matter knowledge}
    \label{fig:Ch6_8_QJsTeachInitialexpect}
\end{figure}

The teachers’ initial expectations abut the course (see Figure. \ref{fig:Ch6_8_QJsTeachInitialexpect}) show the will to address the evolution of quantum theory through the use of meaningful experiments and a simplified, not necessarily rigorous, formal structure. The request was also to have materials at the end of the course that could be used in class.

To produce the competence in the integration of quantum physics content and teaching resources for building a good teaching-learning environment, the collaboration with teachers aimed to foster the reflection about the curriculum design process (see Section \ref{sec:LEDesign}.

In their practice, teachers are used to follow a specific sequence of topics they introduce to students based on the  the table of content of their textbooks \cite{amaldi2020il}, integrated by elements from their teaching experience.

\begin{figure}[hbt!]
    \centering
   \includegraphics[width=\textwidth,height=\textheight,keepaspectratio]{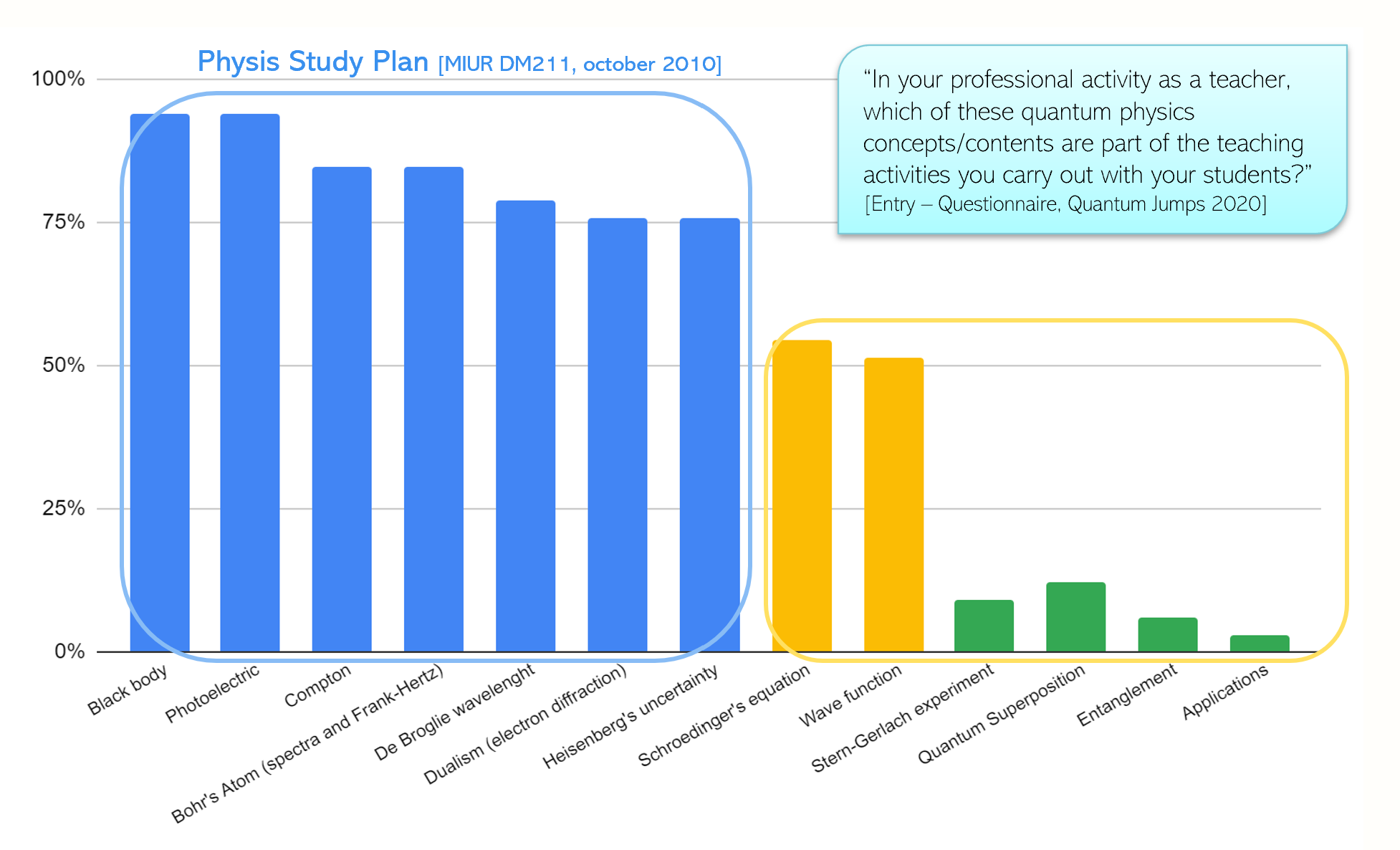}
    \caption{Quantum curriculum: content choices based on the national Study plan \cite{MIUR2010_211} and teachers personal choices about quantum applications}
    \label{fig:Ch6_9_QJsTeachChoices}
\end{figure}

In some cases topics outside the official curriculum are included to give students the opportunity to see some recent applications of quantum ad part of their curriculum.

In their reflections, teachers shared their perplexity about such structure, as non effective in introducing concepts properly. In some cases teachers expressed their concerns about the use of the approach to physics presented in the textbooks. It places too much emphasis on historical aspects, emphasising more narrative than substantive aspects. It also does not provide a context in which the new knowledge can be used in a meaningful way, for example in problem solving. This is a very different way of teaching physics from that used for other topics.

\begin{quote} {\fontfamily{cmtt}\selectfont
    We were used to explain a theory, solving problem \dots We end up with telling stories[QJs20EnQ]
}\end{quote}

In other comments it is pointed out that the approach used by the books hints at the tortuousness that characterised the early years of the development of quantum theory. This, according to teachers, can confuse students, leaving them in doubt as to whether quantum theory is in any way incomplete or fragile, or even confused and contradictory.

\begin{quote} {\fontfamily{cmtt}\selectfont
    [Quantum physics in textbooks is ] a somewhat tortuous path [\dots] I often need to leave something behind (is light a particle or a wave?), to force students to accept and not to understand [QJs20EnQ]
}\end{quote}
Looking at the way quantum physics is usually presented in schools, some teachers believe that it is presented as an adjustment of classical physics. \cite{SusskindQT}

\begin{quote} {\fontfamily{cmtt}\selectfont
    Quantum looks like an hybrid that uses classical methods and quantum rules[QJs20EnQ]
}\end{quote}

The approach used by many textbooks and that is part of teachers practice seems not fully adequate to introduce quantum because leaves teachers' expectations unattended and students puzzled

\begin{quote} {\fontfamily{cmtt}\selectfont
    Lots of questions remain unanswered \dots for example, is light a wave or a particle? And the electron? Duality seems a gross way to fix a problem)[QJs20EnQ]
}\end{quote}

In the PD the idea was to rethink the way teachers can approach the teaching of quantum (see Chapter \ref{ch:3-QApproach}), without destroing the official curriculum.

The official italian physics Study Plan \cite{MIUR2010_211} is structured as a list of topics and a list of skills that are related to those topics and that will be eventually part of the tests or assessments. The combination of those two elements define what students are supposed achieve.

\begin{figure}[hbt!]
    \centering
   \includegraphics[width=\textwidth,height=\textheight,keepaspectratio]{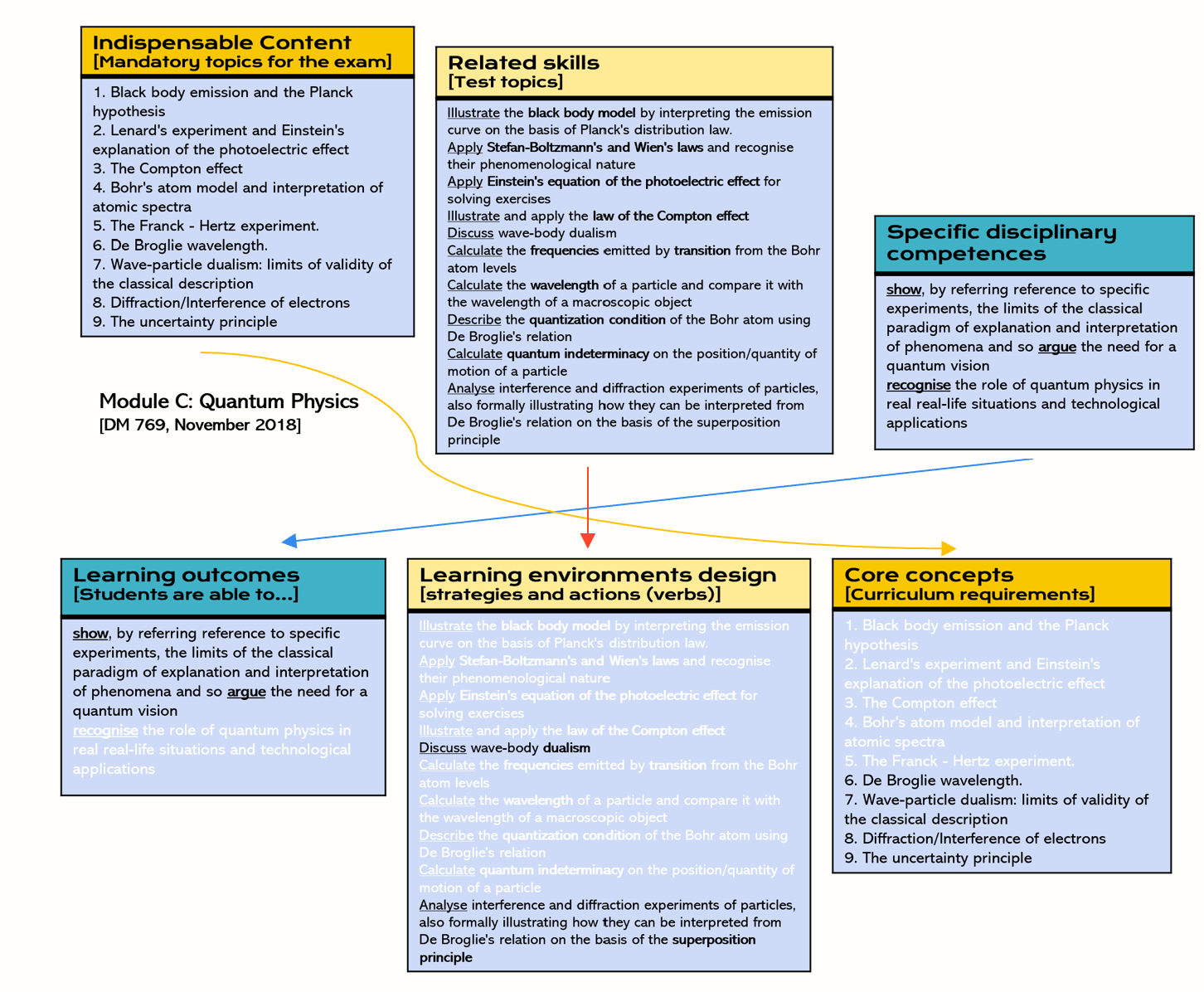}
    \caption{Quantum curriculum proposal. Using a design approach (see Section \ref{sec:LEDesign}, we started from the definition of intended learning outcomes (ILOs) already present in tha curriculum to design learning activities related to some specific indespensable contents}
    \label{fig:Ch6_10_QJsShuffleProposal}
\end{figure}

In a curriculum design framework, we propose to use a new approach that start from the definition of the intended learning outcomes ILOs (see Section \ref{sec:LEDesign}. These can be written adapting the learning goals already present in the study plan. This approach can help the identification of the teaching strategies and methodologies that can support the learning process, including the assessment tools. Finally the contents needed to support that specific process are collected from the list of mandatory topics that already present in the study plan.

In this context the \textit{polite informational axiomatic approach} we define above (see Section \ref{sec:PIAA}) can be used to identify the key concepts and the representations that can be used to design the activities. The central idea around which the design proposal is built is the exploration of the concept of quantum states via the analysis of experiments about qubits to characterize the quantum nature of physical objects.\\

\begin{figure}[hbt!]
    \centering
   \includegraphics[width=\textwidth,height=\textheight,keepaspectratio]{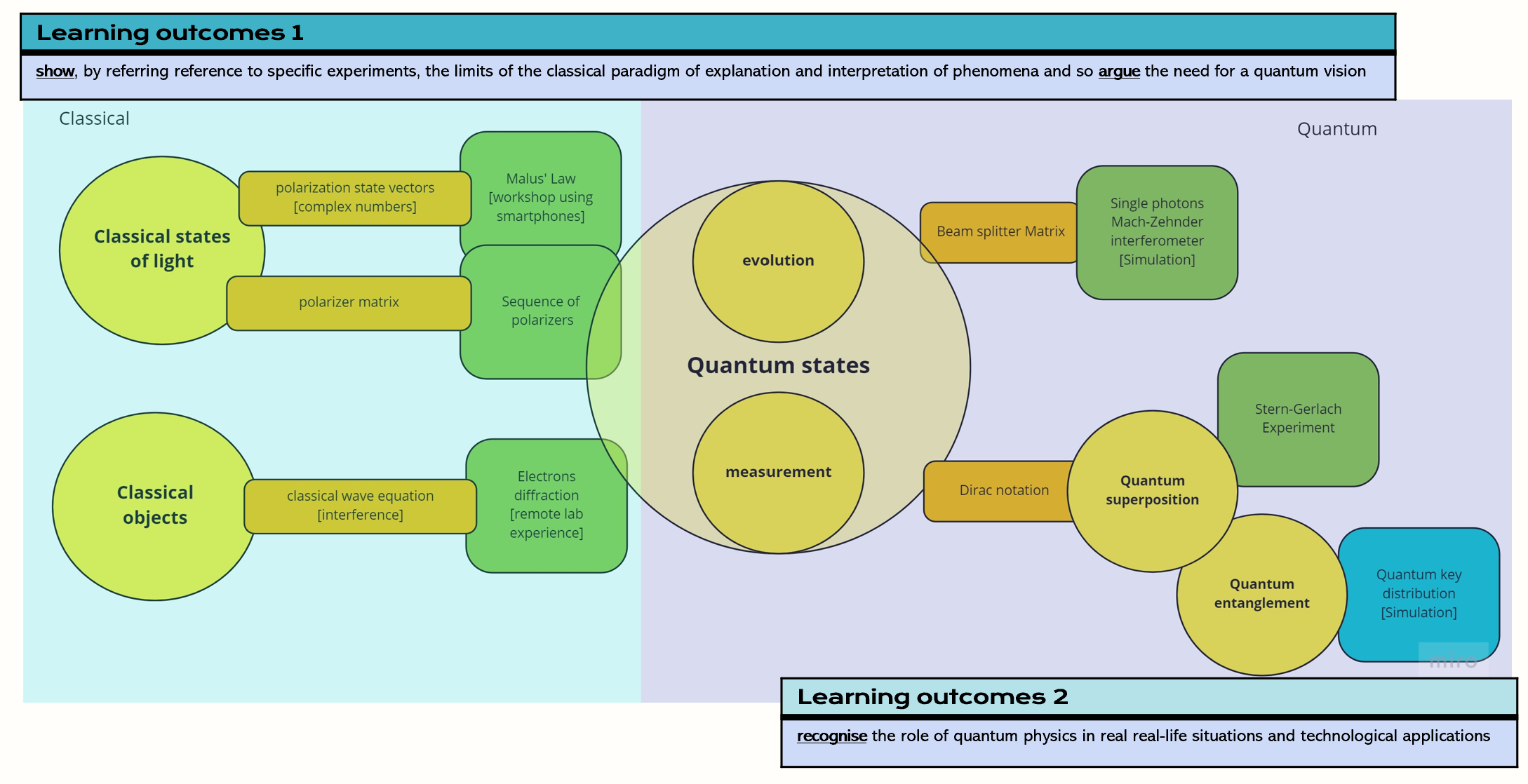}
    \caption{Quantum Jumps PD programme main structure: the core concept of quantum state is explored in different experimental context using the idea of quantum qubit}
    \label{fig:Ch6_12_QJsCoreconcept}
\end{figure}


\subsection{Methods}
The professional development programme is structured in three modules that are linked to specific learning outcomes of the italian physics Study Plan.

\begin{figure}[hbt!]
    \centering
   \includegraphics[width=\textwidth,height=\textheight,keepaspectratio]{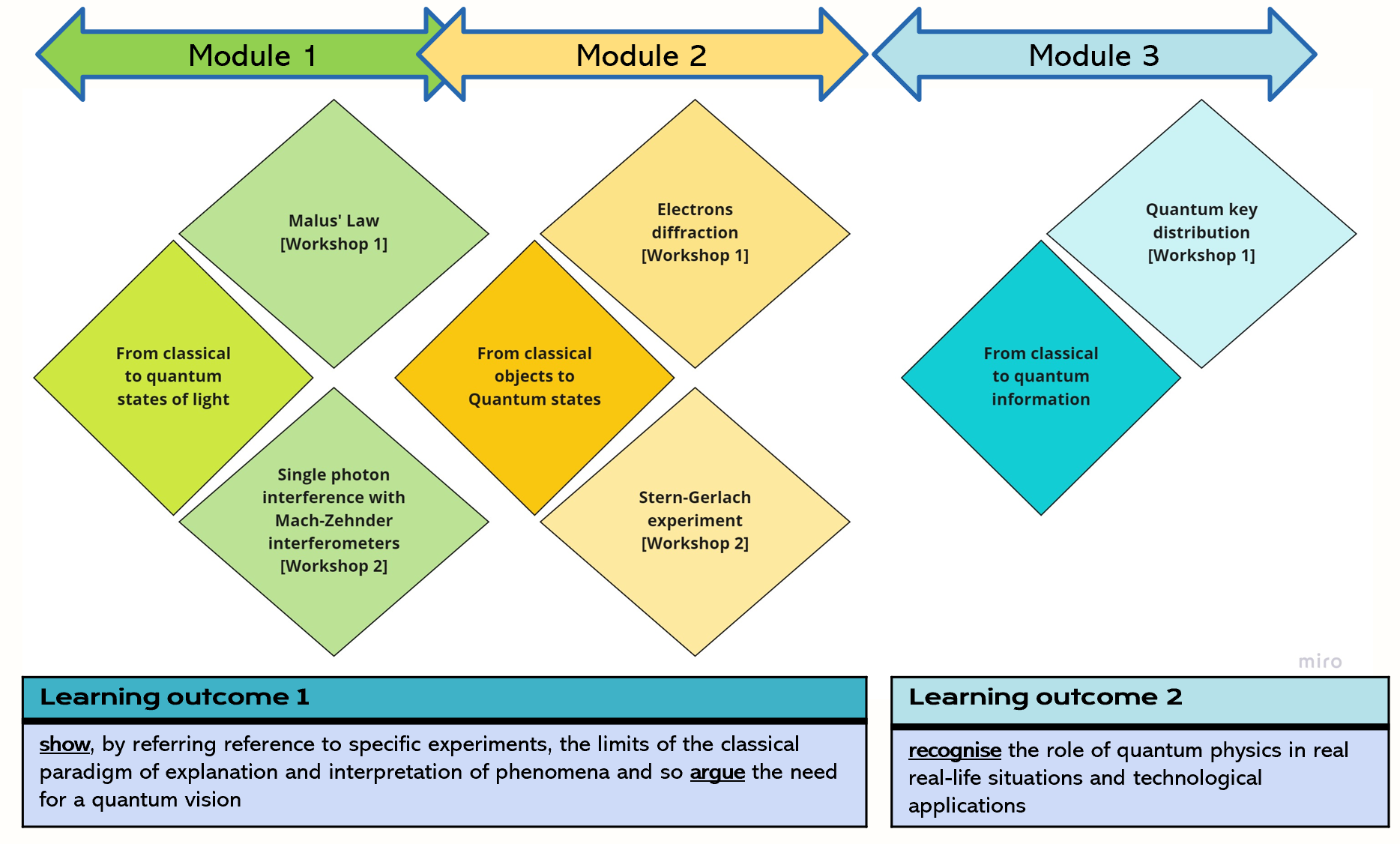}
    \caption{Quantum Jumps Professional development programme modules}
    \label{fig:Ch6_11_QJsModules}
\end{figure}

Each module has a introduction to the main concepts and the representations that can be used in teaching activities to investigate the behaviour of quantum objects. The workshops with learning activities about specific experiments provide the context to use those teaching tools.
Due to the restrictions imposed by the Covid-19 health emergency, all activities took place online. The materials used as well as the recordings of all the interventions were shared with the participants in order to favour teachers' partecipations also in asynchronous mode.

In addition to the questionnaire at the beginning of the course, the answers to a final questionnaire were collected as well as notes on the participants' reflections extracted from the recordings.
At the end of the PD programme, some teachers gave their availability to try out some of the contents of the course in class. Their work was monitored through regular meetings.
At the end, a semi-structured interview was conducted with some of the teachers who participated in the course. The interview protocol is available in the appendix \ref{appx:IntervProtocol}.

\subsection{Teachers'participation}
With the aim of involving a large number of teachers, the PD was promoted through SOFIA (https://sofia.istruzione.it/), a platform of the Italian Ministry of Education through which teachers can select professional development experiences.
\begin{figure}[hbt!]
    \centering
    \includegraphics[width=\textwidth,height=\textheight,keepaspectratio]{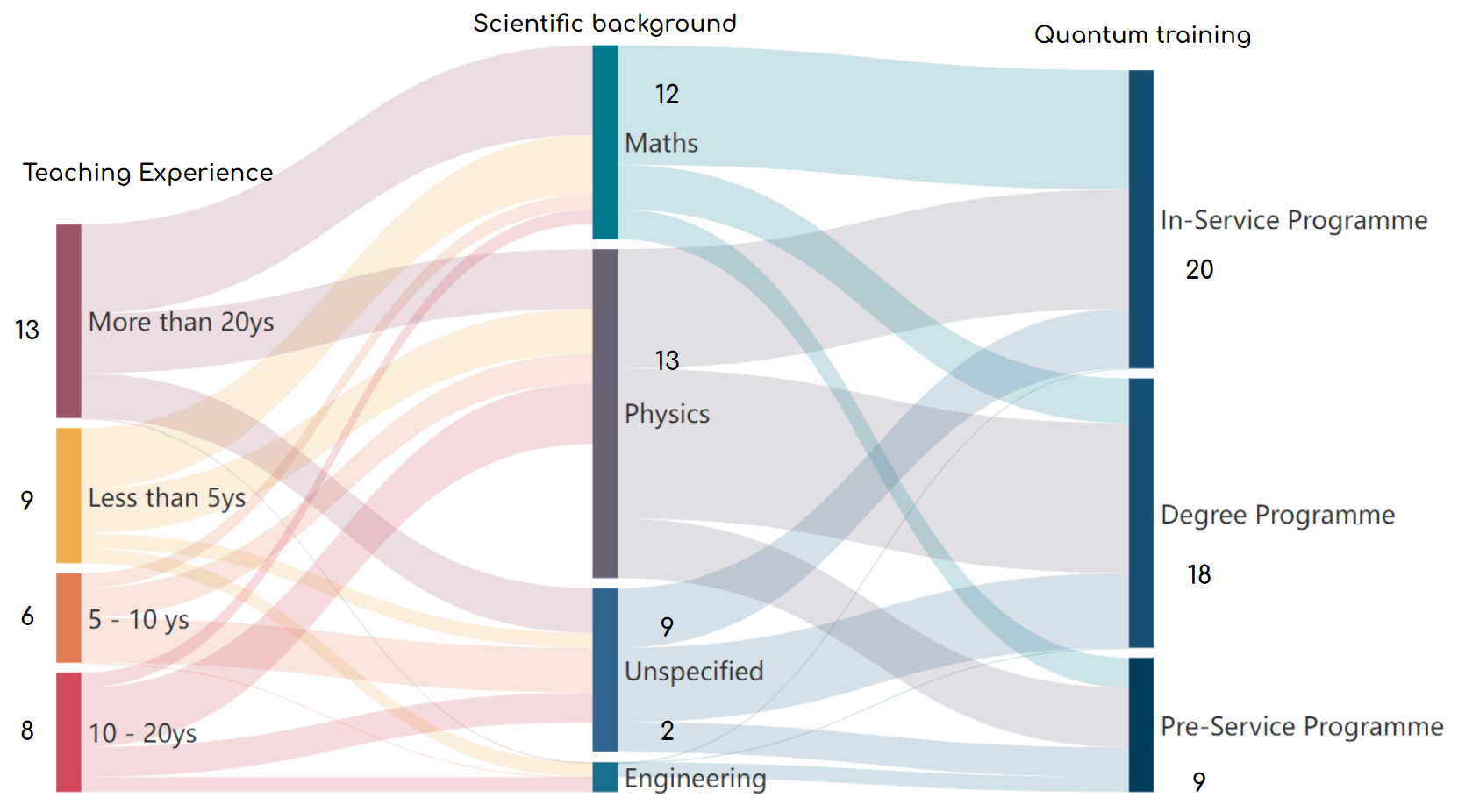}
    \caption{Quantum Jumps Professional development programme: participants. The 36 participant had different teaching experience and scientific backgrounds. We also asked in which context they have been trained in quantum mechanics}
    \label{fig:Ch6_11_QJsModules}
\end{figure}

The teachers have different scientific backgrounds: compared to the previous course (see Figureure \ref{fig:Ch6_3_QSKPDTeachersProfile}) there is a substantial balance between mathematicians and physicists. Although they are all in-service teachers, the teaching experience is very diverse, with a large group of very experienced teachers joined by a group of teachers who have just started their careers. The context in which the teachers were exposed to the content of quantum mechanics is that of professional development programmes. At some points during the course, participants repeatedly stressed that their training in quantum physics took place while they were alreading teaching and often drew on personal readings, i.e. outside a structured training context (see Section\ref{subsec:Results_interviews}).

\subsection{Results}
\begin{figure}[hbt!]
    \centering
    \includegraphics[width=\textwidth,height=\textheight,keepaspectratio]{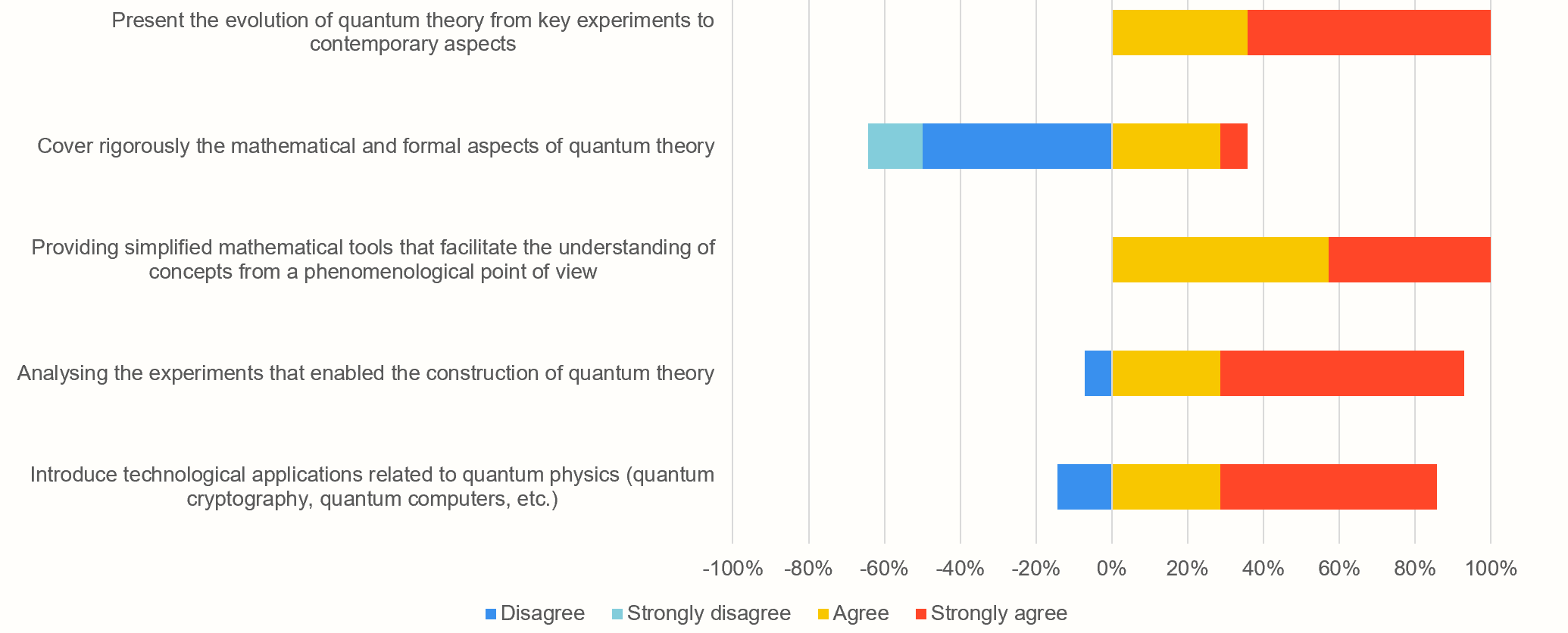}
    \caption{Quantum Jumps Professional development programme: Evaluation of the programme on the three dimensions of PCK. Data from the Quantum Jumps exit questionnaire}
    \label{fig:Ch6_18_ExitExpect}
\end{figure}

Compared to expectations at the beginning of the course (see Figure \ref{fig:Ch6_8_QJsTeachInitialexpect}), a main shift is registered relate to the role of formalism (line 2 in Figure \ref{fig:Ch6_18_ExitExpect}).

At the end of the PD programme, the use of the formal structure of quantum physics is seen as an opportunity to give the explanation more rigour and to prevent the trivialisation that can characterise the purely discursive description of a physical phenomenon (see Section\ref{subsec:QJTEresult1})

Another element reported by the teachers at the end of the PD programme was that they were more inclined to reduce the time devoted to the discussion of aspects of the historical evolution of quantum mechanics in favour of the introduction of conceptual aspects considered more essential to understand quantum phenomena. This attitude was also supported by the contingency linked to the change of procedures for the final state examination. Due to the health emergency, the Ministry of Education has removed the disciplinary written tests from the final examination, leaving as the only form of assessment the production of a paper on the subjects characterising the course of study, which for the scientific high school are Mathematics and Physics \cite{MIUR_Esami21}. This choice relieved the teachers of the tension related to the students' preparation for the exam and allowed some of them to opt for a more in-depth study of the course contents.

\begin{quote} {\fontfamily{cmtt}\selectfont
    This year we [teachers] have more freedom to complete the study programme\dots now I can spend the time I used to spend preparing for the maths test on completing the physics programme at my leisure..[QJs20ObF]
}\end{quote}

\begin{figure}[hbt!]
    \centering
    \includegraphics[width=\textwidth,height=\textheight,keepaspectratio]{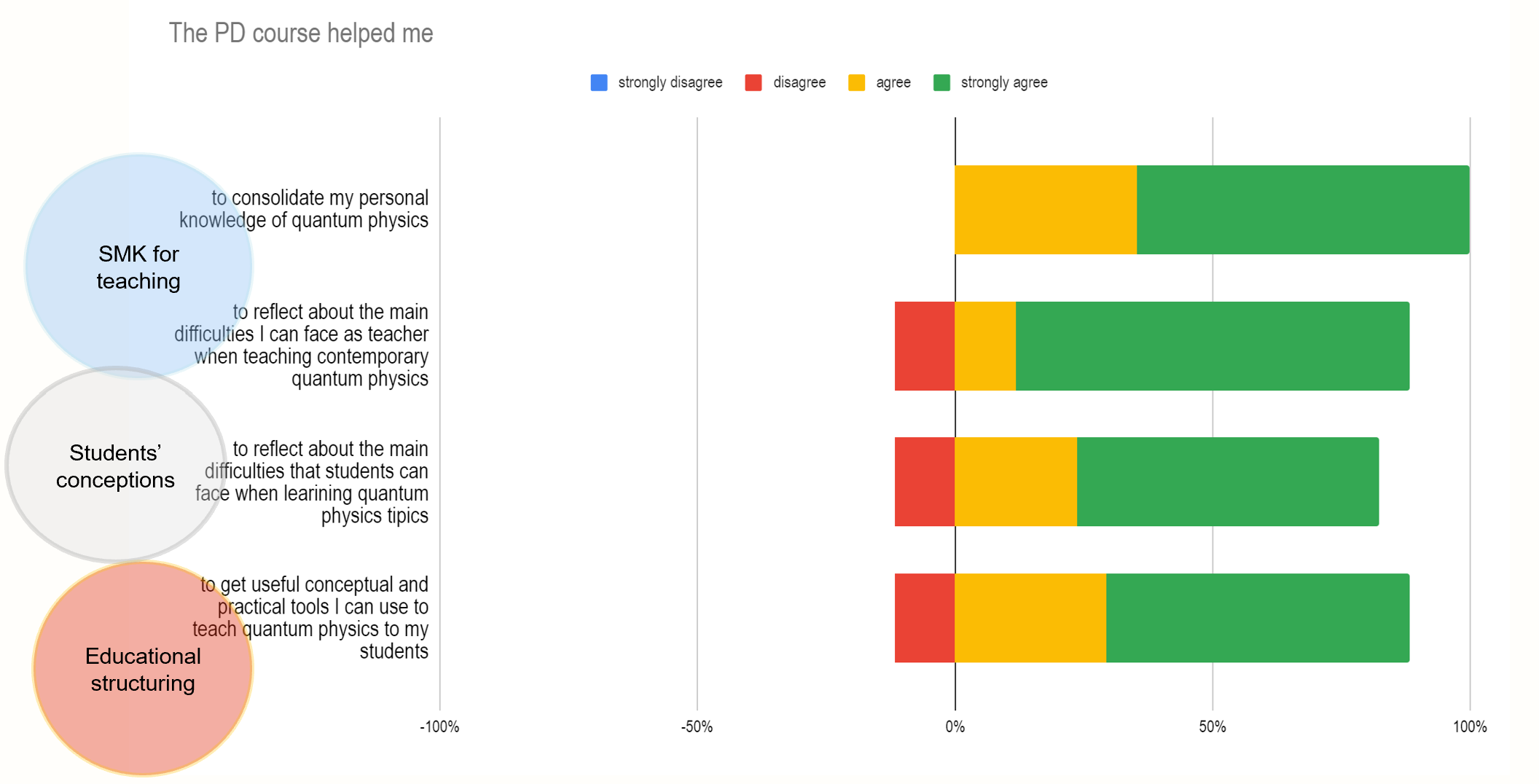}
    \caption{Quantum Jumps Professional development programme: Evaluation of the programme on the three dimensions of PCK. Data from the Quantum Jumps exit questionnaire}
    \label{fig:Ch6_17_PCKeffetc}
\end{figure}

Based on questionnaires results (see Figureure \ref{fig:Ch6_17_PCKeffetc}, the PD seemed to have supported teachers’ reflection about the three PCK dimensions quite effectively. In the following Sections we analyse those results in relation to the specific Research Questions stated at the beginning of this Chapter (see Table \ref{tab:LET_RQIncipit}

\subsubsection{RQ6.3: Informational approach to foster Teachers'PCK development}\label{subsec:QJTEresult1}
The main result of the proposed activities was to continue a reflection on the centrality of the concept of quantum state as a conceptual tool for the interpretation of quantum behaviours.
One choice made in the PD programme was not to enter into issues related to the ontology of the quantum state. \cite{Pusey2012_QstatesReality,Leifer2014_quantumsattereal} but based on the proposed approach (see Section \ref{sec:ChangeP}) try to use the concept of quantum state as a tool to describe behaviours of physical objects that are "\textit{absolutely} impossible, to explain in any classical way, and which has in it the
heart of quantum mechanics" \cite{feynmanLect_Vol3}.\\ 

The teachers took the opportunity to approach the content in a different way. This is to describe the transition from classical objects (particles and waves) to the concept of state as the information we have about all the possible measurement outcomes. The rules of quantum mechanics (see Section \ref{sec:QAxioms}) describe how quantum state evolves and changes inside specific devices or experiments. Measuring the state is necessary to get the information, but quantum measurements changes the state according to Born's rule. The role of quantum measurements and the difference with classical measurement is therefore essential.

The idea of describing what happens to electrons using qubits and simulations is a way of giving students a key to understanding behaviour that would otherwise be strange and bizarre. However, the use of an abstract concept such as a qubit or state must always be linked to something that can be visualized.

\begin{quote} {\fontfamily{cmtt}\selectfont
    It [the qubit] looks like mathematics \dots it is important that the students do not let the physics go \dots that they don't forget that we are talking about real physical entities. \dots When we talk about fields [in classroom] we also ask them to make this effort, but in that case I can show them how to calculate the Lorentz force on a moving charge and then that abstract description makes sense.[QjJs20ObF]
}\end{quote}

\begin{quote} {\fontfamily{cmtt}\selectfont
    I often find myself in difficulty when I want to explain these subjects [of quantum physics], \dots the discourse always seems to be interrupted at the best part, \dots partly because of my inability to understand and explain things. The examples given through the simulations [experiment using sequences of Stern Gerlach apparatus] allows to explain things also by showing them and this makes things much easier.[QJs20ObF]
}\end{quote}

\begin{quote} {\fontfamily{cmtt}\selectfont
    [talking about dualism] it is difficult to change ideas about things without being able to \textit{see} them [QJs20Obf]
}\end{quote}

The doubt expressed by some teachers is that this may not be enough.

\begin{quote} {\fontfamily{cmtt}\selectfont
    is an experiment enough to believe?[QJ20I]
}\end{quote}

The reflection relates to the effectiveness of experiments or simulations shown as sufficient elements to allow the student to establish effective links between what the experiment shows and the theory that explains it \cite{CrouchMazur2004_classroomdemo}.

Regarding the use of a complete formal mathematical representation, teachers remained quite skeptical about the possibilities of students to fully understand it.

\begin{quote} {\fontfamily{cmtt}\selectfont
    Some parts use too much advanced mathematics [QJ20ExQ]
}\end{quote}

The formal structure is the one, however, that provides elements that some teachers consider to give rigour to the discourse.

\begin{quote} {\fontfamily{cmtt}\selectfont
    Certainly for some more "fragile" students certain discourses [i.e. the use of mathematical formalism] can be too complicated. However, the risk of trivialisation is always very high and mathematics certainly makes it possible, with a little effort, to construct a discourse that "stands up" [rigorous, solid, difficult to question]. [QJs20ObF]
}\end{quote}

Explanation through formalism provides a better understanding of the structure of the theory and its internal coherence.

\begin{quote} {\fontfamily{cmtt}\selectfont
    it doesn't make a lot of sense to make [students] do those accounts. But showing them I think gives them the sense that there is a thought behind it, that [the explanation] is not just something you say to justify yourself.[QJs20ObF]
}\end{quote}

In this sense the use of simulations and experiments support the development of teachers' PCK (see Table \ref{tab:PCK_Analysis_GRID}) because it becomes a support to \textit{react flexibly} \cite{Dijk2006_ERTE} to the students' difficulties.

The use of Dirac notation and the idea of the state of a vector that can change in the interaction with the device has been seen as a new opportunity to present the concepts and describe what happen during an experiment.

\begin{quote} {\fontfamily{cmtt}\selectfont
    [In describing what happens] I can think of using what you see from simulations [QJs20ObF]
}\end{quote}

Experiments are valued as central to support the presentation of the concepts. Teachers enjoyed the use of simulations and remote labs as an opportunity to base reasoning around concepts on experimental evidence.

\subsubsection{RQ6.4: Quantum technologies to built learning environments }\label{QJTEresult2}
Some tools to explore quantum technologies as an opportunity to learn quantum physics were shared with teachers. Avoiding technical details, the focus was on an interdisciplinary approach to reflect with teachers on how the exploration of quantum technologies can be used to consolidate the learning of different core concepts of quantum physics.
Module 3 of the course was devoted to the analysis of the educational use of quantum technologies. (see Figureure \ref{fig:Ch6_11_QJsModules}).

For many participants these were new topics. Being able to deal with quantum technologies during the PD Programme was welcomed as interesting both in terms of personal cultural enrichment and as a possibility to included new topics in the lessons with students.

\begin{quote} {\fontfamily{cmtt}\selectfont
I often hear about quantum technologies but I don't have a clear idea of what they are and \dots I would like to understand if they can be used in the classroom with students, also in order to update the curriculum a bit and talk about new things. [QJs20ObF]
}\end{quote}

The evaluation of the proposed activities was overall positive. In particular, the teachers' comments highlight how dealing with applicative aspects can be a good way of involving students, exploiting the interest of some of them in technology.

\begin{quote} {\fontfamily{cmtt}\selectfont
When I talk about quantum mechanics in class, some students sometimes mention quantum computers \dots They probably read about them on the Internet \dots so they are interested in talking about them in class. [QJs20Obf]
}\end{quote}

The choice made in the introduction to quantum technologies was to start from the general problem of understanding the reason for the advantage brought by these technologies in solving certain problems and to link it to the key concepts of quantum physics (e.g. quantum superposition). This choice has been considered as an interesting stimulus also for those students who are not particularly inclined to a more technical study or who struggle to understand what the study of physics is really useful for.

\begin{quote} {\fontfamily{cmtt}\selectfont
Quantum technologies make quantum physics look like something useful and not just a difficult and somewhat nonsensical theory. [QJs20ObF]
}\end{quote}

Focusing on understanding how these technologies work can also reduce the level of abstraction. While remaining on a rather complex conceptual level, the study of how some quantum behaviours (e.g. quantum superposition and quantum measurement) can be used for the distribution of cryptographic keys makes them seem less distant from reality and more meaningful.

\begin{quote} {\fontfamily{cmtt}\selectfont
Now my idea of quantum is different\dots and it is something students can see too \dots Quantum physics is not magic after all \dots I can use it \dots and it is something useful [QJs20ObF]
}\end{quote}

\begin{quote} {\fontfamily{cmtt}\selectfont
With Quantum technologies \dots I can see quantum physics in action [QJs20ObF]
}\end{quote}

The use of the \cite{QuVisSim} simulators also proved to be a useful tool for dealing with quantum technology content once it is properly introduced to students.

\begin{quote} {\fontfamily{cmtt}\selectfont
The use of simulations together with worksheets allows you to get your hands on it. [QJs20ObF]
}\end{quote}

With regard to the inclusion of these topics in class activities, the attitude was generally cautious.

\begin{quote} {\fontfamily{cmtt}\selectfont
The course certainly gave me a lot of ideas \dots However, before using these things with my students I need time \dots I don't feel like using them already, I have to study them myself first \dots for this I think I will use the course [PD] materials. [QJs20ObF]
}\end{quote}



\section{Teaching - learning sequences designed by teachers}\label{sec:Teachersdesign}
After both Quantum Skills and Quantum Jumps PD programmes, groups teachers teamed up with the researchers to design learning activities. The results of this collaboration and the subsequent classroom implementation were very useful to understand how to adapt the teaching approach developed during the PD to a secondary level physics course.\\

In this section we specifically present some of the educational activities designed by the teachers who participated in the course. The aim is to highlight which elements of the educational pathway have been acquired in terms of tools useful for the construction of activities to be brought to the students.
In order to respond to the institutional constraints relating to the objectives of competences envisaged in the secondary school cycle, the various proposals refer to the guidelines provided in the Study Plan for the scientific high school (Liceo Scientifico) \cite{MIUR2010_211} and to the curriculum guidelines for the preparations of final exam written tests (Quadri di riferimento per la redazione e lo svolgimento delle prove scritte) \cite{DM769_18Quadri}. These documents provide guidelines about the specific scientific competences to develop, skills to acquire and topics to be covered. In particular those activities have been focus on the following intended learning outcome:

\begin{quote}
    To be able to show, with reference to specific experiments, the limits of the classical paradigm of explanation and interpretation of phenomena and to be able to argue the need for a quantum vision. \cite{DM769_18Quadri}
\end{quote}

The presentation of teachers' designs aims to highlight how the collaboration between the different members of the learning ecosystem and the introduced physics approach has facilitated the activation of a reflection on the different dimensions of teachers' PCK and the development of instructional design competences.

\subsection{Quantum behaviours: superposition and entanglement} \label{sec:massimo}


One teacher proposed that the transition between classical and quantum physics should be driven by analysis and interpretation of quantum experiments results:

\begin{quote} {\fontfamily{cmtt}\selectfont
\textit{One way to approach the counter-intuitive features of quantum physics is using data-based evidence}[TeachRefl]
}\end{quote}

\begin{figure}[hbt!]
    \centering
    \includegraphics[scale=0.9]{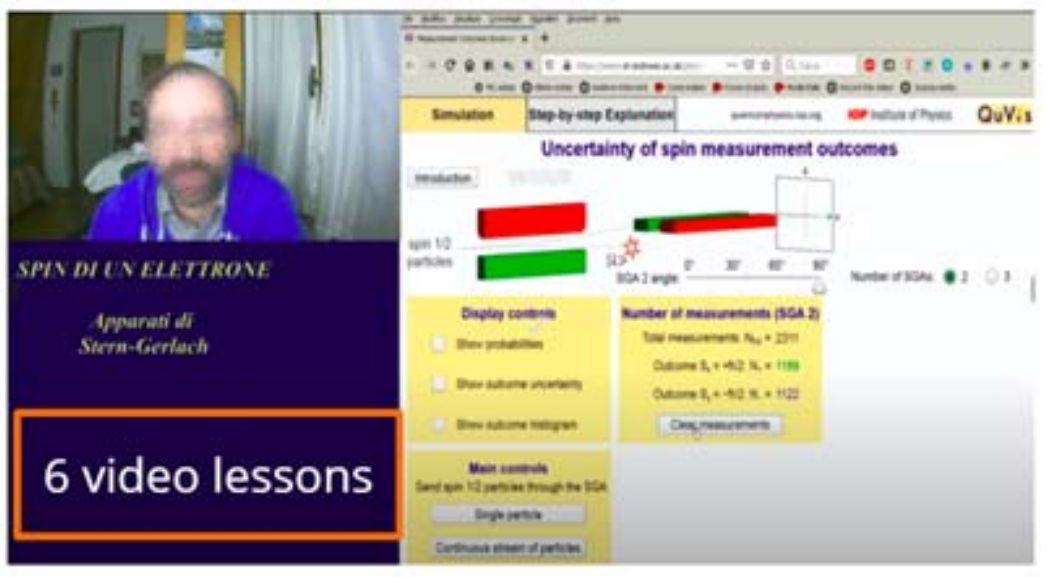}
    \caption{Quantum Skills teachers' materials}
    \label{fig:Ch6_6_QSKPD_TeacherMaterials}
\end{figure}

Teacher created their own set of materials (see Figureure \ref{fig:Ch6_6_QSKPD_TeacherMaterials} ), including video lessons and worksheets and prepared questionnaires for use during online lessons to collect students’ reflections on the experiments.
Due to the need for distance learning, the teacher organised the lessons according to a methodology inspired by the flipped classroom methodology \cite{bergmann2014flipped} \cite{Mazur2009_flipped}.

The teacher first recorded a video lecture which he provided to the students together with other study material, among others the slides of the presentations used during the lectures. Then the students watched and studied the material first, and finally there was a "synchronous" videoconference meeting in which the topics of the lecture were reviewed and deepened. The teacher answered the questions of the students, doubts were clarified and, sometimes, a debate on the topics was attempted.

The assessment of learning was carried out through oral interviews and the presentation of a lab report on the simulated experiments to the teacher. (see Figure \ref{fig:Ch6_6_QSKPD_TeacherMaterials})

The course was designed at the end of the Quantum Skills PD programme in April 2019.

\begin{figure}[hbt!]
    \centering
   \includegraphics[width=\textwidth,height=\textheight,keepaspectratio]{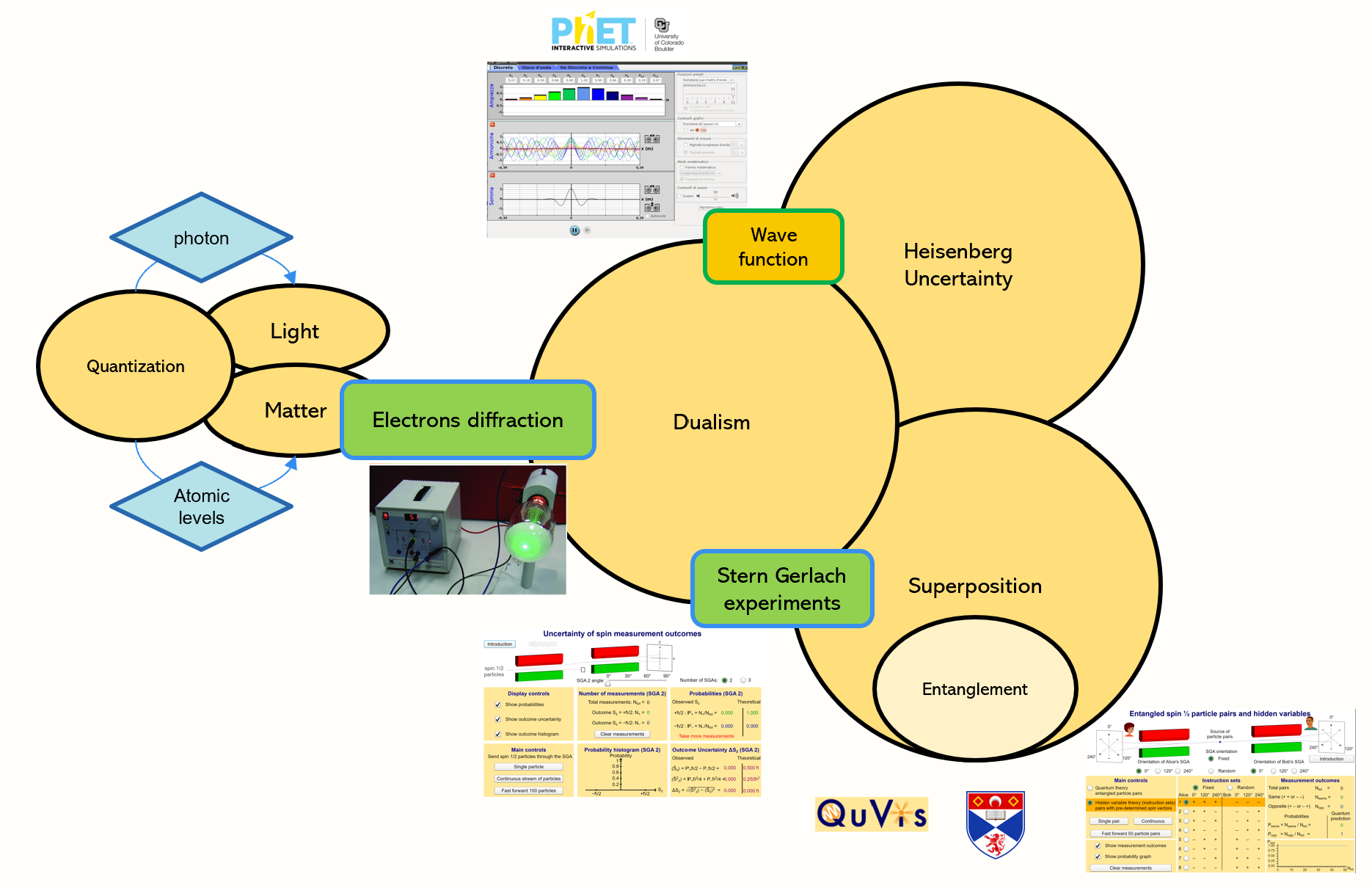}
    \caption{Exploration of quantum behaviour: the general structure of the teaching - learning activites. Starting from the concepts of quantization, the analysis of electrons diffraction leads to the key concepts of dualism, uncertainty and superpostion that are investigated using simulations}
    \label{fig:Ch6_21_TLSmassimo}
\end{figure}

The structure of the teaching-learning activities is shown in Figure \ref{fig:Ch6_21_TLSmassimo}.

The first objective of the activity explained by the teacher is "cultural".
\begin{quote} {\fontfamily{cmtt}\selectfont
It is unthinkable for a student to leave a scientific high school without having a basic understanding of a number of concepts that are the foundations of modern physics.[TeachRefl]
}\end{quote}
This is especially intended for those students who will not undertake scientific studies and therefore will not have the opportunity to deal with this content in the future.\\

The second objective is of a 'methodological-cultural' nature and concerns a reflection on certain aspects of the process of constructing scientific knowledge.
\begin{quote} {\fontfamily{cmtt}\selectfont
I think it is important for students to realise that, apart from personal opinions, in science there are ways of comparing two theories/hypotheses and checking whether a theory is correct or not, whereas the simple "I think so" is not "consistent".[TeachRefl]
}\end{quote}
This type of reflection can easily be extended to other subject areas (not only scientific ones) and is therefore important for all students, regardless of the specific areas in which they will be building their future as citizens.

In the teacher's words the first part of the course is structured as follows
\begin{quote} {\fontfamily{cmtt}\selectfont
   We began by looking at quantization, starting with the quantization of light radiation (black body and photoelectric effect, as per the traditional path) and then moving on to the quantization of the energy levels of the hydrogen atom. From the quantization of light radiation with the introduction of the photon we arrived at the wave-matter dualism for light to the analogous dualism for the electron and the particles constituting the matter we know. \dots The introduction of De Broglie's hypothesis comes after that followed by its experimental verification through the diffraction of an electron beam through graphite and the experience of the double slit. [TeachRefl]
}\end{quote}

According to the teacher, building this parallelism can facilitate the understanding of the concept
\begin{quote} {\fontfamily{cmtt}\selectfont
    A short but delicate step\dots because while students easily "digest" the dual nature of light, it is more difficult to "digest" the fact that the matter we are made of has this dual nature.[TeachRefl]
}\end{quote}

The second part of the course breaks away from the standard approach to organising content by choice. The reflections made during the PD facilitated the teacher's didactic planning, allowing the structure of the teaching learning sequence to be developed around the exploration of certain key concepts. The aim is to highlight the close connections between the elements the teacher considers to be characteristic of the quantum vision of the world.

\begin{quote} {\fontfamily{cmtt}\selectfont
    From this point onwards, the course differed substantially from the "canonical2 one. In fact, the presentation of the three "fundamental nuclei" (wave-particle dualism with its representation through wave packets, Heisenberg's uncertainty principle and superposition of states) was done by highlighting the close connections between them.[TeachRefl]
}\end{quote}

Dualism was introduced by analysing the results of the double-slit experiment. Unfortunately, it was not possible to carry out the experiment in the laboratory, but it was decided to illustrate the experiment during the lessons. (see also Section \ref{sec:carcano})

\begin{quote} {\fontfamily{cmtt}\selectfont
    Showing the empirical evidence becomes fundamental, which is why the biggest regret of the experience was not being able to take the students to the Insubria laboratories where they could "autonomously" carry out the experiment and the accounts necessary to verify De Broglie's hypothesis. In any case, the experiment and the resulting accounts were described and analysed in detail, even without being able to carry out the measurements.
}\end{quote}

Heisenberg's uncertainty was at first introduced using an approach presented in the textbook (\cite{romeni2017fisica}, Chapter 28) in the context of the single-slit experiment.

\begin{figure}[hbt!]
    \centering
   \includegraphics[width=\textwidth,height=\textheight,keepaspectratio]{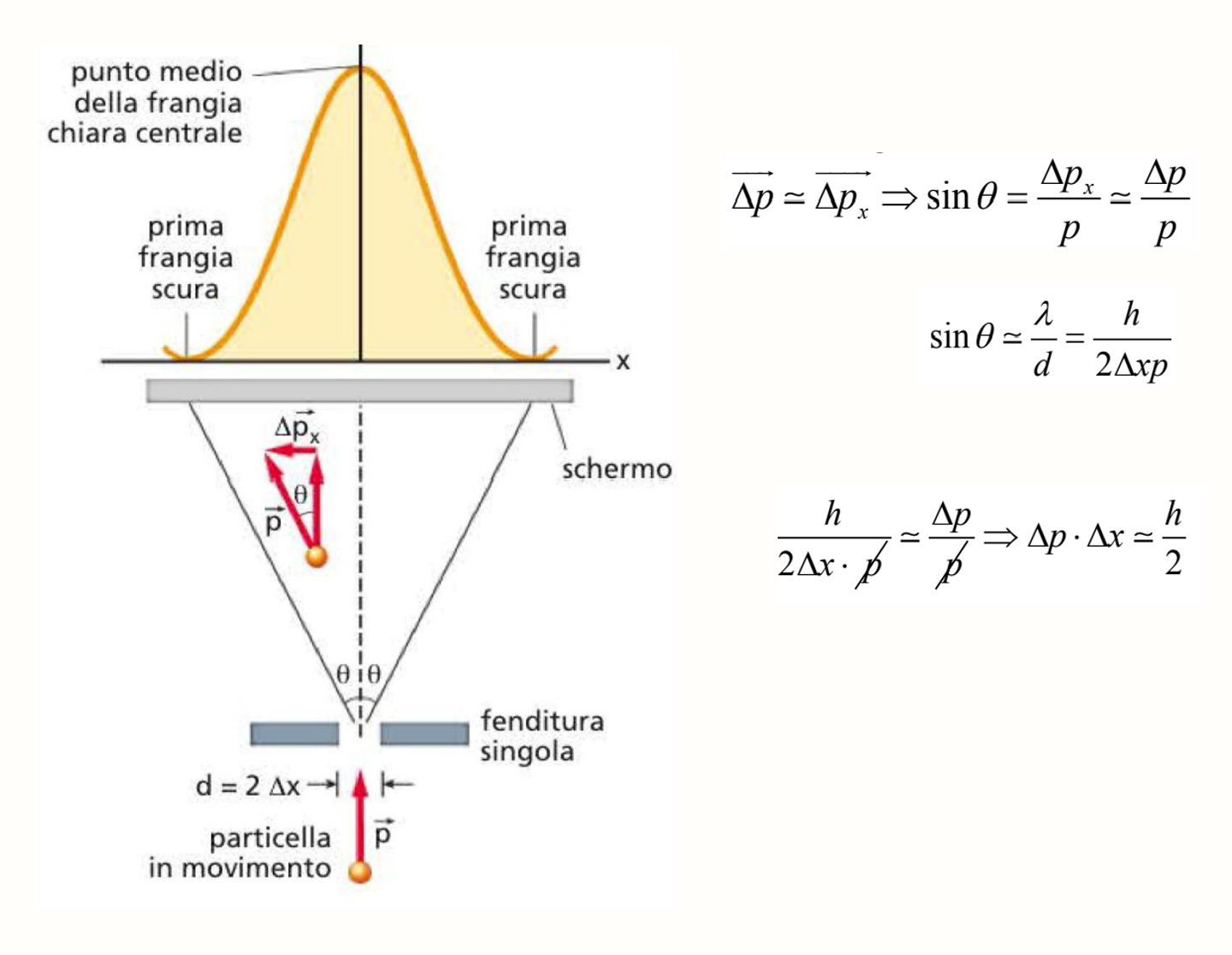}
    \caption{Heisenber uncertainty: The uncertainty relation between position and momentum of the particle representing the light is obtained from the width of the diffraction centre. Image taken from \cite{romeni2017fisica}}
    \label{fig:Ch6_19_HeisenRomeni}
\end{figure}

To highlight the link between Heisenberg's indeterminacy and dualism, the wave function was then introduced as the mathematical object used to represent the quantum object. This parallelism allowed the introduction of a formalism related to wave theories for the structuring of the interpretative model.

The choice was to introduce the wave function starting from a reference to plane waves with which the students are already familiar and to link the amplitude of the wave to the probability of identifying the quantum object. In other words, it is a question of using waves as a \textit{ convenient mathematical tool} for describing the behaviour of these particles \cite{schwichtenberg2019no-nonsense}. 
See also Section \ref{sec:carcano}.
\begin{quote} {\fontfamily{cmtt}\selectfont
   The wave packet is understood as a superposition of plane waves, each with an appropriate amplitude $A_n$, and wavelengths in an interval centred in $\lambda_0$ and of amplitude $2\Delta \lambda$.[TeachRelf]
}\end{quote}

\begin{figure}[hbt!]
    \centering
   \includegraphics[width=\textwidth,height=\textheight,keepaspectratio]{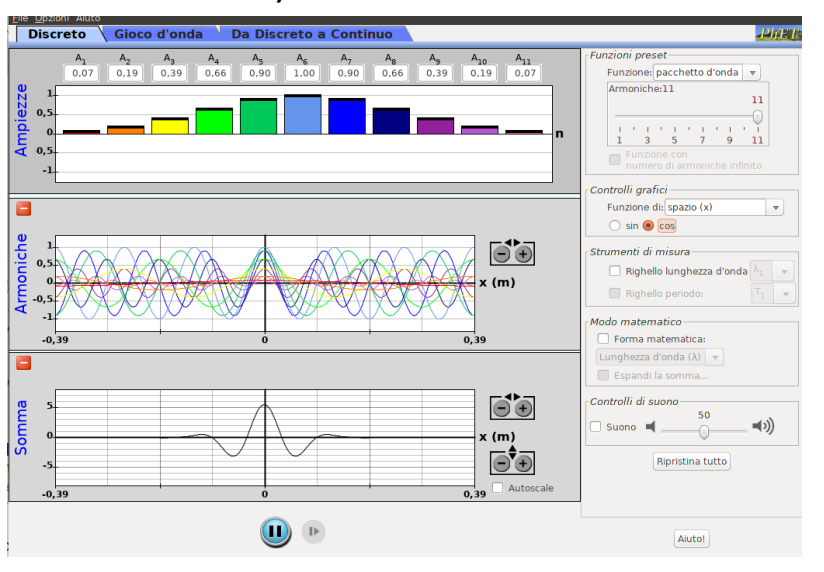}
    \caption{Wave packet as a superposition of waves: in the teacher own words "By means of a simulation it is possible to show how the superposition of several waves (cosine) generates a localised package." \cite{PhetFourier}}
    \label{fig:Ch6_20_HeisenFourierTransfor}
\end{figure}

It was chosen not to deal specifically with Fourier transforms (see Section \ref{sec:carcano}) but to show the idea of superposition through a simulator.
For the introduction of such a representative language the use of a simulator \cite{PhetFourier} has indeed proved essential to support the understanding of the wave representation of the quantum object (see Figure \ref{fig:Ch6_20_HeisenFourierTransfor})

\begin{quote} {\fontfamily{cmtt}\selectfont
    To this end [presentation of the wave packet] the use of an application to calculate and represent the packet both as a discrete sum and as a "sum in the continuum" made it more approachable for the students.[TeachRefl]
}\end{quote}

The last part of the teaching - learning sequence is about superposition and entanglement and has been implemented in collaboration with physics education researcher that participated in two online classroom lessons. The intented learning outcome (ILO) about this last section can be expressed in these terms:


\begin{table}[hbt!]
\centering
\begin{tabular}{cp{12cm}}
\toprule
 & \textbf{ILO}: evaluate the counterintuitive aspects of quantum objects behaviours through the creation of interpretative model of experimental results using the concept of quantum state. \\
 &\textit{At the end of the activities students should be able to:} \\
\midrule
ILO-a & Compare different interpretative model about complex quantum behaviour, such as quantum superposition and entanglement\\
ILO-b & Argue the need of a quantum interpretative paradigm based on the identification of the characteristics and behaviours of physical systems that can be tested using simulations about sequences of Stern Gerlach apparatus.\\
ILO-c & Analyse the relationship between the concept of quantum state and the properties of quantum systems in the measurement process related to experiments using sequences of Stern Gerlach apparatus.\\
ILO-d & Illustrate the behaviour of quantum systems using the concept of quantum state in the experiments using Stern Gerlach apparatus.\\
ILO-e & Predict the outcome of experiments using Stern Gerlach apparatus using the concept of quantum states superposition.\\
ILO-f & evaluate the efficacy of hidden variables hypothesis in the explanation of Bell’s experiment performed using  Stern Gerlach apparatus and support the idea of entanglement based on Bell’s inequalities
\end{tabular}
\caption{ILOs: Quantum behaviours}
\label{tab:ILO_massimo}
\end{table}


Formative assessment strategies include the use of closed and open-ended questionnaires to guide students through the analysis of experiments with Stern Gerlach sequences. These activities aim to monitor students' reflections and support them in exploring the experiments by stimulating the reflection of the experimental data.
The activities are carried out synchronously online via the Google Meet$^{TM}$ platform.

This last activity in the sequence is divided into three parts

\begin{enumerate}
    \item [part 1]: ($\approx$50') introduction to quantum superposition through the analysis of experiments with Stern Gerlach apparatus (see Section \ref{subsec:SGAsims})
    \item [part 2]:(homework) experiment to determine the equation of a state $|\theta \rangle$ prepared in a generic direction $\theta$ using $|\uparrow \rangle$ e $|\downarrow \rangle$ basis
    \item [part 3]: ($\approx$80') study of the EPR problem and Bell's inequality by comparing predictions of the outcomes of experiments made under the hidden variable hypothesis with quantum probabilities.
\end{enumerate}

As these are online lessons, the first and third parts were structured according to the chunked lesson methodology (see Section \ref{sec:online_methodology}).


In the first part, students are introduced to the use of the Stern Gerlach apparatus and to make predictions about the experimental results of specific apparatus sequences (see Section \ref{subsec:SGAsims}).
In small groups created through Rooms \cite{Saltz2020_OnlineLearn}, students are asked to verify their results using the simulator and to justify their answers through short texts that are then shared on an online platform visible to all participants.
This interplay between predictions and results helped students make their assumption about the nature of quantum objects more explicit and identify the coherence between quantum models and experimental results.

One student commented this way:
\begin{quote} {\fontfamily{cmtt}\selectfont
the result of this experiment [spin state measurement using a sequence of three Stern-Gerlach apparatus] is coherent (agrees) with quantum physics, with the superposition principle. [...] The classical interpretation would have foreseen a different outcome [StudentsRefl]
}\end{quote}

In this context, a formal analysis of the experiments (see Section \ref{sec:seqSGA_matrix}) was not proposed for the construction of the theoretical interpretative model, preferring to give the lectures a qualitative slant. The rigour sought was to base conclusions about quantum behaviour on the results of experiments.


The simulations were used to study quantum entanglement within the framework of EPR interpretative model and Bell’s inequalities \cite{Pospiech1999_EPRatHS}.\\
In the second part, using their experience with the Stern-Gerlach simulators, the students were able to frame the quantum state of the electron in a general $\theta$ direction by measuring the outcomes of the simulator.\\

\begin{figure}[hbt!]
    \centering
   \includegraphics[width=\textwidth,height=\textheight,keepaspectratio]{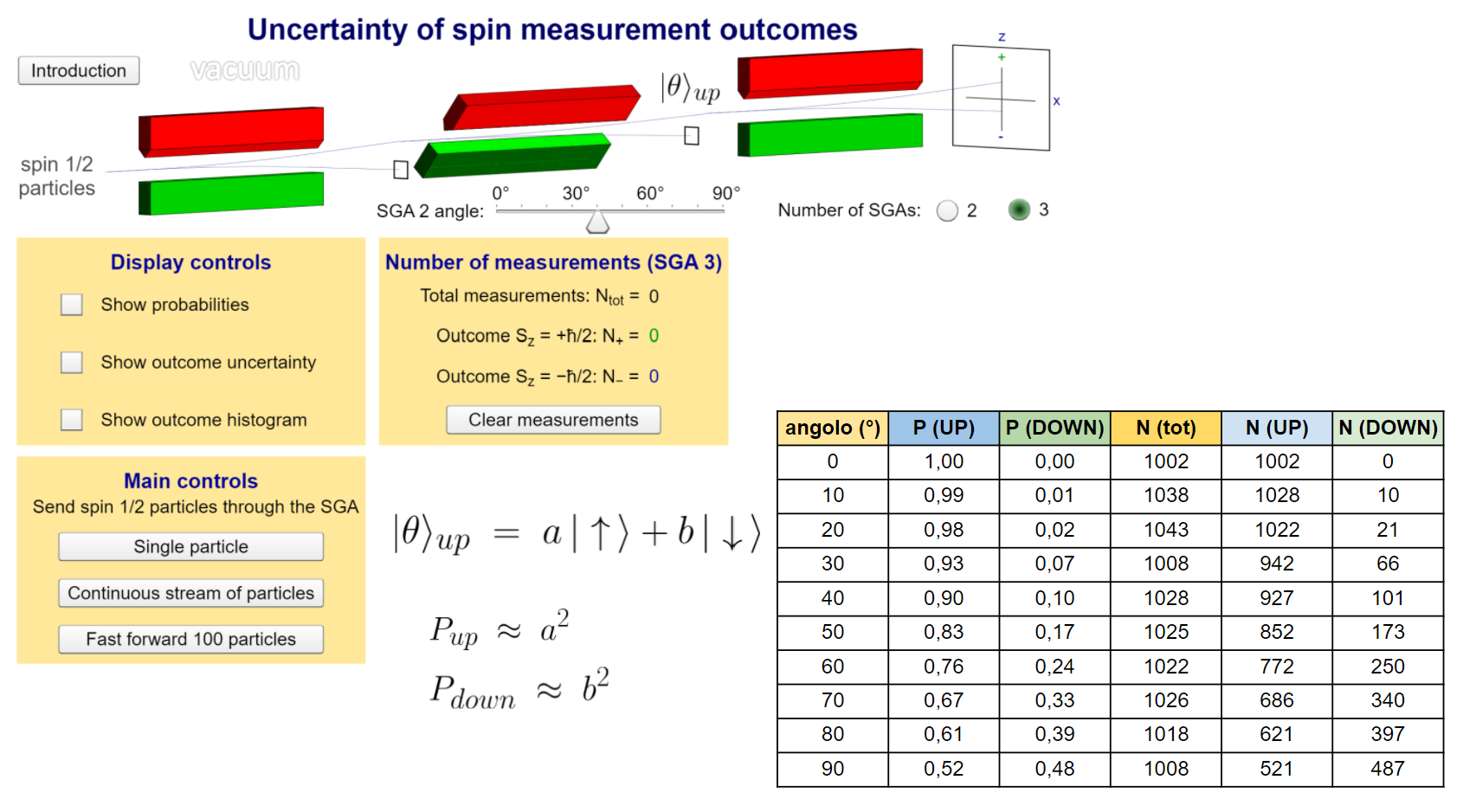}
    \caption{Experimental construction of the output state of an apparatus oriented in the $\theta$ direction. The figure shows the experimental setting used by the students and an example of the data collected. 
    }
    \label{fig:Ch6_23_EntagQuVISData}
\end{figure}

An interesting resource developed by the teacher to introduce Bell's experiment was to use the simulator to reconstruct the electron spin state at any angle. This activity was proposed as homework and the analysis of the results formed the start of the second meeting.
Students are aware that the square value of the coefficients in front of the basis vectors gives the probability of measuring that particular value of the observable associated on that particular state.  Instead of using the vector and matrix representation (sec Sectin \ref{sec:seqSGA_matrix}), students were given the challenge of reconstructing the coefficients of the spin state prepared in the $\theta$ direction out of a Stern Gerlach apparatus oriented along $Z$ (see Figure. \ref{fig:Ch6_23_EntagQuVISData}). The data collected are the number of relative detections, i.e. how many electrons are measured with spin "+1" (UP) or "-1" (DOWN) with respect to the total number of objects detected on the screen (which do not correspond to the total number of objects sent because a part is blocked).

The reconstruction of the expression for the state was done by comparing the qualitative fit of the data obtained with different functions of the angle $\theta$.

\begin{figure}[hbt!]
    \centering
   \includegraphics[width=\textwidth,height=\textheight,keepaspectratio]{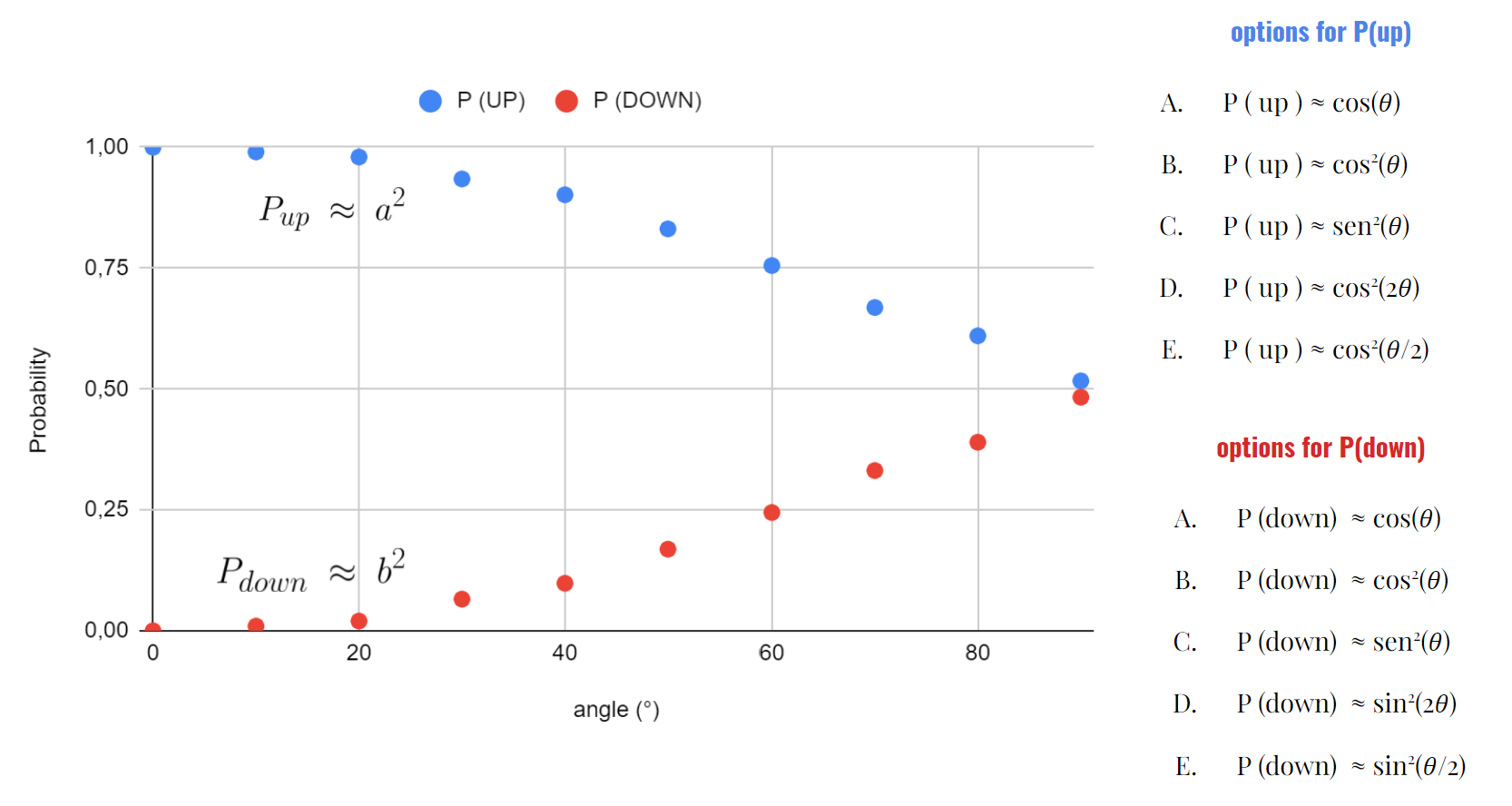}
    \caption{Experimental reconstruction of the coefficients of the output state of an apparatus oriented in the $theta$ direction. The graphs obtained are shown in the figure. In the activity, students have to understand which of the proposed functions can be used to fit the data. In this way it is possible to identify the functions of the angle $theta$ that can be used as coefficients $a$ and $b$ in the expression of $|theta \rangle_{up}$.}
    \label{fig:Ch6_24_EntagQuVISAnaly}
\end{figure}

From the analysis of the data students can recognise the function that describes the probability of detecting a spin UP in relation to the value of the angle $\theta$
\begin{equation}
    P(\mbox{spin UP}) \; = \; \cos^2 \frac{\theta}{2} \qquad
    P(\mbox{spin DOWN}) \; = \; \sin^2 \frac{\theta}{2}
\end{equation}\label{eq:probQuantum}

e quindi ricavare che lo stato UP preparato nella direzione $\theta$ scritto nella base \{$|\uparrow \rangle \,,\, |\downarrow \rangle$\} è dato da


\begin{equation}
    |\theta\rangle_{up} \; = \; \cos \left( \frac{\theta}{2}\right) \, |\uparrow \,\rangle + \sin \left( \frac{\theta}{2}\right) \, |\downarrow \, \rangle
\end{equation}




This activity allows them to build a proper mathematical tool to compare classical predictions of the Bell’s experiments with the quantum ones.\\

In the third part, the concept of entanglement is introduced in the context of the EPR problem.
The EPR \textit{gendaken} experiment considers pairs of entangled spin $\frac{1}{2}$ particles (e.g. electrons) generated in the singlet state
\begin{equation}
    \ \psi \rangle \; = \; \frac{1}{\sqrt{2}} |\uparrow \downarrow \rangle \, + \, \frac{1}{\sqrt{2}} |\downarrow \uparrow \rangle
\end{equation}
The particles are spatially separated and sent to two different observers that can measure the spin components on different basis.

The activity is centred around the comparison of the ability to make experimentally consistent predictions possible using either the hidden variable hypothesis or the probabilities defined by quantum hypotesis.


\begin{figure}[hbt!]
    \centering
   \includegraphics[width=\textwidth,height=\textheight,keepaspectratio]{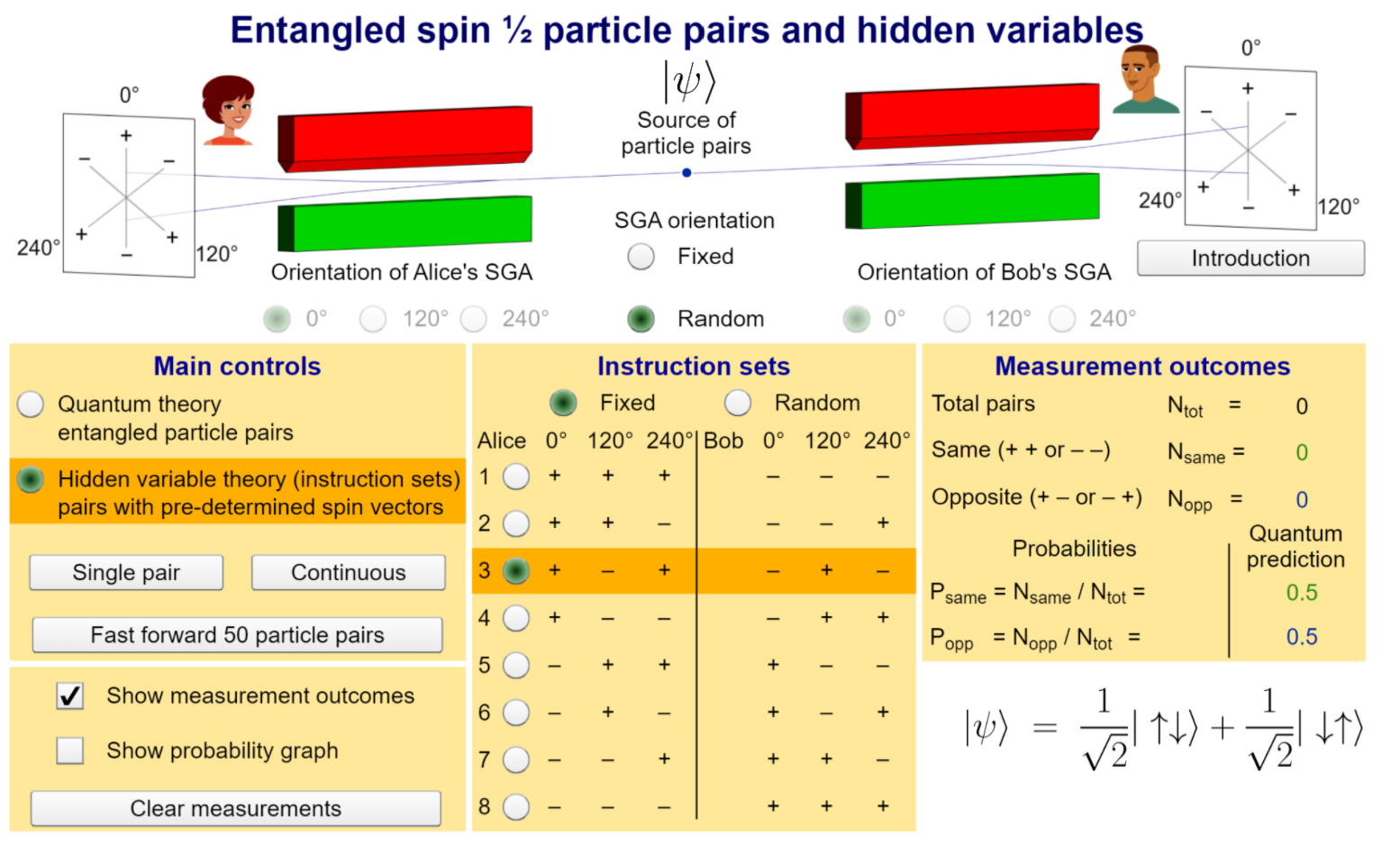}
    \caption{Quvis simulator for the exploration of the hidden variable hypotesis}
    \label{fig:Ch6_24b_HiddenVarSIM}
\end{figure}

\begin{figure}[hbt!]
    \centering
   \includegraphics[width=\textwidth,height=\textheight,keepaspectratio]{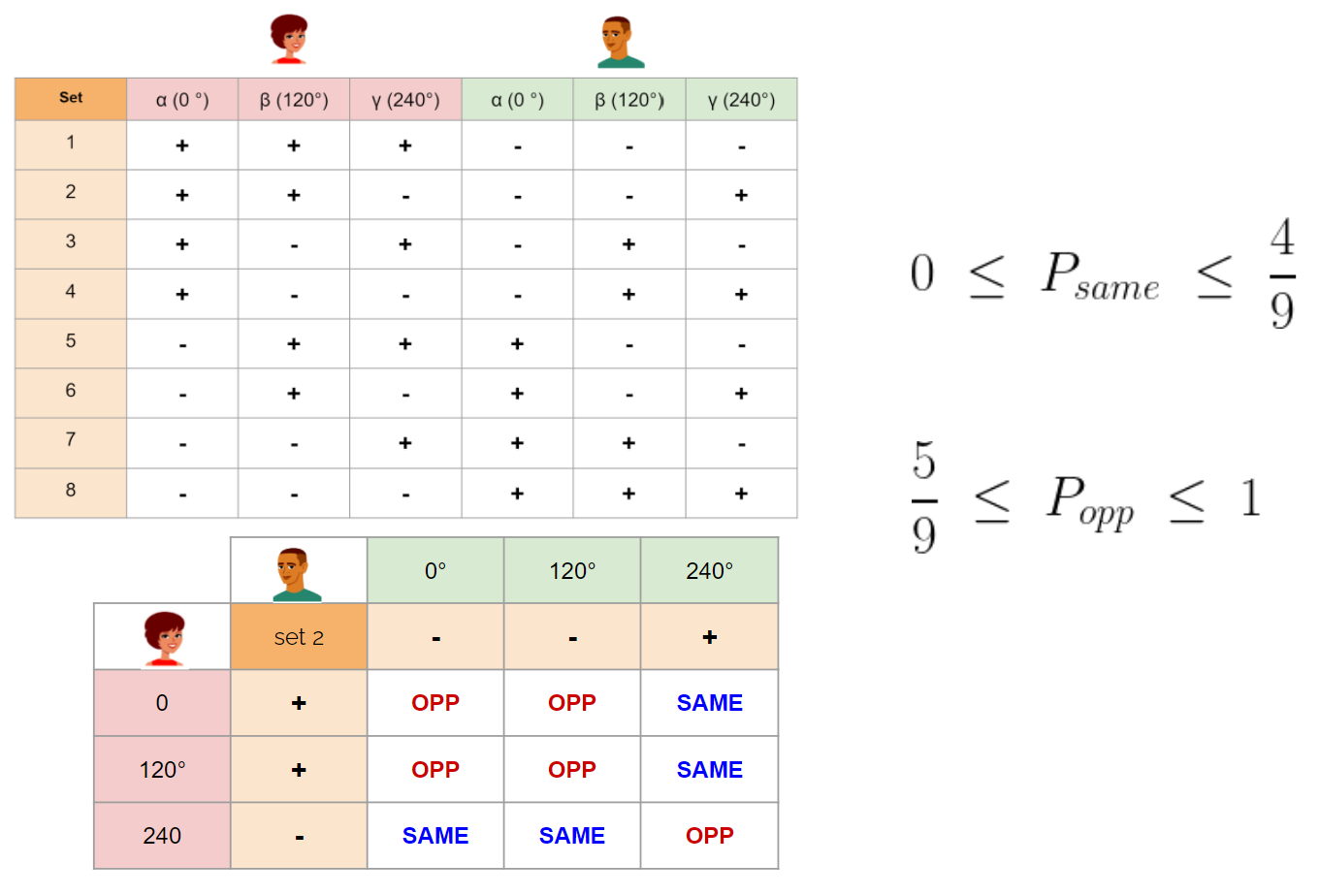}
    \caption{Le previsioni del modello a variabili nascoste si possono ottenere analizzando gli elementi della tabella}
    \label{fig:Ch6_25_HiddenVar}
\end{figure}

With the help of a simulator (see Figure \ref{fig:Ch6_24b_HiddenVarSIM}) and working in small groups, it was possible to guide the students in deriving the predictions of the two models.
It can be verified that the predictions based on hidden variables are (see Figure. \ref{fig:Ch6_25_HiddenVar})

\begin{equation}
    0 \,\leq\,P_{same}\,\leq\,\frac{4}{9} \qquad \frac{5}{9} \,\leq\,P_{opposite}\,\leq\,1
\end{equation}\label{eq:EPRHiddenProb}

These two inequalities are two possible formulations of Bell's inequalities: an inequality that defines conditions that measurement probabilities must meet in the context of the theory of hidden variables (i.e. if we consider \textit{reasonable} the assumptions of "realism" and "locality").

\begin{figure}[hbt!]
    \centering
   \includegraphics[width=\textwidth,height=\textheight,keepaspectratio]{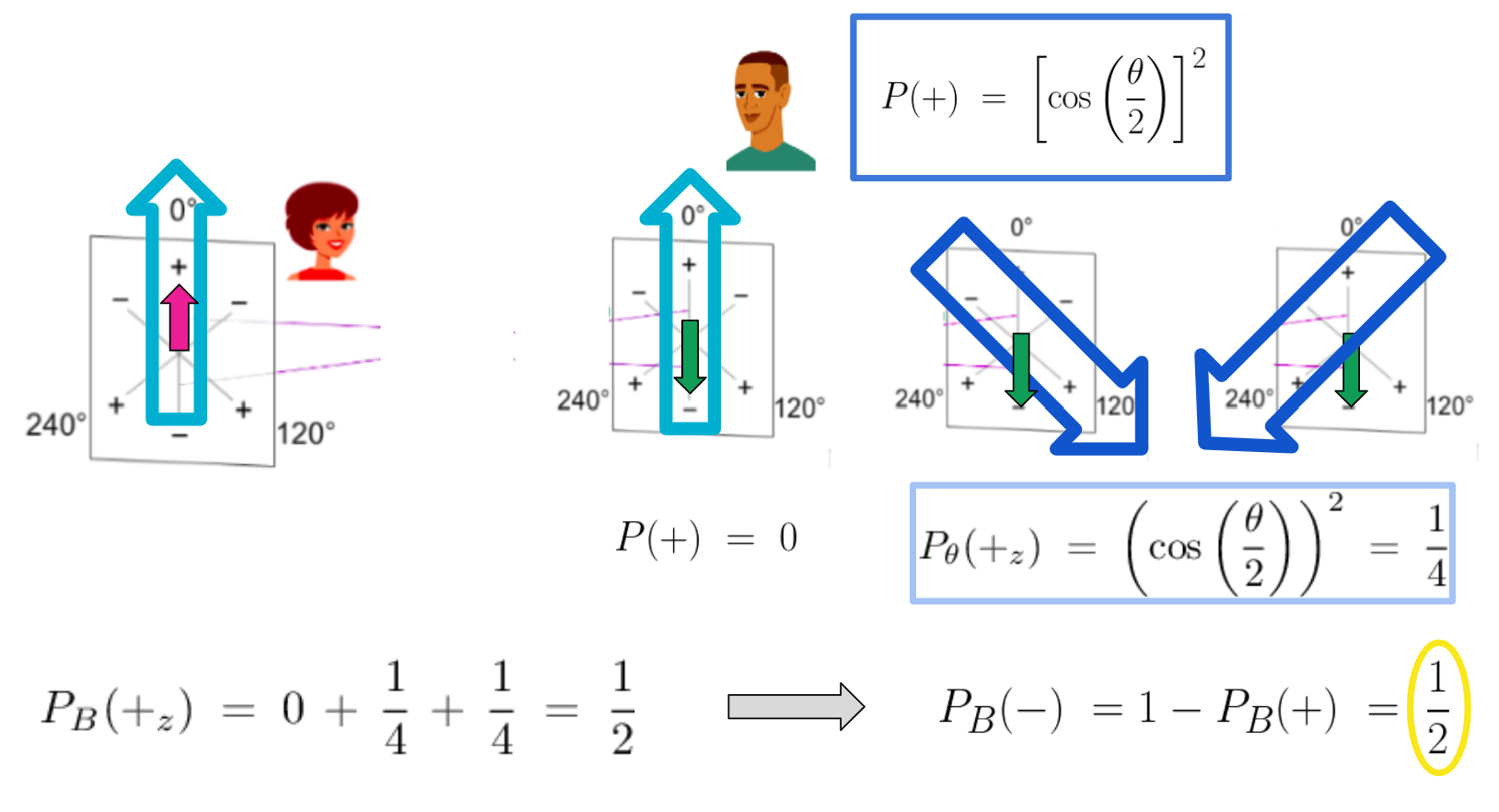}
    \caption{Quantum predictions can be obtained by considering the different (mutually exclusive) measurement situations and the formulas for calculating the probabilities obtained from the analysis of the experiments with the Stern Gerlach sequences}
    \label{fig:Ch6_26_QuantumProb}
\end{figure}

Still referring to what happens in the simulation (see Figureure \ref{fig:Ch6_26_QuantumProb}) and calculating quantum probabilities using equation \ref{eq:probQuantum} we obtain that

\begin{equation}
    P_{same} \; = \; P_{opposite} \; = \; \frac{1}{2}
\end{equation}\label{eq:EPRQuantumProb}


Results in equations \ref{eq:EPRHiddenProb} and \ref{eq:EPRQuantumProb} are not compatible with each other. Using the comparison with experimental data as a reference, attention is drawn to the fact that experiments carried out with quantum systems \cite{kaiser2020tacklingBellLoopholes} \cite{Pan2000_BellLoopholes} give results consistent with quantum predictions stated in Equation \ref{eq:EPRQuantumProb}.


This is a crucial point of arrival in relation to the learning goals set at the beginning of the course.

In one of his post-activity reflections, the teacher added:

\begin{quote} {\fontfamily{cmtt}\selectfont
   I believe it is important for students to realize that... beyond personal opinions... in the scientific field there are ways to compare two theories/hypotheses and to test whether or not a theory is correct… whereas the simple "I think so" is not "Consistent". [TeachRefl]
}\end{quote}


\subsubsection{Results}
As for the general objective of the course, the teacher was satisfied.
\begin{quote} {\fontfamily{cmtt}\selectfont
    from what I heard in the interviews/quizzes, it seems to me that for almost all students [the cultural objective] has been fully achieved.
}\end{quote}
The students were in fact generally able to argue the description of the most counterintuitive aspects of quantum behaviour by making connections with the experimental aspects.
In the words of one student (collected during in-class implementation of activities)

\begin{quote} {\fontfamily{cmtt}\selectfont
    I struggle to visualise, to imagine these things. What does it mean that something is in a state of superposition? Having seen the experiments gives me a sense of reality \dots These things are really happening, even if I can't find a way to simply tell and describe them.
}\end{quote}

As for the second objective, the assessment is positive with respect to the use of experiments for knowledge building.

\begin{quote} {\fontfamily{cmtt}\selectfont
    [the idea of using experiments to built knowledge] seems to have passed for a majority of the students. If the idea that a scientific theory must always be "measured" by experimental verification seems to have passed \dots perhaps the idea that the validity of the theory is also measured by its ability to make predictions that turn out to be correct has been less understood. I fear this concept has remained in the background and is really grasped by not many students.
}\end{quote}

And the teacher was in favour of further applying the introduced approach.

\begin{quote} {\fontfamily{cmtt}\selectfont
    When I do teach [quantum] physics (almost certainly not next year), the approach to quantum physics that I will take will be as similar as possible to the one taken in this "course".
}\end{quote}

The work done by the lecturer shows a development with respect to different aspects related to PCK, showing great awareness of the interaction between the different dimensions (see Table \ref{tab:PCK_Analysis_GRID})


A first element concerns the development of the subject matter knowledge for teaching and the process of identifying the central concepts on which to base the reconstruction of the contents to be proposed to the students. The teacher has worked to create an important synthesis between his teaching experience, his specific disciplinary knowledge and the need to redesign content teaching starting from the identification of some key concepts.

A second aspect concerns aspects of educational structuring.  The teacher was able to exploit his considerable IT skills to build a complete virtual learning environment which proved essential for online activities.
The teacher also knew how to integrate the tool of simulations in a very effective way and in line with the general objectives of the planned course, showing great flexibility in integrating different types of teaching tools (lecture, use of simulators, laboratory experiences). The teacher's knowledge of the difficulties linked to the acquisition of new concepts stimulated the creation of a learning environment in which different types of representations (mathematical, formal and visual) contributed to the students' involvement. Students were stimulated to go beyond previous ideas about the nature of quantum objects and classical ways of describing their behaviour.\\
As a result of the development of the proposal on entanglement, the dialogue between teachers and researchers led to the definition of a proposal on the study of the "robustness" of the hypotheses of hidden variables through the search for a "counterexample".
This activity was not carried out with the students but was designed after the activity following reflections on the results of the experiment.
It involves constructing an inequality that is always valid in the context of the hidden variables. An example is then identified in which the quantum predictions (which are in agreement with the experimental results) show that this inequality is false.
In addition to completing the discussion of Bell's inequalities, this type of argumentation becomes an opportunity to reflect on the methods of scientific argumentation \cite{newton1999placeArgum, osborne2012introArgument,cavagnetto2012importanceArg} and elements of nature of science \cite{Stadermann2020ConnectNOS, dawson2010teachingArgument}.


What is then achieved is the construction of a different formulation of Bell's inequalities from the analysis of the data and the table in the simulator.

\begin{figure}[hbt!]
    \centering
   \includegraphics[width=\textwidth,height=\textheight,keepaspectratio]{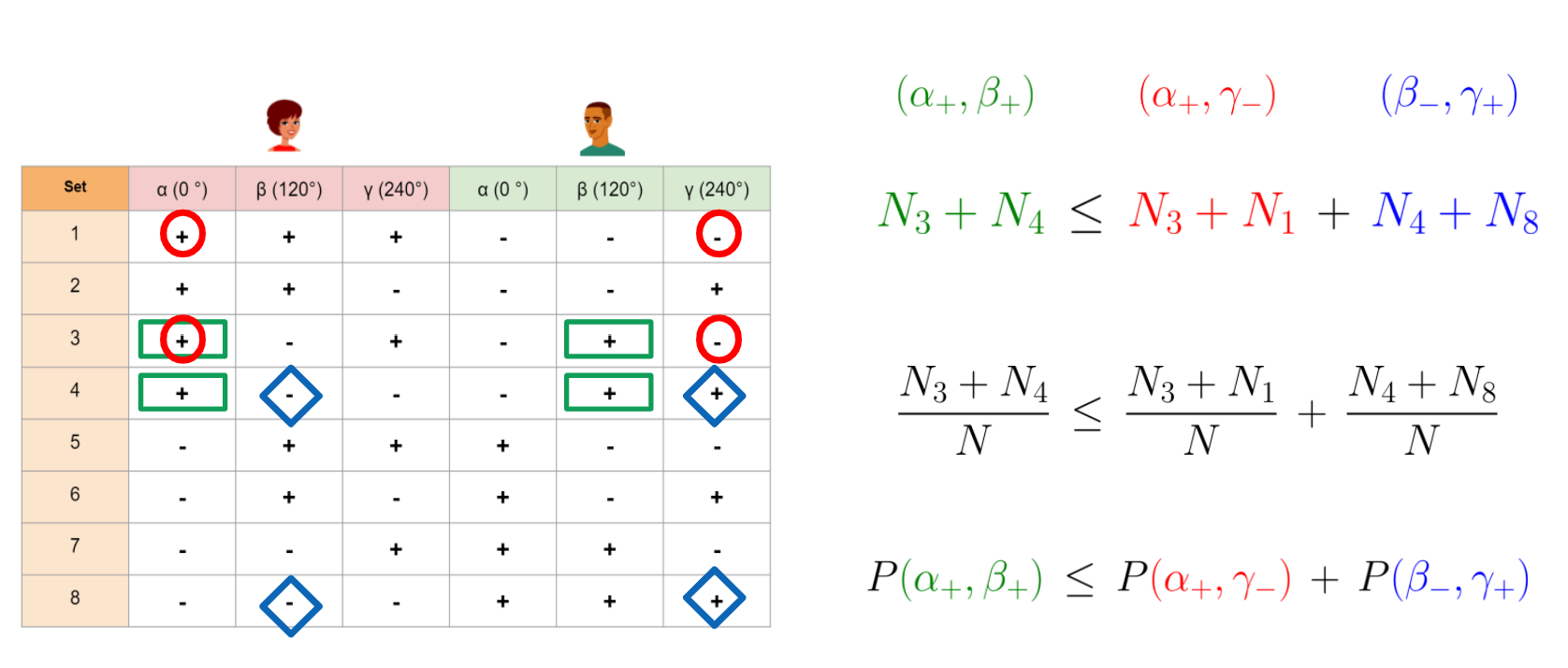}
    \caption{Bell's inequality construction: the inequality is constructed by taking as a reference the table with the sets of counts provided by the hidden variable model.}
    \label{fig:Ch6_27_BuiltBell}
\end{figure}

The construction (see Figure \ref{fig:Ch6_27_BuiltBell}) is done for example considering the case $(\alpha_+, \beta_+)$, i.e. all the cases in which Alice measures $\alpha_+$ and Bob measures $\beta_+$. According to the table, the cases are obtained by summing the counts $N_3$ relative to set 3 and $N_4$ relative to set 4, i.e. $N_3 + N_4$.

\begin{equation}
    (\alpha_+, \beta_+) \longleftrightarrow  N_3 + N_4
\end{equation}

The inequality
\begin{equation}
    N_3 + N_4 \leq N_3 + N_ 1 + N_4 + N_8
\end{equation}\label{eq:N3N1N4N8}

is surely true since $N_1$ and $N_8$ are positive quantities. Still considering the table shown in Figure \ref{fig:Ch6_27_BuiltBell} we see that

\begin{equation}
    (\alpha_+, \gamma_-) \longleftrightarrow  N_3 + N_1 \qquad (\beta_-, \gamma_+) \longleftrightarrow  N_4 + N_8
\end{equation}

Dividing both members of the equation \ref{eq:N3N1N4N8} by the total number of $N$-counts yields the following inequality (eq. \ref{eq:ProbBellalphabetagamma}) between the probabilities of measuring specific pairs of spin values, which is surely true under the assumptions of realism and locality that is assumed to organize the data in the table.



\begin{equation}
   P(\alpha_+,\beta_+) \,\leq \,P(\alpha_+, \gamma_-) \, + \, P(\beta_-, \gamma_+)
\end{equation} \label{eq:ProbBellalphabetagamma}

The formulae relating to the calculation of quantum probabilities in the case of the pair of apparatuses can be derived or verified by an experiment similar to the previous one (part 2 of the path) by choosing fixed orientations in the apparatuses of Alice and Bob.

\begin{equation}
    P(\alpha_+,\beta_-)\,=\,P(\alpha_-,\beta_+)\,=\,\frac{1}{2} \cos^2 \left( \frac{\alpha-\beta}{2}\right)\\
\end{equation}

\begin{equation}
    P(\alpha_+,\beta_+)\,=\,P(\alpha_-,\beta_-)\,=\,\frac{1}{2} \sin^2 \left( \frac{\alpha-\beta}{2}\right)
\end{equation}

\begin{figure}[hbt!]
    \centering
   \includegraphics[width=\textwidth,height=\textheight,keepaspectratio]{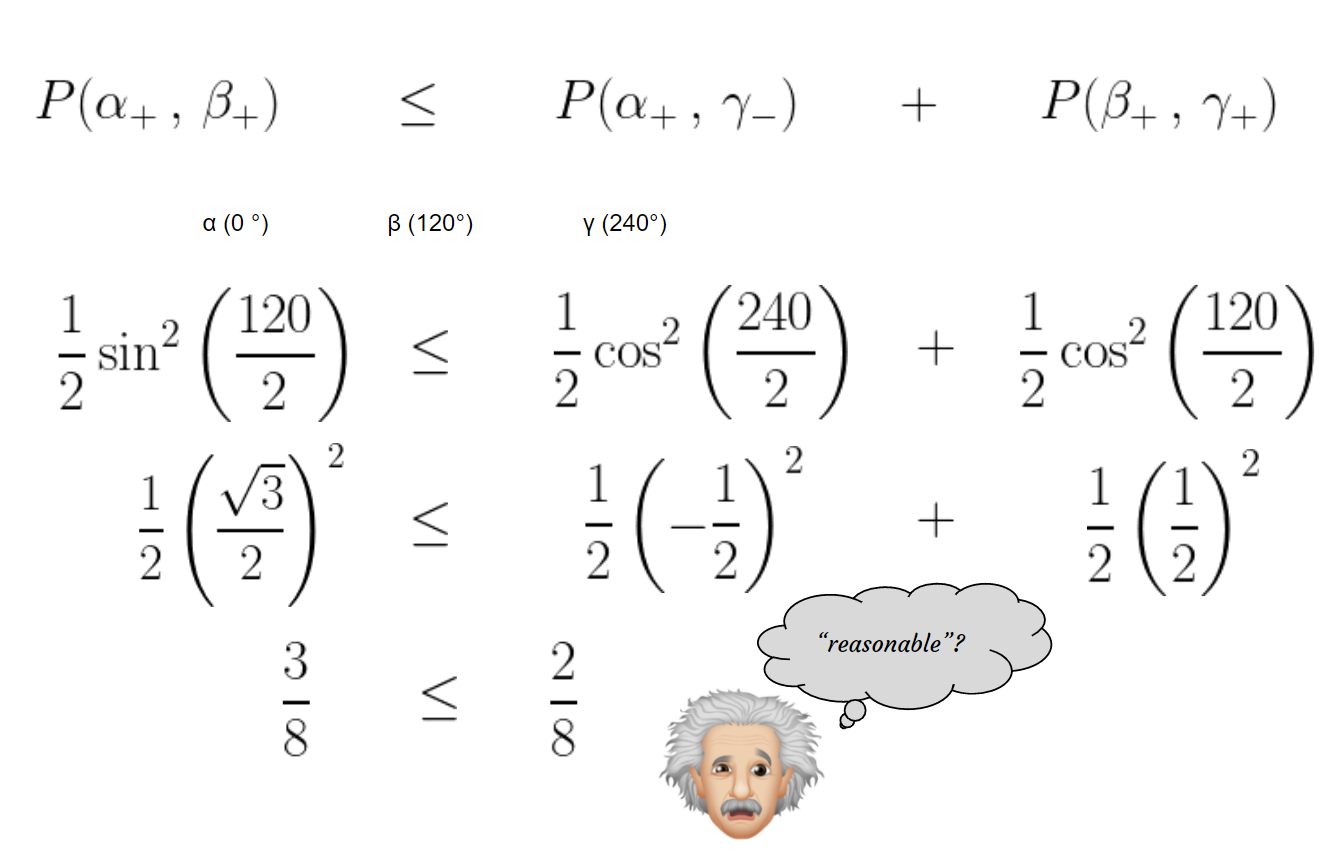}
    \caption{Calculations to show how quantum probabilities, consistent with experiments, violate the inequality constructed under the assumptions of realism and locality inherent in the hidden variable model.}
    \label{fig:Ch6_28_ViolazioneBell}
\end{figure}

A simple calculation made by substituting the values of the angles (see Figure. \ref{fig:Ch6_28_ViolazioneBell}) leads to a false inequality and thus the quantum laws that allow for experimentally consistent predictions do not respect the realism and locality constraints of the hidden variable model.



\subsection{Duality and the nature of quantum objects: from electrons diffraction to quantum states} \label{sec:carcano}

The diffraction of an electron beam proposed in the Quantum Jumps PD  was the “central experiment” \cite{Jones1991_photonMiscon} used by the teachers to start the investigation of the nature of quantum objects. That laboratory experience was already part of most teachers’ physics syllabus and was used to provide students with a direct measurement of the wavelength of the wave associated with the electrons.\\
In this context, the electron can be represented in terms of a quantum state, which provides “the only knowledge or information about some aspect of reality \cite{Pusey2012_QstatesReality}. The qubit approach developed was adapted for the construction of a path to investigate the quantum nature of the electron.
The two teachers who designed the activity started from the knowledge that in science courses students become familiar with different representations of the electron, especially in the study of atomic models. In that context, the dual nature of the electron as an object with both particle and wave behaviour is presented. The aim of the activity is to clarify some aspects of dualism by trying to base the description of the behaviour of the electron on the analysis of experimental evidence. In the words of one of the teachers.

\begin{quote} {\fontfamily{cmtt}\selectfont
the students know what an electron is, in the sense that they consider it to be a fundamental element of the atom. They know about the existence of orbitals and I would like to try to go into the details of this description with them and I would like it to be a discourse based on experiments. [TeachRefl]
}\end{quote}

\begin{figure}[hbt!]
    \centering
   \includegraphics[width=\textwidth,height=\textheight,keepaspectratio]{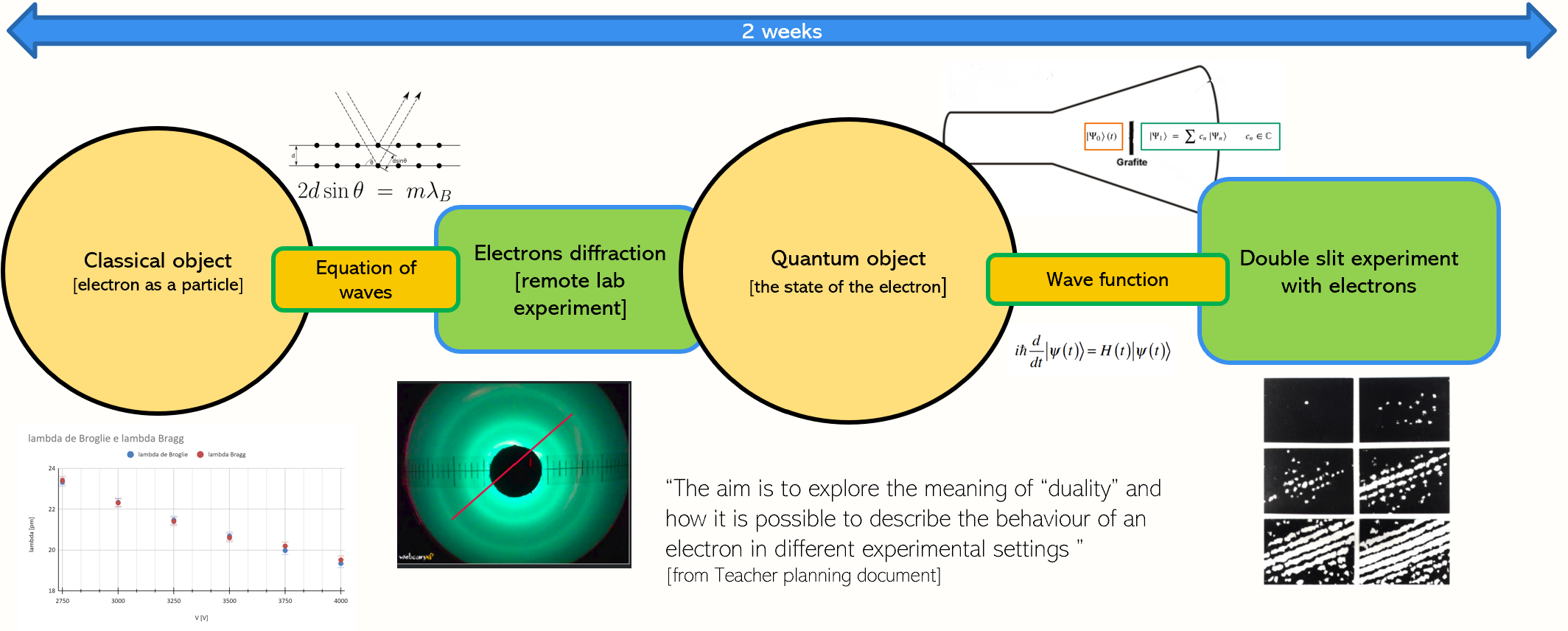}
    \caption{Struttura generale del percorso}
    \label{fig:Ch6_36_carcanoStructure}
\end{figure}

The idea is to focus on the concept of quantum state as an abstract construct that contains information about the object and can be described and manipulated mathematically. Highlighting the central role of the measurement process as an essential element to attribute a property (position, wavelength) to the physical object, the activity allows to show how it is possible to use the concept of state to interpret the results of some experiments with electrons, such as diffraction grid or double slit (see Figure \ref{fig:Ch6_36_carcanoStructure}).

So the general aim is to explore the meaning of “duality” and how it is possible to describe the behaviour of an electron in different experimental settings.


\begin{table}[hbt!]
\centering
\begin{tabular}{cp{12cm}}
\toprule
 & \textbf{ILO}: characterize the quantum nature of the electron in the context of different experimental settings using the concept of quantum state. \\
 &\textit{At the end of the activities students should be able to:} \\
\midrule
ILO-a & describe the results of experiments about electrons using specific language and the wave formalism\\
ILO-b & reflect on the nature of the electron by relating the results of experiments to the concept of the quantum state
\end{tabular}
\caption{ILOs: duality and the nature of quantum objects}
\label{tab:ILO_carcano}
\end{table}



Assessment strategies include writing a lab report on the electron diffraction experience and a series of oral tests on topics related to the lab experience.


The course was designed for two classes of a scientific high school (February 2019) and two classes of a technical institute for chemistry education (April 2020). In the first case, the lessons took place in the school's laboratories. In the second case, the laboratory activities took place online using remote lab equipments, while the discussion of the results was coordinated by the teacher in presence.


In the following paragraphs some strategies designed together with teachers are presented. The analysis of the design process is part of the study of the development of the teachers' PCK dimensions (see Section \ref{sec:ERTE}.\\

\begin{figure}[hbt!]
    \centering
   \includegraphics[width=\textwidth,height=\textheight,keepaspectratio]{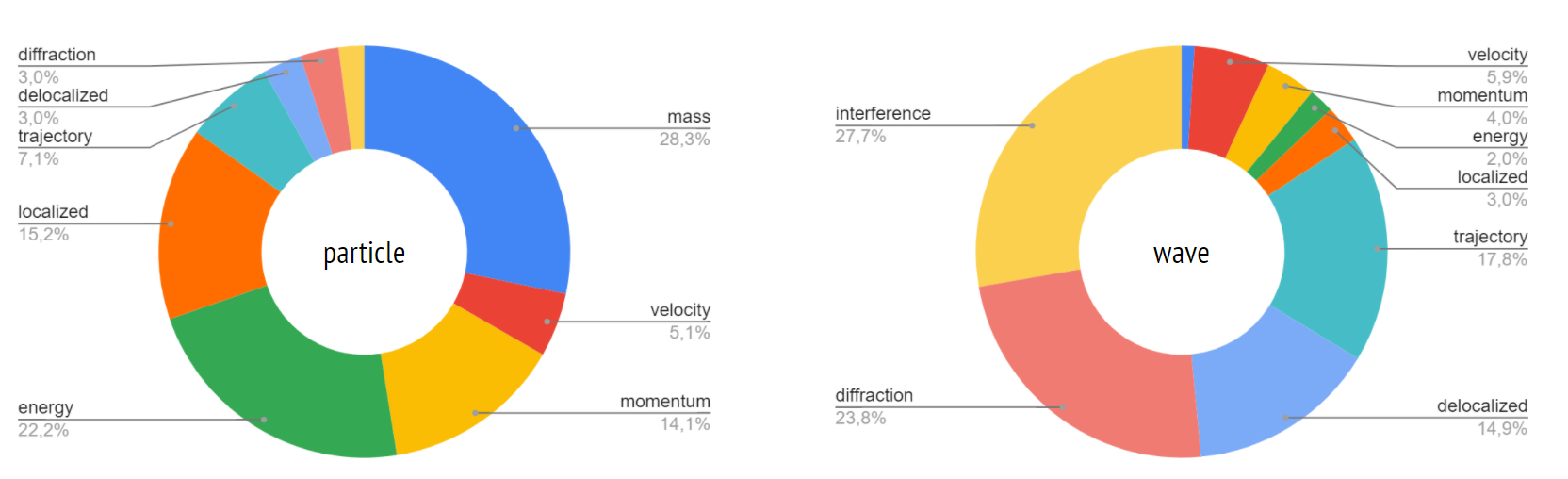}
    \caption{Properties of particles and waves: students had to specify which properties or characteristics are characteristic of particles or waves.}
    \label{fig:Ch6_31a_PartWAve1}
\end{figure}

\begin{figure}[hbt!]
    \centering
   \includegraphics[width=\textwidth,height=\textheight,keepaspectratio]{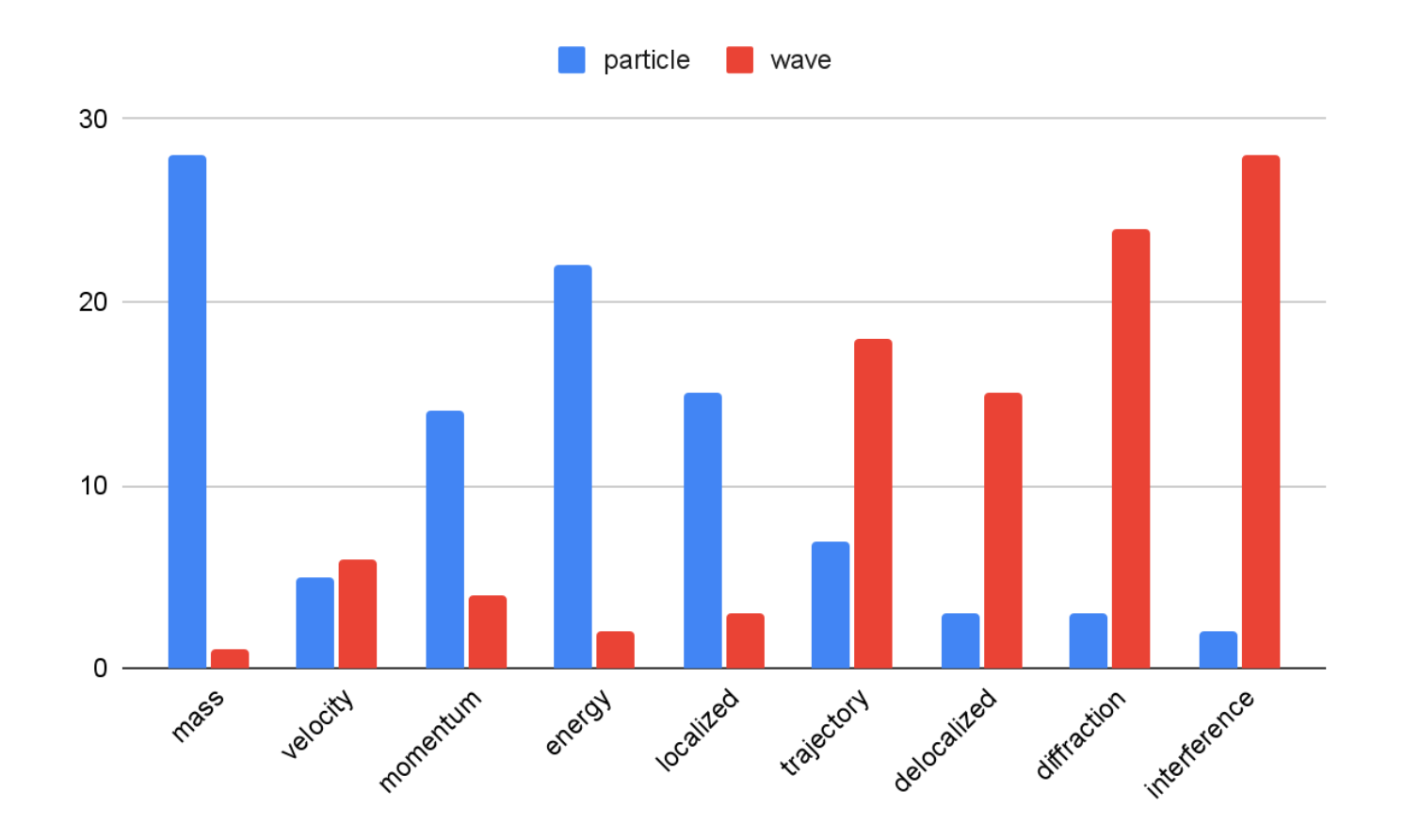}
    \caption{Comparing particles and waves}
    \label{fig:Ch6_31b_PartWave2}
\end{figure}

The first proposed action is an example of how teachers decided to take into account the pre-instrucional knowledge of their students. The proposed activity is an investigation into how students approach the concept of dualism. In particular, the focus is on their knowledge and beliefs about the distinction between particle and wave. By means of a questionnaire to be filled in the days preceding the laboratory activities, the students were asked to list the characteristics of particles and waves, choosing them from a list (see Figureure \ref{fig:Ch6_31a_PartWAve1} and Figure \ref{fig:Ch6_31b_PartWave2}).

The results show that a particle is basically a physical object with mass (28 out of 34) and mechanical energy (22 out of 34). The particle is localised (15 out of 34), i.e. it has a defined position in space and it is possible to define a trajectory of the particle (7 out of 34).
The wave, on the other hand, is mainly characterised by the fact that it can generate interference (28 out of 34) and diffraction (24 out of 34). These are in fact the characteristics that identify waves in many textbooks \cite{romeni2017fisica, amaldi2020il}. Experiments on diffraction and interference are in fact used to establish the wave nature of light. According to the students, it is also possible to determine the trajectory followed by a wave as it propagates through space (18 out of 34) in the same way as it is possible for particles.
At the same time, waves are delocalised (15 out of 34), i.e. they can be at two different points in space at the same time.

\begin{figure}[hbt!]
    \centering
   \includegraphics[width=\textwidth,height=\textheight,keepaspectratio]{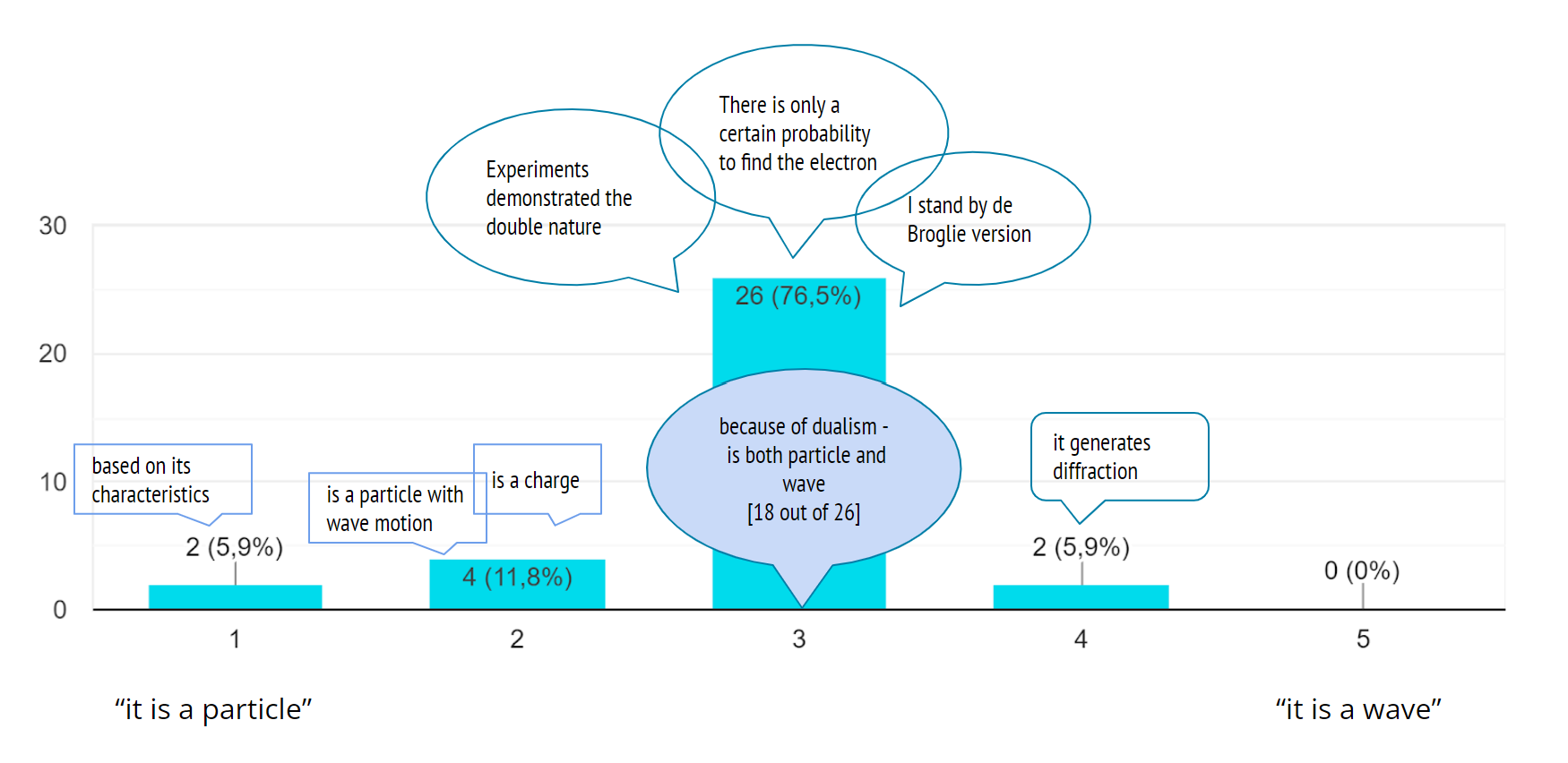}
    \caption{"How can you define an electron": students responses show how they seem to agree about the dual nature of the electron, but that nature is simply stated for the most of them}
    \label{fig:Ch6_30_WhatisElectron}
\end{figure}

Based on these answers, students were then asked to define what an electron is, indicating on a scale whether it is more correct to define it as a particle or a wave, and to justify their choice (see Figure \ref{fig:Ch6_30_WhatisElectron}).
The analysis of the questionnaire data shows that students are already accustomed to describing the nature of the electron as "something dual". This narrative is, however, uncritical as is for light \cite{Henriksen2018_Whatislight}.\\

The teachers agreed that it was necessary to have information on the meaning students give to the different terms before examining the problem of dualism. The data collected on student conceptions was therefore used for the design of the activities. In this way, it is possible to reflect and try to reconstruct the processes or reasoning that students use to attribute different properties to a physical object. The educational objective is therefore to increase the level of complexity of thinking \cite{Webb1997_DOK}, moving from simply remembering an answer to being able to think strategically (see Figure \ref{fig:Ch4_6_DOK}).

To achieve this, the teachers designed the activities by identifying key concepts and relationships.

A first element for the reconstruction of the contents was to structure the interpretation of the phenomenon on the idea that the properties of a physical object can be attributed through the process of measurement. 

In the case of the electron diffraction experiment, the conclusion proposed is that it is possible to assign electrons a wavelength, which is precisely the physical quantity that is measured in the diffraction grating experiment. For a physical object to possess a wavelength means to be delocalised, a wave element. The electron therefore possesses a characteristic or property that in the context of classical physics is proper to waves. This does not mean that the electron \textit{is} a wave. It means that it is possible to say that the electron is delocalised, just like waves.
Therefore both waves and electrins have a characterising element in common: being "delocalised". This characteristic is expressed by the fact that through a measurement it is possible to associate a wavelength to both waves and electrons.

The experiment is part of the course as an opportunity to consolidate the link between the interpretative model and the property associated to the object.
The teachers' proposal to the students is therefore to carry out the diffraction experiment, structuring it around the comparison between the experimental results and the capacity of an intepretative model to generate predictions that are coherent and compatible with these results. On the one hand, the description of the interaction of electromagnetic waves with the crystal lattice is described using the classical wave model. In this context, the Bragg equation allows to derive an expression for the wavelength of the object interacting with the lattice. On the other hand, the de Broglie hypothesis also allows obtaining an expression for the wavelength of the electron.

\begin{figure}[hbt!]
    \centering
   \includegraphics[width=\textwidth,height=\textheight,keepaspectratio]{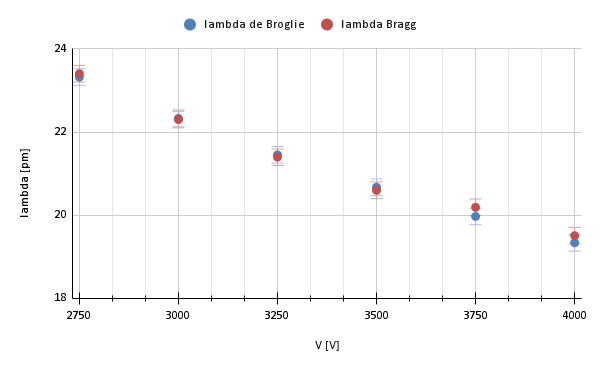}
    \caption{Comparison of wavelength values obtained using de Broglie's relation and the Bragg equation. Experimental results obtained by aggregating the collected students data}
    \label{fig:Ch6_32_ElDiffract}
\end{figure}

The fact that the prediction through de Broglie's hypothesis coincides with the experimental one derived using the wave model serves to justify the idea that the wave formalism is useful for describing the behaviour of electrons. Teachers intention was to support the students in finding the link between interpretative models and the description of the behaviour of the object \cite{schwichtenberg2019no-nonsense}. In the words of a teacher:
\begin{quote} {\fontfamily{cmtt}\selectfont
    We are not saying that the electron is a wave, but that if I want to find a way to describe the behaviour of this particular quantum object and predict the experimental results, I can use the same formalism, the same kind of representation that is used for waves [TeachRefl]
}\end{quote}

The reconstruction process ends with the introduction of the concept of quantum state and its relation to information about the physical object and the measurement of the object's properties. The teachers therefore used an information-based approach to construct an interpretative theoretical framework for the experiment. That involves the use of abstract concepts (such as state) placed in a formal mathematical context. The attention during the design phase was to reduce the use of equations and calculations to a minimum trying to maintain a high level of student participation.

\begin{quote} {\fontfamily{cmtt}\selectfont
    too much maths can kill them [the students] [TeachRefl]
}\end{quote}

The formalism is to be considered as a language to compensate for the limited ability to represent quantum behaviour in terms of familiar representations  \cite{SusskindQT}. In the teachers' intentions, the constant reminders of what students know about waves and the focus on the different meanings of this formalism should help them to develop a way to describe the phenomena.\\

\begin{figure}[hbt!]
    \centering
   \includegraphics[width=\textwidth,height=\textheight,keepaspectratio]{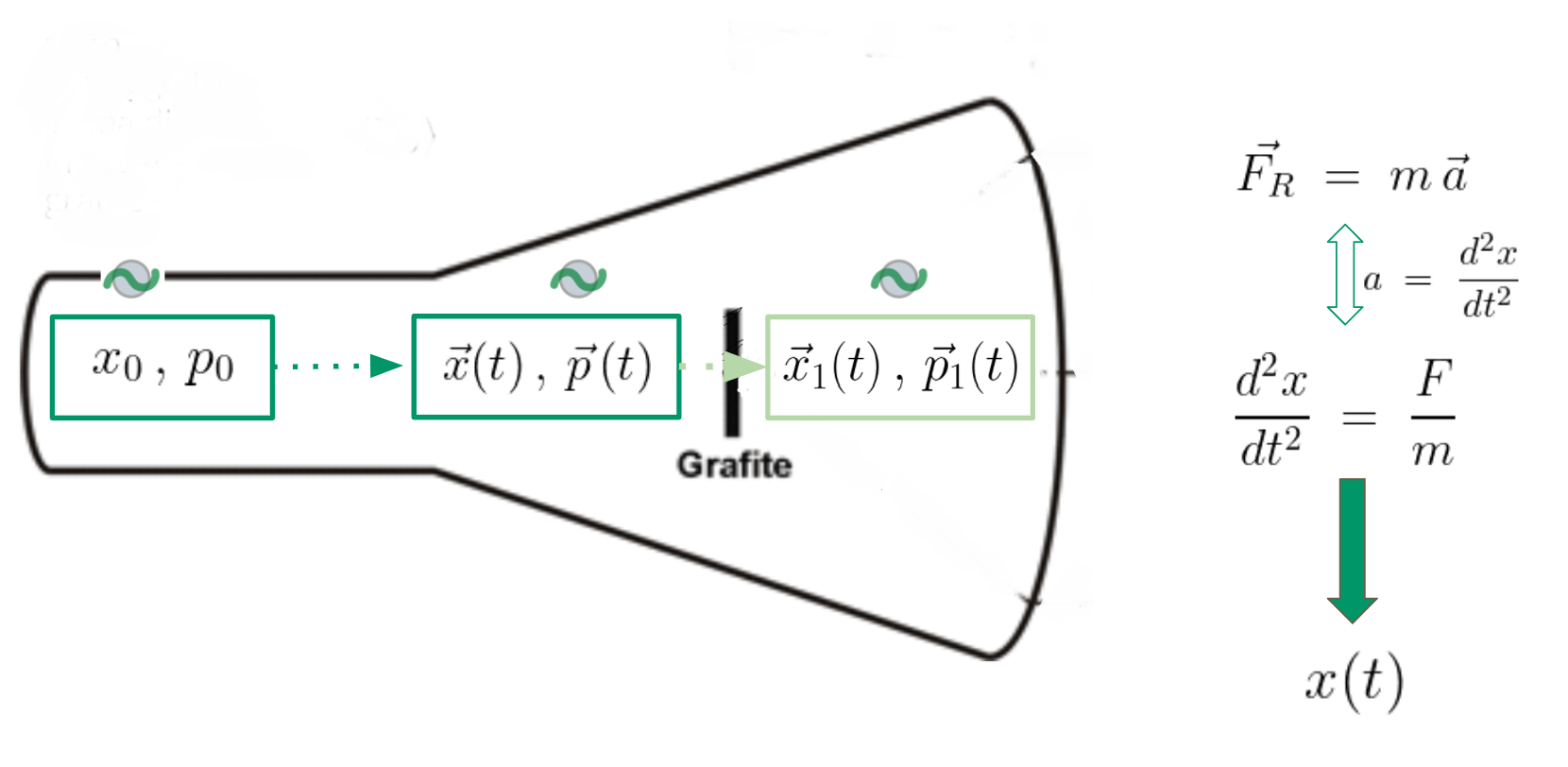}
    \caption{Description of the evolution of a classical state within the apparatus: I can think of the motion of the object as a trajectory whose information about the position of the object can be derived as a solution of the equation of motion}
    \label{fig:Ch6_34_ElEvoluzClassica}
\end{figure}

The first fundamental step in the teachers' reconstruction of the contents is to identify similarities and differences between the description of the object in terms of classical and quantum states.
The choice was to define the state in classical terms as that function which allows deriving information about the object: e.g. where it is located (position) or how it moves (momentum) (see Figure \ref{fig:Ch6_34_ElEvoluzClassica}). This function is obtained as a solution of the classical equation of motion.

The solution of the equation of motion provides the time evolution of the position point by point, the trajectory, and therefore to make deterministic predictions about future positions independently of the execution of a measurement.

The quantum state can also be obtained as the solution of an equation, the Schr\"{o}edinger equation, which was constructed taking into account the de Broglie hypothesis and therefore using the wave formalism. However, its solution, the wave function $|\Psi\rangle (t)$ provides different information. It is still a function that contains the set of information about the object. In this case it expresses the "instructions" to make previsions on the results of a measurement. In this sense the wave function "it is not a description but a prescription" \cite{ball2019beyond}.

\begin{figure}[hbt!]
    \centering
   \includegraphics[width=\textwidth,height=\textheight,keepaspectratio]{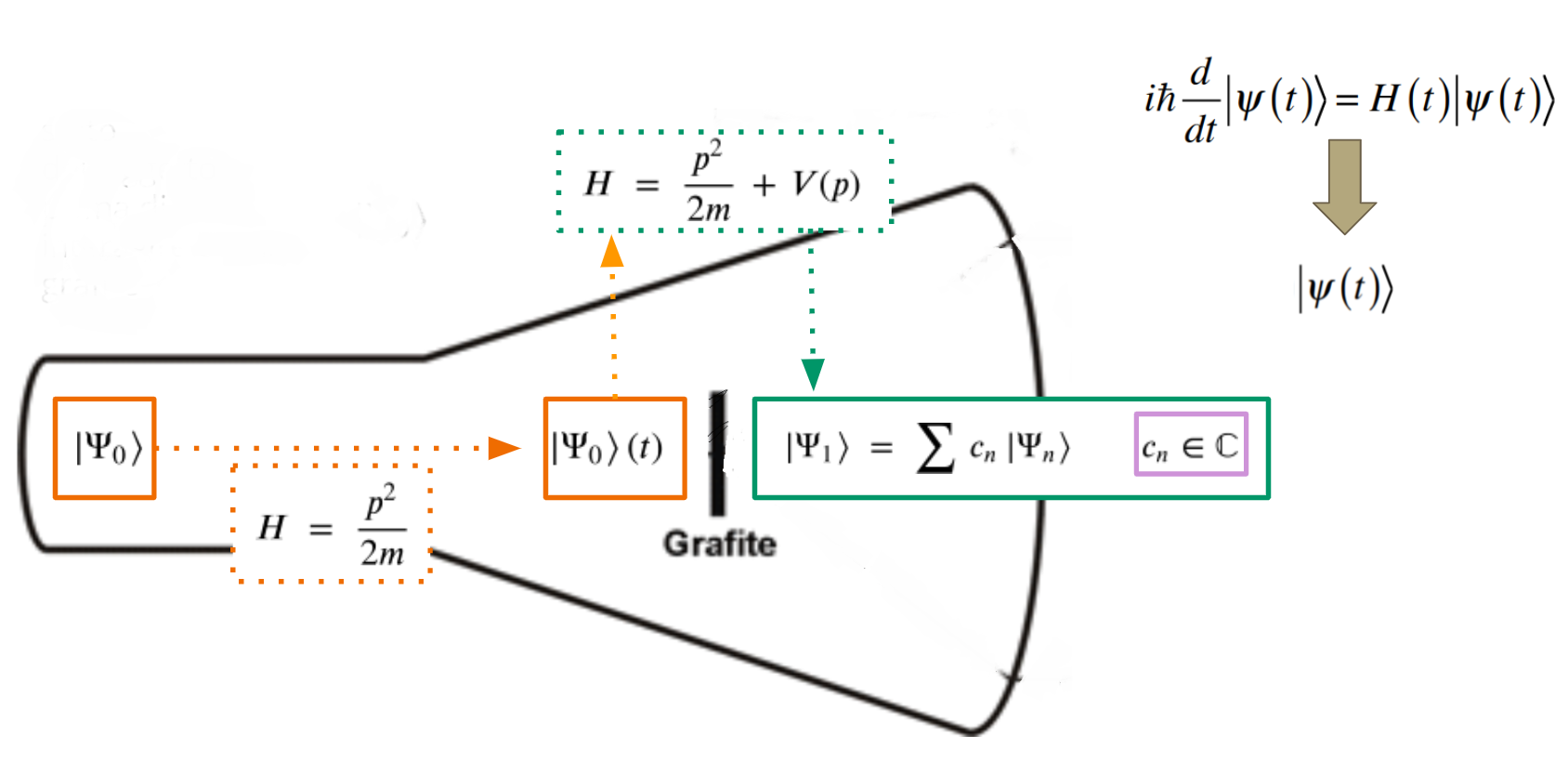}
    \caption{Description of the evolution of a quantum state within the apparatus: as it passes through the graphite, the Hamiltonian includes a potential term describing the interaction with atoms in the lattice modifies the wave function.}
    \label{fig:Ch6_35_ElEvoluzQuantum}
\end{figure}

In order to facilitate the connection with the students' previous knowledge, we intend to exploit their familiarity with the phenomenon of grating diffraction (seen and taken up in detail in preparation for the laboratory experience) \cite{Vokos2000_matterwave}. When a wave passes through a system of slits, the output wave is obtained as a superposition of the wavefronts of the waves emerging from the individual slits. In analogy with Huygens principle for classical waves, the state of the object after it has interacted with the graphite can be described as a superposition of wave packets

\begin{equation} \label{eq:StateSum}
    |\Psi_1\rangle \; = \; \sum c_n |\Psi_n\rangle
\end{equation}

$c_n$ is a complex number. The wave packets have phase differences between one and the other and this generates the diffraction figure that is observed on the screen.


The complex nature of the wave function means that there is no direct correspondence with a real value that unambiguously defines the result of a measurement and therefore the wave function does not correspond directly to a physical object.


The second point concerns the relationship between the state (complex function that contains all the information about the quantum object) and the measure (what can be said about the quantum object).
The proposal is to introduce Born's rule. In this sense the measures of the properties are real numbers and are derived by calculating the amplitudes of these functions in analogy to what is done with waves. In that classical case the square of the amplitude corresponds to the intensity and provides a measure of one property (energy) of the wave.
In this case instead, the modulus of the square of the amplitude gives the probability of finding (measuring) the electron in a certain position at the instant in which the measurement is taken. (see Equation \ref{eq:StateSumBORNrule} ).

\begin{equation} \label{eq:StateSumBORNrule}
    |\Psi_1|^2\; = \; \sum_n \sum_m c_n c_m^* \langle \Psi_n |\Psi_m\rangle
\end{equation}

In the expression there are mixed terms (double products) $c_n c_m^*$ which contain phases that modulate the "intensities", understood as the probability of detecting the electron, and which explain the diffraction/interference pattern.
At the moment of the measurement for example of the position of the electron, it will be detected at a precise point in space. In this way it is possible to extract information about the physical object. Thus, when the measurement is taken, the electron is localised, similarly to a particle.

\begin{quote} {\fontfamily{cmtt}\selectfont
    A key point to be made explicitly is that measurement is always localised [TeachRefl]
}\end{quote}

Since this is a central point in the reconstruction, the teachers have chosen to ask the students to consider what would happen if one electron at a time was sent at the lattice. This is a formative assessment strategy which aims to stimulate in the students the process of reorganising the concepts presented in class, increasing the level of complexity \cite{Webb1997_DOK}. Referring to the Depht of Knowledge (DOK) model, the aim is to promote a shift from the DOK-1 level ("remembering and reproduction") to the DOK-2 level ("skills and concepts") (see Figure \ref{fig:Ch4_6_DOK}).

\begin{figure}[hbt!]
    \centering
   \includegraphics[width=\textwidth,height=\textheight,keepaspectratio]{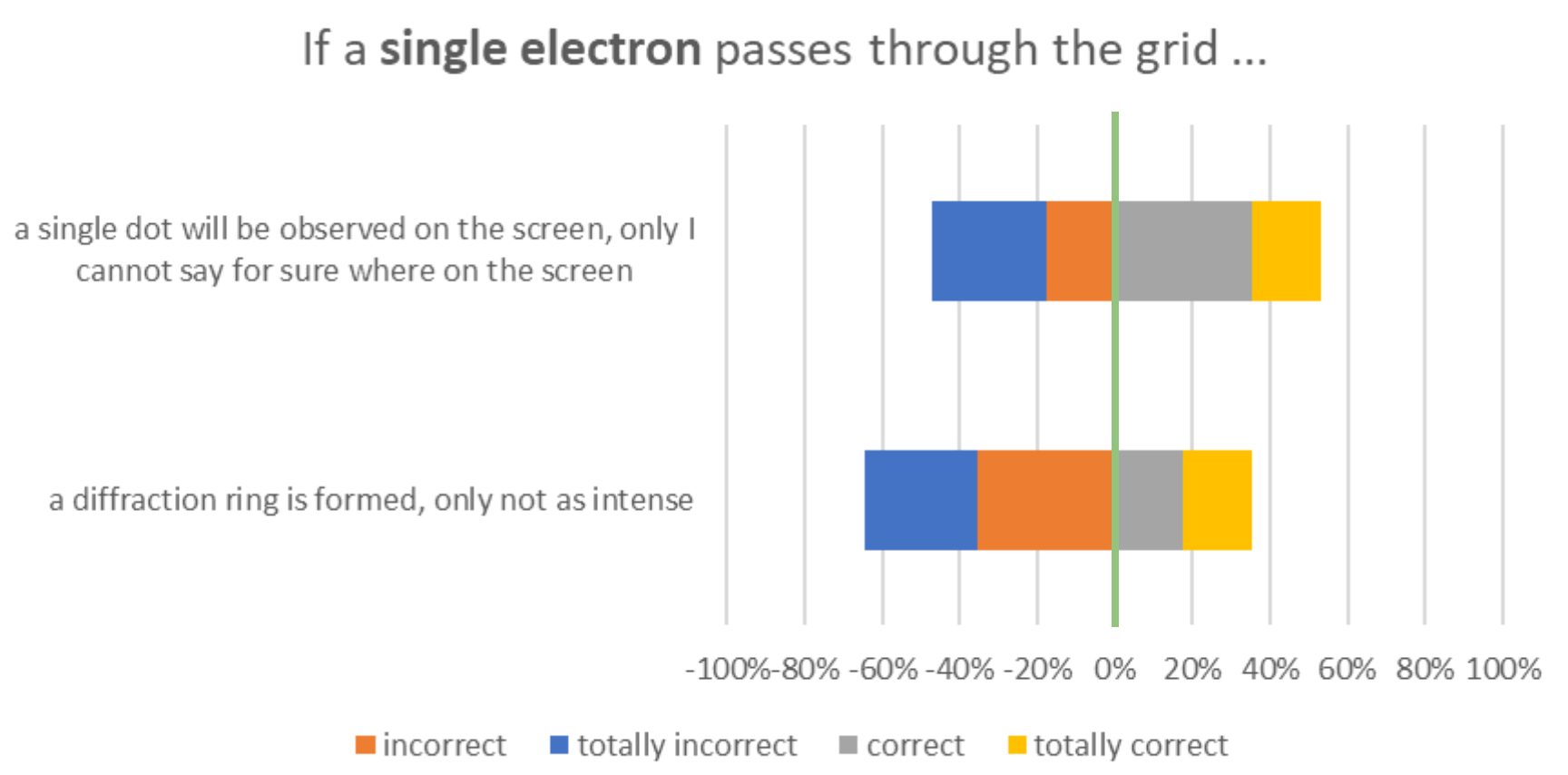}
    \caption[Students'interpretation of sigle electron diffraction]{Interpretation of diffraction figure: Students should specify on a scale of 1 to 4 the level of correctness of each of the sentences presented in representing when the diffraction experiment with a single electron is observed. The first sentence is essentially correct}
    \label{fig:Ch6_37_DiffractInterpretFIG}
\end{figure}

The results (see Figure \ref{fig:Ch6_37_DiffractInterpretFIG}) show that a fair number of students were able to apply what was covered in the course. In the discussion following the presentation, it was concluded that when the electron interacts with the detector (screen) after passing through the graphite lattice, the detector exchanges energy with the electron. This exchange corresponds to the emission of light that can be detect ("seen") at a specific point on the screen.
The interaction between electron and screen occurs at a single point. As confirmed in one of the students comment during the lesson

\begin{quote} {\fontfamily{cmtt}\selectfont
If we send one electron at a time, the diffraction pattern does not appear all at once \dots but slowly forms as the individual electrons arrive. [StudentsRefl]
}\end{quote}

The electrons therefore hit the screen at different points following a particular distribution (described by the wave function). It is therefore a statistical process.

In order to push this process to a higher level of complexity, the teachers chose to apply what they had learned in a different experimental context, such as the double slit experiment with electrons. The students showed that they knew the outcome of the experiment, namely that the electron beam generates the diffraction figure. However, most of the group were unable to identify the type of argument consistent with the interpretation of the phenomenon presented.

\begin{figure}[hbt!]
    \centering
   \includegraphics[width=\textwidth,height=\textheight,keepaspectratio]{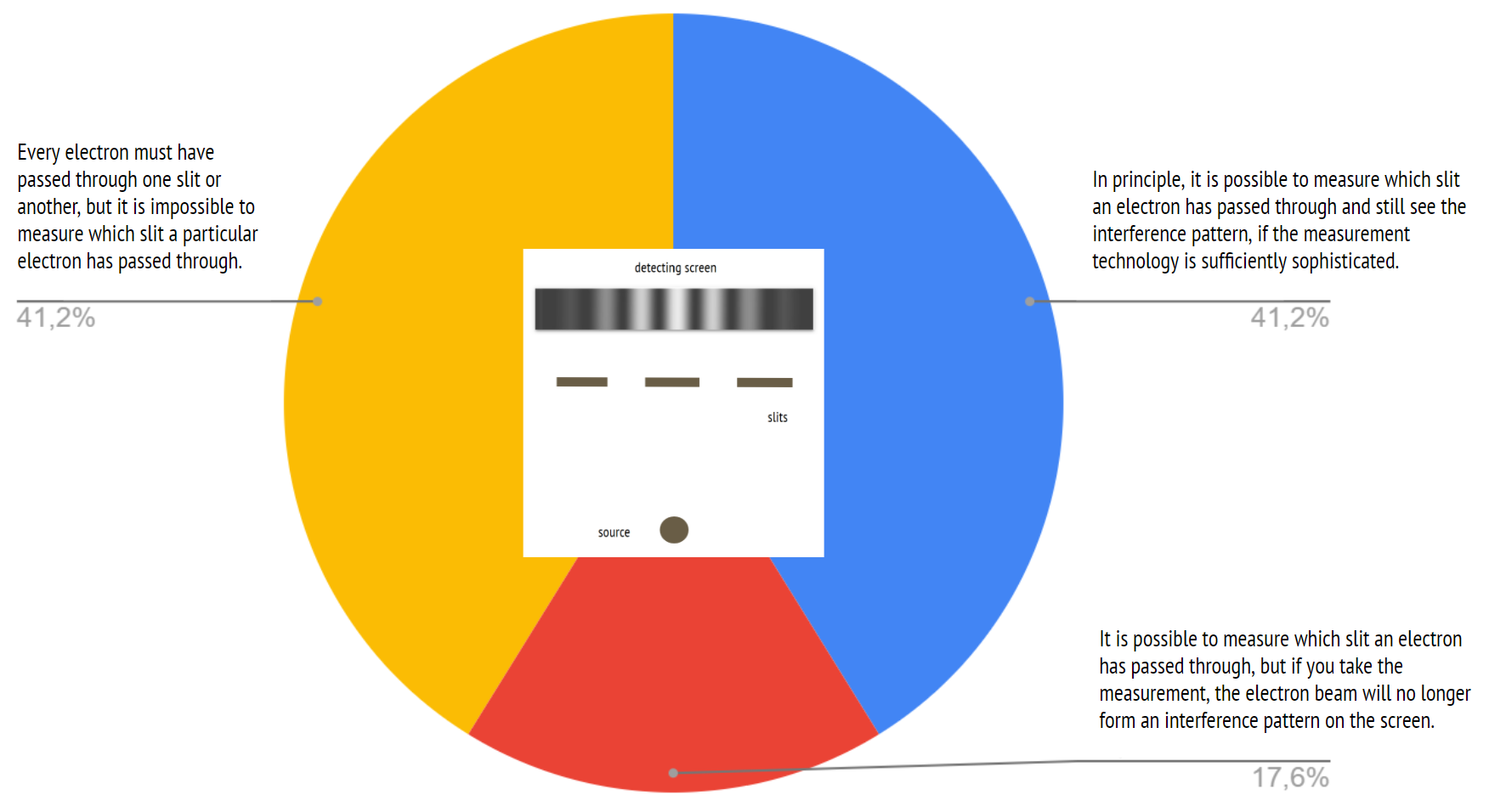}
    \caption{Interpretation of diffraction figure: analysis of students' answers to the questionnaire administered at the end of the course. The figure shows the alternatives proposed and the percentages of the answers selected}
    \label{fig:Ch6_38_DoubleSlitFIG}
\end{figure}

\begin{figure}[hbt!]
    \centering
   \includegraphics[width=\textwidth,height=\textheight,keepaspectratio]{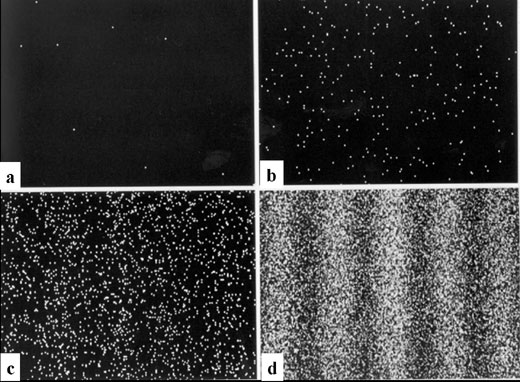}
    \caption[Single electron interference pattern]{Single electron events build up to from an interference pattern in the double-slit experiments. The number of electron accumulated on the screen. (a) 8 electrons; (b) 270 electrons; (c) 2000 electrons; (d) 160,000. The total exponsure time from the beginning to the stage (d) is 20 min. Source: Hitachi Global Research https://www.hitachi.com/rd/research/materials/quantum/doubleslit/index.html}
    \label{fig:Ch6_39_Tonomura}
\end{figure}

These results are not to be understood as an evaluation of learning but as a monitoring of the students' reconstruction processes. The data collected helped to trigger moments of reflection during the lesson involving the students starting from their answers \cite{MazurPeer1997}. Watching the video of the Tonomura experiment \cite{TonomuraVideo} then made it possible to show how the statistical nature of the process could be generalised to another very similar experimental context. This allowed the coherence between the proposed theoretical model and the experimental results to be made explicit. (see Figure \ref{fig:Ch6_39_Tonomura} )

\subsubsection{Results} 

With regard to the course developed, the teachers indicated as a positive element the level of enjoyment and participation of the students. Having integrated the exploration of elementary physical phenomena with laboratory experiences certainly stimulated the curiosity of the students, who participated actively.
\begin{figure}[hbt!]
    \centering
   \includegraphics[width=\textwidth,height=\textheight,keepaspectratio]{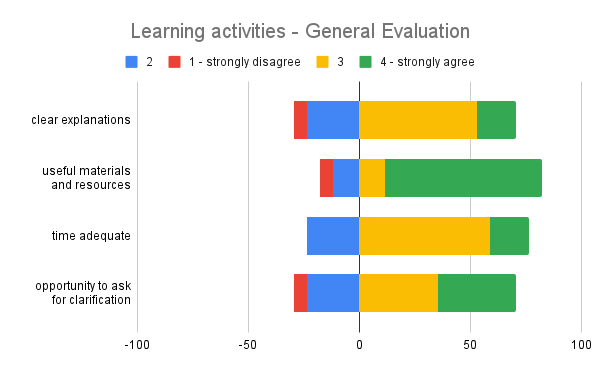}
    \caption{Percorso sul dualismo: valutazione partecipazione alle lezioni. Risposte raccolte su un campione parziale del partecipanti N = 17 su un totale di 34}
    \label{fig:Ch6_45_CarcanoEvaluation}
\end{figure}

The teachers report that the topics, despite their difficulty, stimulated students' reflection on several aspects. On the one hand, the close relationship between the performance of experiments and the construction of models

\begin{quote} {\fontfamily{cmtt}\selectfont
   The structure of the lessons made it possible to see both aspects: the experimental part and the theoretical part. In this sense the course is complete. [TeachRefl]
}\end{quote}

On the other hand, the course has sparked debate on fundamental questions related to the image of reality that arises from the study of the quantum world.

\begin{quote} {\fontfamily{cmtt}\selectfont The students were very involved in the discussion of the fundamentals and the interpretation \dots Some of them have had their chakras opened!
[TeachRefl]
}\end{quote}

\begin{quote} {\fontfamily{cmtt}\selectfont Some of the students then took up in class what they had seen in the lessons. They told me that they wished they had more time to ask questions, that the time was a bit too compressed.
[TeachRefl]
}\end{quote}

An important element to be added to the treatment of quantum physics, according to the lecturers, are therefore aspects relating to the process of constructing physical theories and, more generally, aspects linked to the Nature of Science (NOS). \cite{Stadermann2020ConnectNOS}.

\begin{quote} {\fontfamily{cmtt}\selectfont These [NOS] [NOS] aspects that are interesting for the students and that always enter into the discussion of quantum physics \dots for teachers is not easy to handle them \dots
[TeachRefl]
}\end{quote}

This is an opportunity for teachers to possibly involve other colleagues from other departments (e.g. Philosophy, Literature).


With respect to the development of PCK, a first element to be noted was the choice to build the course from the students' knowledge. The teaching experience (PCK source) supported by the use of questionnarires allowed them to make the right anticipations about the way in which students can approach the explanation of quantum phenomena. The lecturers reflected at length on the correct way to approach the behaviour of quantum objects from the students' mental images.

\begin{quote} {\fontfamily{cmtt}\selectfont When you talk about dualism, the students come with something already in their heads \dots it's not something new for them. This is an advantage in the sense that they have an idea \dots they know there is a problem, but it is also a disadvantage because you don't know what they really mean \dots and their ideas can lead them astray.
[TeachRefl]
}\end{quote}





The main criticism is that the time available (3 lessons of about 60 minutes each) is considered by the students too short to assimilate new and difficult concepts. In the words of one student
\begin{quote} {\fontfamily{cmtt}\selectfont
   [The course was] very interesting \dots I would have preferred to do the course in more than three hours in order to do it more calmly and to deepen some aspects. [StudentsRefl]
}\end{quote}


\section{Listening to teachers}\label{sec:TLE_focus_interview}
Between May and June 2021, two meetings were organised with those teachers who in different ways chose to participate in the proposed activities. A group of 4 teachers who participated in the Professional Development programme (PD) Quantum Jumps gave their consent to conduct individual interviews. After a few weeks, an online meeting was organised between the teachers who carried out classroom activities using some of the materials proposed during the PD Quantum Jumps and other teachers who participated in the activities carried out with the students during the Spring School introduction to quantum technologies (see Section \ref{sec:SumSch_QT}). Of the ten teachers who participated, two had also participated in the interviews.

\subsection{Semi structured interviews}\label{subsec:Teachrs_Interv}
As for the interviews, a protocol was prepared in order to detect some useful elements to understand if the PD helped the development of teachers' PCK with regard to the three dimensions defined in Section \ref{sec:PCKdimGrid}.
The structure is similar to that proposed in other studies of teachers' PCK done using the ERTE model \cite{Dijk2008_ERTEevolution}. In our case, these are semi-structured interviews aimed at collecting teachers' reflections by creating a context that allows them to "articulate their PCK". The design of the protocol made it possible to make explicit the relationships between the different dimensions of PCK and the questions put to the teachers in the different parts of the protocol (see Table \ref{tab:PCKintervGrid1}). The definition of the themes for the analysis have been caried out in two phases (see Table \ref{tab:LET_RQInterviewThemes})

\begin{landscape}
\begin{table}[hbt!]
\begin{center}
\centering
\begin{tabular}{p{5cm}p{1cm}p{1cm}p{1cm}p{1cm}p{1cm}}
& \multicolumn{1}{c}{Part 1} & \multicolumn{1}{c}{Part 2} &
\multicolumn{1}{c}{Part 3} & \multicolumn{1}{c}{Part 4} & \multicolumn{1}{c}{Part 5}\\
& \multicolumn{1}{c}{Background} & \multicolumn{1}{c}{Experience} &
\multicolumn{1}{c}{Educational choices} & \multicolumn{1}{c}{Scenario Questions} & \multicolumn{1}{c}{Anticipation}\\
\hline
PCK source & \multicolumn{1}{c}{x} & \multicolumn{1}{c}{x} &
\multicolumn{1}{c}{} &\multicolumn{1}{c}{} & \multicolumn{1}{c}{}\\
\hline
Subject matter knowledge for teaching & \multicolumn{1}{c}{} & \multicolumn{1}{c}{x} &
\multicolumn{1}{c}{x} &\multicolumn{1}{c}{x} & \multicolumn{1}{c}{x}\\
\hline
Students' pre-instructional conception & \multicolumn{1}{c}{} & \multicolumn{1}{c}{x} &
\multicolumn{1}{c}{x} &\multicolumn{1}{c}{x} & \multicolumn{1}{c}{x}\\
\hline
Educational structuring & \multicolumn{1}{c}{} & \multicolumn{1}{c}{x} &
\multicolumn{1}{c}{x} &\multicolumn{1}{c}{x} & \multicolumn{1}{c}{x}\\
\end{tabular}
\caption{Semi-structured interview protocol: alignment between interview questions and PCK dimensions}
\label{tab:PCKintervGrid1}
\end{center}
\end{table}
\end{landscape}

\pagestyle{empty}
\begin{landscape}
\begin{table}[hbt!]
\centering
\begin{tabular}{p{4cm}p{9cm}p{9cm}}
\multicolumn{3}{c}{\textbf{Themes}} \\
\hline
\multicolumn{1}{c}{\textit{PCK Dimensions}}&
\multicolumn{1}{c}{\textit{Phase 1}} & \multicolumn{1}{c}{\textit{Phase 2}} \\
\hline
&\\
\textbf{\textit{PCK sources}} &  Subject matter knowledge&\\
& teaching experience&\\
& Workshops&\\
& Memories&\\
&\\
\hline
& \\
\textbf{\textit{Subject Matter Knowledge for teaching}}
&  Use of subject matter knowledge & Personal lack of subject matter knowledge\\
& Flexibility in handling daily complexity&Transition from classical to quantum paradigms\\
& Reaction to unanticipated situations&\\
& \\
\hline
& \\
\textbf{\textit{Students pre instructional conceptions}}
& Awareness of students' prior-knowledge & Level of mathematical knowledge\\
& Ability to follow students conceptions & Link between concepts and experiments\\
&\\
\hline
& \\
\textbf{\textit{Educational structuring}}
& Use of subject matter knowledge representations & Use of simulations\\
& Design competences & Time constrains\\
& Familiarity with different teaching methodologies &\\
\end{tabular}
\caption{Themes from interviews}
\label{tab:LET_RQInterviewThemes}
\end{table}
\end{landscape}
\pagestyle{plain}

In a first phase, a deductive procedure was used to identify in the recordings those passages that could be traced back to one of the three dimensions of PCK already defined (see Section \ref{sec:PCKdimGrid}). The passages identified in this way were transcribed so as to be able to report the words used, maintaining in particular the use of specific terms, analogy and metaphors.
In the second phase an inductive type procedure was used for the identification of particularly relevant and recurrent themes, transversal to the different interviews. Also in this case a transcription of the most relevant portions of the recording was made.
The set of themes thus obtained was organised by identifying the connections with the different dimensions of the PCK. Some of the elements that emerged from the analysis in relation to the research questions are presented in Table \ref{tab:LET_RQIncipit}.

The different dimensions of PCK examined are strongly intertwined. The PCK dimensions haave been used as a guide, focusing on those aspects that seem most relevant to understand if and how the activities carried out with the teachers have contributed to their development.
Regarding the reflection on PCK sources (see Table \ref{tab:LET_RQInterviewThemes}), it should be specified that all the teachers who participated in the interviews have a background of mathematical studies (to understand the contribution of teachers with different backgrounds see the next section \ref{subsec:Teachrs_Gdicsussion}).
For some of them this is experienced as a limitation, which at times can hinder work with students.

\begin{quote} {\fontfamily{cmtt}\selectfont
   Many things are new to me, I have never studied them at university and I realise that when I am teaching I understand them \dots and this creates strong doubts in my mind, and I often wonder if the students do not understand because of me, \dots because deep down I also do not understand these things. [TeachInterv1]
}\end{quote}

For others, coming from the "world of mathematicians" it is also seen as an opportunity to pay greater attention and care of the communication processes.

\begin{quote} {\fontfamily{cmtt}\selectfont
   As a mathematician I think that those who write physics books take a lot of things for granted \dots from my university days, as a mathematician studying physics I found that I asked myself a lot of questions and I realise that when I teach physics I take care of every single step [TeachInterv2]
}\end{quote}

The perception of "having gaps" in one's subject matter knowledge was compensated by the activation of some self-education tools, such as the purchase of books and the participation in training courses. This kind of activity reinforces the reflection around one's professional practice and contributes to the construction of a stock of useful experiences.

\begin{quote} {\fontfamily{cmtt}\selectfont
   I have bought a lot of books and many of them are on quantum physics \dots I have also participated in various training courses on these subjects [modern physics] \dots and I must say that every time I take something useful with me and I try to incorporate it into what I propose later on. [TeachInterv3]
}\end{quote}

The reflection on some elements of Subject Matter Knowledge for Teaching is expressed through the explanation of the difficulties teachers have in designing and implementing teaching activities. These difficulties are linked to the effort needed to follow the students' reasoning, in the choice of concepts and examples to propose and in the management of the timing in relation to the specific curricular objectives.

The motivation for teachers to tackle the conceptual complexity of modern physics stems from the importance of including these topics in students' education. Aware of the limits imposed by the characteristics of school practice, the relevance is not so much linked to the need for formative training on specific knowledge and skills.\\

\begin{quote} {\fontfamily{cmtt}\selectfont
   Specific aspects, details and technicalities are not important at high school level \dots and we are not prepared enough to do this and I don't think this is what needs to happen at school. [TeachInter1]
}\end{quote}

Rather, the relevance is related to the cultural impact that these topics have for a citizen's scientific education. It is one of the objectives of the physics curriculum and is not necessarily linked to the possibility that students then go on to study STEM subjects or want to become part of the future quantum technology workforce.

\begin{quote} {\fontfamily{cmtt}\selectfont
   Just to quote an anecdote \dots, during a national conference, representatives of associations for the teaching of physics and of the Ministry [of Education] told us to leave teaching quantum physics alone, that those who want to will study it at the University, \dots because that's where you can study it [quantum physics] properly \dots in secondary school you are only able to teach mechanics \dots Apart from the fact that I'm not sure that in high school you teach mechanics well [chuckles] this for me is nonsense \dots 95\% of the students will not do physics, so [the course of quantum physics at school] is the only chance they will have to hear about these subjects \dots is an opportunity that cannot be missed [TeachInterv2]
}\end{quote}

Quantum topics have a great impact on students and are an opportunity to stimulate learning about scientific thinking.

\begin{quote} {\fontfamily{cmtt}\selectfont
   These are topics [quantum physics] that inspire students, fascinate them \dots think of "Schr\"{o}dinger's cat" and certainly it has more appeal than the principles of thermodynamics or other topics they have to do in school. [TeachInterv2]
}\end{quote}

\begin{quote} {\fontfamily{cmtt}\selectfont
   [Quantum physics] opens your mind, it's an opportunity to deal with fundamental, fascinating topics. I am convinced that students are also fascinated by these topics. [TeachInterv4]
}\end{quote}

\begin{quote} {\fontfamily{cmtt}\selectfont
   These topics [quantum physics and technologies] resonate with the kids \dots open the door to the future and provide a breath of fresh air. \dots [TeachInterv1]
}\end{quote}

With regard to the difficulties of teaching the main concepts of quantum physics in lessons with students, one aspect that emerged across the board concerns the ways in which students can be supported in the transition from the classical to the quantum world.\\

\begin{quote} {\fontfamily{cmtt}\selectfont
   One care to be taken is to link classical physics to quantum physics by trying as much as possible to avoid "jumping around" \dots and where possible using classical physics to explain \dots  [TeachInterv2]
}\end{quote}

The difficulty arises from not being able to "see things2. Aware of not having the possibility of knowing quantum objects through a sensory-mediated intuition of reality \cite{SusskindQT}, students must be supported and guided through a number of "stratagems".

\begin{quote} {\fontfamily{cmtt}\selectfont
   I have to remove this sense that it's just "magic", this attitude that <<if I don't see it \dots I don't believe it>> \dots then I have to provide a "practical" motivation, I have to put them in a context \dots in front of a problem where it is necessary to introduce something new to understand \dots [The students] have to feel that what I have at my disposal] is not enough for me \dots a bit like the $\sqrt{-1}$ and the complex numbers in mathematics! [TeachInterv4]
}\end{quote}

Mathematics is then a possible way of thinking about things that cannot be seen. From this data some elements of reflection emerge, followed by the introduction of the axiomatic approach (see Section \ref{sec:PIAA}) seen in the Professional Development programme.

The starting point cannot be axiomatic in the strict sense of the term.
\begin{quote} {\fontfamily{cmtt}\selectfont
a purely axiomatic approach, the existence and necessity of axioms must be demonstrated.
}\end{quote}

The use of axioms is not, however, something totally new for students. Already in their early years, the approach to mathematics involves the introduction of formal postulates and structures. This can certainly help the introduction of formalism.

\begin{quote} {\fontfamily{cmtt}\selectfont
Already in the first year of high school they are told about the "commutative property" and they accept it, as they are used to do with mathematics \dots because for them that ["obey to rules"] \textit{is} mathematics. Then maybe when they see the same property in another context, for example in vector calculus, then it makes sense. [TeachInterv2]
}\end{quote}


The starting point for physics is necessarily different and always coincides with the description of the physical phenomenon. It is from there that the introduction of the formal structure on which the axioms are built can arise and be justified.

\begin{quote} {\fontfamily{cmtt}\selectfont
Mathematicians like me must always remember that "physics provides beautiful examples". As mathematicians sometimes we "get lost" and talk about very abstract and sensible things. If we want to introduce formalism, we can use physics.
We start with spin and we show that it is not what the science teacher explained to you [chuckles]\dots We can then shows that the linear algebra "is born from the spin" \dots in the sense that linear algebra already existed before \dots but only there [in the physical phenomenon] it takes on meaning \dots is a bit like what Heisenberg did \dots his greatness: "I don't know a part of mathematics and so I reinvent it" \dots Physics provides a context.[TeachInterv2]
}\end{quote}

The problem is then reversed: not to insert or add mathematics into the discourse of physics, but to make it arise from physics itself and make it to be the necessary tool for understanding.

\begin{quote} {\fontfamily{cmtt}\selectfont
In the end the Pauli matrices were born from Pauli [TeachInterv2]
}\end{quote}

Where the necessary specific mathematical tools are lacking, they can be introduced as and when needed by creating meaningful links between subject programmes. The selection of what mathematical tools are needed is closely related to the particular subject matter.

\begin{quote} {\fontfamily{cmtt}\selectfont
Especially if the physics teacher does not also teach mathematics, it happens that the physics teacher has "to wait" for the mathematics teacher \dots by teaching them together I can instead introduce what I need when I want. [TeachInterv2]
}\end{quote}

A further support tool for students is the use of simulations. Introduced during the Professional Development programme, they are considered extremely useful in helping students to develop one of the curricular skills, namely reflection on the limitations of the classical interpretative paradigm \cite{DM769_18Quadri}.

Simulations can reduce, but not eliminate, the perceived gap between theory and physical reality.

\begin{quote} {\fontfamily{cmtt}\selectfont
Students are always a bit sceptical\dots that if the experiment is "simulated", then it is "fake". [TeachInterv3]
}\end{quote}

It is, however, a very functional tool for achieving teaching objectives.

\begin{quote} {\fontfamily{cmtt}\selectfont
When I talk about quantum mechanics I feel as if I were telling a story without formalism, but I remain convinced that the mathematics [of quantum mechanics] is not feasible, above all because of a question of time, and then they will see if they want to \dots The simulations at least make it possible to show that the things I am telling are not fairy tales, but are linked to reality. [TeachInterv3]
}\end{quote}

The use of simulators was also seen as an opportunity to structure lessons differently, following the pace of the students.

\begin{quote} {\fontfamily{cmtt}\selectfont
I used the simulations by working alongside the students, commenting together on the different steps using the worksheet \dots a lot of questions emerged \dots the class was very active and I also learned a lot [TeachInterv1]
}\end{quote}

The reflections collected allow us to see how the teachers were able to use what they experienced during the Professional Development Course as an opportunity to reflect on their professional practice. The different dimensions of the PCK and their interconnections have made it possible to identify some specific areas of development. In particular, the influence of the interrelation of subject matter knowledge for teaching and students' pre-instructional conceptions on the process of educational structuring is revealed (see Section \ref{sec:ERTE}).
Teachers related their personal difficulties in teaching quantum physics to those they attribute to students' pre-instructional conceptions of design elements, including appropriate use of formalism and simulations.


\subsection{Group discussion}\label{subsec:Teachrs_Gdicsussion}
The meeting was designed as an opportunity for teachers to discuss the conditions they consider necessary for the implementation of quantum physics teaching with the approach used in the PD Quantum Jumps and in the summer schools for students. Among the 10 teachers involved in the discussion, a subgroup of 4 teachers had tried in their classrooms to use some of the resources presented in the course during Module 1 and Module 2 (see Figure \ref{fig:Ch6_11_QJsModules}). A second subgroup made up of the remaining 4 teachers includes those who had actively participated in the Spring School - Introduction to Quantum technology and at the end of which had expressed interest in continuing to participate in dedicated teacher activities.
Two teachers from the first subgroup and one teacher from the second subgroup had also already participated in the interviews.

\begin{figure}[hbt!]
    \centering
     \includegraphics[width=\textwidth,height=\textheight,keepaspectratio]{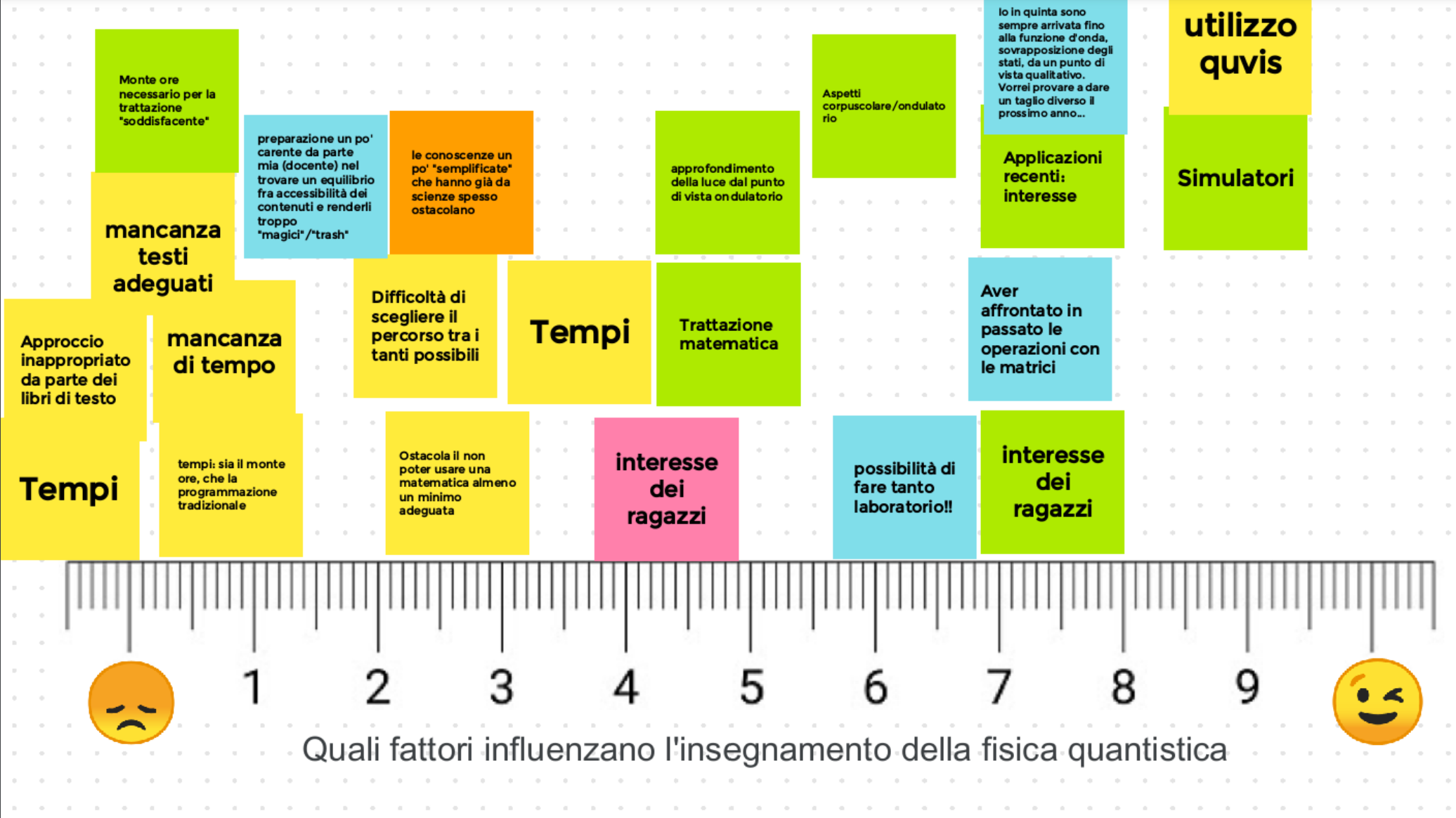}
    \caption[Teachers' discussion about teaching quantum physics]{Central ideas about the factors that hinder and facilitate the practice of teaching quantum physics. The group of 10 teachers worked online using Google's Jamboard$^{\footnotesize{TM}}$ platform.}
    \label{fig:Ch6_41_TEGroupDiscussion}
\end{figure}

The discussion lasting about one hour was centered on their experience in teaching quantum physics. The idea was to create an environment in which the two subgroups could compare and possibly support each other in evaluating effective practices for classroom work. The start of the activities coincided with a task about the identification of what the teachers consider as facilitating or hindering the teaching of quantum physics(see Figure \ref{fig:Ch6_41_TEGroupDiscussion})

From the recording of the meeting, the interventions considered most useful for the evaluation of the teachers' PCK development were extrapolated. The themes identified were integrated with those that had already emerged during the interviews. Among them is the theme of "time", linked to the problem of teaching design in relation to the limits imposed by the national curriculum. Then there is the discovery of the use of simulations, which responds to the need for a "practical" approach to the study of quantum mechanics.

The discussion revealed some critical issues in physics teaching practice. Each participant described his or her personal experience in this regard and how they have managed to adapt their professional action.
One challenge is to find the balance between the desire to present interesting aspects of scientific development to one's students and the need to respect the timetable and the curriculum.
There are many possible choices of topics to be covered and therefore it is important to select from the many approaches the one that best touches the essential conceptual nodes.

\begin{quote} {\fontfamily{cmtt}\selectfont
 The difficulty in choosing which learnimg activity to do is linked to the many different proposals I have heard in the different courses I have attended or in the textbooks. \dots Since I have to be careful to do everything I need to do [for the fonal exam], the starting point for me is to understand how to select the most important concepts and how to tie them together. [TeachRefl]
}\end{quote}

The motivation to teach and to define effective methodologies stems from the interest shown by the students. Participants report a good level of involvement in their class groups. However, interest tends to wane as complexity increases.

\begin{quote} {\fontfamily{cmtt}\selectfont
It is certainly easy to arouse their [students] curiosity, to hook them, but then this interest is lost and even frustration prevails because things are difficult \dots they do not understand. [TeachRelf]
}\end{quote}

In order to manage this process properly, support tools must be introduced. Among these, simulations become an opportunity to carry out those experimental activities considered essential for the acquisition of basic concepts.

\begin{quote} {\fontfamily{cmtt}\selectfont
Doing a lot of laboratory activity has a positive effect \dots Using the simulations they [the students] are a bit more "in the centre" \dots [TeachRefl]
}\end{quote}

Time constraints limit the possibility of addressing the topics adequately. So does the lack of an adequate textbook. The resources available to teachers use an approach that is considered inadequate.

\begin{quote} {\fontfamily{cmtt}\selectfont
   The textbook is a reference point, for us [teachers] and for them [students]. \dots But contents are introduced badly, lots of "talkie" and little substance.  [TeachRefl]
}\end{quote}

\begin{quote} {\fontfamily{cmtt}\selectfont
  I tried to introduce more popular books for the students to read, which would support understanding. The approach is too different, you don't go into enough detail. [TeachRefl]
}\end{quote}

A central point of reflection concerned the role played by mathematics in connection with quantum physics. Generally regarded as an obstacle if it is presented too rigorously, it is nonetheless a distinguishing feature that is difficult not to include in a comprehensive course

\begin{quote} {\fontfamily{cmtt}\selectfont
  Without the mathematical part, the whole argument risks falling into banality. [TeachRefl]
}\end{quote}

In attempting to characterise the appropriate type of formalism, a number of thoughts have emerged as to what kind of mathematics might be suitable for high school learning activities.

A first feature is to use those mathematical topics with which students are familiar.
\begin{quote} {\fontfamily{cmtt}\selectfont
   There is no simple or difficult mathematics per se. There is, however, a mathematics that students are used to. [TeachRefl]
}\end{quote}

Anticipating some mathematical topics that are known to be useful for the study of quantum mechanics can therefore facilitate its teaching. If this is not possible, the introduction of specific calculation tools (e.g. vectors or matrices) should be considered.
In this case, however, it is a question of using mathematical concepts at the level of pure representation. A rigorous and structured introduction of the concepts as one would do in a mathematics course is not intended. Instead, the idea is to use the mathematical language as a thinking tool.

\begin{quote} {\fontfamily{cmtt}\selectfont
   The formalism [Dirac notation] is a way of writing in mathematical language what can be told to them using the words \dots to describe what they see in the simulator [TeachRefl]
}\end{quote}

A second related aspect that may in fact facilitate the understanding of the formalism is whether it is clearly related to the physical phenomenon under consideration.
The relationship between mathematics and physics is through the use of a formula. In general, however, this way of using mathematics is not conducive to learning physics, a kind of "spell to cast" that is functional for solving exercises.

\begin{quote} {\fontfamily{cmtt}\selectfont
    An expression in which plug in numbers in to get the numerical result of a problem, this is how generally students see maths in relation to physics [TeachRefl]
}\end{quote}

Instead, a process of modelling phenomena through mathematical language should be stimulated with students. Mathematics can therefore be used to built equations that describe the relationship between variables and so describe the behaviour of a physical system.

The fact that students then use mathematics is no guarantee that they understand the relationship between the various concepts. A final reflection was therefore devoted to the way in which quantum physics courses are evaluated. It is not possible to rely on batteries of exercises on various levels as in other subjects (e.g. mechanics).

\begin{quote} {\fontfamily{cmtt}\selectfont
   The formulas are few and always the same \dots for example to calculate the wavelength of a tennis ball dots and then that is not the heart of the matter [TeachRefl]
}\end{quote}

The assessment criteria should be structured around the students' ability to describe experiments using the concepts they have learned, such as the idea of superposition.
During the discussion, some of the results of the experiments were shared with the participants (see Section \ref{sec:massimo} or \ref{sec:carcano}). Those who participated in the interview proposed to share the scenario questions used during the interview (see Appendix \ref{appx:IntervProtocol}).

\begin{quote} {\fontfamily{cmtt}\selectfont
   I don't like closed-ended tests very much, but those questions \dots give me the impression that their learning can be measured \dots of course they [students] have to explain to me why they made those choices. [TeachRefl]
}\end{quote}

\begin{quote} {\fontfamily{cmtt}\selectfont
   If my students were able to answer all those questions correctly, I would be satisfied. [TeachRefl]
}\end{quote}

With regard to the possibility of using quantum technologies in physics education, some participants showed interest in reworking and proposing in class what they had seen using the simulators and materials provided.
The use of the worksheets presented together with the simulators was considered as a concrete opportunity for students to carry out specific work to consolidate their acquired knowledge.

\begin{quote} {\fontfamily{cmtt}\selectfont
  Next year I want to try to use the simulators to talk about cryptography \dots I thought it was a very interesting activity, with many possible connections \dots students can understand and use the simulators also by themselves \dots The strong point I think is that [the simulations] can guide the students in the activity \dots [TeachRefl]
}\end{quote}




\section{Conclusions}\label{sec:TLEconclusions}
The activities with teachers were an opportunity to work synergistically on both strategic lines of the project. In fact, teachers can play an important role in the educational context of the high school as a link between the world of physics research and the educational and training needs of students. This is a complex role in which different levels relating to the processes of elaboration of didactic contents and design of learning activities that take into account the specific context of application overlap.
The aim of the project was to understand how to facilitate this role by acting both on the development of some specific competences related to PCK and on the reflection about design strategies.

The actions carried out have allowed the creation of different contexts in which teachers have been able to discuss and reflect together. 
The sharing made it possible to highlight the ways in which each teacher is able to manage different aspects of subject matter for teaching. Particularly evident to the introduction of quantum physics is the importance of reflection on students' pre-conceptions. It becomes the instrument to accompany them in the transition from the classical to the quantum world.
The opportunity provided to take part in summer school activities alongside their students allowed them to experience the criticalities and potentialities of the approach presented. This gave them the opportunity to reflect on the factors that can hinder or enhance their teaching activities.\\

The collaboration between teachers and researchers also provided the teachers with some useful tools for their personal teaching design. In the various activities proposed, "ingredients" were provided (e.g. the use of simulators) which each teacher then made their own. The coherence between the resources provided and the approach to quantum mechanics presented facilitated the identification of the fundamental concepts to be presented (e.g. quantum state and quantum measure) and promoted teachers' reflection on how to structure their lessons in order to grasp the essential elements of the theory.\\

In this sense, the negative perception of the role of mathematics has also been partially eroded.
Although there is still a strong scepticism about the rigorous use of formalism, the idea that it can at least become a tool to support the description of physical phenomena has permeated. The particular nature of the physics teaching staff in Italy, populated for the most part by mathematicians, and the setting of the physics curriculum, closely linked to that of mathematics, can certainly facilitate reflection on the relationship between the disciplines. \\

\chapter{\label{ch:7-LEStudents} Learning environments for students}







In this chapter some examples of learning environments designed for extracurricular activities with students are presented. The basic elements of the design are the same as those used for teachers' training. The activities are related to the introduction of some core concepts of quantum physics through the informational approach described in Chapter \ref{ch:3-QApproach}.

In the Section \ref{sec:SumSch_light} the attention is focused on the activities related to the study of polarisation and the nature of light, according to the approach of which was introduced in Section \ref{sec:qubit_polarization} and Section \ref{subsec:Polar_MathInterplay}. In Section \ref{sec:SumSch_QT} activities more closely related to quantum technologies, in particular tpo  cryptographic protocols are analysed.

These activities were carried out outside the regular curricular school practice and involved self-selected students, i.e. students who showed their teachers an interest in the course topics and therefore chose to participate. Not all students were specifically interested in physics, but they are oriented towards scientific and STEM disciplines, specifically engineering and computer science.

Referring to the research questions defined in the Table \ref{tab:ProjFrame}, activities with students are mainly focused on understanding how the informational approach can help them to understand the relationships between the principles of quantum physics, the nature of quantum objects (RQ5) and quantum technologies (RQ7).

\section{From classical to quantum nature of light}\label{sec:SumSch_light}
The transition to a quantum view of reality can be built through an investigation of the nature of light. Answering the question "what is light?" can hide many pitfalls from a didactic point of view \cite{Henriksen2018_Whatislight}, and the introduction of concepts such as the photon does not necessarily help to simplify the confusion about what should be the correct way to describe the behaviour of light at the fundamental level (see Section \ref{sec:SingPhot}).
Following the didactic approach presented in the Section \ref{sec:qubit_polarization}, the course has been structured in order to identify which elements of the informational approach can help students to construct a representation of the quantum behaviour of light that is consistent with the results of the experiments.

The research questions used to design the activities with the students are the same as those used for the work with the teachers. In this case, however, we tried to see if the informational approach could help students to build a theoretical framework useful to describe the nature of quantum objects. 

\begin{table}[hbt!]
\centering
\begin{tabular}{cp{12cm}}
 & \textbf{Quantum behaviours}\\
 & \textit{How the polite informational axiomatic approach can help students to explore the nature of quantum objects}\\
\hline
\rule[-4mm]{0mm}{1cm}
RQ5.1 & How is it possible to characterise the quantum nature of physical objects using qubits to encode information?\\
 & \\
 \midrule
 \rule[-4mm]{0mm}{1cm}
RQ5.2 & Through which type of representation is it possible to describe the process by which the information associated with quantum objects can be transformed inside a physical apparatus and extracted through a measurement?\\
 & \\
 \midrule
 \rule[-4mm]{0mm}{1cm}
RQ5.3 & Which experimental contexts are significant for the study of how quantum objects can be used to encode information?
\end{tabular}
\caption{Research questions: Learning Environment for students about the characterization of quantum objects}
\label{tab:RQ_SLE_Qnature}
\end{table}

\begin{figure}[hbt!]
    \centering
   \includegraphics[width=\textwidth,height=\textheight,keepaspectratio]{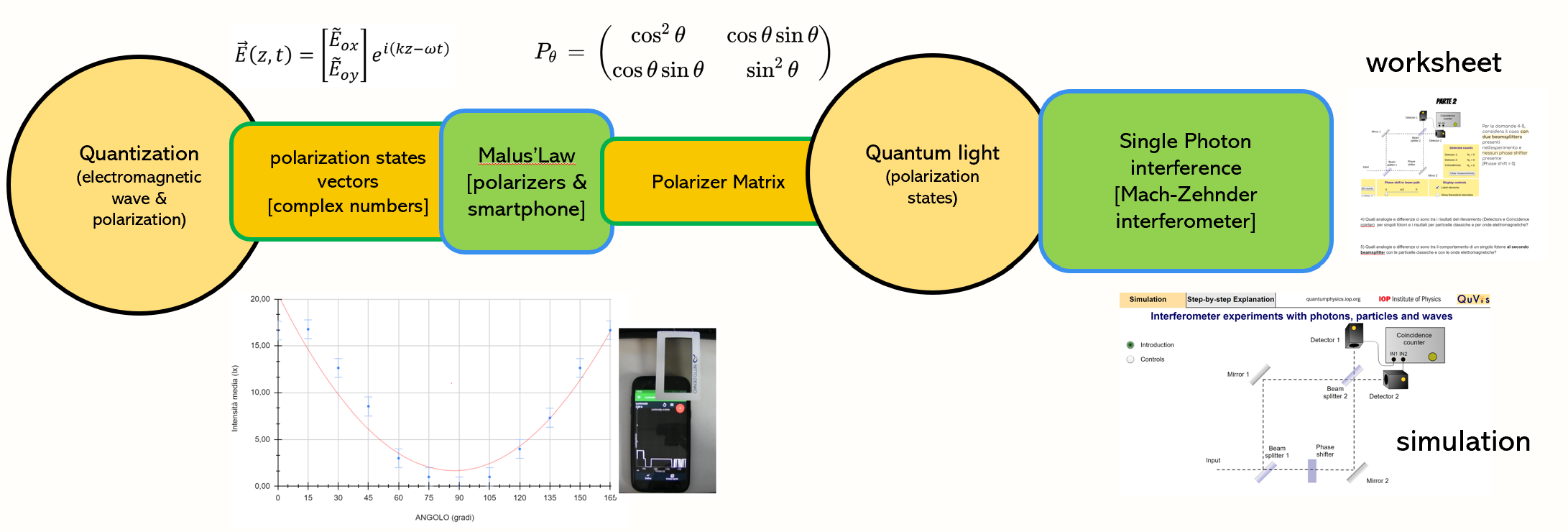}
    \caption{Teaching - learning sequence about the nature of light}
    \label{fig:Ch7_1_TLSNatureLight}
\end{figure}

The general aim of the teaching learning activity is to use the concept of quantum states to explore the limits of classical model of light and introduce a quantum model of light based on the concept of photon (see Section \ref{sec:SingPhot}).\\

\begin{table}[hbt!]
\centering
\begin{tabular}{cp{12cm}}

 & \textbf{ILO}: investigate the nature of light through the analysis of classical and quantum states in specific experimental setups (sequences of polarizars and Mach-Zehnder interferometer). \\
 & \\
 &\textit{At the end of the activities students should be able to:} \\
\midrule
\rule[-4mm]{0mm}{1cm}
ILO-a & describe the light polarization states and polarizers in terms of vectors and matrices.\\
\rule[-4mm]{0mm}{1cm}
ILO-b & apply the mathematical representation of light states to predict the outcomes of polarizers sequences experiments in terms of light intensity.\\
\rule[-4mm]{0mm}{1cm}
ILO-c &  characterize the behaviours of light using different physical variables in the analysis of experiments perfomed with Mach-Zehnder simulators.\\
\rule[-4mm]{0mm}{1cm}
ILO-d & compare interpretative models of light in the context of single photon experiments using a simulator Mach-Zehnder interferometer
\end{tabular}
\caption{ILOs: from classical to quantum nature of light}
\label{tab:ILO_light}
\end{table}


The activities we describe here took place during the International Summer School of Optics organised by the Department of Science and High Technology (DiSAT) of the University of Insubria in June 2021. This is an intensive week of study for upper secondary students dedicated to the study of optics. The high-profile theoretical and experimental skills of the members of DiSAT's Quantum Optics group allow students to study classical and quantum models of light, and to participate in photonics experiments in the laboratory. This initiative has been running since 2003 and until 2019 has also allowed participants to take part in workshops at the Joint Laboratory of Optics at Palacky University in Olomouc, Czech Republic. 2021's edition took place online due to the health emergency and involved 54 students from various schools in northern Italy.

\begin{figure}
     \centering
     \begin{subfigure}[hbt!]{\textwidth}
         \centering
         \includegraphics[width=\textwidth]{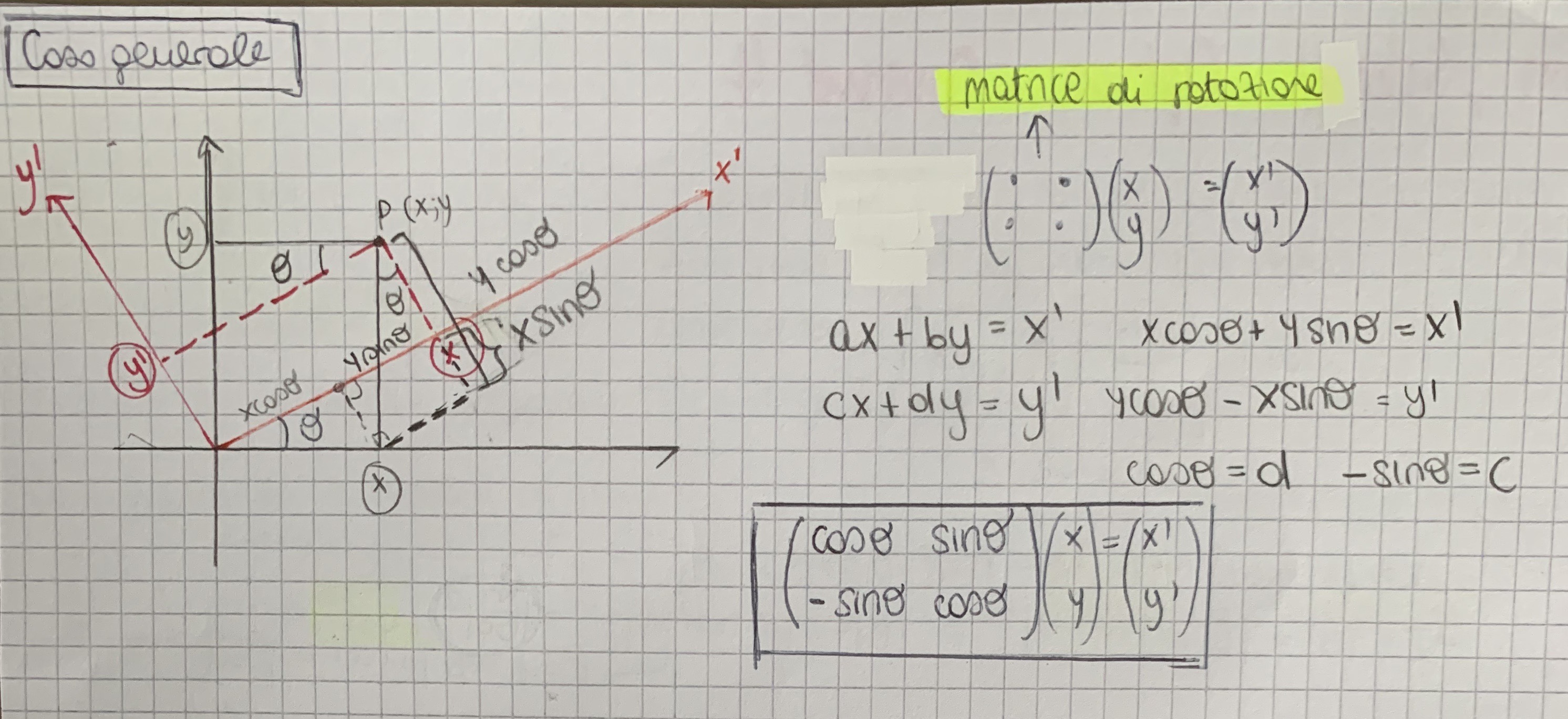}
     \end{subfigure}
     \hfill
     \begin{subfigure}[hbt!]{\textwidth}
         \centering
         \includegraphics[width=\textwidth]{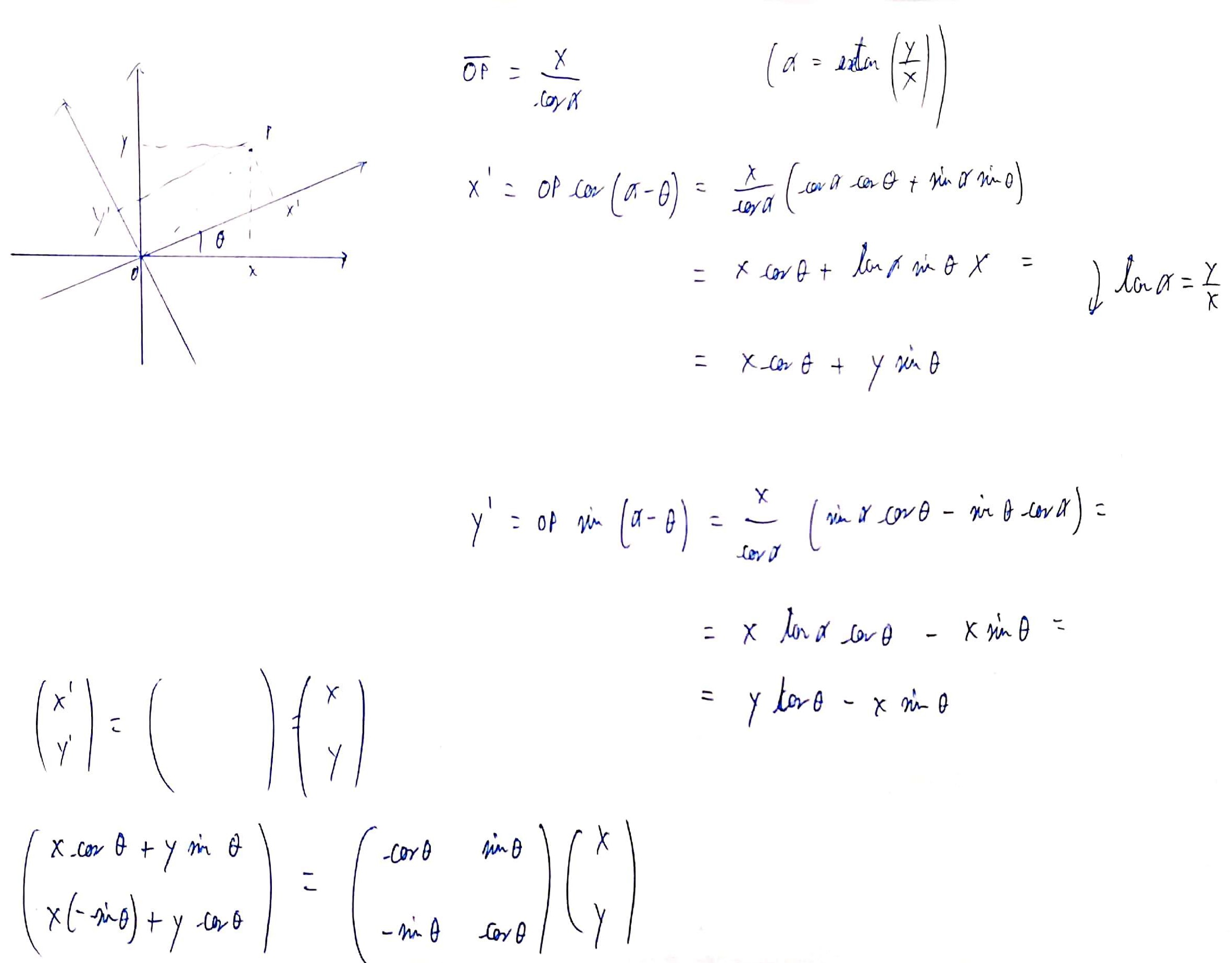}
     \end{subfigure}
      \caption{Polarizer matrix task: using basic trigonometric rules presented by the instructor (see graph on the left of each note) students worked out the coefficients of the rotation matrix}
     \label{fig:ch7_StMatrix}
\end{figure}

The most important aspect of the evaluation of the activities on the study of polarisation concerns the possibilities offered by the use of mathematical formalism for the exploration of physical phenomena. The effort in the design phase was to use mathematical concepts familiar to the students, such as the basic rules of trigonometry or complex numbers. Not all students had mastered basic elements of linear algebra, but given the students' familiarity with the concept of vector and the simplicity of the required rules for calculation with matrices, it was possible to integrate the missing knowledge in a short time and with good results.\\
In the course of extra-curricular activities, students fully appreciated the possibility of exploring the deep connections between mathematics and physics. 
During the small group activities on using the polarizer matrix (see Section  \ref{subsec:Polar_MathInterplay}), students helped each other figure out how to properly use the vector - matrix product rule that had been previously introduced to them (see Figure  \ref{fig:ch7_StMatrix}). By the end of the activity and with very little guidance, they were able to solve the problems in the worksheets. The students seemed to enjoy being able to use such a mathematical tool. In one of the students comment during the activity:

\begin{quote} {\fontfamily{cmtt}\selectfont
    When you get it… it is easy \dots you just plug the numbers in [the matrix] and you are able to know what happens next.[StudentsRefl]
}\end{quote}

\begin{figure}
     \centering
     \begin{subfigure}[b]{0.6\textwidth}
         \centering
         \includegraphics[width=\textwidth]{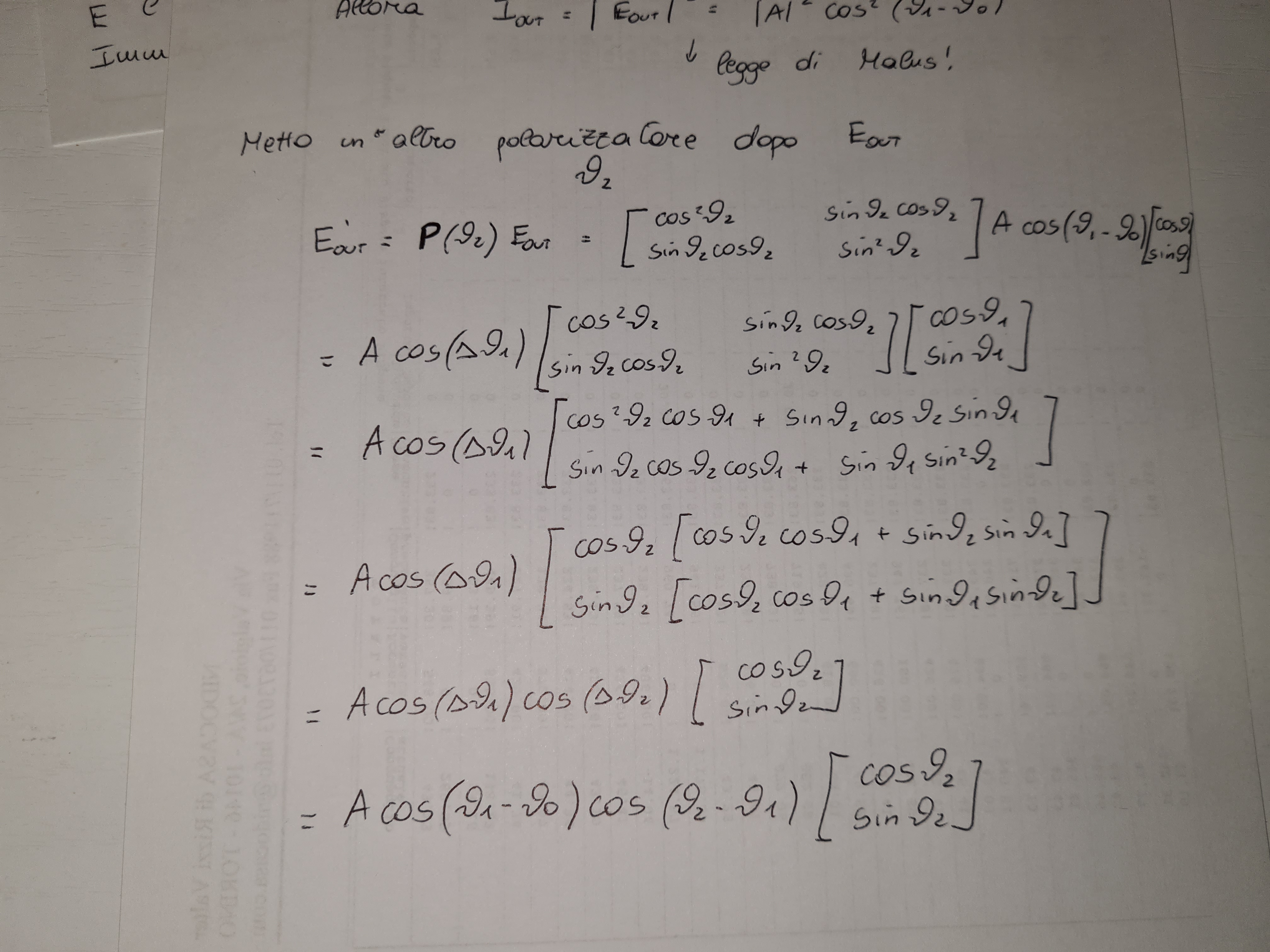}
     \end{subfigure}
     \hfill
     \begin{subfigure}[b]{0.6\textwidth}
         \centering
         \includegraphics[width=\textwidth]{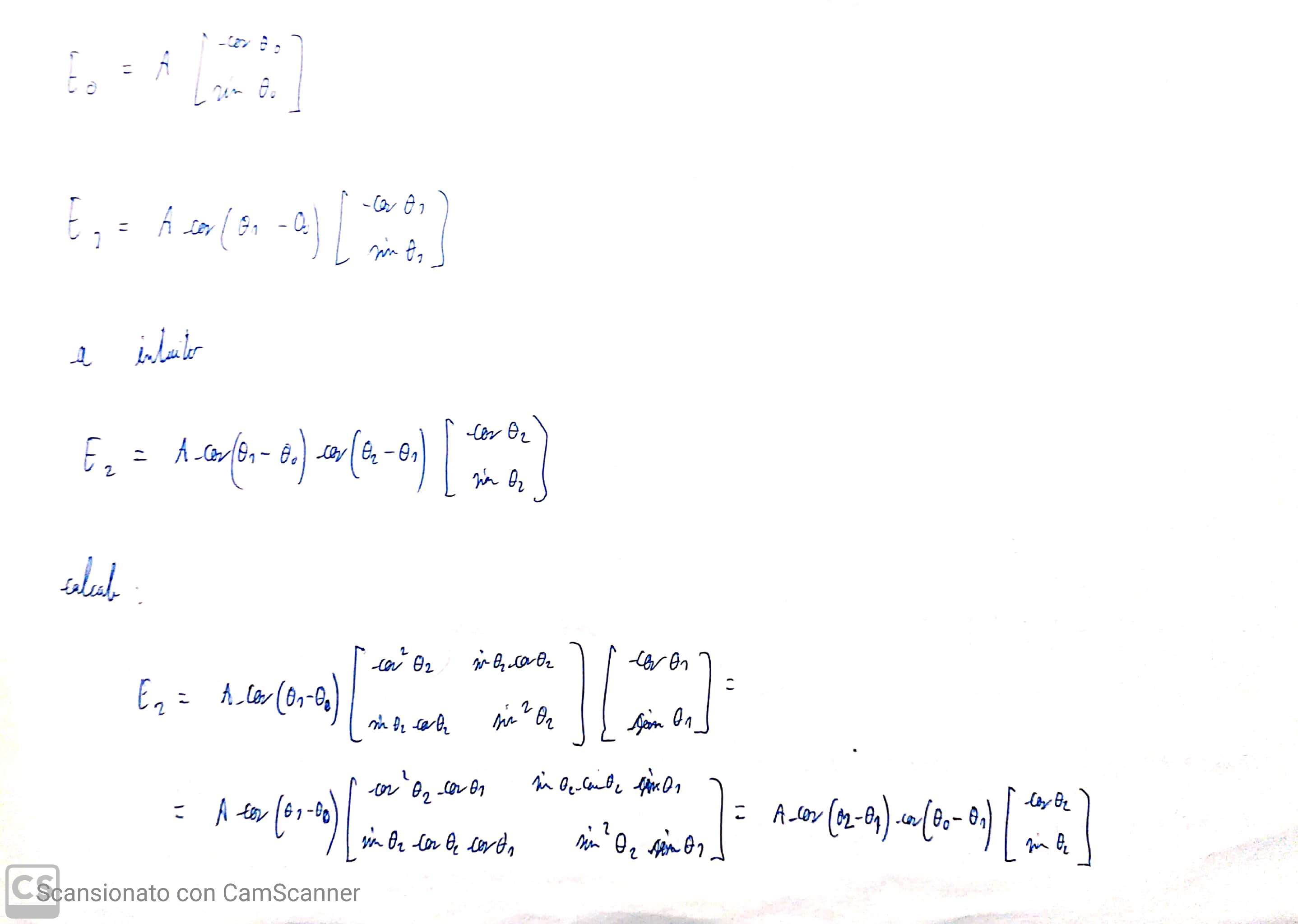}
     \end{subfigure}
     \hfill
     \begin{subfigure}[b]{0.6\textwidth}
         \centering
         \includegraphics[width=\textwidth]{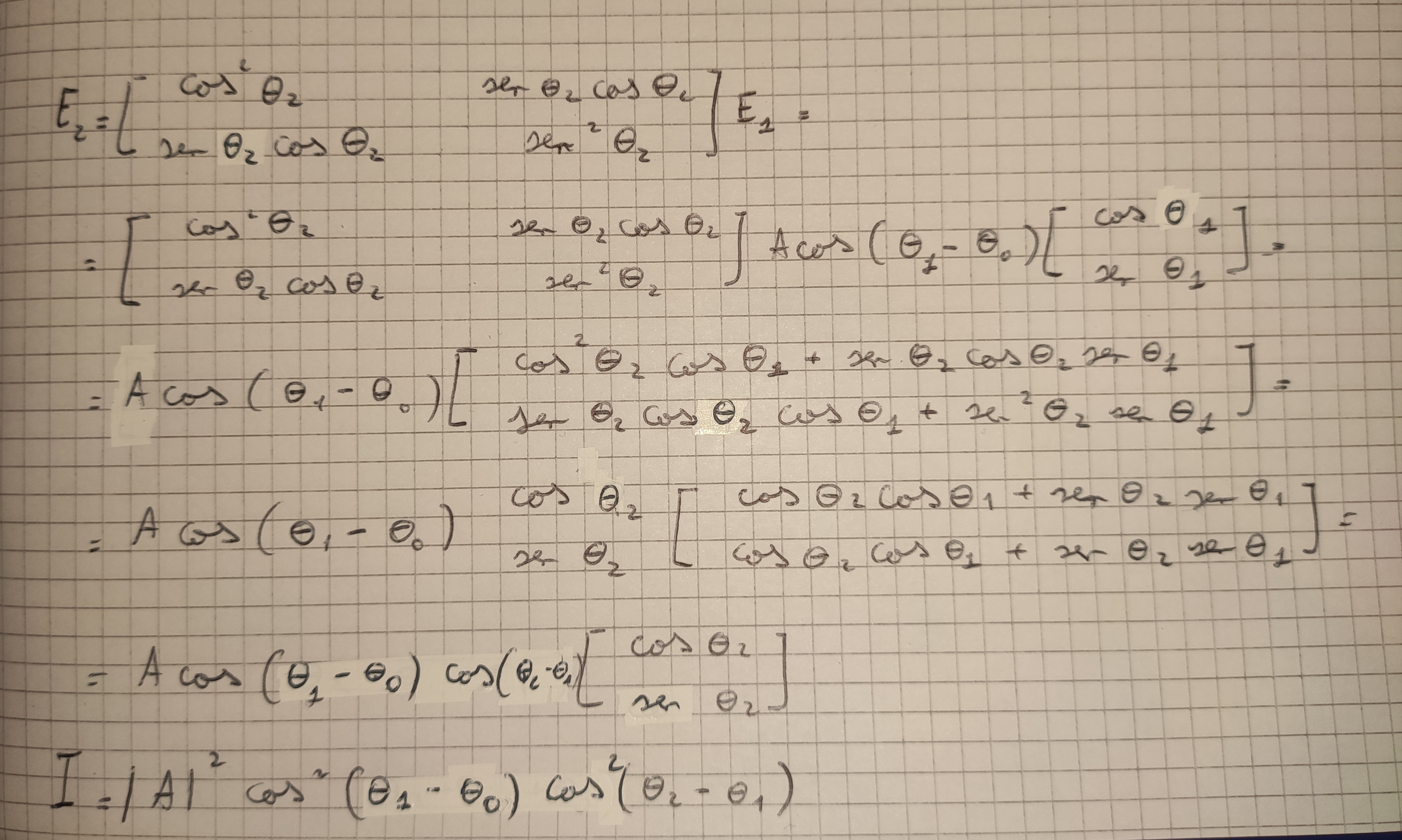}
     \end{subfigure}
        \caption{Polarizers sequence matrices task: after developing a mathematical description of Malus' law in terms of vectors and matrices, students were asked to generalize their findings with a generic sequence of polarizers.}
        \label{fig:ch7_StPolEx}
\end{figure}

As for ILO-a and ILO-b, when working on the experimental activities, students were able to apply the mathematical formalism to model how light states change in the sequence of polarizers. The use of matrices and trigonometry helped them to predict the intensity outcome that was compared with experimental measurements (see Figure \ref{fig:ch7_StMatrix} and Figure \ref{fig:ch7_StPolEx})

\begin{quote} {\fontfamily{cmtt}\selectfont
    I didn't think that matrices could be used for this \dots I only studied them in mathematics [course]. [StudentsRefl]
}\end{quote}

\begin{quote} {\fontfamily{cmtt}\selectfont
    I believe that using maths can help you learn physics \dots we always see them together at school anyway. [StudentsRefl]
}\end{quote}

Developing and consolidating familiarity with these mathematical tools in such physical context makes it easy to add them to the body of knowledge that students can develop. These tools are indeed very useful for the exploration of quantum systems representing qubits (see Section \ref{sec:seqSGA_matrix}). The knowledge of the vector state representing the information of the quantum object together with the possibility to represent the action of the devices by matrices can be useful to describe the behaviour of quantum objects in the context of quantum technologies.
The ease with which students have acquired this kind of representational language suggests that it can also be generalised to other contexts.

\begin{figure}[hbt!]
    \centering
   \includegraphics[width=\textwidth,height=\textheight,keepaspectratio]{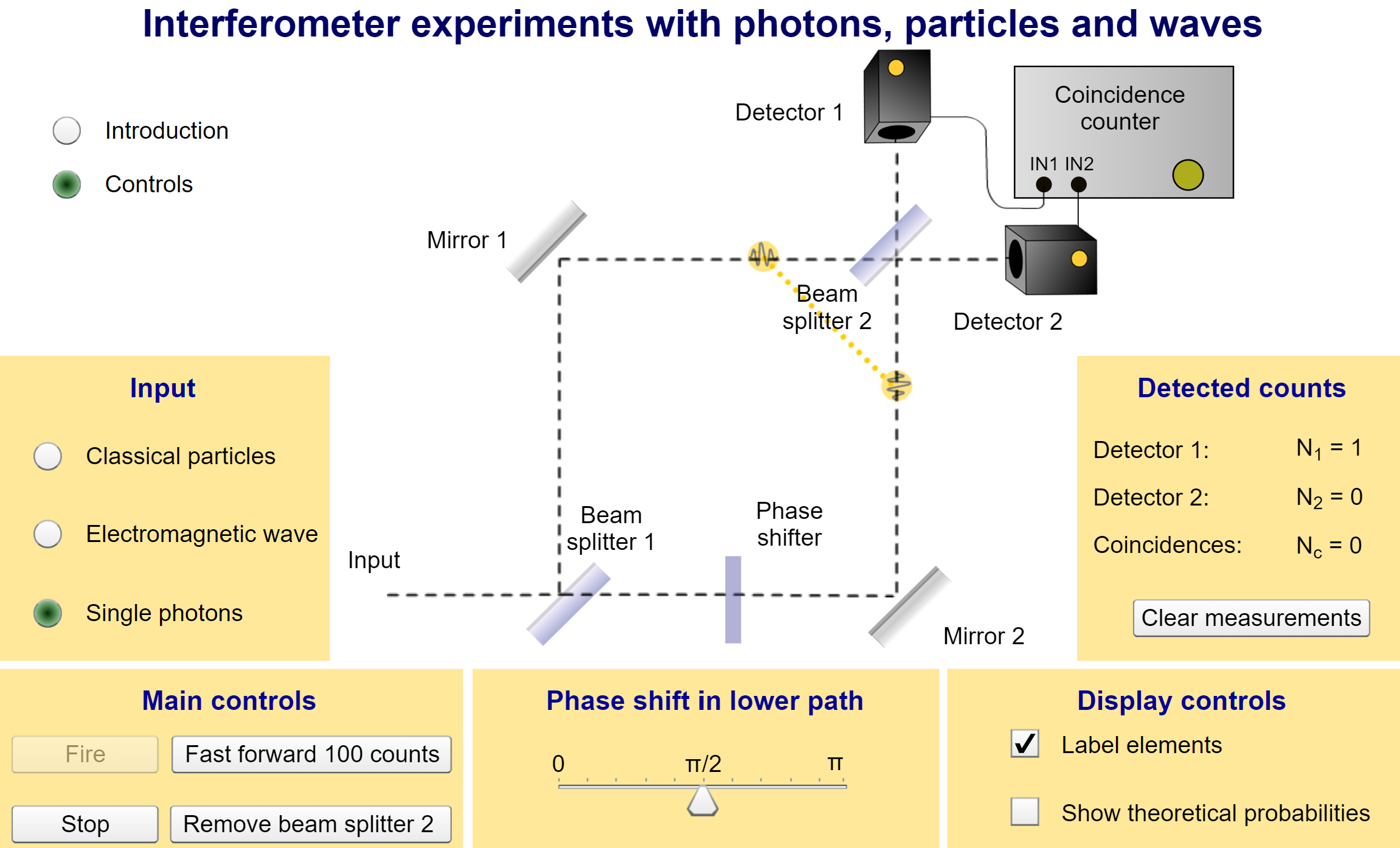}
    \caption[QuVis simulation: single photon]{Mach-Zehnder interferometer simulator \cite{QuVisSim}. Students can compare the behaviour of light using particle, wave or single photon models}
    \label{fig:Ch7_18_QuvisSingleP}
\end{figure}

The second part of the activity deals more specifically with the comparison of descriptive models of light (see Figure \ref{fig:Ch7_18_QuvisSingleP}).

Through the simulator students can compare the behaviour of light inside a Mach-Zehnder interferometer using three different models of light: particle, wave and single photon. The inquiry process is guided through the gradual introduction of some elements, such as the use of a second beam splitter or a tool to create a $\delta$ phase shift between the two paths. The student's attention is drawn to the information that can be gathered about the light in the measurement process, which is done in terms of counts on the detectors and in the number of coincidences.
The experience made with the simulator allowed the students to compare in small groups using a worksheet as a guide. For detailed description of the use of the simulation, see appendix \ref{appx:QuVisMZI}).\\

As for ILO 3 and ILO4, thanks to the use of simulators, students were able to identify the elements useful to define analogies and differences between the interpretative models of the behaviour of a single photon made using the concepts of wave and particle. Among these, the attention to be paid to the number of coincidences and the impact on the measurements (counts and coincidences) that the choice of the value of the phase difference that can be introduced between the two paths has.

\begin{quote}{\fontfamily{cmtt}\selectfont
In the case of a single beam splitter, the photons behave like classical particles since passing through the beam splitter [the single photon] does not split but is alternately directed in one of the two directions and then being detected by one detector at a time without coincidence.[StudentsRefl] 
}\end{quote}

\begin{quote}{\fontfamily{cmtt}\selectfont
    For the single photon it is observed that, increasing the phase shift, an increasing frequency with which detector 1 detects the photon is observed. Starting with a phase shift greater than $\frac{pi}{2}$ a reversal of the trend is observed, in fact more photons are recorded by detector 1. With a phase shift of $\pi$ in we obtain the opposite situation to the null phase shift in which photons are recorded only by detector 1. In the latter case the photon is transmitted by both beam splitters or reflected by both. The same happens in the case of electromagnetic waves, the difference being that with electromagnetic waves coincidences are recorded, so the wave is split by the second beam splitter. In electromagnetic waves no coincidence is detected with a phase shift of 0 and $\pi$. [StudentsRefl]
}\end{quote}

The final question required the students to make a synthesis of the different experimental situations. The analysis of the answers was done through the rubric in Table \ref{tab:LES_Grid_NatureLight}. With this instrument it was intended to detect whether the students were able to characterise the nature of the single photon as something that could not be simply ascribed to the wave or particle model and thus highlight the need to identify a different model.\\


\begin{table}[ht]
\centering
\caption{Grid to analyse the answers about the nature of single photons}
\label{tab:LES_Grid_NatureLight}
\begin{tabular}[t]{c>{\raggedright}p{0.1\linewidth}>{\raggedright\arraybackslash}p{0.5\linewidth}}
\toprule
score & level & description\\
\midrule
0 & absent &  no answer provided\\
& &\\
1& limited & statement with no explanation or no link to physical concepts\\
& &\\
2& partial & limited or generic use of physical concepts to support or justify the answer\\
& &\\
3 & complete & explanation provided is built on clear reference to meaningful and appropriate physical concepts\\
\bottomrule
\end{tabular}
\end{table}

\begin{figure}[hbt!]
    \centering
   \includegraphics[width=\textwidth,height=\textheight,keepaspectratio]{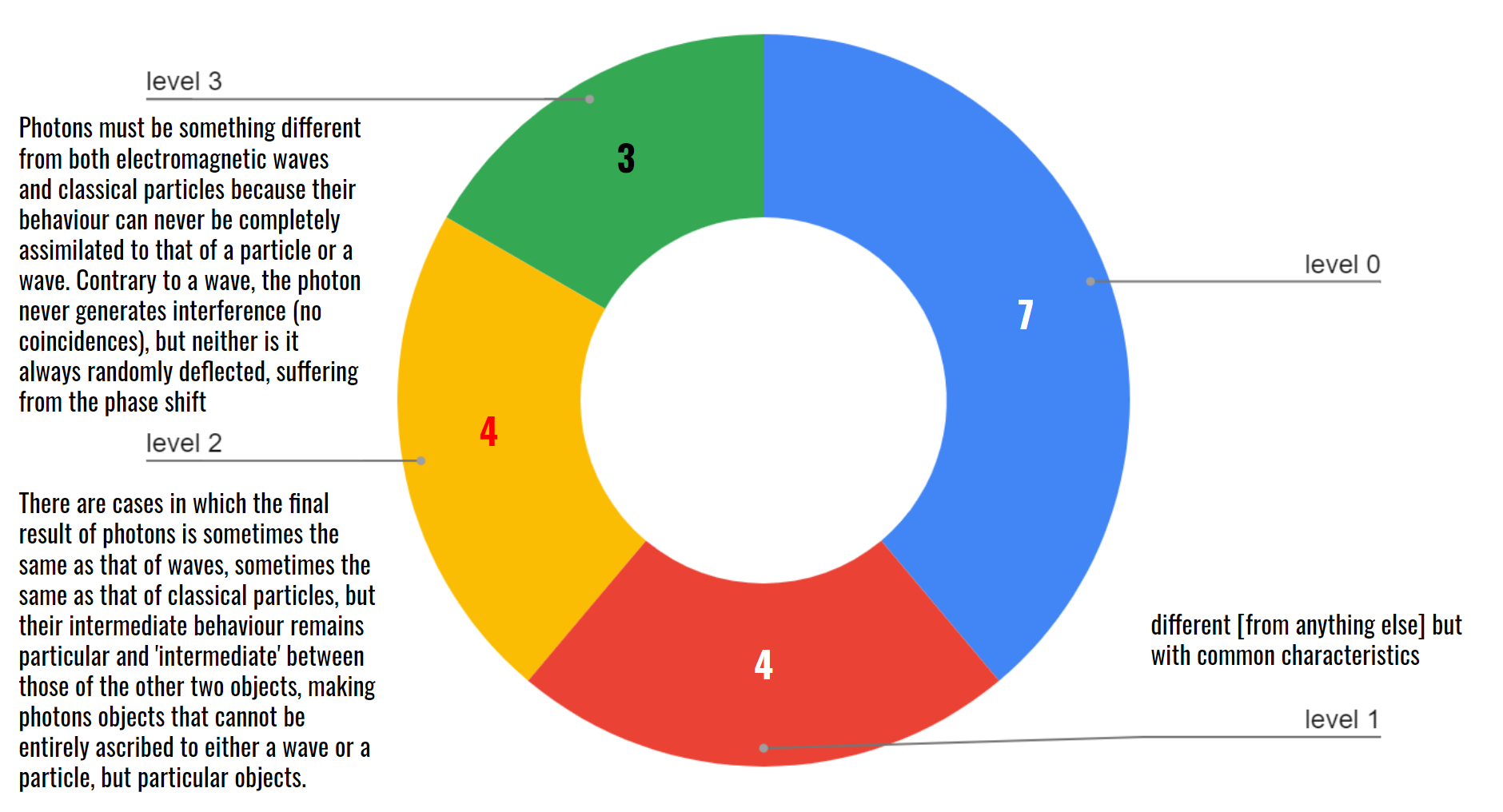}
    \caption[Students' responses about the nature of photons]{Results of the analysis of N = 18 students responses about the nature of photons. In the Figure there are some examples of students'responses. The level is defined in Table \ref{tab:LES_Grid_NatureLight}}
    \label{fig:Ch7_4_QuestPhoton}
\end{figure}

The question is (see Appendix \ref{appx:QuVisMZI}, question 11): "\textit{In what sense do they (single photon) behave like classical particles?  In what sense do they behave like classical waves?  Was there a case where the photons acted just like classical particles; or acted just like electromagnetic waves? Or must photons be something different to both electromagnetic waves and classical particles?}"\\
The task was completed in groups of three as an opportunity for students to compare and reflect. A total of 18 questionnaires were collected and the levels achieved are depicted in Figure (\ref{fig:Ch7_4_QuestPhoton}). 

It is therefore necessary a definition in which the dual nature of light is brought forward not as a simple statement, but critically argued from experimental evidence. These reflections made the presentation of the interpretative model using the concept of state more meaningful for the students. The abstract nature of this concept at this point can be justified by the impossibility of finding something classical that can adequately represent the photon.

\begin{quote}{\fontfamily{cmtt}\selectfont
    [photons] are not assimilated to either [particles or waves], so they must be something different. [StudentsRefl]
}\end{quote}




\section{Quantum technologies for high school students}\label{sec:SumSch_QT}


As discussed in Chapter \ref{ch:3-QApproach}, the Second Quantum Revolution does not only concern the production of new technologies, but it is mainly about the generation of a new vision of reality as we know it. So we designed learning activities to engage students (and their teachers) in the exploration of the quantum realm through the analysis of quantum technologies.\\

In general, the study of quantum technologies is proposed following the introduction of the concepts and fundamental principles that describe the behaviour of quantum objects through the evolution of interpretative models originating in the classical world.
Technologies are seen as the result of the application of scientific knowledge gained in a given field of research to the creation of tools to perform specific tasks. The technologies are then presented after introducing the theoretical basis for understanding how they work.
In the specific case of quantum technologies, in many textbooks \cite{amaldi2020il,romeni2017fisica} the applications of the first quantum revolution (lasers, leds, etc.) find space at the end of the course. The curriculum \cite{MIUR2010_211} envisages the construction of a model of the atom that is coherent with a series of experimental results as the culmination of the study of quantum physics. The applications of the theory to the microscopic world derived from such a model are often left to the discretion of the teacher.

The choice made in the design of the educational activities was to \textit{reverse the approach} and introduce the study of quantum physics starting from the problem of how information can be encoded in a physical system and the difference between choosing a classical (bits) or a quantum system (qubits) (see Section \ref{sec:PIAA}).
Reversing the approach and focusing on the informational process opens the possibility to introduce the axioms that are the tenets of quantum theory (see Section \ref{sec:QAxioms}). 

The research questions are specifically about how the intertwining between maths, physics and computer science can support student understanding of quantum computation.

\begin{table}[hbt!]
\centering
\begin{tabular}{cp{12cm}}
 & \textbf{Explore the relation between quantum physics and quantum technologies}\\
 & \textit{Interdisplinary nature of quantum technologies}\\
\hline
\rule[-4mm]{0mm}{1cm}
RQ7.1 & How does the interdisciplinarity between mathematics, computer science and physics can support students’ exploration of the tenets of quantum physics
and their relation with quantum technologies?
\end{tabular}
\caption{Research questions: Learning Environment for students about Quantum Technologies}
\label{tab:RQ_SLE_QT}
\end{table}


These activities were carried out in the context of the "Late Summer School - Introduction to Quantum Technologies" organised within the Department of Science and High Technology (DiSAT) of the University of Insubria as part of the activities of the Piano Lauree Scientifiche project of MIUR.
During the last few years, three Schools have been held with different groups of students from different school districts. Generally the students involved have a strong interest in science subjects, especially physics, mathematics and computer science.
The design and implementation of the activities took place thanks to the fundamental collaboration of the research groups in physics education of the Universities of Bologna and Pavia. The construction of a collaborative network was essential to adapt the choices of the approach to the specific learning environment of the Summer School.

An interesting aspect is that some teachers, many of whom also teach the course participants, also participated in the course. The teachers experienced this occasion mainly as a moment of personal training linked to their interest in the topics. Some of them described the experience as a moment of professional reflection, sharing with the organisers their thoughts about the criteria for transferring the contents into their lessons. This created a learning environment involving different elements of the learning ecosystem related to quantum physics. The dynamic interaction and mutual contamination facilitated the identification of critical points in the process and highlighted its strengths. As far as the feedback from the teachers on these activities is concerned, please refer to the Section \ref{sec:TLE_focus_interview}.

Using the concept of quantum qubit, we introduced the main features of quantum physical systems processes, that is state preparation, evolution and measurement and we reconceptualized them in terms of input information, processing, and output, triggering an analogy between the physical and the computational perspectives. Going then into the substance of how the quantum logic gates, circuits, and algorithms work, we switched to the mathematical perspective that intertwined with the other.

\begin{figure}[hbt!]
    \centering
   \includegraphics[width=\textwidth,height=\textheight,keepaspectratio]{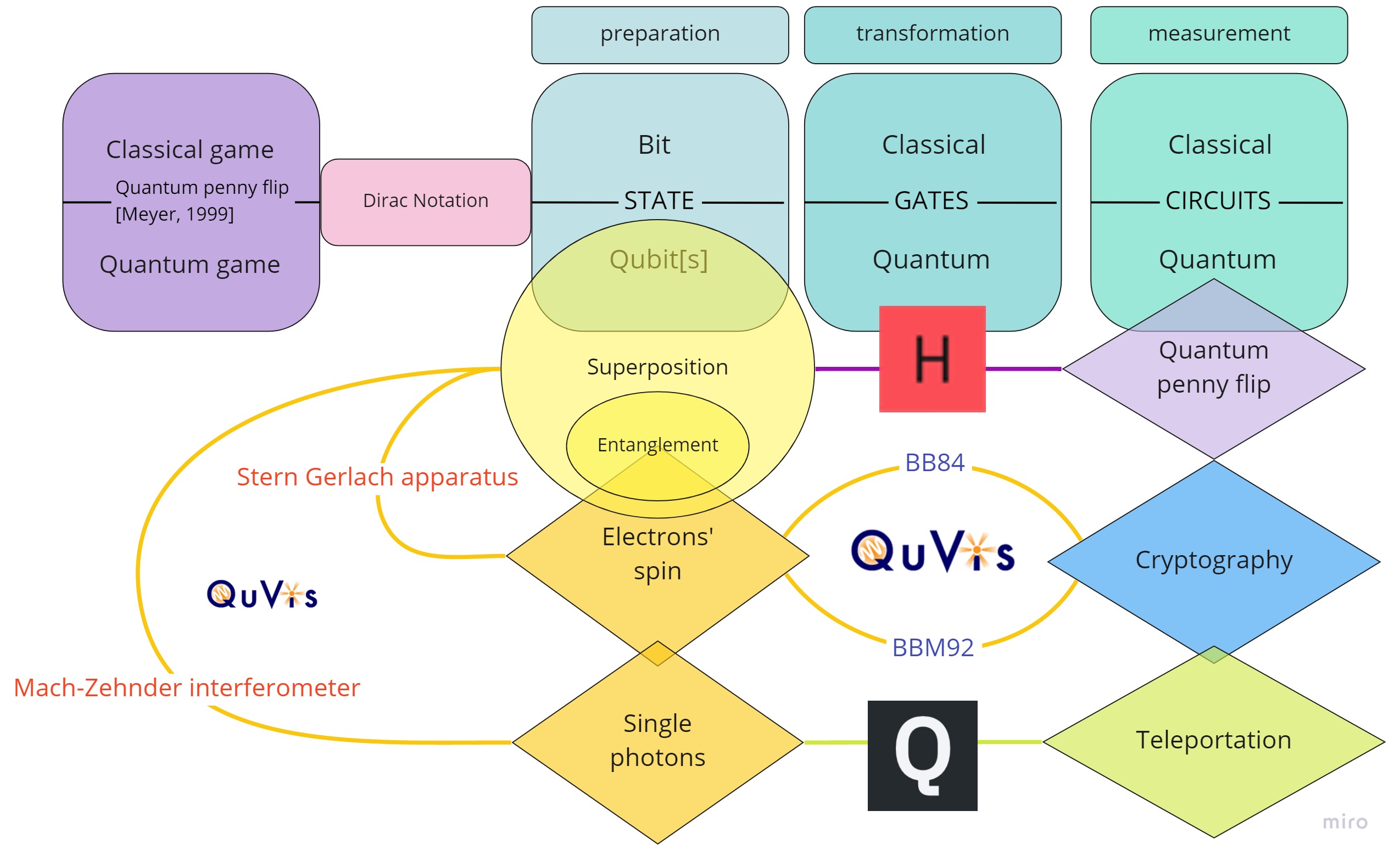}
    \caption[Summer School - Introduction to Quantum technologies, general strutcture]{The key concept of qubit is explored both studying the physical objects that can represent it and using it as the information resouce in a computational structure}
    \label{fig:Ch7_5_TLSQTSumSch}
\end{figure}

An example of a general sequence implemented during the Summer School is in fig.\ref{fig:Ch7_5_TLSQTSumSch}
With the penny flip game we start the exploration of what makes a Quantum Computer different from a Classical Computer. We gradually introduce different mathematical representations (in particular the Dirac Notation) as a tool to analyse the computational structure of quantum description of the behaviour of quantum states both in its interaction with experimental devices and inside a circuit. Using an informational approach, every concept is explored from a phenomenological point of view in an experimental context (Stern Gerlach apparatus and and Mach-Zehnder interferometers) and also as a key element for the functioning and structuring of a quantum algorithm using the IBM composer. In this sense, Quantum technologies are an opportunity to study the interplay between physics, logic and computer science in the making of a new description of reality based on quantum theory.

\begin{figure}[hbt!]
    \centering
   \includegraphics[width=\textwidth,height=\textheight,keepaspectratio]{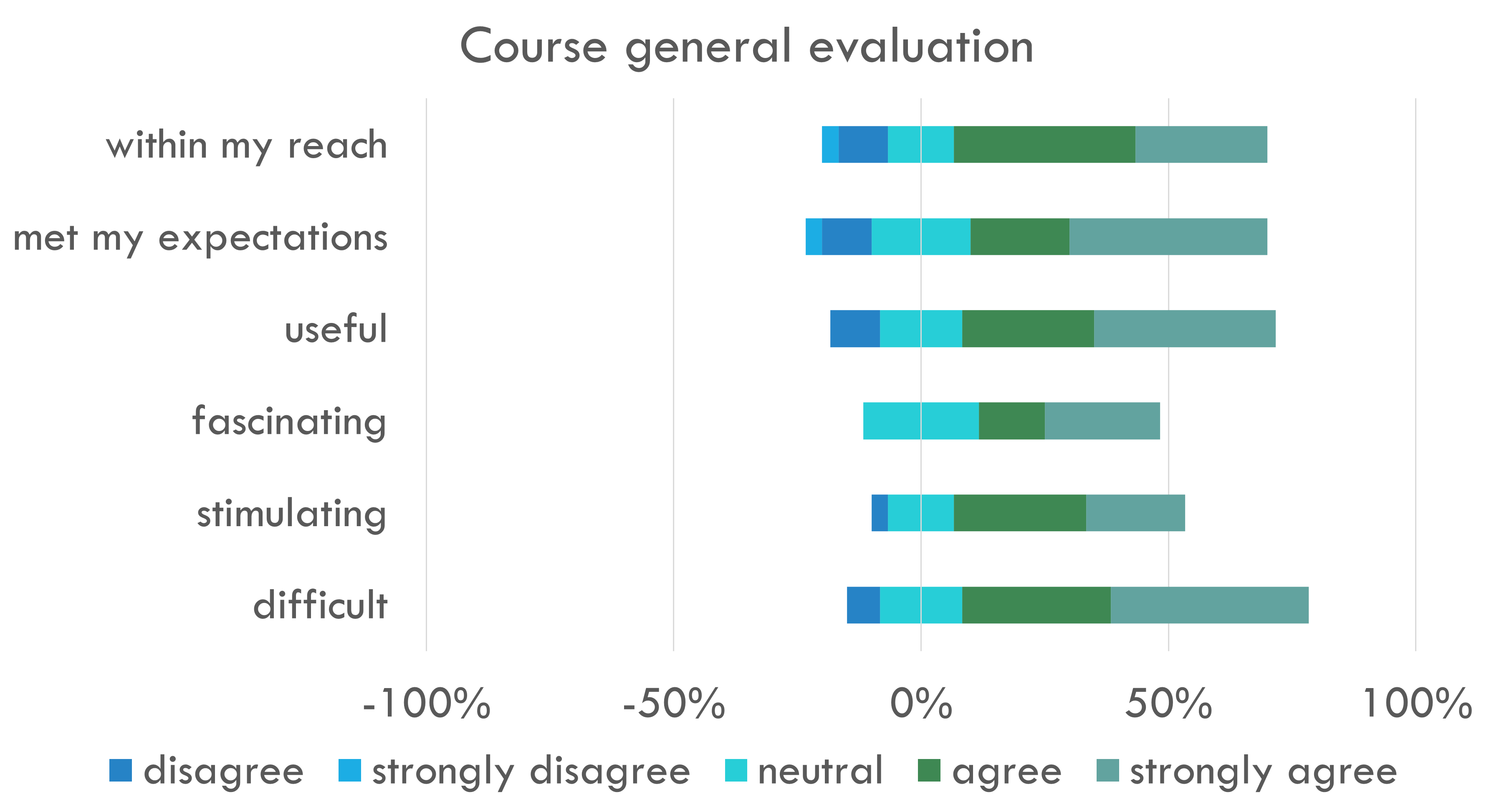}
    \caption[Summer School - Introduction to Quantum technologies, students'evaluation of the course]{ Students'responses in the questionnaire about the evaluation of the Summer School - Introduction to Quantum technologies, September 2020 edition}
    \label{fig:Ch7_6_QTSumSchEvalu}
\end{figure}

Students found the course challenging in a positive way (see Figure \ref{fig:Ch7_6_QTSumSchEvalu}. Even though it was hard, the activities were stimulating and useful.


As for RQ7.1 (see Table \ref{tab:RQ_SLE_QT}) in the collected questionnaires and interviews, students showed considerable awareness of the role of different the disciplines. Students shared interesting reflections on the nature of quantum processes they have been able to reconstruct thinking about the different activities.

At the end of the Summer School one students sent an email to share some of his reflection about the course. Like other participants, his participation in the course was motivated by a curiosity about the topics rather than a specific interest in physics

\begin{quote} {\fontfamily{cmtt}\selectfont
 When I enrolled in the course I had no certainty about how it would be or whether I would like it, in fact I enrolled almost by chance but since the "quantum" theme has always intrigued me I thought it was time to find out something about it and take the opportunity to go deeper. 
Before the course started I had some doubts because even though I attend the scientific high school (in a few days I'll be starting my fifth year at the Savona high school with a specialisation in applied sciences), so it's quite obvious that I'm interested in scientific subjects, physics is my least favourite of all the scientific subjects, and knowing that the course would be full of it I was afraid that I might find it uninteresting, perhaps even not very useful, given that physics has never been one of my possible options for university and it's not even one of my favourite subjects. [StudentsRefl]
}\end{quote}

Participation also had a significant and profound impact at the level of study interests

\begin{quote} {\fontfamily{cmtt}\selectfont
Despite my doubts I still wanted to take part and was surprised to find interest and pleasure in discovering the topics of quantum mechanics and it unsettled me because I am basically an indecisive person, plus I have to re-evaluate physics among my options for university. [StudentsRefl]
}\end{quote}

The student then shared some reflections on the approach to quantum physics, in particular on how the analysis of quantum technologies has shed light on certain aspects of physical theory

\begin{quote} {\fontfamily{cmtt}\selectfont
Of course, to say that it was easy to understand would be absolutely untrue, but even if it is not easy, quantum mechanics is feasible, in the sense that it can be understood and is not at all detached from reality. This course has been really shocking because I have always heard quantum physics being talked about as a world apart (and it is a bit) and detached from reality, but it is not like that at all, on the contrary it is much more attentive to reality and how it is interpreted than any branch of physics I have ever dealt with.[StudnetsRefl]
}\end{quote}

The contents presented in the Summer School also stimulated links to other disciplines

\begin{quote}{\fontfamily{cmtt}\selectfont
I found a lot in common between quantum physics and Kant's thinking, at least as a view of reality, and just as I enjoyed this philosopher I also enjoyed the topics seen during the course. [StudentsRefl]
}\end{quote}


As for ILO1 and ILO2, the approach used helped students to understand how essential and strong the interplay between physics, mathematics and computer science is to generate the fabric of quantum technologies. In one of participants words

\begin{quote}{\fontfamily{cmtt}\selectfont
    What helps most to understand what quantum computing is the way in which the three disciplines mentioned [mathematics, physics and computer science] interact with each other: the use and study of certain physical phenomena, the way in which they can be manipulated through simple mathematical commands which can then in turn be exploited in the creation of a computer algorithm. [StudentsRefl]
}\end{quote}

A first consequence is that the definition of what a quantum computer is has been enriched by considerable connections with physics: from being a faster and more powerful computer, it has become a technology that uses quantum objects (qubits) and physical laws (the superposition principle) to process information.

\begin{figure}
     \centering
     \begin{subfigure}[hbt!]{0.6\textwidth}
         \centering
         \includegraphics[width=\textwidth]{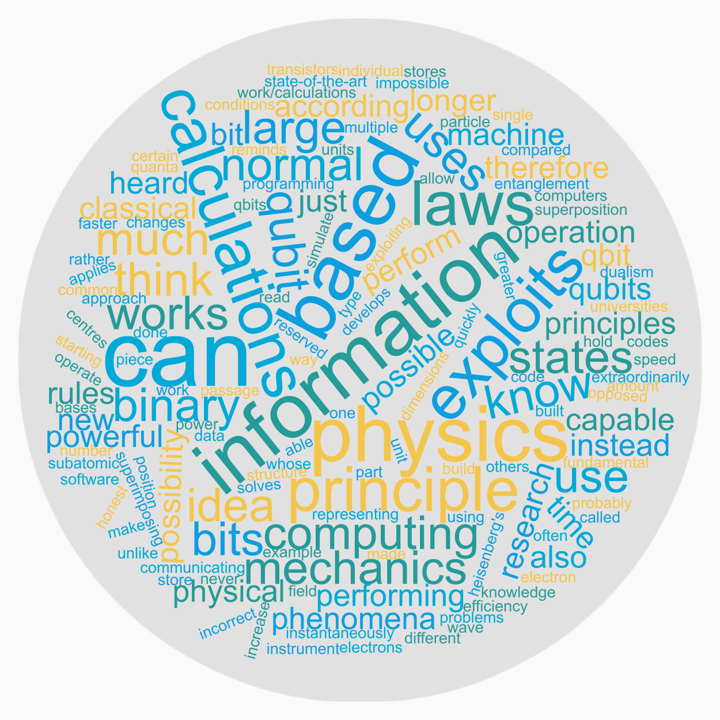}
         \caption{"Can you define what a quantum computer is?": pre Summer School}
     \end{subfigure}
     \hfill
     \begin{subfigure}[hbt!]{0.6\textwidth}
         \centering
         \includegraphics[width=\textwidth]{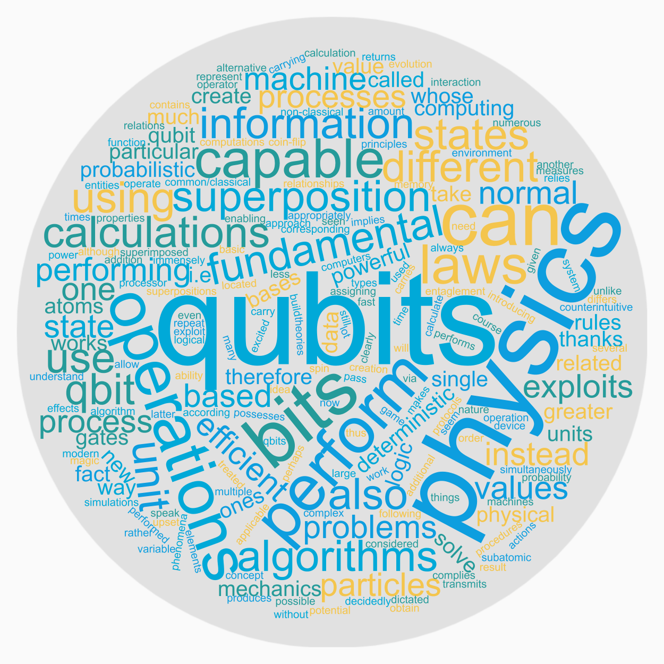}
         \caption{"Can you define what a quantum computer is?": post Summer School}
     \end{subfigure}
        \caption[Defitions of quantum computers]{Quantum computers: elements to characterize a quantum computer before and after the Summer School. Physical concepts such as qubit and superposition are now the key words to define the nature of quantum computers}
        \label{fig:ch7_7_QCDefinitions}
\end{figure}

In particular mathematics emerged as a powerful learning tool.
\begin{quote}{\fontfamily{cmtt}\selectfont
The rigour of mathematical formalism has helped me to interpret quantum processes correctly and without misunderstanding. [StudentsRefl]    
}\end{quote}

Students have been able to identify the relazione tra computazione classica e quantistica ed il modo in cui nuovo in cui quest'ultima definisce processi con cui è possibile estrarre informazione sulla realtà attraverso la manipolazione di oggetti quantistici.

\begin{quote}{\fontfamily{cmtt}\selectfont
    The basic difference between classical and quantum computation lies in the minimum unit of information in a processor that forms the computer: the qubit is capable of storing much more information than the classical bit. [StudentsRefl]
}\end{quote}

\begin{quote}{\fontfamily{cmtt}\selectfont
    “[…] the bit is either 0 or it is 1, even during the computation phase; the qubit […] before measurement can take on superposition values, which are neither 0 nor 1, and this greatly facilitates the computation phase, allowing the machine to consider several states at the same time […] the measurement destroys all doubt, making the qubit "collapse" and reducing it to the form of a classical bit: only 0 and 1 can indeed be measured. [StudentsRefl]
}\end{quote}

\subsubsection{Quantum cryptography}\label{subsec:crypto}
One example of activities that have been designed and implemented with different group of students are related to the exploration of quantum cryptographic protocols.
This is an extremely interesting technological context from an educational point of view \cite{BessonMalgieri2018}. The physical principles underlying the processes of construction and distribution of cryptographic keys (BB84 and BBM92 protocols) directly refer to the tenets and axioms of quanutm physics  (see Section \ref{sec:PIAA} and Section \ref{sec:ApprPillars}).
An activity that focuses on understanding how to build a secure cryptographic protocol has the dual advantage of raising awareness of the potential of quantum technologies, and of creating a learning context in which to deepen understanding of quantum superposition and quantum measurement.
The simulations used are those developed by the QuVis project at St.Andrews Univesity \cite{QuVisSim}. The worksheet related to the simulation have been adapted to the purpose of the research.

The prerequisites for the course are fundamentally linked to the familiarity with the use of the Stern-Gerlach apparatus and its simulation (see Section \ref{subsec:SGAsims}).
As part of their prior-knowledge, student should be able to:
\begin{itemize}
    \item Predict the experimental outcomes (measurements) using different sequences of Stern – Gerlach apparatus
    \item Interpret the results using the concept of quantum state and quantum superposition
\end{itemize}
In a previous workshop students should have had the opportunity to use Stern-Gerlach apparatus (SGA) to make predictions about SGA experimental result and compare them with actual result using the simulator. This is to introduce the idea of quantum state (see Section \ref{sec:ChangeP}) and the concept of quantum superposition. The interpretative model is presented using Dirac notation. 

The paths presented below were implemented during extracurricular activities carried out with groups of students. The work on the BB84 protocol was carried out both during the Spring School, introduction to quantum technologies in April 2020, in which 24 students took part, and in one of the PCTO's in-depth courses on quantum information, in which 34 students took part. The work on the BBM92 protocol was carried out with a group of 17 students during the last Summer School in September 2021. All the courses took place online. All groups have previously carried out a course introducing the concept of qubits and are familiar with the results and operation of the Stern-Gerlach apparatus.\

\begin{table}[hbt!]
\centering
\begin{tabular}{cp{12cm}}
\toprule
 & \textbf{ILO}: Analysing how the quantum characteristics of qubits underpin the process of generating and distributing cryptographic keys obtained using Stern-Gerlach devices. \\
 &\textit{At the end of the activities students should be able to:} \\
\midrule
ILO-a & characterize the problem of key distribution in relation to the generation of a secure key\\
ILO-b & relate the process of generating quantum key to the quantum nature of qubits, in terms of quantum superposition, entanglement and the probabilistic nature of quantum measurement\\
ILO-c & evaluate the impact of eavesdropping in the key distribution process using probability
\end{tabular}
\caption{ILOs: Quantum behaviours}
\label{tab:ILO_Cripto}
\end{table}

The chosen methodology refers to the guided inquiry \cite{FraserMazur2014_Bridging}. After a first moment ("starter") that serves as a link with the previous course modules, the students can work in small groups carrying out short tasks.  These moments allow the participants to compare notes and evaluate the validity of their conclusions through peer-to-peer dialogue \cite{MazurPeer1997} and the use of the simulator. Students are invited to share what they have learned through an online platform (Google Forms or Wooclap) and to discuss the main results in plenary sessions.
This sequence is carried out in each of the two parts into which each course is divided:
\begin{itemize}
    \item \textbf{part 1} Key distribution \textit{without} eavesdropping
    \item \textit{part 2} Key distribution \textit{with} eavesdropping
\end{itemize}
Each part has been designed with reference to the chunked lesson methodology (see Section \ref{sec:online_methodology}) to facilitate student participation in online teaching. 
It starts with a starter activity (about 10') followed by an explanation part (about 20') that is punctuated by a series of questions that students can answer online and that are discussed in plenary. The next phase involves carrying out a series of tasks (about 30') in small groups created within online rooms. Students can discuss with each other and fill in the online form together to collect their answers. Finally there is a plenary discussion around the different answers provided by the groups as an opportunity to clarify and consolidate learning.\\
The purpose of the starter exercises is to reintroduce some knowledge about specific aspects of the behaviour of quantum objects. The concept of quantum state, the superposition principle and entanglement are presented again, and their relationship with the representation given by Dirac's notation is reinforced.

\begin{figure}
     \centering
     \begin{subfigure}[hbt!]{\textwidth}
         \centering
         \includegraphics[width=\textwidth]{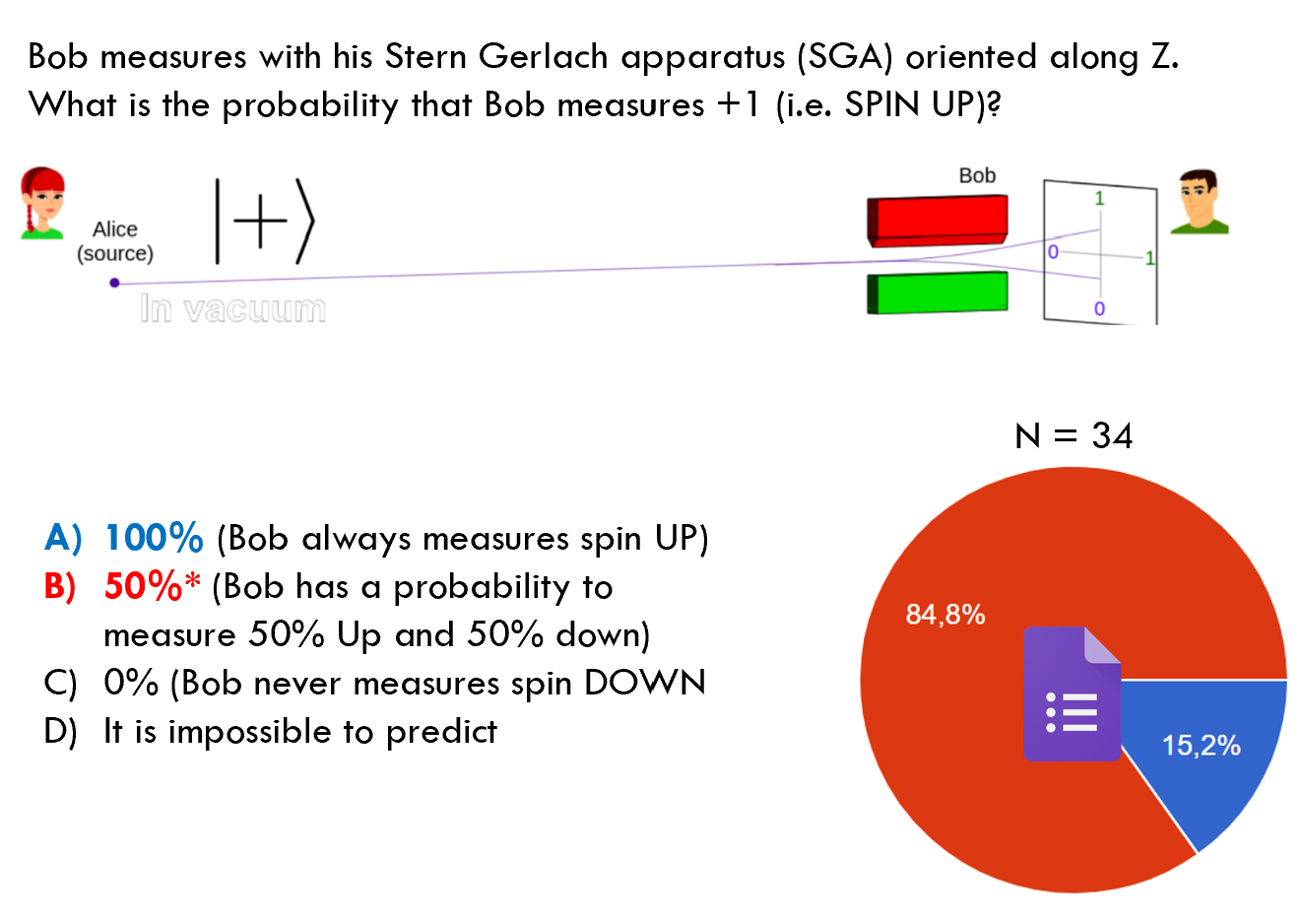}
         \caption{Starter question: superposition, N = 34 upper secondary students}
     \end{subfigure}
     \hfill
     \begin{subfigure}[hbt!]{\textwidth}
         \centering
         \includegraphics[width=\textwidth]{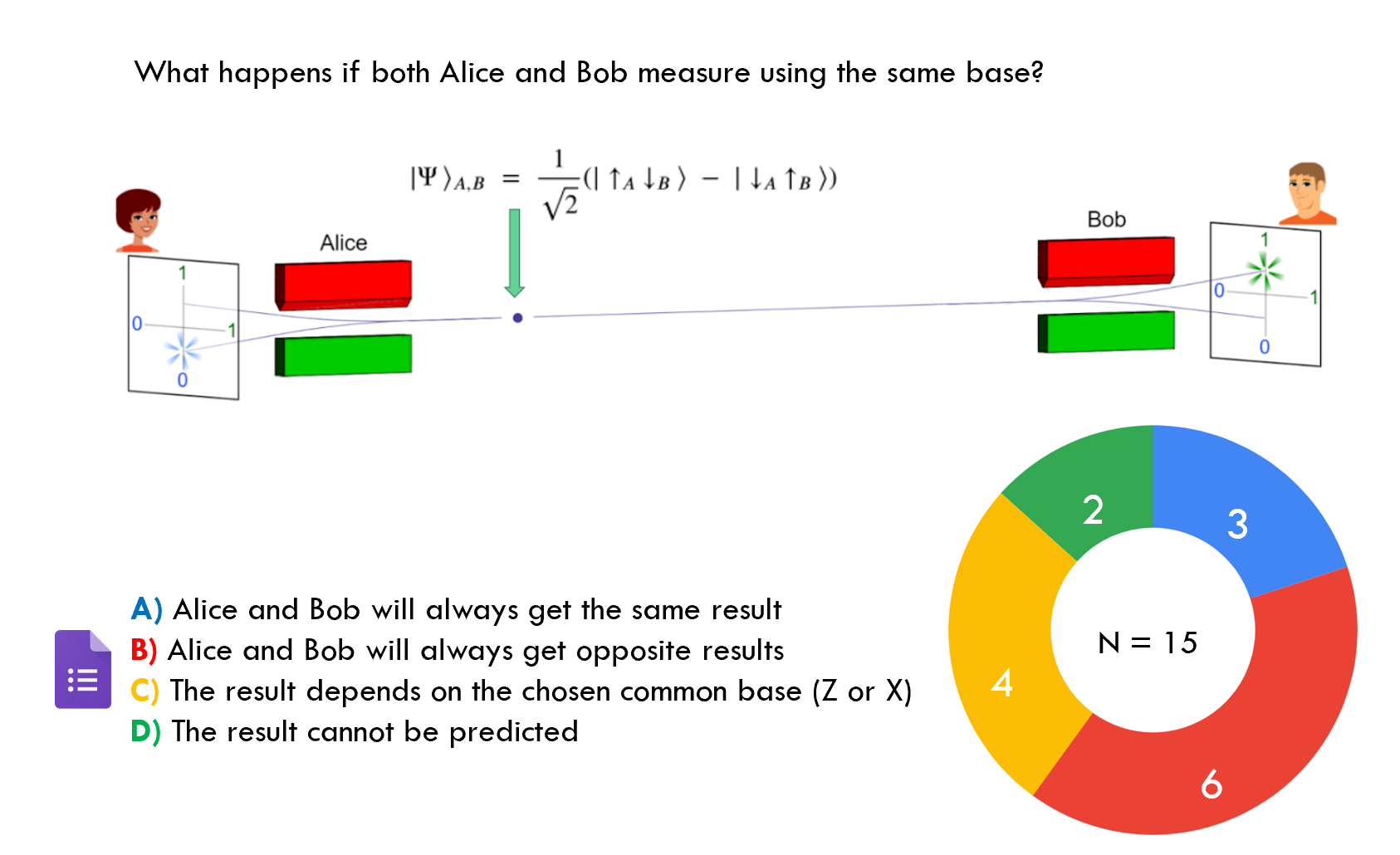}
         \caption{Starter question: entanglement, N = 15 upper secondary students}
     \end{subfigure}
        \caption[Starter questions: cryptography]{Started questions as part of priok-knowledge assessment}
        \label{fig:ch7_11_CryptoStarters}
\end{figure}

\begin{figure}
     \centering
     \begin{subfigure}[hbt!]{1\textwidth}
         \centering
         \includegraphics[width=\textwidth]{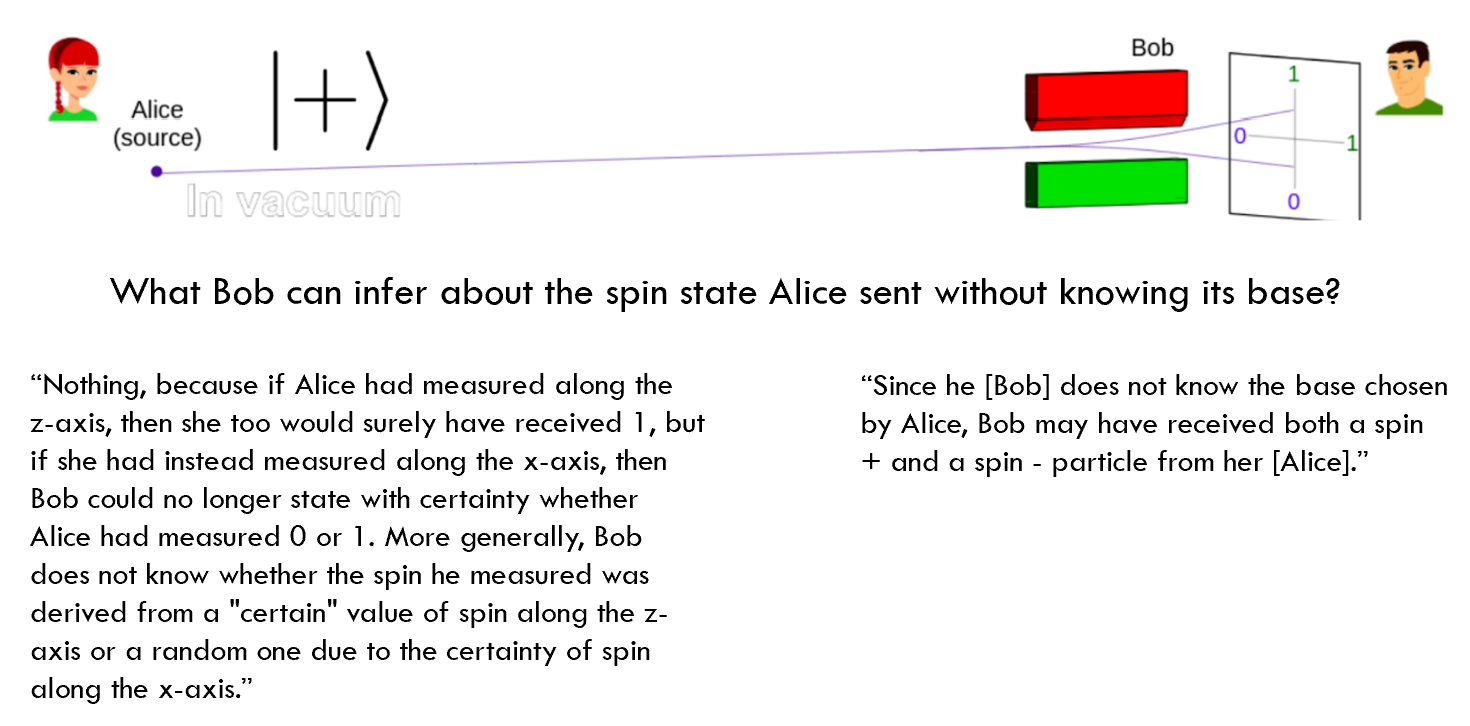}
         \caption{}
     \end{subfigure}
     \hfill
     \begin{subfigure}[hbt!]{1\textwidth}
         \centering
         \includegraphics[width=\textwidth]{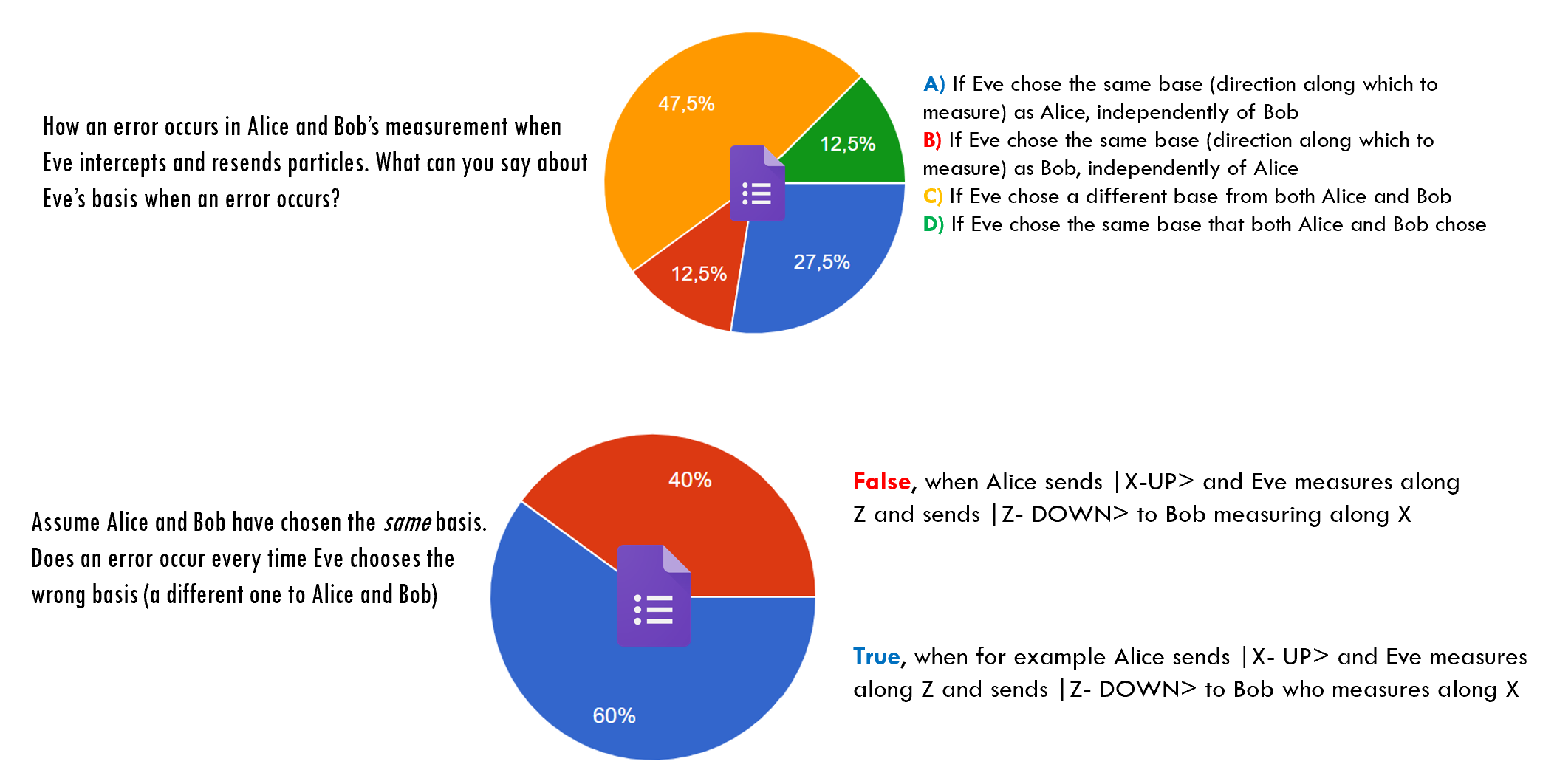}
         \caption{upper secondary students}
     \end{subfigure}
        \caption[Scenario questions: BB84 cryptography]{Scenario questions with examples of students answers: these questions can be used to evaluate how students car relate the process of generating quantum key to the quantum nature of qubits}
        \label{fig:ch7_12_BB84Scenario}
\end{figure}

\begin{figure}
     \centering
     \begin{subfigure}[hbt!]{1\textwidth}
         \centering
         \includegraphics[width=\textwidth]{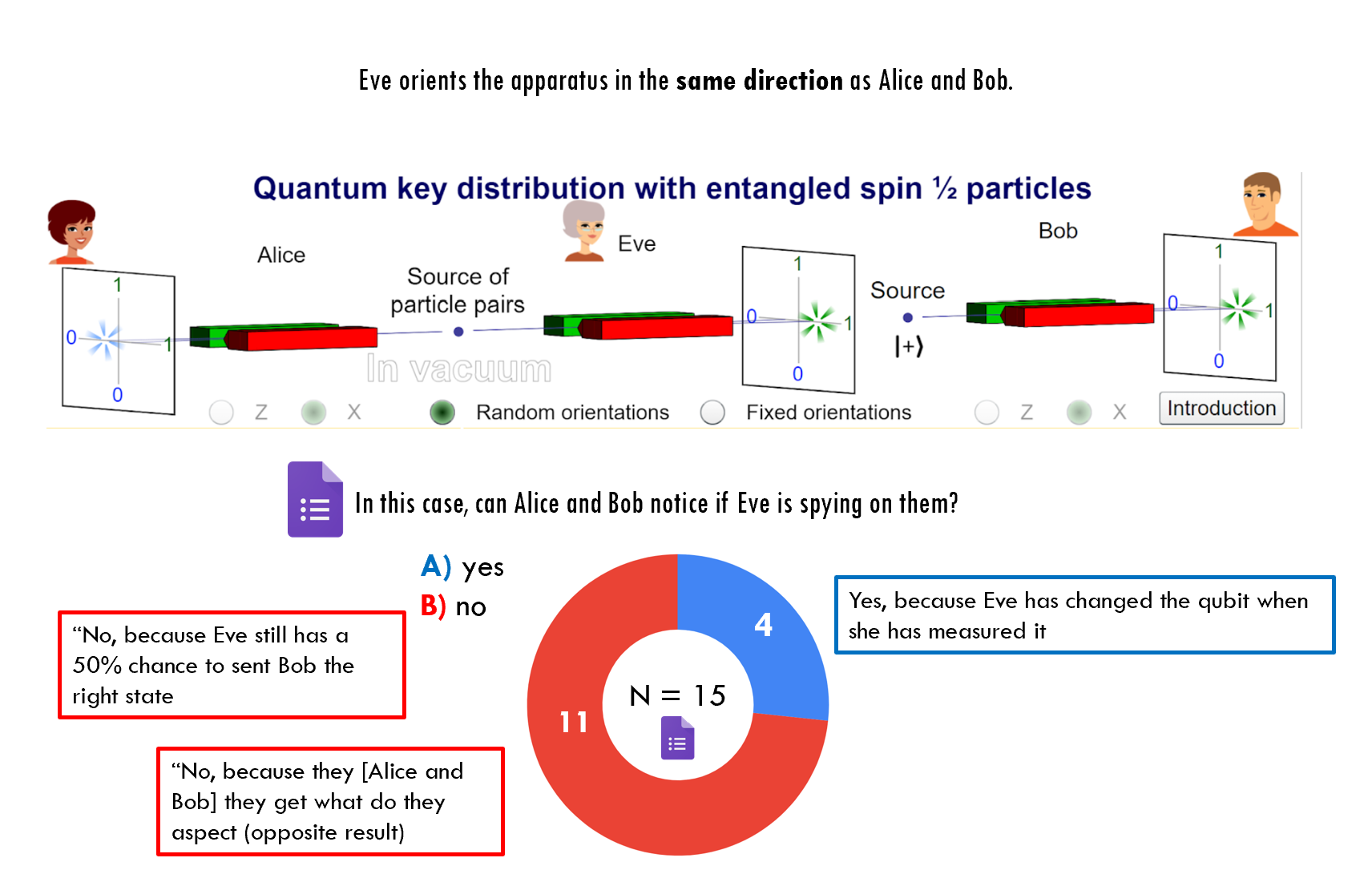}
         \caption{}
     \end{subfigure}
     \hfill
     \begin{subfigure}[hbt!]{1\textwidth}
         \centering
         \includegraphics[width=\textwidth]{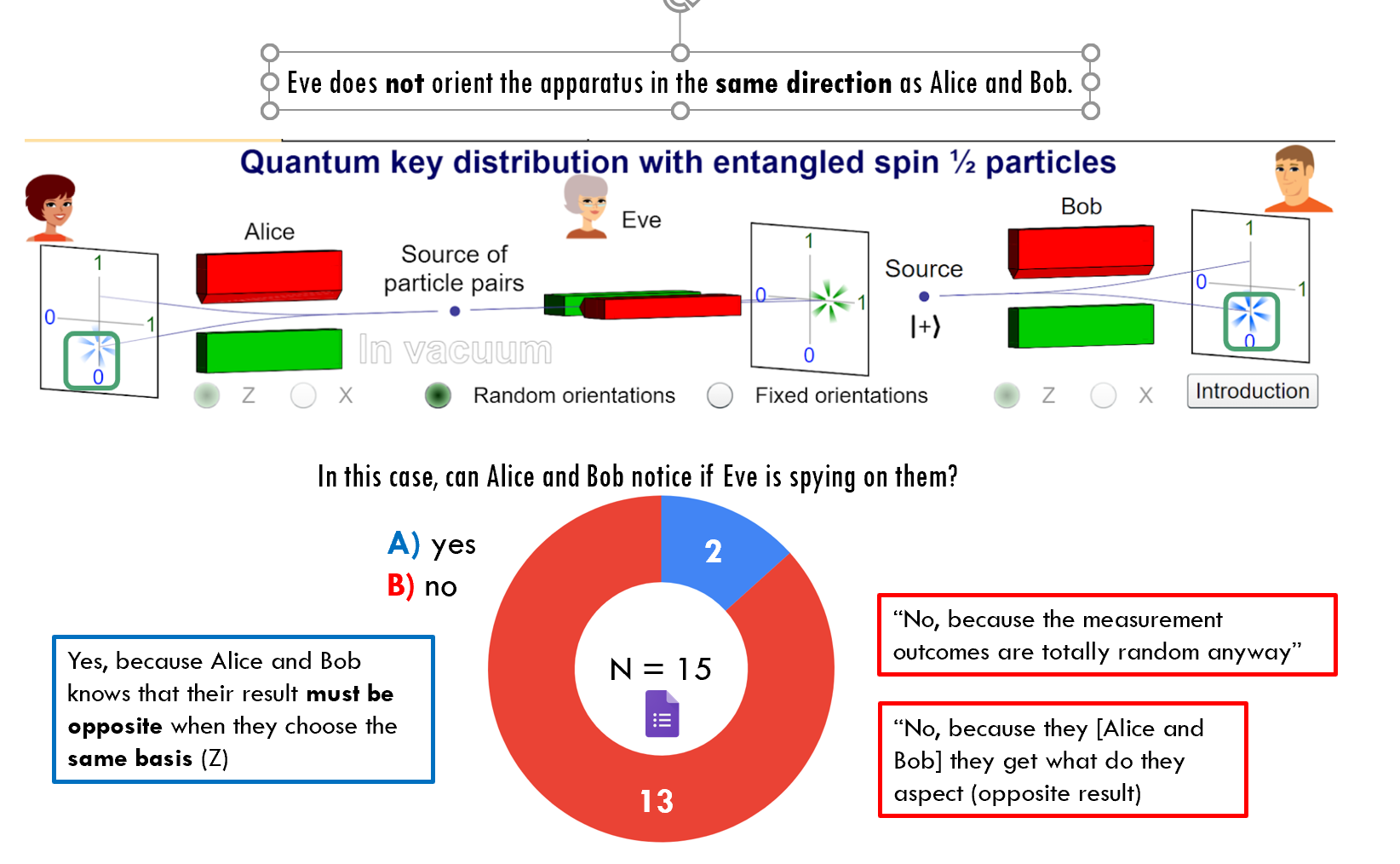}
         \caption{upper secondary students}
     \end{subfigure}
        \caption[Scenario questions: BBM92 cryptography]{Scenario questions with examples of students answers: in this case the knowlege about the characteristics of entangled states (totally random and perfectply anti-correlated) should help students justify their answers}
        \label{fig:ch7_14_BBM92Scenario}
\end{figure}

The central part of the activity with the students revolves around exploring the process in which the key is distributed. For both the BB84 and BBM92 protocols, the tasks use scenario questions \cite{Dijk2008_ERTEevolution} that are intended to create an opportunity for students to reflect (see Figure \ref{fig:ch7_12_BB84Scenario} and Figure \ref{fig:ch7_14_BBM92Scenario}). They are asked to make predictions by anticipating the results of the measurements. After a moment of discussion within the group or in plenary, the predictions are tested with the simulator.

\begin{figure}
     \centering
     \begin{subfigure}[hbt!]{0.6\textwidth}
         \centering
         \includegraphics[width=\textwidth]{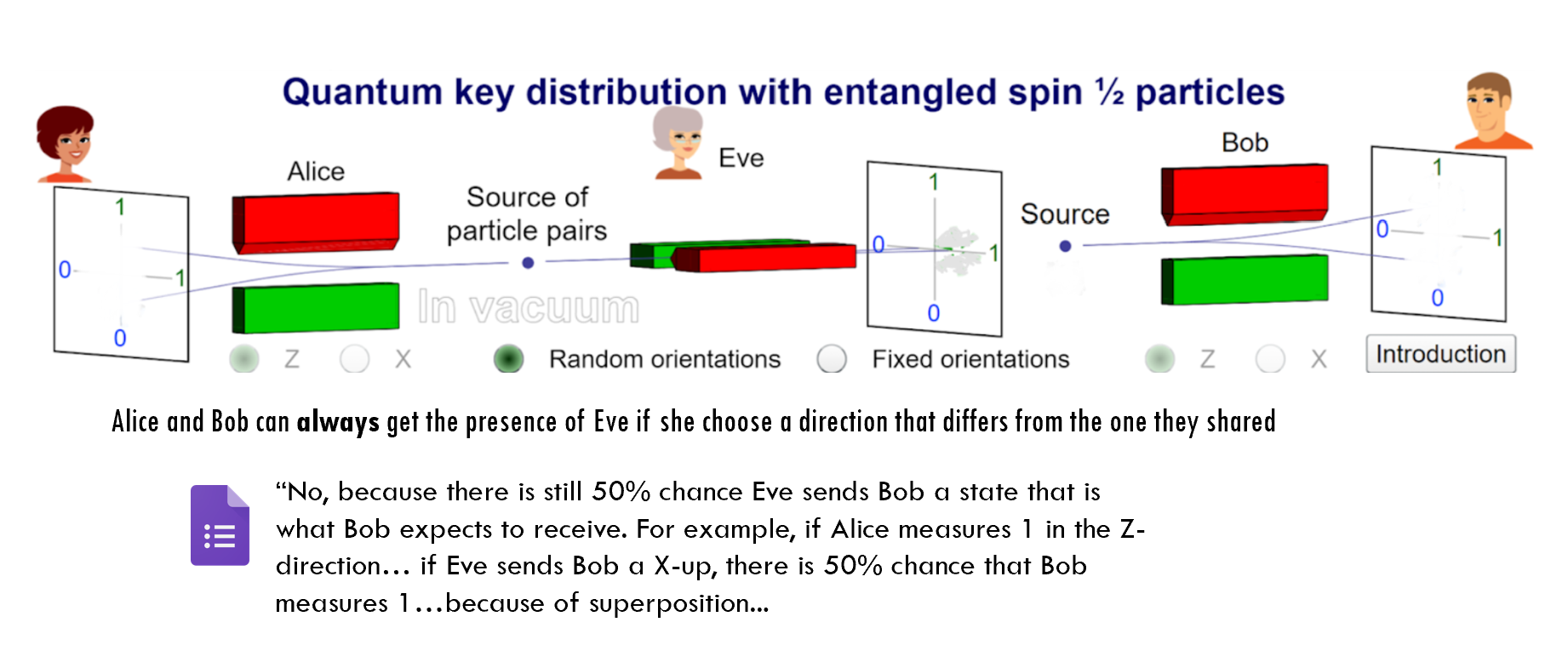}
         \caption{}
     \end{subfigure}
     \hfill
     \begin{subfigure}[hbt!]{0.6\textwidth}
         \centering
         \includegraphics[width=\textwidth]{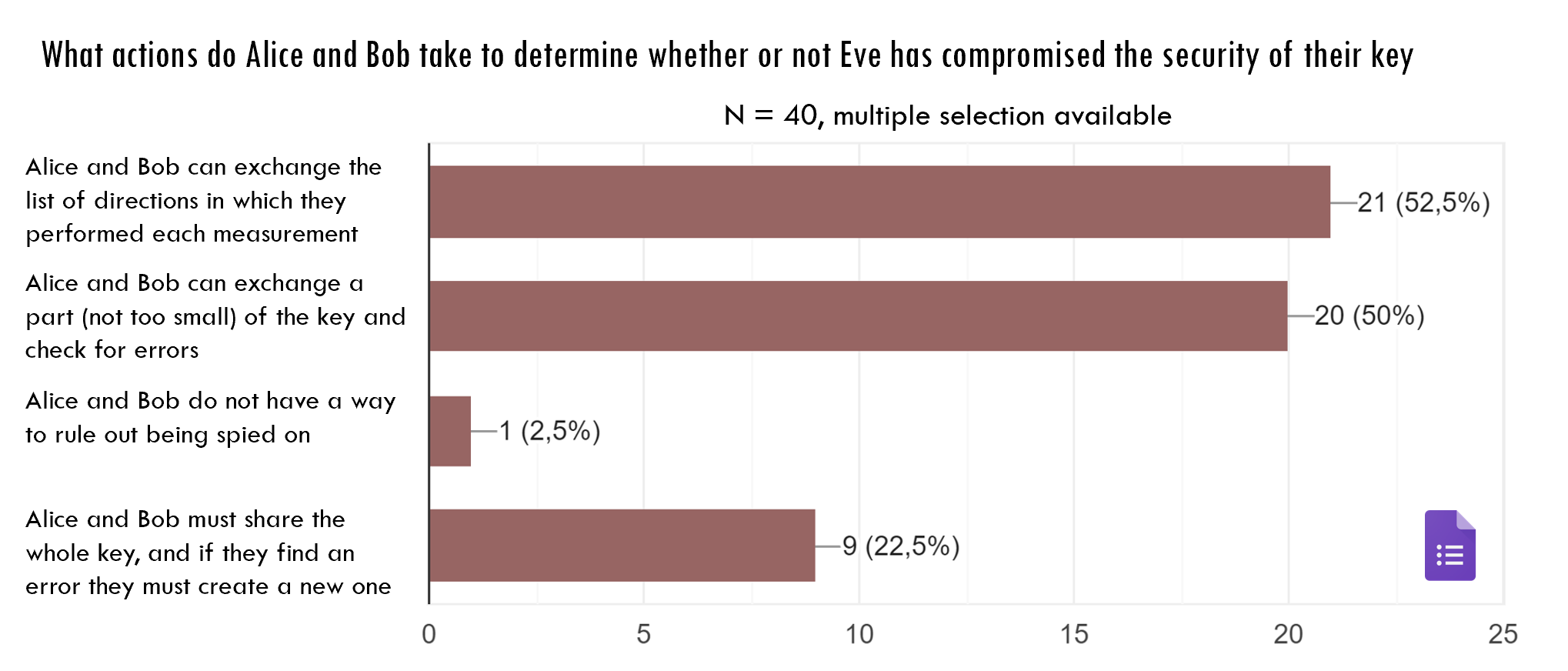}
         \caption{upper secondary students}
     \end{subfigure}
        \caption[Open questions]{Open questions: students should be able to justify their answer about the security of the protocol}
        \label{fig:ch7_15_OpenQuest}
\end{figure}

Specific questions (see Figure \ref{fig:ch7_15_OpenQuest} have been prepared to enable students to evaluate the impact of the eavesdropper. In this case students should justify their answer making clear reference to the quantum properties of quantum objects and probability

The use of the simulator is very effective in supporting students in analysing the different elements of the key distribution process.


The analysis of the texts produced by the students shows that they are able to identify which elements characterising quantum behaviour can be used for the construction of the key. 

\begin{quote}{\fontfamily{cmtt}\selectfont
One bit of the key is generated when Bob and Alice use the same base and get the same result. If "random orientation" is selected, more measurements need to be taken as the choice of orientation is random, so not always the same or different. 
In the case of "fixed orientation", on the other hand, they will never get a key if they have chosen a different base [StudentsRefl]
}\end{quote}

This type of activity was evaluated very positively by the students, as it helped them understand how quantum physics can become a technological resource and not just an abstract or counterintuitive theory.

\section{Conclusions}
The activities described in this chapter made it possible to test the use of the informational approach (see Section \ref{sec:ApprPillars}) to introduce the tenets of quantum physics. The activities were proposed to groups of students who self-selected for their specific interest in the contents of the courses proposed to them. This factor certainly made it possible to deal with the different contents more easily than would be the case in a curricular classroom context. The specific level of preparation of this group of students is certainly more homogeneous and the interest in the topics has favoured the engagement of the students in the course
This particular context made it possible to highlight some useful elements for the design of regular curricular activities.\\

As for the research questions RQ5 concerning the characterisation of quantum behaviour (Table \ref{tab:RQ_SLE_Qnature}), the use of the informational approach proved useful in providing students with representative and descriptive tools for quantum objects. Using the simulator, students were faced with the situation of having to choose which interpretative model was consistent with the observations. The conclusions were then constructed by highlighting the link between the quantum state and the measurement process.
Recognition of the impossibility of being able to completely relate the behaviour of a single photon to classical representations (particle or wave) can be used at the educational level to make room for a new type of representation. The description of the behaviour of quantum objects can thus be expressed in terms of state-related information and measurement-related properties. \\

As for the research question RQ7 related to the use of quantum technologies, the students showed great interest and involvement. Tackling the conceptual problems of quantum mechanics by exploring how quantum technologies use those same concepts and principles to function spontaneously generated a meaningful context. Although the problem was to understand the process of distributing cryptographic keys, the students' interest was often directed towards the way in which "strange" quantum behaviour can become a resource for achieving something unique.
As is often the case in physics education, the identification of a rich context offers many possibilities for constructing particularly meaningful learning activities.

\chapter{\label{ch:8-Conclusions}Conclusions} 
This chapter evaluates the impact of the actions carried out in relation to the various strategic lines. Finally, some critical aspects of the project are pointed out and  may constitute possible indications for the redefinition of new actions or new strategic lines.

\section{Strategy 1: Promote cohesion in the learning ecosystem}




The strategic choice to strengthen the cohesion between the different members of the learning ecosystem (researchers, teachers and students) has allowed the creation of different environments in which to stimulate sharing and dialogue.
In particular, it was important to provide opportunities for teachers to play an active role as "researchers" rather than simply being passive users of a training course. 

The possibility offered to co-design some teaching activities has in fact allowed to identify useful teaching strategies and actions to be tested in practice. The effort of the education researchers was not to replace the teachers, but to provide "ingredients" that they could adapt to their own objectives and teaching skills.

This type of relationship is a delicate aspect of building an effective collaboration between the research world and educational institutions. Legitimized by some Ministry guidelines that calls for universities to conduct training interventions in schools, a process of "formative delegation" has often characterized the way schools and universities collaborate together. The result has been an impoverishment of the relationship that has gradually consolidated in guaranteeing the carrying out of mandatory activities without allowing an effective exchange of knowledge between the two realities. 
Co-designing activities have therefore facilitated a transfer of skills, in both senses. The researchers could get a more precise picture of the scenario in which the teaching of modern physics takes place and have the opportunity to overcome some prejudices about the constraints and the organization of teaching at the high school level.
Teachers had the opportunity to confront with extremely competent professionals that offered useful indications to overcome some criticalities in the teaching of quantum physics. In particular devoting time to revise and reflect on the central content of contemporary physics is not always feasible in the schools.
The fact that some teachers felt able to rework and propose autonomously some ideas and materials pursued in the courses is an encouraging result in this sense.\\

Another action that yielded positive spin-offs is to have offered high school teachers the opportunity to participate in some activities together with their students. During the Summer Schools the teachers combined the reflection on the pre-conception of their students with the evaluation of possible actions of educational structuring. In this way it was possible to support the development of two fundamental dimensions of teachers' PCK. The reflections collected during and at the end of the activities show a significant evolution of the teachers' thoughts about the possibilities that can be built for the teaching of quantum physics, in particular about the role that the educational use of quantum technologies can play in it.\\

A different aspect to evaluate is my personal place within the learning ecosystem in relation to this strategic line. 
In the interaction with researchers in physics, my position has basically been that of a student. The very high profile of the researchers I interacted with (primarily my tutor) allowed me to rigorously investigate the disciplinary content aspects of my research that proved essential to the construction of the approach (see Chapter \ref{ch:3b-Qubits}). At the same time, in this role I was able to observe some of the dynamics present in the way physics researchers approach teaching. The deep and detailed knowledge of physics that is evident in the way content is organized for teaching is generally directed towards the construction of rigorous arguments that allow them to construct logically coherent lectures. The educational implementation is sometimes "cursed" by that same knowledge \cite{schwichtenberg2019no-nonsense} and have difficulties in balancing the transfer process with the capability of students to digest it.
When designing learning ecosystems for secondary schools, the timing and methodologies to be chosen play an important role. If not accompanied by careful design of formative feedback activities, there is a risk that even elegantly chiseled lessons will fill the notebooks of high school students without adequate critical reflection on their part.
The specific training for teaching physics at secondary level has allowed me to bring my contribution in the management of these dynamics. In particular in the definition of intended learning outcomes consistent with the specific objectives of secondary school and in the promotion of diversified methodologies that could take into account the different levels of preparation and knowledge of students.\\

When the construction of networks becomes a startegic choice, the process of "self-generation networks" is facilitated \cite{Capra2017}. Another fundamental result for the design and implementation of the pathways for students was in fact the creation and consolidation of a network with other researchers in education, in particular from the University of Bologna and the University of Pavia. The sharing of knowledge and the exchange of skills constituted the fabric that support the creation of many activities and proposals that emerged during the course of the research. Besides being an opportunity for mutual enrichment, this result confirms how innovation also passes through openness to the experiences of other groups allowing the definition of common research paths and facilitating the generation of new ideas.\\





\section{Strategy 2: Regenarating approach to quantum physics}\label{sec:ConclSTR2}
The aim of the second strategic line was to to initiate a reflection on the criteria and practices used to design learning activities about quantum physics for secondary school students.\\

The introduction of the informational approach and the co-design actions allowed to experiment and develop coherent educational design tools. Constrained by the alleys imposed by the completion of topics on the final exam \cite{DM769_18Quadri}, teachers had finally the opportunity to rethink how to prioritize and center educational objectives over content. Having insisted during the co-design activities on the necessity to revisit the "why-s" quantum physics is taught and not so much on "what-s", the broader and deeper reflections have led to the creation of learning environments coherent with shared and specific educational objectives.

Linked with the historical evolution of quantum concepts or to the different national curricula, the existing teaching proposals in the lively Italian and international community have been a shining example to follow. The daring intention expressed with Strategy 2 was not to create an "additional" approach to be added to what already existed.  Instead, it was to explicitly identify "revolving points" (revolution) in order to move towards something new and open new possibilities. This action of rupture was intended to free energies and to catch those aspects of quantum theory that can be consider genuinely revolutionary in the approach with reality and therefore better linked with the characteristics of the second quantum revolution.
Obviously we didn't want to reinvent the wheel, and therefore what we propose recovers experiences already matured in other contexts. The desire was to introduce a new frame of meaning that would stimulate among those involved in teaching a deep reflection on why today it makes sense to teach quantum physics.

The participants in the various actions have perceived the breath of fresh air brought by the desire to talk about physical reality through "information". Suspended between an "application" level (that of technologies) and a more "conceptual" one (related to the reality of the qubit) the proposed way to interpret the behavior of quantum objects through the concept of state has brought to light the interdisciplinary nature of the second quantum revolution. Understanding the qubit is in fact not only a matter of analysis of a physical phenomenon, but needs a new way of thinking about reality and "new foundation of logic" \cite{SusskindQT}

The main activation concerned the relationship between mathematics and physics in the teaching of quantum physics.
Necessary for the construction of the axiomatic structure of the theory, the rigorous mathematical formalism that characterizes the quantum physics is seen by teachers either as an insurmountable obstacle or even as a sufficient reason not to talk about quantum physics to high school students.

In particular this obstacle shows its true colors when it is necessary to introduce useful mathematical tools (such as vectors and matrices) in the study of qubits.
The proposed approach, while simplifying the mathematical level necessary to deal with many of the essential aspects of quantum nature (such as superposition or uncertainty relation) cannot avoid asking students to multiply a 2x2 matrix by a two-dimensional vector (see Section \ref{Ch3.1_ApproachShape}).


However, physics teachers often find themselves in the situation of having to use specific mathematical tools throughout the course of the physics curriculum. The problem lies perhaps in "what" elements need to be introduced.
Compared to other types of curriculum organization at the international level, in which mathematics and physics generally belong to different "departments", in the Italian context there is a strong relationship between the two disciplines. This relationship is explicit not only in the curricular objectives but also in the teaching practical organization, where physics teachers is often also the mathematics teacher. This does not mean that the teaching of mathematics and physics are equivalent since there are specific and different objectives and learning outcomes. The fact of the matter is that this type of structuring of the curriculum and of the teachers within the departments can favor the interplay between the disciplines. As also made explicit during the work with teachers (see Section \ref{subsec:Teachrs_Interv}), this feature can become an opportunity to introduce the mathematics that is needed.
Returning, then, to the mathematics for qubits, the basic elements of linear algebra are explicitly provided in the mathematics curriculum precisely in relation to physics.

\begin{quote}
    [The student] will study the concepts of vector, linear dependence and independence, scalar and vector products in the plane and in space, and the elements of matrix calculations. He/she will also deepen the understanding of the fundamental role that the concepts of vector and matrix algebra have in physics \cite{MIUR2010_211} 
\end{quote}

The often-collected observation that these elements of linear algebra "are not in the school programmes" therefore appears to be groundless.

Nevertheless, taking into account the difficulties connected to its implementation (time, different level in student understanding), the approach introduced wants to provide the context in which the objective mentioned above can be achieved, integrating the two disciplines in a fluid way .

In the process of designing learning environments, the introduction of specific mathematical tools useful for understanding phenomena was not forced. The same stratagem that teachers stated they use with their students to make them understand something new (see Subsection \ref{subsec:Teachrs_Interv} was applied. That is introducing a context where it is clear that that specific piece of mathematics makes it easier to understand the problem.\\

Another important element in the work done on the approach has been its construction through constant and continuous interaction between physics researchers, educational researchers, and high school teachers. The joint effort has allowed the identification of central themes (see Section \ref{sec:ApprPillars}) and phenomenological contexts (see Chapter \ref{ch:3b-Qubits}) in which the nature of quantum behaviors could emerge without the need to use complex mathematics.
This is a procedure of "creative optimization" and is one of the main examples given with respect to the development of subject matter for teaching in its relation to educational structuring.




In summary, the conditions for bringing the second quantum revolution into high schools are therefore linked to two processes related to the defined strategies.
The first concerns the promotion of co-design opportunities between researchers and teachers: this is a fundamental interaction if it is to have an impact on educational practice. The second concerns the development of an approach that allows the use of concepts linked to the nature of physical objects that are consistent with contemporary developments in theory.

\section{Future perspectives}


A critical point of the project is that it did not adequately integrate among the design elements the phase of assessment of participants' learning, especially in the courses designed for students.
This is an aspect also pointed out by the teachers (see Section \ref{sec:TLE_focus_interview}) and which is essential for the validation of the impact that the proposed approach can have in secondary school regular classroom activities.
A possible future strategy linked to the assessment process should include, among its actions, the identification of effective criteria for measuring the role played by the specific choices characterising the educational approach (see Section \ref{sec:ApprPillars}) in achieving the specific learning objectives.
Such actions should allow not only the measurement of the acquisition of certain specific skills, but also the monitoring and support of a learning process related to the appropriation of specific concepts \cite{Levrini2015_Appropriation}.
In this sense, a reflection on what it means to "understand" \cite{wiggins2005understanding} quantum physics can become a useful starting point for design.\\

The specific learning environments on quantum technologies should then be tested on classes composed not of self-selected students (as happened in the Summer Schools) but of students with different levels of interest and attitude towards science (as happened with the classes in the course presented in Section \ref{sec:carcano}).

The evaluation of the effectiveness of the learning enviroment designed for teachers should then be tested also with non-self-selected teachers group. The teachers who have participated in the various courses, especially the most active and collaborative ones, are in fact characterised by their own personal drive for research and innovation in their professional practice. A study to be developed with a more heterogeneous sample and possibly not of in-service teachers could provide important hints for the validation of the approach and the possibility that it could become part of what happens in the classroom.\\

Another missing element that has been brought in to enhance the quantum physics courses is the importance of  making central aspects of nature of science as part of the central reflection for teachers \cite{Stadermann2020ConnectNOS}. On a number of occasions, both teachers and students have constructed reflections that have touched on aspects relating precisely to the way in which science constructs knowledge. These are dimensions of cultural enrichment central to the educational process that could be integrated into the approach without necessarily distorting its meaning.\\

A final element concerns the enhancement of the "breaking elements" that characterized the construction of the approach presented (see Section \ref{sec:ConclSTR2}). Those who participated in the various training courses, whether students or teachers, were asked to make a great effort to break away from more "familiar" and "classically reassuring" visions. In particular for teachers the leap was more difficult, because it required a rethinking of their teaching practice also at methodological level.
Faced with the complexity of the contents proposed, it was often easy for teachers to barricade themselves in the trenches of traditional teaching. The non-negotiable central role of the "moments of content transfer" characterised by an almost exclusive use of lecturing that was in some cases justified by anecdotes and personal experiences. Experience is an essential source for teachers' PCK, but it risks introducing dangerous bias if not moderated by the contributions of research in teaching.

Evidence-based Physics Education Research has validated the effectiveness of specific non-traditional teaching practices, "Physics education research seeks to provide both instructors and students with strategies to maximise successful and lasting learning experiences within their limited time." \cite{FraserMazur2014_Bridging}

In this sense, it is considered appropriate to provide teachers with research skills in education so that they can become the "authors" of their own discoveries.

Although the dominant narrative may say otherwise, innovations and revolutions do not come from the action of an gifted individuals but generally start from small groups and communities, where the seeds of change are protected and can have time to grow. In order to continue the cohesion-building strategy, it is necessary to continue to provide opportunities to build tools for experimenting, making mistakes, reflecting and creating.
Students are asked to reflect on the value of experimentation for the development of scientific knowledge. In the context of competence development, this is not only to be understood in a disciplinary sense (e.g how to do an experiment in a laboratory) but as a general mindset. The orientation is to see scientific methods as an aptitude for knowledge of reality which can therefore be spent in any context, even not specifically scientific. In the same way, it can be useful to create research communities for teachers, where they can exercise research methods in their own profession.\\
And so continue to sow the seeds of the revolutions that the world of education constantly needs.

\appendix
\chapter{Experiments with Stern Gerlach apparatus}\label{appx:SGA}

\begin{figure}[hbt!]
    \centering
     \includegraphics[width=\textwidth,height=\textheight,keepaspectratio]{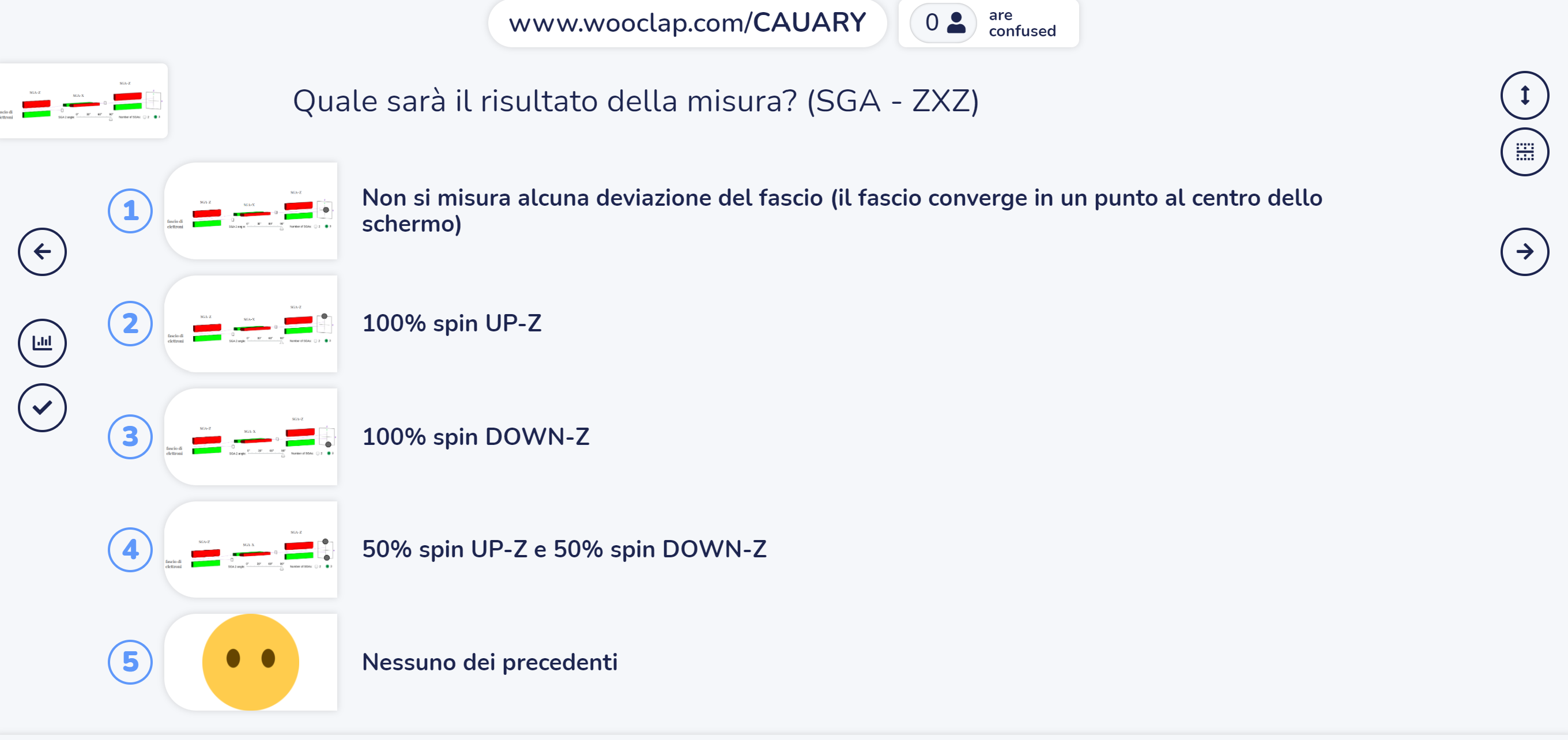}
    \caption[SGA experiments - Wooclap]{Wooclap have been used to collect participant predictins of experimental outcomes}
    \label{fig:AppSGA_wooclap}
\end{figure}

In this appendix, we report the questions on Stern Gerlach apparatus sequences proposed in the activities with teachers and students. The aim is to collect all  participants predictions about the outcome of the experiment shown in the diagram. Four different options are available for each experiment. Participants prediction can be collected using Wooclap$^{TM}$ (see Figure \ref{fig:AppSGA_wooclap}). After all the predictions are collected, participants can check the consistency of their predictions with the results of the simulation \cite{QuVisSim}.\\

\begin{figure}
     \centering
     \begin{subfigure}[hbt!]{\textwidth}
         \centering
         \includegraphics[width=\textwidth]{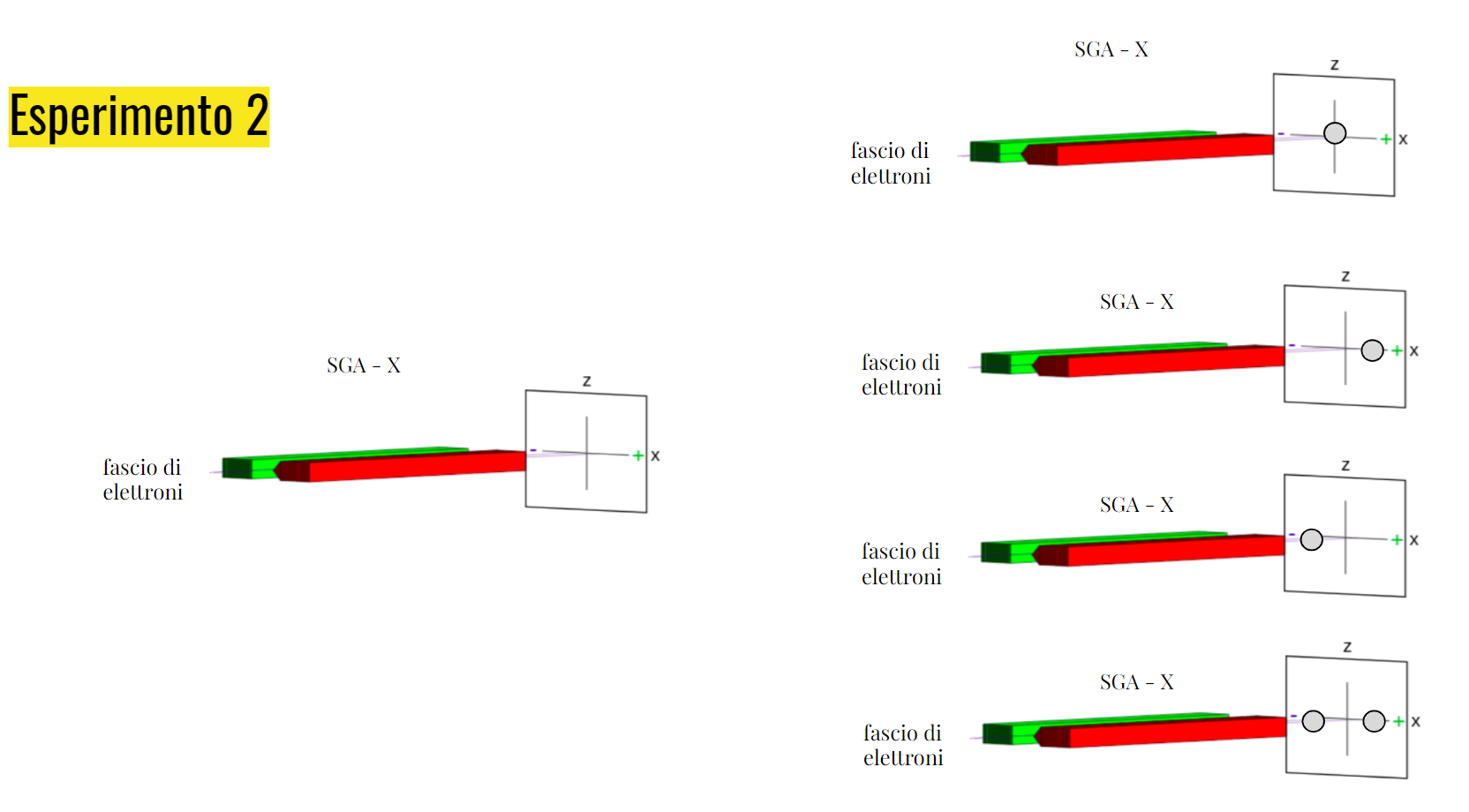}
     \end{subfigure}
     \hfill
     \begin{subfigure}[hbt!]{\textwidth}
         \centering
         \includegraphics[width=\textwidth]{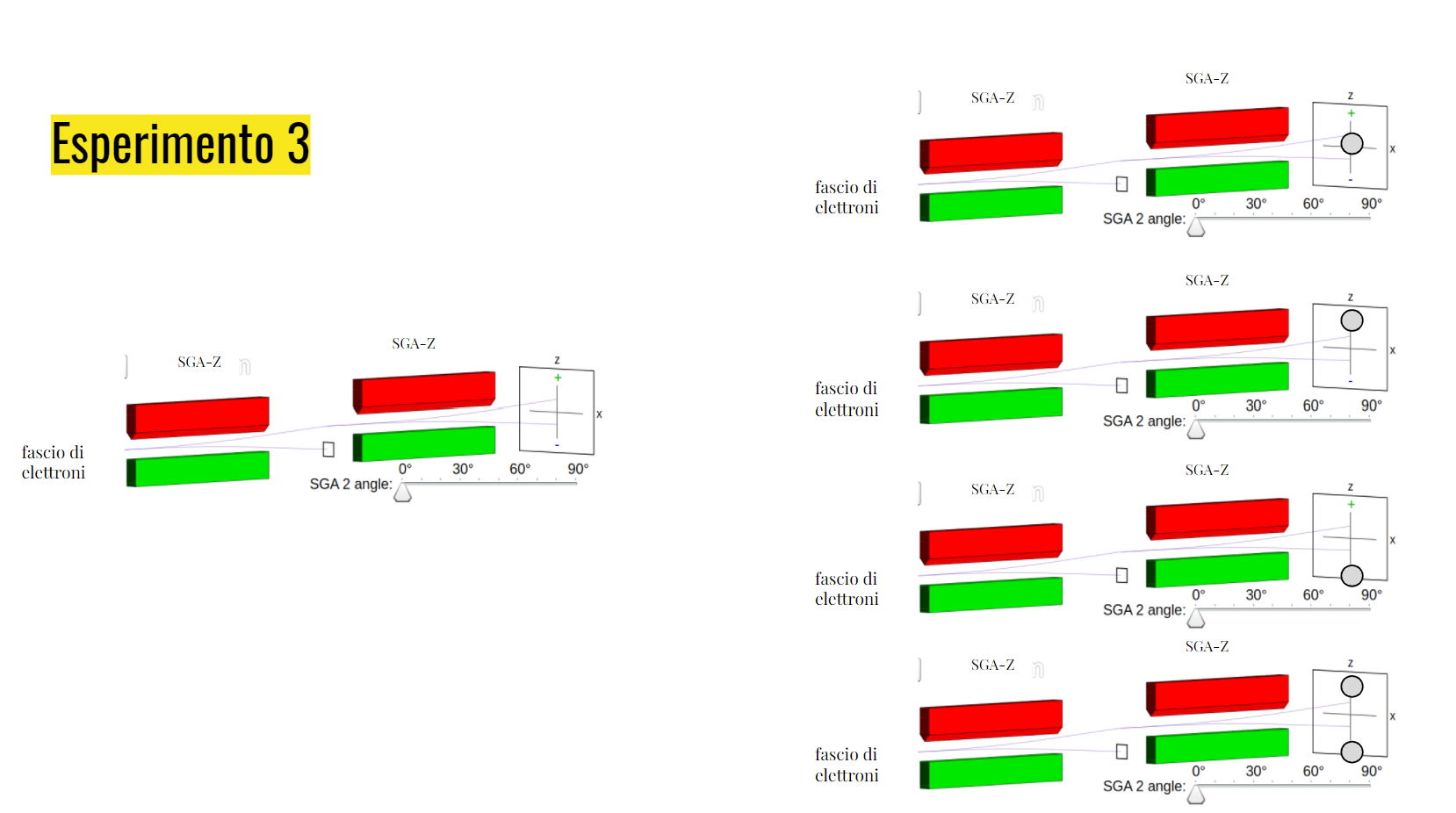}
         \caption{$y=3sinx$}
         \label{fig:App_SGA1}
     \end{subfigure}
     \caption{Experiments with sequences of Stern Gerlach apparatus. Part 1. All the images have been adapted from the QuVis project website}
     \label{fig:App_SGA1}
\end{figure}

\begin{figure}
     \centering
     \begin{subfigure}[hbt!]{\textwidth}
         \centering
         \includegraphics[width=\textwidth]{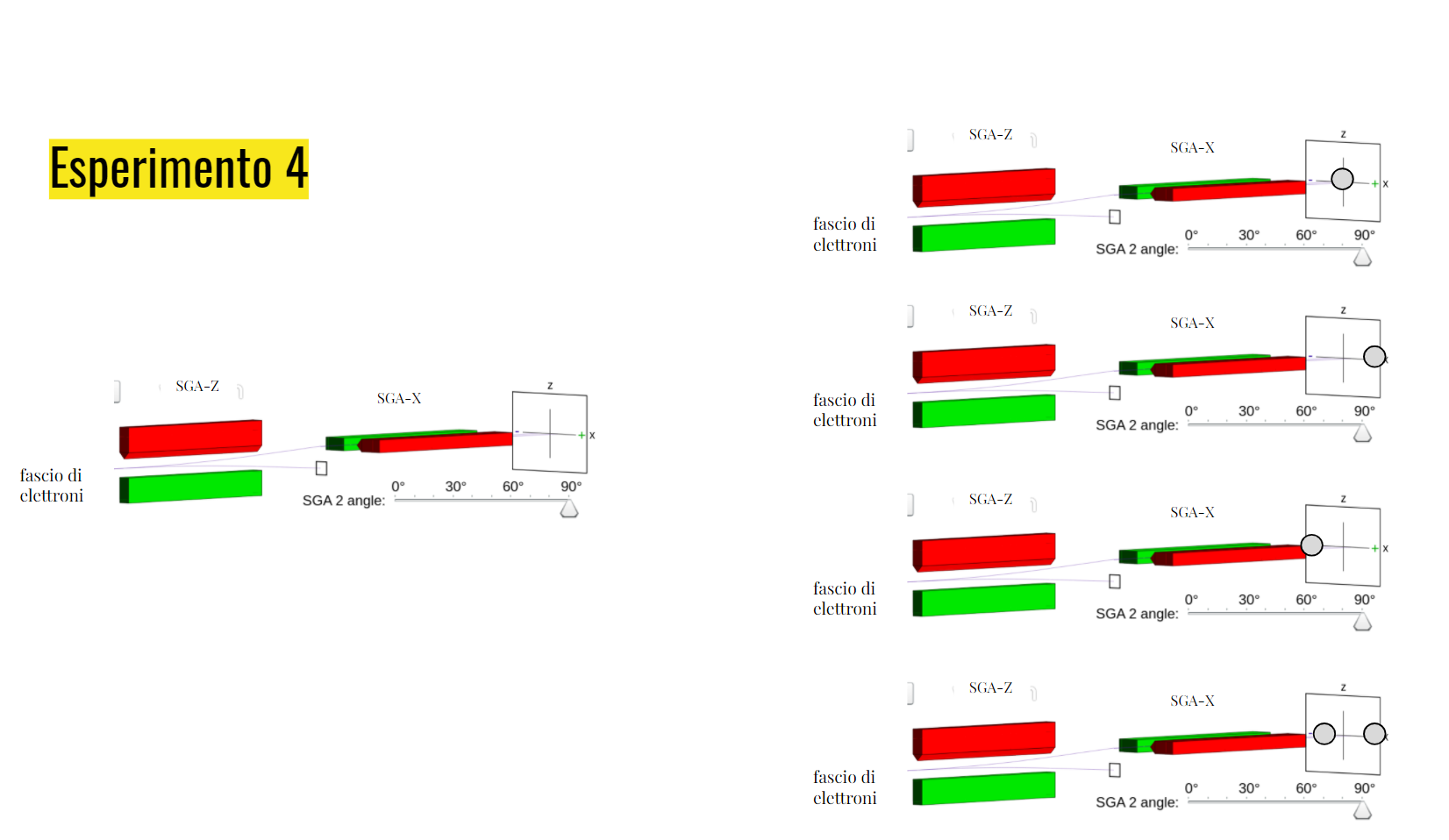}
     \end{subfigure}
      \hfill
     \begin{subfigure}[hbt!]{\textwidth}
         \centering
         \includegraphics[width=\textwidth]{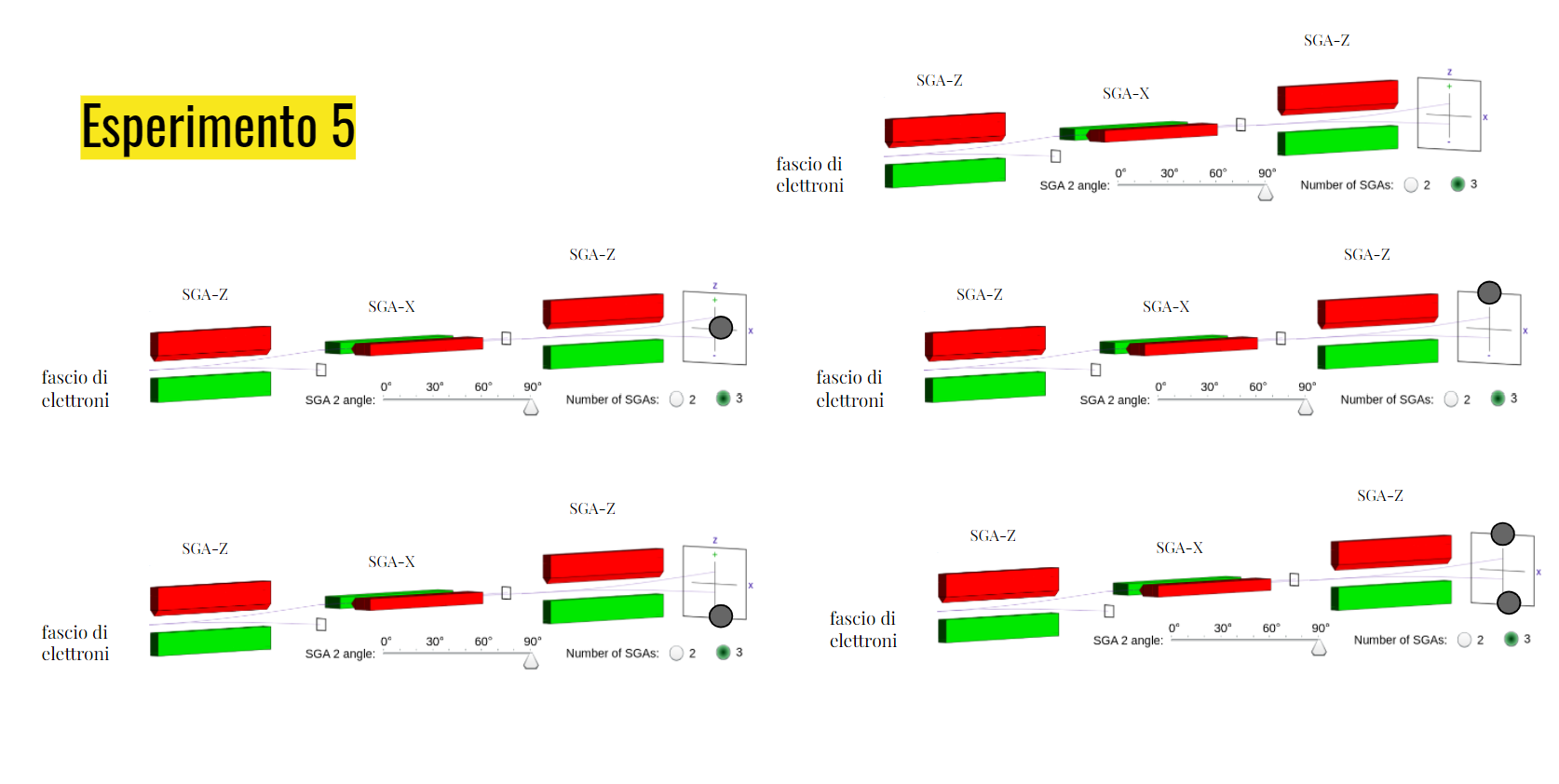}
     \end{subfigure}
      \caption{Experiments with sequences of Stern Gerlach apparatus - part 2. All the images have been adapted from the QuVis project website}
     \label{fig:App_SGA2}
\end{figure}
\chapter{Interferometer worksheet}\label{appx:QuVisMZI}

In this appendix we present the worksheet used for the inquiry activity on light patterns. The questions and diagrams are taken from the QuVis project website \cite{QuVisSim}.

\begin{figure}[hbt!]
    \centering
     \includegraphics[width=\textwidth,height=\textheight,keepaspectratio]{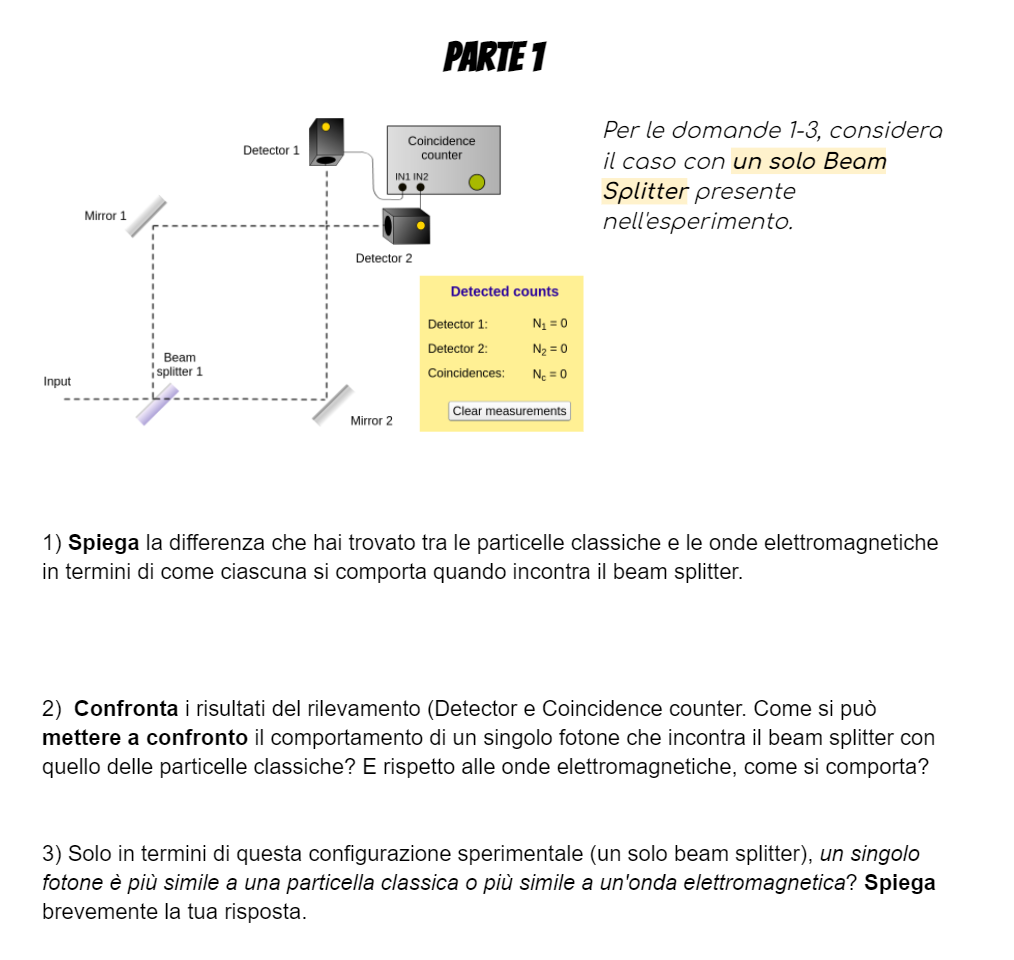}
    \caption{Mach-Zehnder Interferometer: one beam splitter}
    \label{fig:AppMZI1}
\end{figure}

\begin{figure}[hbt!]
    \centering
     \includegraphics[width=\textwidth,height=\textheight,keepaspectratio]{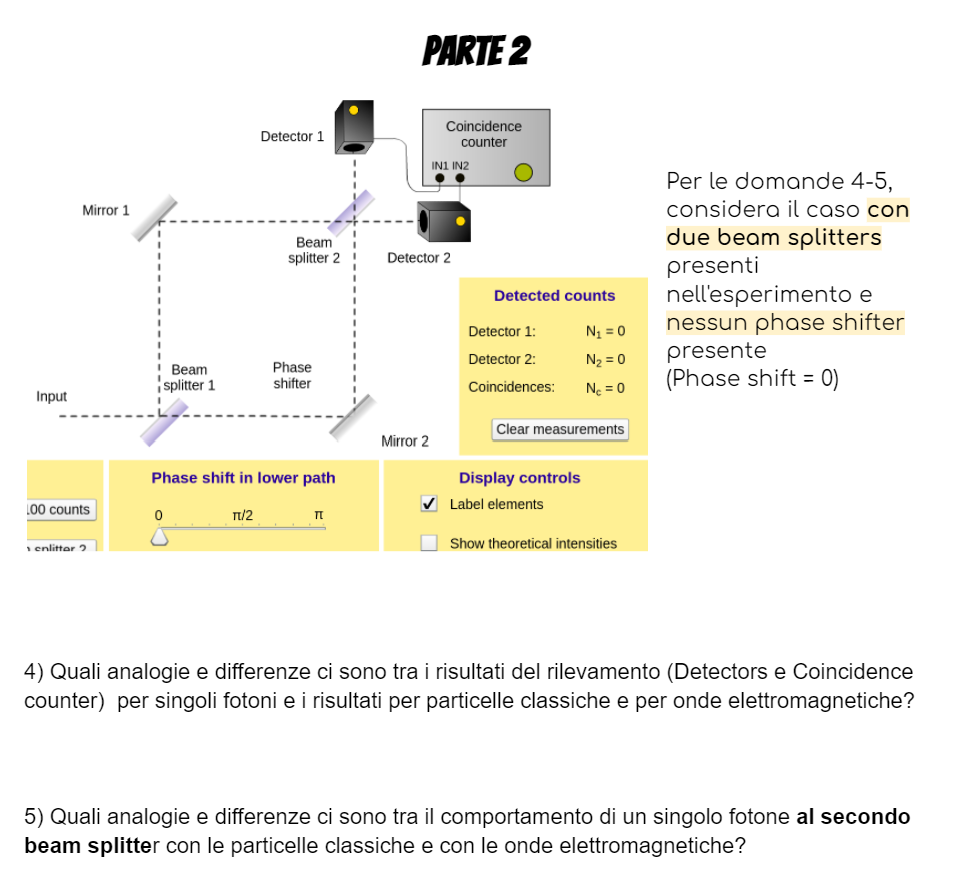}
    \caption{Mach-Zehnder Interferometer: two beam splitter}
    \label{fig:AppMZI2}
\end{figure}

\begin{figure}[hbt!]
    \centering
     \includegraphics[width=\textwidth,height=\textheight,keepaspectratio]{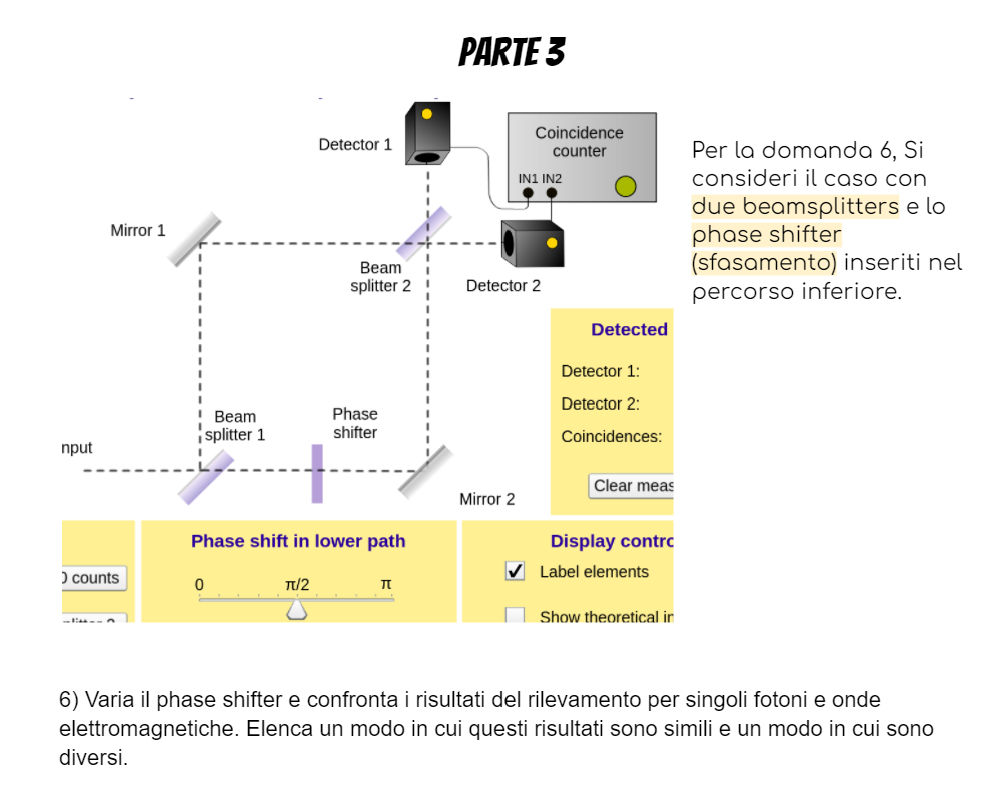}
    \caption{Mach-Zehnder Interferometer: two beam splitter and phase shifter}
    \label{fig:AppMZI3}
\end{figure}
\chapter{Semi structured interview protocol}\label{appx:IntervProtocol}

Interview protocol: A PCK-study on contemporary quantum physics

\textbf{Part 1}: Questions concerning your background
\begin{flushright}
\textit{Teaching experience is a relevant element of in-service teachers' PCK.}
\end{flushright}
\begin{enumerate}
    \item What education/training have you received? (subjects; level; teacher training courses)
    \item What teaching experience do you have? (number of years; level; subjects)
\end{enumerate}

\textbf{Part 2a}: Questions concerning teaching quantum physics - organization
\begin{flushright}
\textit{In this part we discuss relevant elements about the relation between teachers' subject matter knowledge for teaching and subject matter representation. It is particularly relevant to understand what is the role and relevance of teaching quantum in teachers' practice.}
\end{flushright}

\begin{itemize}
    \item  In which grade(s) do you teach quantum physics (as a separate topic or integrated in other themes)?
    \item How much time do you spend on teaching quantum physics in the different grades?
    \item What status has the teaching of quantum theory within the physics education for you?
    \item Which literature will be used by the students during the lessons?
    \item Which literature (books, journals) do you use for the preparation of your instruction?
\end{itemize}

\textbf{Part 2b}: Questions concerning teaching quantum physics - learning goals
\begin{flushright}
\textit{In this part we discuss relevant elements about teachers' subject matter knowledge for teaching. In particular what are their learning goal and how do they design their activities in relation to the core concepts}
\end{flushright}

\begin{enumerate}
    \item What do you intend the students to learn about quantum physics, what are the important concepts, or core ideas?
    \item Why is it important for students to know these core ideas?
\end{enumerate}

\textbf{Part 3}: Questions concerning the instruction on the topic of quantum physics core ideas

\begin{enumerate}
    \item What are the difficulties and challenges connected with teaching this idea?
    \item What do you know about students’ thinking that influences your teaching of this idea?
    \item What particular cases, phenomena, situations, experiments, people, events can be
used to make the content in question interesting, worth asking questions about,
accessible, comprehensible, conceivable for the students at their level and grade?
(Have you sometimes tried out other strategies for teaching this core idea?)
\end{enumerate}

\textbf{Part 4}: Scenario questions
You present the following questions, 1–6, to a student who has been taught the core concepts of quantum physics (quantization, photoelectric effect, duality, uncertainty, etc)  but failed to acquire a basic understanding of quantum physics. Questions are taken from Quantum Physics Conceptual Survey (QPCS) \footnote{https://www.physport.org/assessments/assessment.cfm?A=QPCS)}

\begin{enumerate}
    \item What answers do you expect the student to give, given that he or she does not understand quantum physics theory? (What would the good answer be?)
    \item How would you react to the answer that was given by the student?
\end{enumerate}

\begin{figure}[hbt!]
    \centering
     \includegraphics[width=\textwidth,height=\textheight,keepaspectratio]{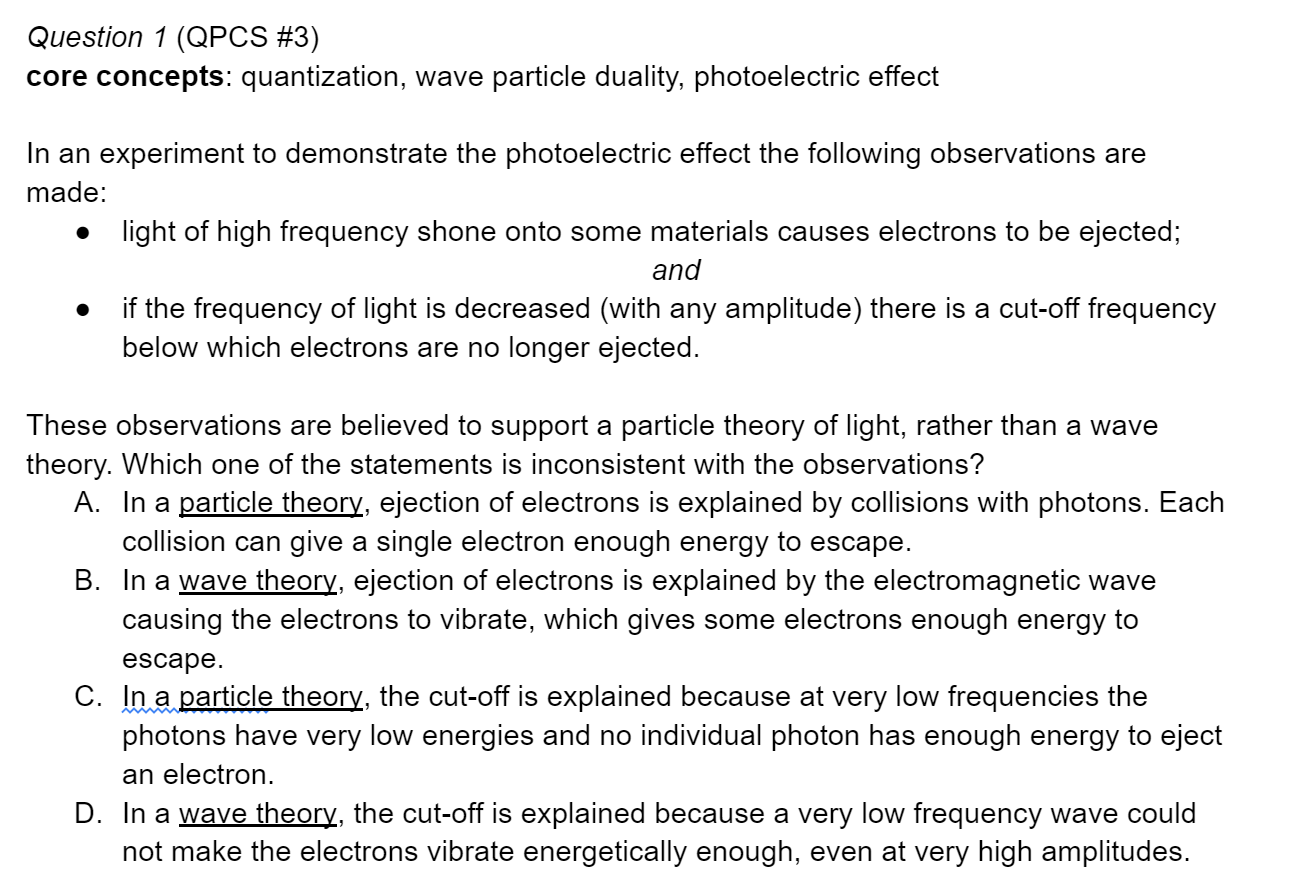}
    \caption{Scenario questions: QPCS number 3}
    \label{fig:scenario1}
\end{figure}

\begin{figure}[hbt!]
    \centering
     \includegraphics[width=\textwidth,height=\textheight,keepaspectratio]{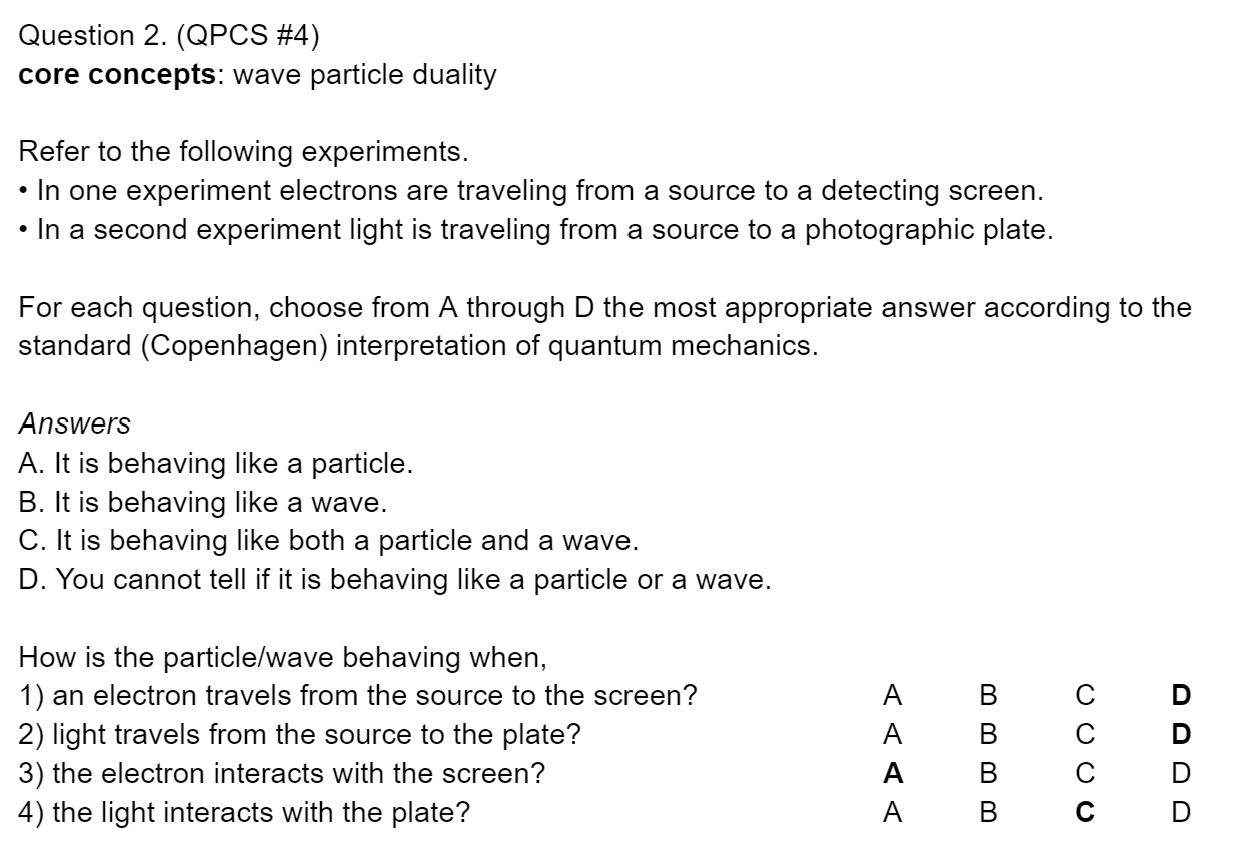}
    \caption{Scenario questions: QPCS number 4}
    \label{fig:scenario2}
\end{figure}

\begin{figure}
     \centering
     \begin{subfigure}[hbt!]{\textwidth}
         \centering
         \includegraphics[width=\textwidth]{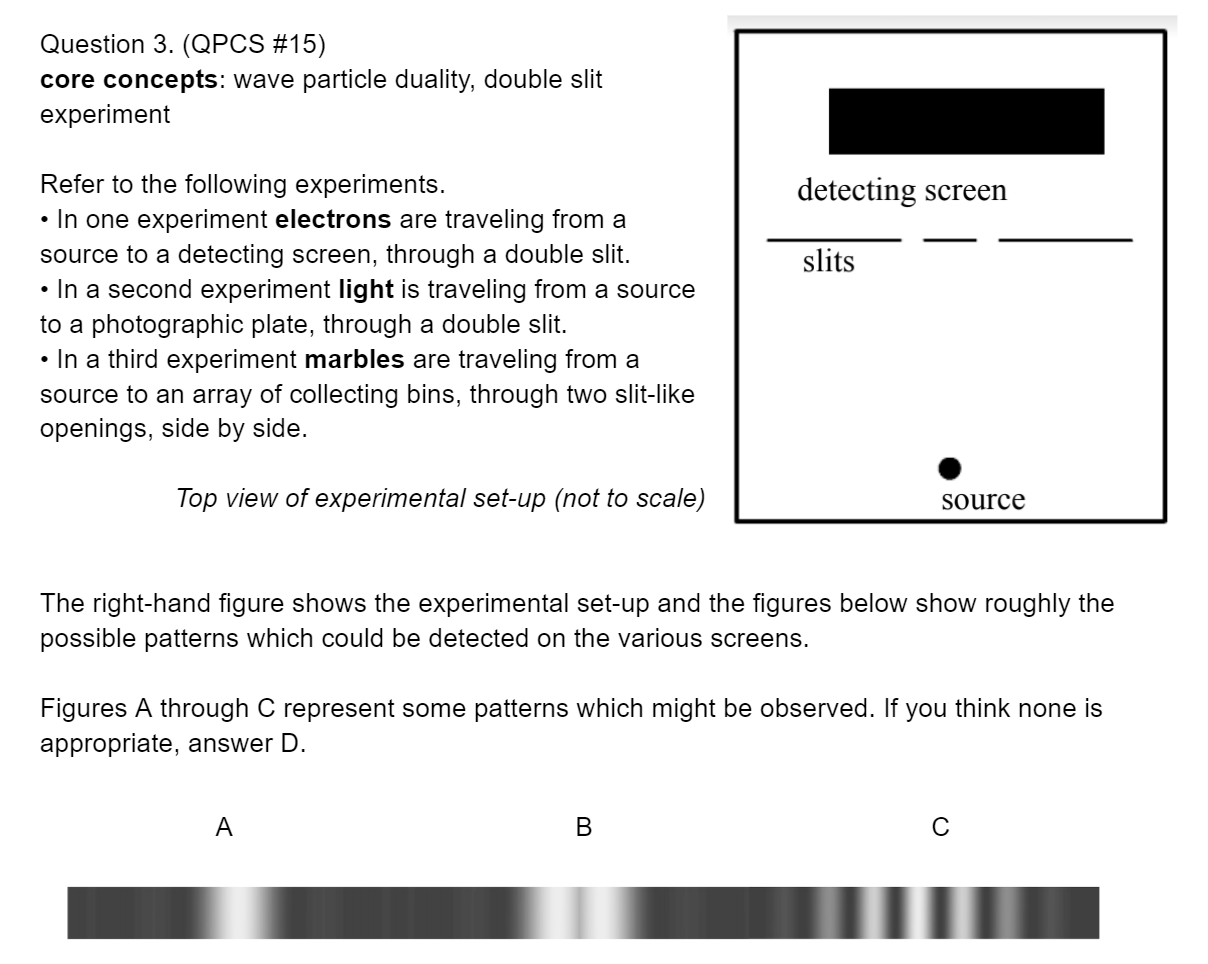}
     \end{subfigure}
     \hfill
     \begin{subfigure}[hbt!]{\textwidth}
         \centering
         \includegraphics[width=\textwidth]{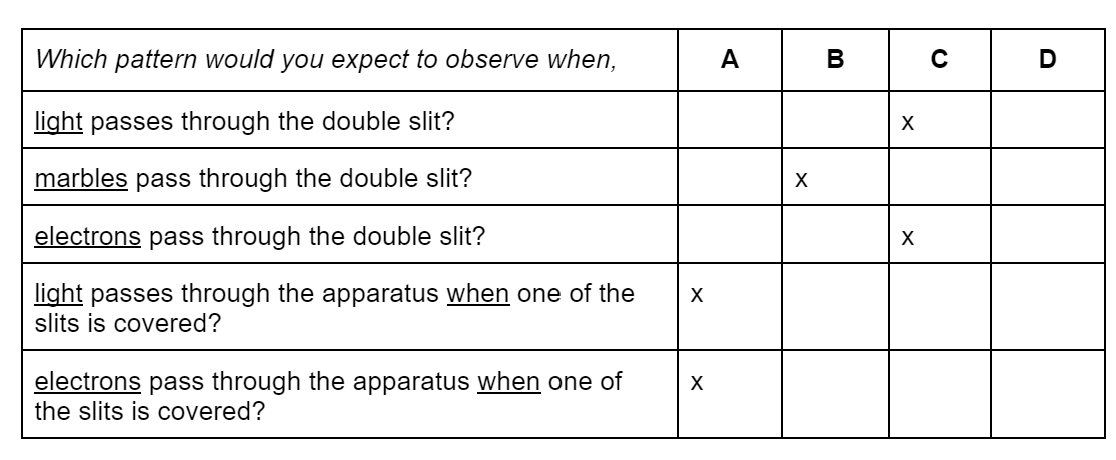}
         \caption{$y=3sinx$}
         \label{fig:App_SGA1}
     \end{subfigure}
     \caption{Scenario questions: QPCS number 15}
     \label{fig:scenario3}
\end{figure}

\begin{figure}[hbt!]
    \centering
     \includegraphics[width=\textwidth,height=\textheight,keepaspectratio]{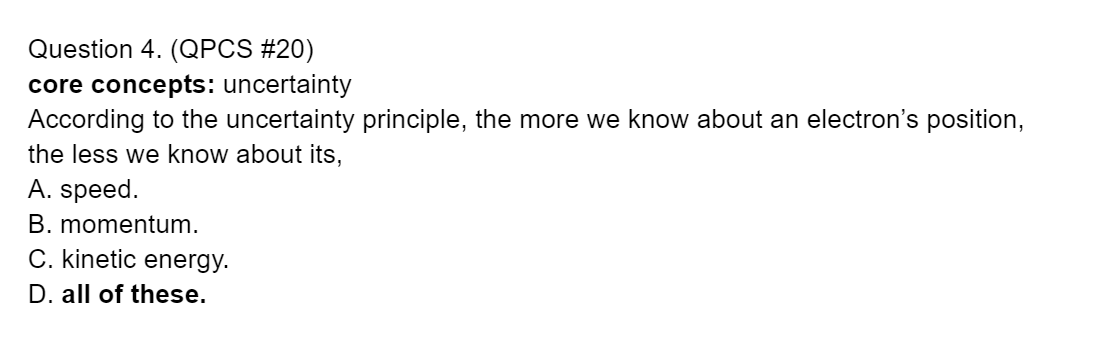}
    \caption{Scenario questions: QPCS number 20}
    \label{fig:scenario4}
\end{figure}

\begin{figure}[hbt!]
    \centering
     \includegraphics[width=\textwidth,height=\textheight,keepaspectratio]{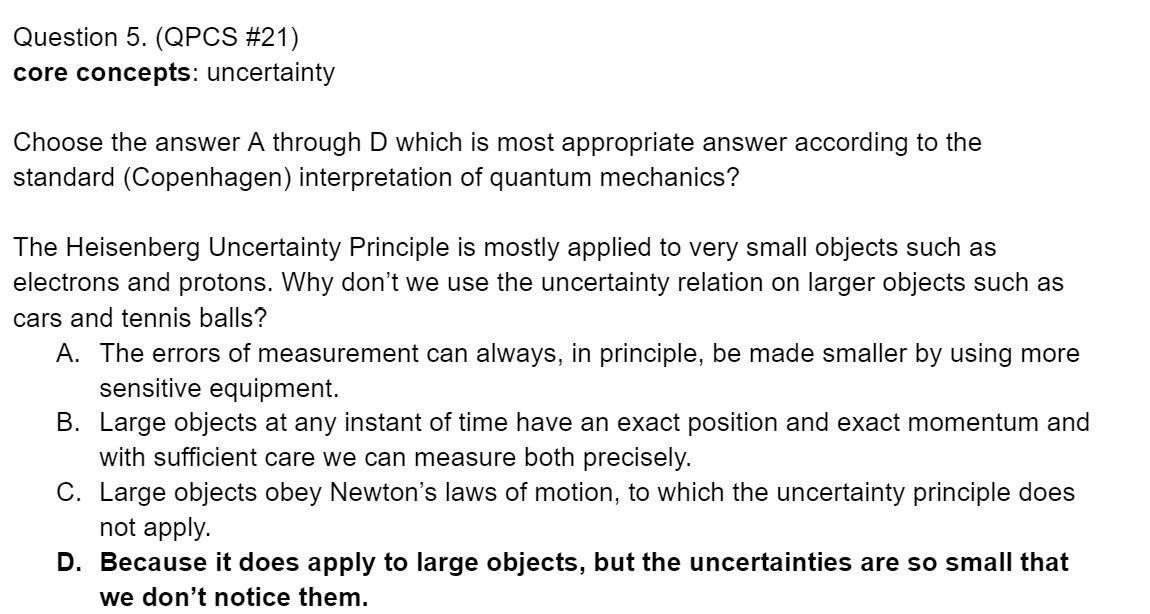}
    \caption{Scenario questions: QPCS number 21}
    \label{fig:scenario5}
\end{figure}

The scenario question are in Figures: \ref{fig:scenario1}, \ref{fig:scenario2}, \ref{fig:scenario3}, \ref{fig:scenario4}, \ref{fig:scenario5}

\textbf{Part 5}: Closing questions
\begin{enumerate}
    \item Can you recall remarkable – critical or fruitful – incidents during your teaching of quantum?
    \item Can you mention items concerning the teaching of quantum evolution that you find important but have not been discussed during this interview?
\end{enumerate}

\chapter{List of Publications}\label{appx:SGA}

F. Pallotta, A.Parola and M.Bondani, Developing scientific competencies: a collaboration between High School teachers and Physics researchers to create experiment-based learning activities J. Phys.: Conf. Ser. 1929 012010
https://doi.org/10.1088/1742-6596/1929/1/012010\\

Le potenzialità di Jupyter nella didattica della fisica - The potential of Jupyter in physics education
Authors: Filippo Pallotta, Davide Passaro, Claudio Sutrini
Giornale di Fisica, issue 4, year 2021, pp. 445-459
DOI: 10.1393/gdf/i2022-10419-y

3. C.Sutrini, D.Passaro and F.Pallotta (2021) The potential of using the jupyter notebook in physics education: experimentation for high school students.
Il Nuovo Cimento (in press)\\

4. I.Testa, M.Malgieri, M.Bondani et al.(2021)  Investigating upper secondary students’ epistemic views and plausibility judgements about quantum physics: the role of physics identity, perception of competency, and engagement in extracurricular activities on quantum technologies (in press)\\

F.Pallotta, M. Bondani (2021) Experimental and mathematical tools to investigate light polarization, Giornale di Fisica (submitted)

\addcontentsline{toc}{chapter}{Bibliography}
\bibliography{main.bib}        

\begin{thebibliography}{100}

\bibitem{amaldi2020il}
Ugo Amaldi.
\newblock {\em Il nuovo Amaldi per i licei scientifici.blu}.
\newblock Zanichelli, Bologna, 2020.

\bibitem{anderson2014_BloomREv}
Lorin Anderson.
\newblock {\em A taxonomy for learning, teaching, and assessing : a revision of
  Bloom's}.
\newblock Pearson, Essex, 2014.

\bibitem{QuVisSim}
St.Andrews~University Antje~Kohnle.
\newblock Quvis, the quantum mechanics visualization project, 2009.

\bibitem{baggott2011Story}
J.~E. Baggott.
\newblock {\em The quantum story : a history in 40 moments}.
\newblock Oxford University Press, Oxford England New York, 2011.

\bibitem{Baily2015}
Charles Baily and Noah~D. Finkelstein.
\newblock Teaching quantum interpretations: Revisiting the goals and practices
  of introductory quantum physics courses.
\newblock {\em Physical Review Special Topics - Physics Education Research},
  11(2), Sep 2015.

\bibitem{ball2019beyond}
Philip Ball.
\newblock {\em Beyond weird : why everything you thought you knew about quantum
  physics is... different}.
\newblock Vintage Books, London, 2019.

\bibitem{Barenco95Gates}
Adriano Barenco, Charles~H. Bennett, Richard Cleve, David~P. DiVincenzo, Norman
  Margolus, Peter Shor, Tycho Sleator, John~A. Smolin, and Harald Weinfurter.
\newblock Elementary gates for quantum computation.
\newblock {\em Phys. Rev. A}, 52:3457--3467, Nov 1995.

\bibitem{benenti2019principles}
Giuliano Benenti.
\newblock {\em Principles of quantum computation and information : a
  comprehensive textbook}.
\newblock World Scientific Publishing Co. Pte. Ltd, Singapore Hackensack, NJ,
  2019.

\bibitem{bergmann2014flipped}
J.~Bergmann and A.~Sams.
\newblock {\em Flipped Learning: Gateway to Student Engagement}.
\newblock Flipped Learning Series. International Society for Technology in
  Education, 2014.

\bibitem{BessonMalgieri2018}
Ugo Besson and Massimiliano Malgieri.
\newblock {\em Insegnare la fisica moderna : proposte e percorsi didattici}.
\newblock Carocci, Roma, 2018.

\bibitem{biddle2007QualRes_Quotes}
Karen Biddle.
\newblock {\em Composing qualitative research}.
\newblock Sage, Thousand Oaks, Calif, 2007.

\bibitem{biggs1982SOLO}
John Biggs.
\newblock {\em Evaluating the quality of learning : the SOLO taxonomy
  (structure of the observed learning outcome}.
\newblock Academic Press, New York, 1982.

\bibitem{Biggs2003}
John Biggs.
\newblock {\em Teaching for Quality Learning at University}.
\newblock 01 2003.

\bibitem{biggs2011teaching}
John Biggs.
\newblock {\em Teaching for quality learning at university : what the student
  does}.
\newblock McGraw-Hill/Society for Research into Higher Education/Open
  University Press, Maidenhead, 2011.

\bibitem{Bitzenbauer2021_herald}
Philipp Bitzenbauer.
\newblock Effect of an introductory quantum physics course using experiments
  with heralded photons on preuniversity students' conceptions about quantum
  physics.
\newblock {\em Phys. Rev. Phys. Educ. Res.}, 17:020103, Jul 2021.

\bibitem{Bitz21_fostering}
Philipp Bitzenbauer.
\newblock Practitioners' views on new teaching material for introducing quantum
  optics in secondary schools.
\newblock {\em Physics Education}, 56:055008, 06 2021.

\bibitem{Bitzenbauer2021_Teachers}
Philipp Bitzenbauer.
\newblock Practitioners' views on new teaching material for introducing quantum
  optics in secondary schools.
\newblock {\em Physics Education}, 56(5):055008, jun 2021.

\bibitem{Black1998_Assessment}
Paul Black and Dylan Wiliam.
\newblock Assessment and classroom learning.
\newblock {\em Assessment in Education: Principles, Policy and Practice},
  5(1):7--74, 1998.

\bibitem{bloom1974taxonomy}
Benjamin Bloom.
\newblock {\em Taxonomy of educational objectives : the classification of
  educational goals}.
\newblock D. McKay, New York, 1974.

\bibitem{Boe2021_HeinsenWave}
Maria~Vetleseter Boe and Susanne Viefers.
\newblock Secondary and university students' descriptions of quantum
  uncertainty and the wave nature of quantum particles.
\newblock {\em Science {\&} Education}, Nov 2021.

\bibitem{Bondani2014_SinglePhoton}
Maria Bondani.
\newblock Single-photon interference experiment for high schools.
\newblock {\em Proceedings of SPIE - The International Society for Optical
  Engineering}, 9289, 07 2014.

\bibitem{borko1996a}
H.~Borko and R.~Putnam.
\newblock Learning to teach.
\newblock In D.~Berliner and R.~Calfee, editors, {\em Handbook of Educational
  Psychology}, page 673–708. MacMillan, New York.

\bibitem{born1989atomic}
Max Born.
\newblock {\em Atomic physics}.
\newblock Dover Publications, New York, 1989.

\bibitem{Bouchee21}
T.~Bouchée, L.~de~Putter~Smits, M.~Thurlings, and B.~Pepin.
\newblock Towards a better understanding of conceptual difficulties in
  introductory quantum physics courses.
\newblock {\em Studies in Science Education}, 0(0):1--20, 2021.

\bibitem{Braun2006ThemAnalys}
Virginia Braun and Victoria Clarke.
\newblock Using thematic analysis in psychology.
\newblock {\em Qualitative Research in Psychology}, 3:77--101, 01 2006.

\bibitem{BrookesEtkina2007}
David~T. Brookes and Eugenia Etkina.
\newblock Using conceptual metaphor and functional grammar to explore how
  language used in physics affects student learning.
\newblock {\em Phys. Rev. ST Phys. Educ. Res.}, 3:010105, May 2007.

\bibitem{Brumfiel2012_Heisen}
Geoff Brumfiel.
\newblock Quantum uncertainty not all in the measurement.
\newblock {\em Nature}, Sep 2012.

\bibitem{bruner1977}
Jerome Bruner.
\newblock {\em The process of education}.
\newblock Harvard University Press, Cambridge, Massachusetts, 1977.

\bibitem{bruner1996_EduCulture}
Jerome Bruner.
\newblock {\em The culture of education}.
\newblock Harvard University Press, Cambridge, Mass, 1996.

\bibitem{Capra2017}
Fritjof Capra and Ove~Daniel Jakobsen.
\newblock A conceptual framework for ecological economics based on systemic
  principles of life.
\newblock {\em International Journal of Social Economics}, 44(6):831--844, Jan
  2017.

\bibitem{cavagnetto2012importanceArg}
Andy Cavagnetto and Brian Hand.
\newblock The importance of embedding argument within science classrooms.
\newblock In {\em Perspectives on scientific argumentation}, pages 39--53.
  Springer, 2012.

\bibitem{Adami97OpticGates}
N.~J. Cerf, C.~Adami, and P.~G. Kwiat.
\newblock Optical simulation of quantum logic.
\newblock {\em Phys. Rev. A}, 57:R1477--R1480, Mar 1998.

\bibitem{EU_keycompetences2018}
European Commission, Sport Directorate-General~for Education, Youth, and
  Culture.
\newblock {\em Key competences for lifelong learning}.
\newblock Publications Office, 2019.

\bibitem{Corsiglia2020Spinfirst}
Giaco Corsiglia, Tyler Garcia, Benjamin Schermerhorn, Gina Passante, Homeyra
  Sadaghiani, and Steven Pollock.
\newblock Characterizing and monitoring student discomfort in upper-division
  quantum mechanics.
\newblock pages 92--97, 09 2020.

\bibitem{eu-2006/962/EC}
{Council of European Union}.
\newblock Council recommendation on key competences for lifelong learning
  ({EU}) no oj l 394, 30.12.2006, 2016.
\newblock \newline\url{http://data.europa.eu/eli/reco/2006/962/oj}.

\bibitem{CrouchMazur2004_classroomdemo}
Catherine Crouch, Adam~P. Fagen, J.~Paul Callan, and Eric Mazur.
\newblock Classroom demonstrations: Learning tools or entertainment?
\newblock {\em American Journal of Physics}, 72(6):835--838, June 2004.

\bibitem{crystal2008think}
David Crystal.
\newblock {\em Think on my words: exploring Shakespeare's language}.
\newblock Cambridge University Press, Cambridge, 2008.

\bibitem{dawson2010teachingArgument}
Vaille~Maree Dawson and Grady Venville.
\newblock Teaching strategies for developing students’ argumentation skills
  about socioscientific issues in high school genetics.
\newblock {\em Research in Science Education}, 40(2):133--148, 2010.

\bibitem{Denzin2013_QualReas}
Norman Denzin.
\newblock {\em The landscape of qualitative research}.
\newblock SAGE Publications, Los Angeles, 2013.

\bibitem{dirac1981_QMprinciples}
P.~A.~M. Dirac.
\newblock {\em The principles of quantum mechanics}.
\newblock Clarendon Press, Oxford England, 1981.

\bibitem{DiVincenzo2000QCphysicalImplement}
David~P. DiVincenzo.
\newblock The physical implementation of quantum computation.
\newblock {\em Fortschritte der Physik}, 48(9-11):771–783, Sep 2000.

\bibitem{Duit2012_MER}
Reinders Duit, Harald Gropengiesser, Ulrich Kattmann, Michael Komorek, and Ilka
  Parchmann.
\newblock {\em The Model of Educational Reconstruction – a framework for
  improving teaching and learning science}, page 13–37.
\newblock 01 2012.

\bibitem{DoringIsham2010}
A.~Döring and C.~Isham.
\newblock “what is a thing?”: Topos theory in the foundations of physics.
\newblock {\em Lecture Notes in Physics}, page 753–937, 2010.

\bibitem{Dur2013qubitlearn}
W.~Dür and S.~Heusler.
\newblock What we can learn about quantum physics from a single qubit, 2013.

\bibitem{Eilks2013_relevanceSE}
Ingo Eilks, Marc Stuckey, Avi Hofstein, and Rachel Mamlok-Naaman.
\newblock The meaning of ‘relevance’ in science education and its
  implications for the science curriculum.
\newblock {\em Studies in Science Education}, 34:1--34, 07 2013.

\bibitem{Eldh2020Quote}
Ann~Catrine Eldh, Liselott Årestedt, and Carina Berterö.
\newblock Quotations in qualitative studies: Reflections on constituents,
  custom, and purpose.
\newblock {\em International Journal of Qualitative Methods},
  19:1609406920969268, 2020.

\bibitem{EPSRC21_QTreport}
Engineering and Physical Sciences~Research Council.
\newblock Quantum technologies public dialogue report, 2021.

\bibitem{feynmanLect_Vol3}
Richard Feynman.
\newblock {\em The Feynman lectures on physics}.
\newblock Basic Books, a member of the Perseus Books Group, New York, 2011.

\bibitem{Ford1966keyideas}
Kenneth~W. Ford.
\newblock The key ideas of quantum mechanics.
\newblock {\em The Physics Teacher}, 4(8):361--367, 1966.

\bibitem{Fox_QuantumIntro}
Mark Fox.
\newblock {\em {Quantum optics: an introduction}}.
\newblock Oxford master series in atomic, optical, and laser physics. Oxford
  Univ. Press, Oxford, 2006.

\bibitem{FraserMazur2014_Bridging}
James Fraser, Anneke Timan, Kelly Miller, Jason Dowd, Laura Tucker, and Eric
  Mazur.
\newblock Teaching and physics education research: Bridging the gap.
\newblock {\em Reports on progress in physics. Physical Society (Great
  Britain)}, 77:032401, 03 2014.

\bibitem{QBism2014}
Christopher~A. Fuchs, N.~David Mermin, and Rüdiger Schack.
\newblock An introduction to qbism with an application to the locality of
  quantum mechanics.
\newblock {\em American Journal of Physics}, 82(8):749–754, Aug 2014.

\bibitem{Gerlach1922}
W.~Gerlach and O.~Stern.
\newblock Der experimentelle nachweis des magnetischen moments des silberatoms.
\newblock {\em Zeitschrift f{\"u}r Physik}, 8(1):110--111, Dec 1922.

\bibitem{gerry2005introQOptics}
C.~C. Gerry.
\newblock {\em Introductory quantum optics}.
\newblock Cambridge University Press, Cambridge, UK New York, 2005.

\bibitem{ghirardi2015_carte}
Gian Ghirardi.
\newblock {\em Un'occhiata alle carte di Dio : gli interrogativi che la scienza
  moderna pone all'uomo}.
\newblock Il saggiatore, Milano, 2015.

\bibitem{Giulini2008}
Domenico Giulini.
\newblock Electron spin or “classically non-describable two-valuedness”.
\newblock {\em Studies in History and Philosophy of Science Part B: Studies in
  History and Philosophy of Modern Physics}, 39(3):557–578, Sep 2008.

\bibitem{Greenbank_QualResValues}
Paul Greenbank.
\newblock The role of values in educational research: the case for reflexivity.
\newblock {\em British Educational Research Journal}, 29(6):791--801, 2003.

\bibitem{Guba1994_Trust}
Egon~G. Guba and Yvonna~S. Lincoln.
\newblock {\em Competing paradigms in qualitative research.}, pages 105--117.
\newblock Handbook of qualitative research. Sage Publications, Inc, Thousand
  Oaks, CA, US, 1994.

\bibitem{Hadzidaki2008_HeisnMicr}
Pandora Hadzidaki.
\newblock The heisenberg microscope: A powerful instructional tool for
  promoting meta-cognitive and meta-scientific thinking on quantum mechanics
  and the nature of science.
\newblock {\em Science \& Education}, 17(6):613--639, 2008.

\bibitem{Hattie2016_LearningStrategies}
John A.~C. Hattie and Gregory~M. Donoghue.
\newblock Learning strategies: a synthesis and conceptual model.
\newblock {\em npj Science of Learning}, 1(1):16013, Aug 2016.

\bibitem{heisenberg1927a}
W.~Heisenberg.
\newblock The physical content of quantum kinematics and mechanics.
\newblock Originally Published: Z. Phys.,.

\bibitem{Henriksen2018_Whatislight}
Ellen Henriksen, Carl Angell, Arnt~Inge Vistnes, and Berit Bungum.
\newblock What is light?: Students’ reflections on the wave-particle duality
  of light and the nature of physics.
\newblock {\em Science and Education}, 27, 03 2018.

\bibitem{Jones1991_photonMiscon}
D~G~C Jones.
\newblock Teaching modern physics-misconceptions of the photon that can damage
  understanding.
\newblock {\em Physics Education}, 26(2):93--98, mar 1991.

\bibitem{kaiser2020tacklingBellLoopholes}
David~I. Kaiser.
\newblock Tackling loopholes in experimental tests of bell's inequality, 2020.

\bibitem{Kaplan2020_PauliPEP}
Ilya~G. Kaplan.
\newblock The pauli exclusion principle and the problems of its experimental
  verification.
\newblock {\em Symmetry}, 12(2), 2020.

\bibitem{Kennard1927Hein}
E.~H. Kennard.
\newblock Zur quantenmechanik einfacher bewegungstypen.
\newblock {\em Zeitschrift f{\"u}r Physik}, 44(4):326--352, Apr 1927.

\bibitem{Kohnle15_QuVIS_2states}
Antje Kohnle, Charles Baily, Anna Campbell, Natalia Korolkova, and Mark
  Paetkau.
\newblock Enhancing student learning of two-level quantum systems with
  interactive simulations.
\newblock {\em American Journal of Physics}, 83, 06 2015.

\bibitem{QUVIS_Kohnle2017_Qkeydistrib}
Antje Kohnle and Aluna Rizzoli.
\newblock Interactive simulations for quantum key distribution.
\newblock {\em European Journal of Physics}, 38(3):035403, Mar 2017.

\bibitem{Krathwohl2002_BloomRevised}
David~R. Krathwohl.
\newblock A revision of bloom's taxonomy: An overview.
\newblock {\em Theory Into Practice}, 41(4):212--218, 2002.

\bibitem{Kim2017}
K.~Krijtenburg-Lewerissa, H.J. Pol, A.~Brinkman, and W.R. van Joolingen.
\newblock Insights into teaching quantum mechanics in secondary and lower
  undergraduate education.
\newblock {\em Physical Review Physics Education Research}, 13(1), Feb 2017.

\bibitem{kuhn1957Copernican}
Thomas Kuhn.
\newblock {\em The Copernican revolution : planetary astronomy in the
  development of Western thought}.
\newblock Harvard University Press, Cambridge, 1957.

\bibitem{Leifer2014_quantumsattereal}
Matthew~Saul Leifer.
\newblock Is the quantum state real? an extended review of $\psi$-ontology
  theorems.
\newblock {\em Quanta}, 3(1):67, Nov 2014.

\bibitem{Levrini2015_Appropriation}
Olivia Levrini, Paola Fantini, Giulia Tasquier, Barbara Pecori, and Mariana
  Levin.
\newblock Defining and operationalizing appropriation for science learning.
\newblock {\em Journal of the Learning Sciences}, 24(1):93--136, 2015.

\bibitem{loudon2000_QthoeryLight}
Rodney Loudon.
\newblock {\em The quantum theory of light}.
\newblock Oxford University Press, Oxford New York, 2000.

\bibitem{maccone2008}
Lorenzo Maccone.
\newblock {\em Fisica moderna : meccanica quantistica, caos e sistemi
  complessi}.
\newblock Carocci, Roma, 2008.

\bibitem{magnason2020TimeWater}
Andri Magnason.
\newblock {\em On time and water}.
\newblock Serpent's Tail, London, UK, 2020.

\bibitem{maknouz2021_Chuncked}
Dany Maknouz.
\newblock {\em La lezione segmentata : ritmata, varia, integrata}.
\newblock Zanichelli, Bologna, 2021.

\bibitem{Malgieri2017}
Massimiliano Malgieri, Pasquale Onorato, and Anna De~Ambrosis.
\newblock Test on the effectiveness of the sum over paths approach in favoring
  the construction of an integrated knowledge of quantum physics in high
  school.
\newblock {\em Phys. Rev. Phys. Educ. Res.}, 13:010101, Jan 2017.

\bibitem{Maries2020_WhichPath}
Alexandru Maries, Ryan Sayer, and Chandralekha Singh.
\newblock Can students apply the concept of
  {\textquotedblleft}which-path{\textquotedblright} information learned in the
  context of mach{\textendash}zehnder interferometer to the double-slit
  experiment?
\newblock {\em American Journal of Physics}, 88(7):542--550, July 2020.

\bibitem{MarshmanSing2015_QdiffFrame}
Emily Marshman and Chandralekha Singh.
\newblock Framework for understanding the patterns of student difficulties in
  quantum mechanics.
\newblock {\em Phys. Rev. ST Phys. Educ. Res.}, 11:020119, Sep 2015.

\bibitem{MazurPeer1997}
Eric Mazur.
\newblock {\em Peer instruction : a users manual}.
\newblock Upper Saddle River, N.J. : Prentice Hall, [1997]
  {\textcopyright}1997, [1997].
\newblock Includes bibliographical references and index.;System requirements:
  IBM compatible PC; disks can be used on computers running Windows, Mac OS,
  DOS, Sun OS 4.13 or 4.14, Solaris 2.3 and 2.4, or HP-UX 9.03.

\bibitem{Mazur2009_flipped}
Eric Mazur.
\newblock Farewell, lecture?
\newblock {\em Science}, 323(5910):50--51, 2009.

\bibitem{mcintyre2016quantum}
D.H. McIntyre, C.A. Manogue, and J.~Tate.
\newblock {\em Quantum Mechanics}.
\newblock Pearson, 2016.

\bibitem{Mermin2003CbitQbit}
N.~David Mermin.
\newblock From cbits to qbits: Teaching computer scientists quantum mechanics.
\newblock {\em American Journal of Physics}, 71(1):23--30, 2003.

\bibitem{Michelini2021}
Marisa Michelini and Alberto Stefanel.
\newblock A path to build basic quantum mechanics ideas in the context of light
  polarization and learning outcomes of secondary students.
\newblock {\em Journal of Physics: Conference Series}, 1929(1):012052, may
  2021.

\bibitem{MIUR2010_211}
{Ministero dell'Istruzione, dell'Università e della Ricerca}.
\newblock Decreto decreto 7 ottobre 2010, n. 211 {Indicazioni nazionali
  riguardanti gli obiettivi specifici di apprendimento concernenti le attivita'
  e gli insegnamenti compresi nei piani degli studi previsti per i percorsi
  liceali}, 2010.
\newblock
  \newline\url{https://www.gazzettaufficiale.it/eli/id/2010/12/14/010G0232/sg}.

\bibitem{Mitchell2016_QOptImpatien}
Morgan~W. Mitchell.
\newblock {\em Quantum optics for impatient}.
\newblock ICFO - Institut de Ciencies Fotoniques, Castelldefels (Barcelona),
  2007.

\bibitem{DM1392007}
{MIUR}.
\newblock Regolamento recante norme in materia di adempimento dell’obbligo di
  istruzione, 2007.
\newblock
  \newline\url{https://archivio.pubblica.istruzione.it/normativa/2007/dm139_07.shtml}.

\bibitem{DM769_18Quadri}
{MIUR}.
\newblock Quadri di riferimento per la redazione e lo svolgimento delle prove
  scritte e griglie di valutazione per l'attribuzione dei punteggi" per gli
  esami di stato del secondo ciclo di istruzione, 2018.
\newblock
  \newline\url{https://www.miur.gov.it/-/esami-di-stato-del-secondo-ciclo-di-istruzione-a-s-2018-2019-d-m-769-del-26-novembre-2018}.

\bibitem{MIUR_Esami21}
{MIUR}.
\newblock Ordinanza ministeriale 53 del 03/03/2021 {Esami di Stato nel secondo
  ciclo di istruzione per l’anno scolastico 2020/2021}, 2021.
\newblock
  \newline\url{https://www.miur.gov.it/documents/20182/5407202/OM-Esami+di+Stato+nel+secondo+ciclo+di+istruzione+per+l-anno+scolastico+20202021.pdf/087431c4-6103-202c-29c1-26f250044730?t=1614865421465}.

\bibitem{PLSWeb}
MIUR.
\newblock Progetto lauree scientifiche, 2021.

\bibitem{Modir2020}
Bahar Modir, John~D. Thompson, and Eleanor~C. Sayre.
\newblock Students' epistemological framing in quantum mechanics problem
  solving.
\newblock {\em Phys. Rev. Phys. Educ. Res.}, 13:020108, Aug 2017.

\bibitem{Modir2019}
Bahar Modir, John~D. Thompson, and Eleanor~C. Sayre.
\newblock Framing difficulties in quantum mechanics.
\newblock {\em Phys. Rev. Phys. Educ. Res.}, 15:020146, Dec 2019.

\bibitem{moraga2020relevance}
Tania~S. Moraga-Calderón, Henk Buisman, and Julia Cramer.
\newblock The relevance of learning quantum physics from the perspective of the
  secondary school student: A case study, 2020.

\bibitem{Muller2021penny}
Rainer Müller and Franziska Greinert.
\newblock Playing with a quantum computer.
\newblock 2021.

\bibitem{Muller2002}
Rainer Müller and Hartmut Wiesner.
\newblock Teaching quantum mechanics on an introductory level.
\newblock {\em American Journal of Physics}, 70(3):200--209, 2002.

\bibitem{newton1999placeArgum}
Paul Newton, Rosalind Driver, and Jonathan Osborne.
\newblock The place of argumentation in the pedagogy of school science.
\newblock {\em International Journal of science education}, 21(5):553--576,
  1999.

\bibitem{nielsen_chuang_2010}
Michael~A. Nielsen and Isaac~L. Chuang.
\newblock {\em Quantum Computation and Quantum Information: 10th Anniversary
  Edition}.
\newblock Cambridge University Press, 2010.

\bibitem{TLTasmaniaILO}
University of~Tasmania.
\newblock Teaching and learning website, how to write ilos, 1999.

\bibitem{oreskes2019why}
Naomi Oreskes.
\newblock {\em Why trust science}.
\newblock Princeton University Press, Princeton, New Jersey, 2019.

\bibitem{osborne2012introArgument}
J~Osborne, A~MacPherson, A~Patterson, E~Szu, and MS~Khine.
\newblock Introduction of argumentation.
\newblock {\em Perspectives on scientific argumentation: Theory, practice and
  research. Dordrecht}, 2012.

\bibitem{Osborne2001_WhatScienceTeach}
Jonathan Osborne, Mary Ratcliffe, Sue Collins, Robin Millar, and Richard
  Duschl.
\newblock What should we teach about science? a delphi study.
\newblock 01 2001.

\bibitem{Pan2000_BellLoopholes}
Jian-Wei Pan, Dik Bouwmeester, Matthew Daniell, Harald Weinfurter, and Anton
  Zeilinger.
\newblock Experimental test of quantum nonlocality in three-photon
  greenberger--horne--zeilinger entanglement.
\newblock {\em Nature}, 403(6769):515--519, Feb 2000.

\bibitem{PauliSpin}
Wolfgang Pauli.
\newblock Exclusion principle and quantum mechanics, nobel lecture delivered on
  december 13th 1946 for the 1945 nobel prize in physics., 1946.

\bibitem{pauli1964}
Wolfgang Pauli.
\newblock {\em Collected Scientific Papers: By Wolfgang Pauli. Edited by R.
  Kronig and VF Weisskopf}.
\newblock London, Sydney, 1964.

\bibitem{PhetFourier}
{PhET Colorado,University of Colorado Boulder}.
\newblock Fourier: Making waves, 2019.
\newblock
  \newline\url{https://phet.colorado.edu/en/simulations/fourier-making-waves}.

\bibitem{Pospiech1999_EPRatHS}
Gesche Pospiech.
\newblock Teaching the epr paradox at high school?
\newblock {\em Physics Education}, 34(5):311–316, Sep 1999.

\bibitem{Pusey2012_QstatesReality}
Matthew~F. Pusey, Jonathan Barrett, and Terry Rudolph.
\newblock On the reality of the quantum state.
\newblock {\em Nature Physics}, 8(6):475--478, Jun 2012.

\bibitem{Rembach2016_SOLO}
Lauren Rembach and Laura Dison.
\newblock Transforming taxonomies into rubrics: Using solo in social science
  and inclusive education.
\newblock {\em Perspectives in Education}, 34:68--83, 05 2016.

\bibitem{TonomuraVideo}
Hitachi~Global Research.
\newblock Interference pattern (with narration), 2018.

\bibitem{Robertson1929_Hein}
H.~P. Robertson.
\newblock The uncertainty principle.
\newblock {\em Phys. Rev.}, 34:163--164, Jul 1929.

\bibitem{robinson2015creative}
Ken Robinson.
\newblock {\em Creative schools : the grassroots revolution that's transforming
  education}.
\newblock Viking, New York, 2015.

\bibitem{Rodriguez2017_SGA}
E~Ben{\'{\i}}tez Rodr{\'{\i}}guez, L~M~Ar{\'{e}}valo Aguilar, and E~Piceno
  Mart{\'{\i}}nez.
\newblock A full quantum analysis of the stern{\textendash}gerlach experiment
  using the evolution operator method: analyzing current issues in teaching
  quantum mechanics.
\newblock {\em European Journal of Physics}, 38(2):025403, January 2017.

\bibitem{romeni2017fisica}
Claudio Romeni.
\newblock {\em Fisica e realta.blu}.
\newblock Zanichelli, Bologna, 2017.

\bibitem{Sadaghiani2016}
Homeyra Sadaghiani.
\newblock Spin first vs. position first instructional approaches to teaching
  introductory quantum mechanics.
\newblock pages 292--295, 12 2016.

\bibitem{Sadaghiani2015}
Homeyra Sadaghiani and James Munteanu.
\newblock Spin first instructional approach to teaching quantum mechanics in
  sophomore level modern physics courses.
\newblock pages 287--290, 12 2015.

\bibitem{sagan2013cosmos}
Carl Sagan.
\newblock {\em Cosmos}.
\newblock Ballantine, New York, 2013.

\bibitem{SakuraiModern}
Jun~John Sakurai.
\newblock {\em {Modern quantum mechanics; rev. ed.}}
\newblock Addison-Wesley, Reading, MA, 1994.

\bibitem{Saltz2020_OnlineLearn}
Jeffrey Saltz and Robert Heckman.
\newblock Using structured pair activities in a distributed online breakout
  room.
\newblock {\em Online Learning}, 24(1), 2020.

\bibitem{sancassani2019progettare}
Susanna Sancassani.
\newblock {\em Progettare l'innovazione didattica}.
\newblock Pearson, MIlano Torino, 2019.

\bibitem{Scerri1995_Pauli}
Eric~R. Scerri.
\newblock The exclusion principle, chemistry and hidden variables.
\newblock {\em Synthese}, 102(1):165--169, 1995.

\bibitem{SCHILDKAMP2020_FormAssessment}
Kim Schildkamp, Fabienne~M. {van der Kleij}, Maaike~C. Heitink, Wilma~B.
  Kippers, and Bernard~P. Veldkamp.
\newblock Formative assessment: A systematic review of critical teacher
  prerequisites for classroom practice.
\newblock {\em International Journal of Educational Research}, 103:101602,
  2020.

\bibitem{Schneider2010qm4beginners}
Mark~B. Schneider.
\newblock Quantum mechanics for beginning physics students.
\newblock {\em The Physics Teacher}, 48(7):484--486, 2010.

\bibitem{Schrodinger1952}
E.~Schrodinger.
\newblock Are there quantum jumps ?
\newblock {\em British Journal for the Philosophy of Science}, 3(11):233--242,
  1952.

\bibitem{schwichtenberg2019no-nonsense}
Jakob Schwichtenberg.
\newblock {\em No-nonsense quantum mechanics : a student-friendly
  introduction}.
\newblock No-Nonsense Books, Karlsruhe, Germany, 2019.

\bibitem{Shuell1986CongitiveLearning}
Thomas~J. Shuell.
\newblock Cognitive conceptions of learning.
\newblock {\em Review of Educational Research}, 56(4):411--436, 1986.

\bibitem{Shulman1987_PCK}
Lee Shulman.
\newblock {Knowledge and Teaching:Foundations of the New Reform}.
\newblock {\em Harvard Educational Review}, 57(1):1--23, 01 2011.

\bibitem{Singh2008_QStudDiff}
Chandralekha Singh.
\newblock Student understanding of quantum mechanics at the beginning of
  graduate instruction.
\newblock {\em American Journal of Physics}, 76(3):277--287, 2008.

\bibitem{Singh2015}
Chandralekha Singh and Emily Marshman.
\newblock Review of student difficulties in upper-level quantum mechanics.
\newblock {\em Physical Review Special Topics - Physics Education Research},
  11(2), Sep 2015.

\bibitem{sousa2017Brain}
David Sousa.
\newblock {\em How the brain learns}.
\newblock Corwin, a Sage Publishing Company, Thousand Oaks, California, 2017.

\bibitem{Stadermann2019_Curricula}
H.~K.~E. Stadermann, E.~van~den Berg, and M.~J. Goedhart.
\newblock Analysis of secondary school quantum physics curricula of 15
  different countries: Different perspectives on a challenging topic.
\newblock {\em Phys. Rev. Phys. Educ. Res.}, 15:010130, May 2019.

\bibitem{Stadermann2020ConnectNOS}
H.K.E. Stadermann.
\newblock {\em Connecting Secondary School Quantum Physics and Nature of
  Science: Possibilities and challenges in curriculum design, teaching, and
  learning}.
\newblock PhD thesis, University of Groningen, January 2022.

\bibitem{SusskindQT}
Leonard Susskind.
\newblock {\em Quantum Mechanics, Theoretical minimum}.
\newblock Penguin Books, London, 2015.

\bibitem{tomlinson2006DiffInstr_UndbyDes}
Carol Tomlinson.
\newblock {\em Integrating differentiated instruction and understanding by
  design : connecting content and kids}.
\newblock Association for Supervision and Curriculum Development, Alexandria,
  Va, 2006.

\bibitem{turchi2007MADIT}
Gianpiero Turchi.
\newblock {\em M.A.D.I.T. ; Manuale per la metodologia di analisi dei dati
  informatizzati testuali}.
\newblock Aracne, Roma, 2007.

\bibitem{Dijk2008_ERTEevolution}
Esther van Dijk.
\newblock Teachers’ views on understanding evolutionary theory: A pck-study
  in the framework of the erte-model.
\newblock {\em Teaching and Teacher Education (2009)}, 25:259--267, 09 2008.

\bibitem{Dijk2006_ERTE}
Esther van Dijk and Ulrich Kattmann.
\newblock A research model for the study of science teachers’ pck and
  improving teacher education.
\newblock {\em Teaching and Teacher Education (2007)}, 23:885--897, 05 2006.

\bibitem{Velentzas2011_HeisenMicrosc}
Athanasios Velentzas and Krystallia Halkia.
\newblock The `heisenberg's microscope' as an example of using thought
  experiments in teaching physics theories to students of the upper secondary
  school.
\newblock {\em Research in Science Education}, 41(4):525--539, Aug 2011.

\bibitem{Vokos2000_matterwave}
Stamatis Vokos, Peter Shaffer, Bradley Ambrose, and Lillian C.~McDermott.
\newblock Student understanding of the wave nature of matter: Diffraction and
  interference of particles.
\newblock {\em American Journal of Physics - AMER J PHYS}, 68, 07 2000.

\bibitem{Webb1997_DOK}
Norman Webb.
\newblock Criteria for alignment of expectations and assessments in mathematics
  and science education. research monograph no. 6.
\newblock 01 1997.

\bibitem{wiggins2005understanding}
Grant Wiggins.
\newblock {\em Understanding by design}.
\newblock Association for Supervision and Curriculum Development, Alexandria,
  VA, 2005.

\bibitem{Singh2009_SGA}
Guangtian Zhu, Chandralekha Singh, Mel Sabella, Charles Henderson, and
  Chandralekha Singh.
\newblock Students’ understanding of stern gerlach experiment.
\newblock {\em AIP Conference Proceedings}, 2009.

\bibitem{Zuccarini2019}
Giacomo Zuccarini.
\newblock Analyzing the structure of basic quantum knowledge for instruction.
\newblock {\em American Journal of Physics}, 88(5):385–394, May 2020.

\end{thebibliography}
\bibliographystyle{plain}  

\end{document}